\documentclass[useAMS,usenatbib,a4paper]{mn2e}
\usepackage{graphicx,natbib}
\pdfoutput=1
\usepackage{rotating}
\usepackage{subfig}
\usepackage{lscape}
\usepackage{amssymb}
\usepackage[pass]{geometry}

\title[Infrared spectroscopy of eruptive variable protostars from VVV]{Infrared spectroscopy of eruptive variable protostars from VVV}
\author[C. Contreras Pe\~{n}a et al.: Infrared spectroscopy of eruptive variables]{C. Contreras Pe\~{n}a$^{1,4,2}$,\thanks{E-mail:cecontrep@gmail.com (CCP)} P. W. Lucas$^{2}$,  R. Kurtev$^{3,4}$, D. Minniti$^{1,7}$, A. Caratti o Garatti$^{6}$, \newauthor 
F. Marocco$^{2}$, M.A. Thompson$^{2}$, D. Froebrich$^{5}$, M. S. N. Kumar$^{2}$,  W. Stimson$^{2}$, \newauthor
C. Navarro Molina$^{4,3}$, J. Borissova$^{3,4}$, T. Gledhill$^{2}$ and R. Terzi$^{2}$ \\
$^{1}$Departamento de Ciencias Fisicas, Universidad Andres Bello, Republica 220, Santiago, Chile\\
$^{2}$Centre for Astrophysics Research, University of Hertfordshire, Hatfield, AL10 9AB, UK\\
$^{3}$Instituto de F\'{i}sica y Astronom\'{i}a, Universidad de Valpara\'{i}so, ave. Gran Breta\~{n}a, 1111, Casilla 5030, Valpara\'{i}so, Chile\\
$^{4}$Millennium Institute of Astrophysics, Av. Vicuna Mackenna 4860, 782-0436, Macul, Santiago, Chile\\
$^{5}$Centre for Astrophysics and Planetary Science, University of Kent, Canterbury CT2 7NH, UK\\
$^{6}$Dublin Institute for Advanced Studies, School of Cosmic Physics, Astronomy \& Astrophysics Section,
31 Fitzwilliam Place, Dublin 2, Ireland \\
$^{7}$Vatican Observatory, V00120 Vatican City State, Italy}
\begin{document}

\date{\today}

\pagerange{\pageref{firstpage}--\pageref{lastpage}} \pubyear{2002}

\maketitle

\label{firstpage}

\begin{abstract}
In a companion work (Paper I) we detected a large population of highly variable Young Stellar Objects (YSOs) in the Vista Variables in the Via Lactea (VVV) survey, typically with class I or flat spectrum spectral energy distributions and diverse light curve types. Here we present infrared spectra (0.9--2.5 $\mu$m) of 37 of these variables, many of them observed in a bright state. The spectra confirm that 15/18 sources with eruptive light curves have signatures of a high accretion rate, either showing EXor-like emission features ($\Delta$v=2 CO, Br$\gamma$) and/or FUor-like features ($\Delta$v=2 CO and H$_{2}$O strongly in absorption). Similar features were seen in some long term periodic YSOs and faders but not in dippers or short-term variables. The sample includes some dusty Mira variables (typically distinguished by smooth Mira-like light curves), 2 cataclysmic variables and a carbon star. In total we have added 19 new objects to the broad class of eruptive variable YSOs with episodic accretion. Eruptive variable YSOs in our sample that were observed at bright states show higher accretion luminosities than the rest of the sample. Most of the eruptive variables differ from the established FUor and EXor subclasses, showing intermediate outburst durations and a mixture of their spectroscopic characteristics. This is in line with a small number of other recent discoveries. Since these previously atypical objects are now the majority amongst embedded members of the class, we propose a new classification for them as MNors. This term (pronounced emnor) follows V1647 Ori, the illuminating star of McNeil's Nebula.
\end{abstract}


\section{Introduction}\label{vvv:intro}


The incidence and theory of episodic accretion in YSOs, and the associated eruptive variability, is one of the biggest remaining gaps 
in our understanding of star formation. While the number of photometrically and spectroscopically verified eruptive variable YSOs 
remains small, the topic has attracted increasing attention from observers and theorists in recent years, alongside a more general 
effort to improve our understanding of lower level accretion variations and photometric variability in YSOs at optical, near infrared 
and mid-infrared wavelengths \citep[e.g.,][]{2012Megeath, 2013Findeisen, 2014Cody, 2014Rebull, 2015Rice, 2015Wolk}. If episodic accretion
is common in normal pre-main-sequence (PMS) stars, this could solve two long-standing puzzles: 
the typical luminosities of YSOs are found to be lower than expected from theoretical models \citep[the so called ``luminosity problem'', see e.g.][]{1990Kenyon,2009Evans,2012Caratti}. This problem can be solved if in fact YSOs spend most of their lifetimes in quiescent states with low accretion rates and gain most of their mass in short-lived accretion bursts. The second problem relates to the wide scatter about the best fitting isochrone 
seen in the Hertzsprung-Russell (HR) diagrams of pre-main-sequence clusters. In this case the final position of the star in HR diagrams would strongly depend on the YSO's accretion history. Moreover, the lasting effects of episodic accretion would have 
to be taken into account not only in stellar evolution theory \citep{2009Baraffe,2012Baraffe} but also in the theory of planet formation
and planet migration.

This paper forms the second part of our investigation of eruptive variability in the VISTA Variables in the Via Lactea (VVV) survey
\citep{2010Minniti,2012Saito}, the first panoramic time domain survey of the Milky Way at infrared wavelengths. In the first part (Contreras 
Pe\~{n}a et al. 2016, hereafter Paper I) we described the discovery of 816 high amplitude infrared variables ($\Delta K_{\rm s}>1$~mag), 
in the mid-plane section of the VVV disc region, a well studied region in the 4th Galactic quadrant 
(longitude $295^{\circ}<l<350^{\circ}$, latitude $-1.1^{\circ}<b<1.1^{\circ}$. 
Most of these variables were located in star formation regions (SFRs) and have very red spectral energy distributions (SEDs), consistent with YSO classification. 
We estimated that about half of the 
816 are YSOs and we made a preliminary 
classification of each into one of the following categories: eruptive, fader, dipper, short timescale variable (STV), long term periodic variable (LPV-YSO), eclipsing binary and 
Mira-like variable (LPV-Mira). The Mira-like variables were distinguished by their relatively smooth, approximately sinusoidal light curves, 
as distinct from
long term periodic variable YSOs with significant lower level scatter on short timescales, superimposed on a higher amplitude
long term periodic variation. The eruptive variables typically displayed outburst durations of 1 to 4 years but in some cases the
eruptions were much shorter or lasted for $>$4 years. We posited that variables with eruptive light curves were mainly YSOs undergoing
episodic accretion and we further suggested that this mechanism (or variations on it) cause the variability of some of the faders and 
long term periodic variables. The dippers (sources with relatively brief but deep drops in the light curve) would more naturally
be explained by variable extinction, although we note that the near-infrared colour variability in many sources in this class does not follow the reddening path. The short term variables (STVs) typically had lower amplitudes and had a bluer distribution
of SEDs than the rest, so the variability in most cases is more likely due to rotational modulation by bright or dark spots on the 
stellar photosphere, or periodic extinction (e.g. by a warped inner disc), rather than episodic accretion.

Spectroscopic follow up remains essential to test our initial conclusions,
in part because the sparsely sampled VVV light curves and SEDs alone are
insufficient to classify individual sources reliably. Moreover, we have only 
2 epochs of colour variability data in the $JHK_{\rm s}$ filters (supplemented by 4 
epoch WISE$+$NEOWISE mid-IR data in some cases) so while the evidence suggested 
that variable extinction is not the main cause of variability in the eruptive
systems, spectroscopic confirmation is required.

Most eruptive variable YSOs have historically been classified as either FUors (after FU Orionis) or EXors (after EX Lupi). FUors have 
long duration outbursts (typically decades but at least 10 years) whereas EXors typically have outburst durations of weeks to months, 
with a maximum of 1.5 years (Herbig 2008). The classical view is that FUors are at an earlier evolutionary stage and belong to the 
transitional phase between class I and class II objects, whilst EXors are associated with instabilities in the discs of class II 
objects. However, both classes of young eruptive variables were originally defined from optical detections in nearby SFRs 
\citep{1977Herbig, 1989Herbig}, which tends to exclude younger YSOs which have higher accretion rates but are too deeply embedded 
in circumstellar matter to be observed at visible wavelengths. 

FUors display strong CO absorption in the 1st overtone bands at 2.3--2.4~$\mu$m and they often show broad H$_2$O absorption bands
from 1--2.5~$\mu$m also. These absorption bands are thought to arise in the 
circumstellar accretion disc (perhaps 1~au from the star \citep{2014Connelley}). The disc is heated in the mid-plane by viscous 
accretion and absorption occurs in the cooler surface layers. FUors typically lack emission lines in their infrared spectra. By 
contrast, EXors show strong HI recombination lines in emission. During outburst the 1st overtone CO bands are in emission also but 
they are seen in absorption during quiescent periods. The HI lines are attributed to infalling matter close to the star and they 
often display $\sim$100 km s$^{-1}$ line widths. The CO emission is thought to arise in the inner part of the accretion disc, which 
has a temperature inversion due to external heating by the star. Na I and Ca I emission lines are also sometimes seen in EXors during 
outburst, their behaviour correlating closely with CO \citep{2009Loren}. CO, Na I and Ca I absorption lines in classical EXors arise 
in the stellar photosphere. This of course requires that the stellar photosphere is not too heavily veiled by the disc, or hidden by high extinction,
as often occurs for normal class I YSOs, \citep[e.g.][]{1996Greene}. 
The accretion events associated with the outbursts of eruptive variables are thought to drive strong outflows \citep{1997Reipurth,2013Magakian}. FUors have been associated with Herbig-Haro objects \citep{1997Reipurth} and drive molecular outflows \citep{1994Evans}. FU Orionis itself, although having massive winds, does not drive a molecular outflow \citep{1994Evans} which could be explained by the lack of organization of magnetic fields required for jet launching and collimation in the binary scenario of \citet{2004Reip} or by previous outflows having cleared away the remnant envelope, leading to a lack of swept-up material that would otherwise constitute a molecular outflow \citep{1994Evans}. However, classical EXors do not show evidence of molecular shocks or jets in the near-infrared (H$_{2}$ or [Fe II] emission) at the sensitivity level of the spectra of \citet{2009Loren}\footnote{We note that PV Cep does show H$_{2}$ emission and it is part of the EXor sample of \citet{2009Loren}. However this object has characteristics that differ from classical EXors \citep[see e.g.][]{2013Caratti}}.


The VVV candidates were initially selected from their variability over only $\sim$2 years (2010 to 2012), a timescale that favours 
detection of EXors over FUors. However, with VVV we are sensitive to optically obscured YSOs with a substantial envelope, i.e. systems
at an earlier evolutionary stage than most EXors. The VVV candidates are in fact almost all optically invisible at the depth of 
panoramic surveys such as VPHAS+ \citep{2014Drew}. This is partly because most of them have intrinsically redder SEDs 
than most of the classical EXors and FUors and in part 
because of interstellar extinction towards the relatively distant SFRs (a few kpc) in which these objects are located 
(see Section \ref{vvvsec:spec}). Owing to their earlier evolutionary status, it is likely to be difficult to detect absorption features that arise 
in the stellar photosphere. In addition, the greater distance of the VVV YSOs biases them towards higher luminosity and mass. This may affect
the relative prominence of spectroscopic features relating to the accretion luminosity of the disc and those related to the 
stellar photosphere \citep[whether seen directly or in reprocessed radiation, see][]{1991Calvet}.

Prior to our work with VVV, very few deeply embedded eruptive variables were known. The known examples were OO Ser, V2775 Ori, HOPS 383, V723 Car and 
GM Cha \citep{1996Hodapp, 2007Persi,2011Caratti,2015Safron,2015Tapia}, now supplemented by our two recent discoveries from the UKIDSS Galactic Plane 
Survey: GPSV3 in Serpens OB2 and the isolated object GPSV15 \citep{2014Contreras}, as well as the recent discovery of CX330 \citep{2016Britt}. CX330 is an isolated very red source in the direction of the Galactic bulge with extreme
(even unique) photometric and spectroscopic properties. As with GPSV15, an eruptive YSO interpretation appears to fit the data best, though we
should consider the possibility that these unusual isolated variables
are more evolved sources or previously unknown types. To this list we might also add
V1647 Ori \citep{2009Aspin} and a few other objects that are somewhat less obscured (A$_V \sim 10$). These embedded objects,
along with a few other less embedded systems, show a mixture of the properties of FUors and EXors.
This has given rise to the scenario where, if we think of the outburst phenomena as a continuum, these new discoveries may instead 
represent a ``connection'' between FUors and EXors, with FUors representing one extreme (long duration) and EXors being the short 
duration end of the same phenomenon. In this sense the different types of outbursts are produced by variations of one or more of the 
parameters involved in the instabilities that give rise to the outburst \citep[][]{2006Gibb,2007Fedele,2009Aspin}.
          
In this paper we present the results of infrared spectroscopic follow up of a subset of the 816 high amplitude variables. The principal aims of this work are firstly to validate our claim in Paper I to have 
discovered a large population of eruptive variables, with increasing incidence towards earlier stages of evolution, and secondly
to explore the characteristics of these embedded systems.

The paper is divided as follows: in section \ref{vvvsec:spec} we describe the selection of the sample and present its photometric properties, including 70~$\mu$m data from {\it Herschel} Hi-Gal. We also review the near-infrared colour variability using two epochs of VVV data, presented previously in Paper I for the full photometric sample. Section \ref{vvv:sec_datared} describes the near-infrared spectroscopic observations and data reduction. In section \ref{vvv:sec_fire} we analyse the general spectroscopic characteristics of the sample, focusing on objects that are found to be likely YSOs. We then analyse these properties according to their light curve classification from Paper I and establish which objects likely belong to the eruptive variable class. In section \ref{vvv:sec_erupclass} we derive accretion luminosities and discuss the classification of our sample in the context of the established subclasses of eruptive variable YSOs (FUors, EXors) and find that most members of our sample cannot be described by the classical scheme. Finally, in section \ref{vvv:sec_summary} we present a summary of our analysis. Individual analyses of objects in our sample are presented in Appendices \ref{vvv:specvar} through \ref{vvv:sec_evolved}. Light curves and spectra of the whole spectroscopic sample are presented in Appendix \ref{apen2}.


\section{The spectroscopic sample}\label{vvvsec:spec}

\begin{figure}
\resizebox{0.99\columnwidth}{!}{\includegraphics{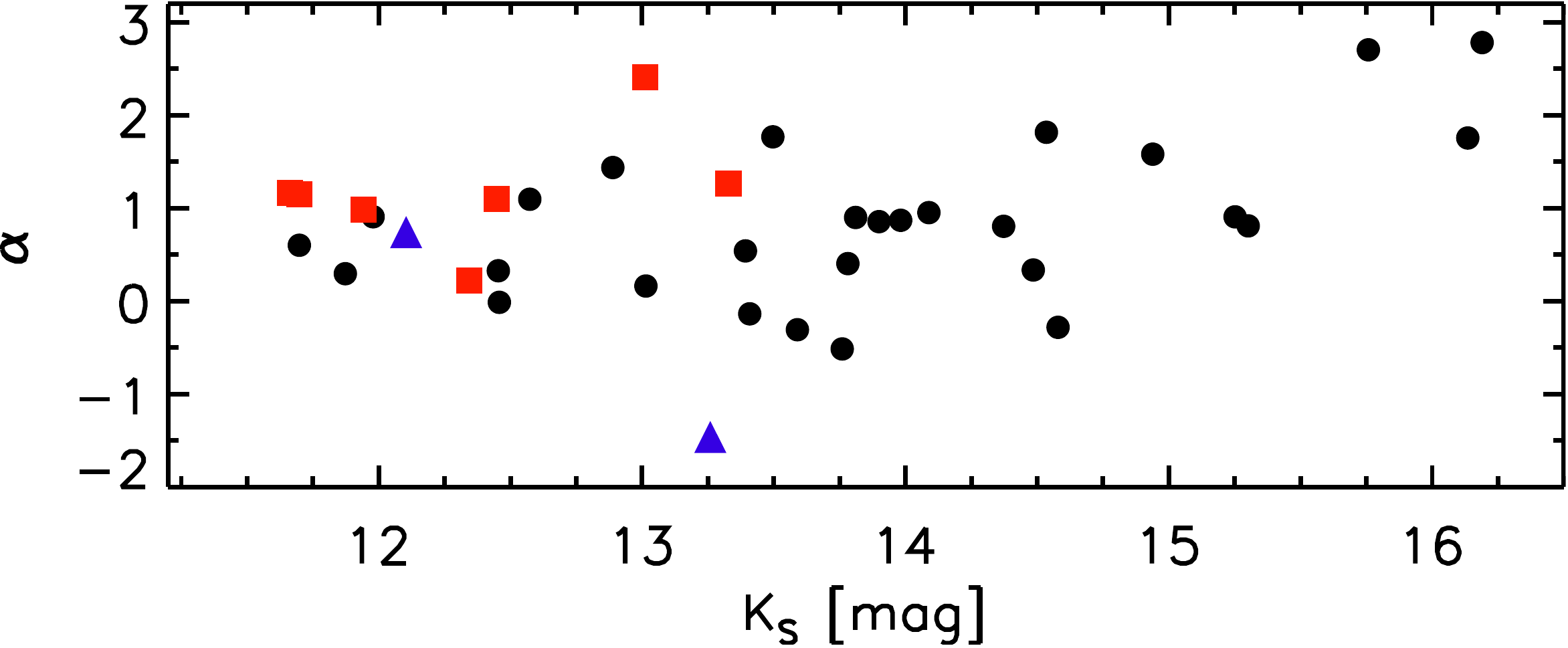}}\\
\resizebox{0.99\columnwidth}{!}{\includegraphics{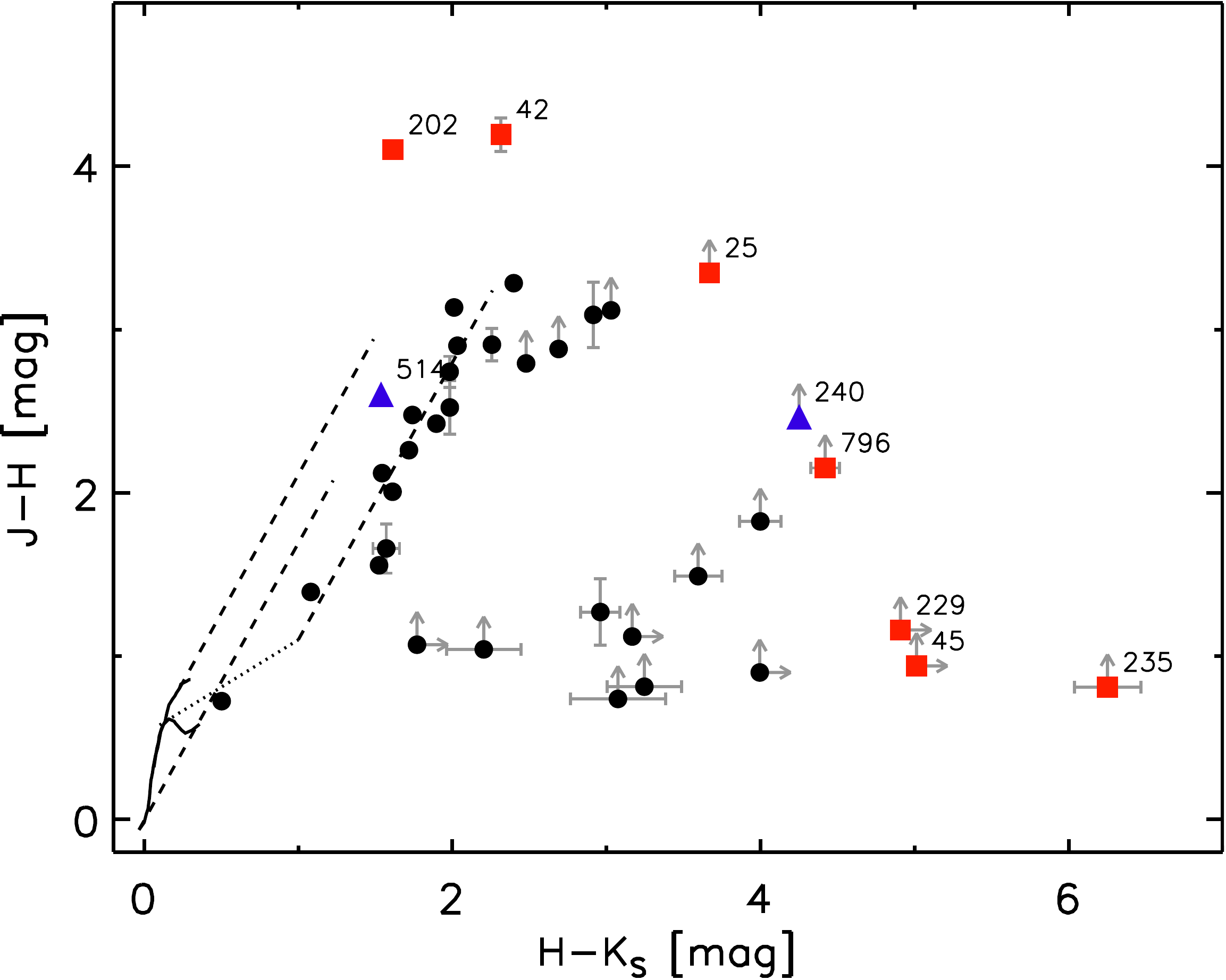}}\\
\resizebox{0.99\columnwidth}{!}{\includegraphics{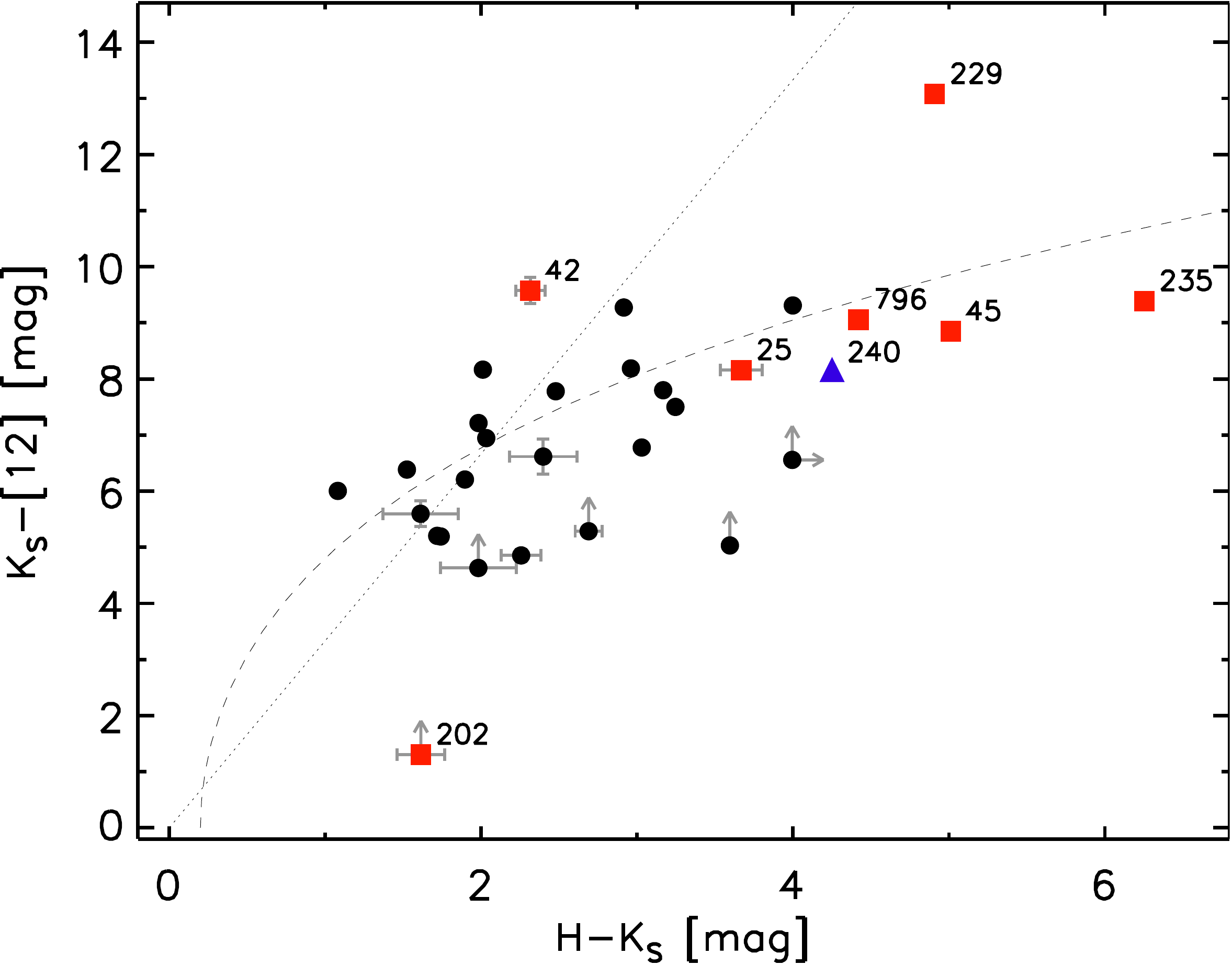}}
\caption{(top) $\alpha$ vs $K_{\rm s}$ for the objects of the VVV spectroscopic sample.  (middle) $(H-K_{\rm s})$, $(J-H)$ colour-colour diagram. (bottom) $K_{\rm s}-[12]$ vs $(H-K_{\rm s})$ for VVV objects. Magnitudes at 12 $\mu$m are from WISE $W3$ filter. Error bars are only shown for objects with significant errors. In all plots red squares mark objects that are likely asymptotic giant branch (AGB) stars, whilst blue triangles mark the two objects that are likely novae. The bottom panel shows the fiducial lines for the sequences of Carbon-rich  (dot-dashed line) and Oxygen-rich AGB stars \citep[dashed line, arising from][]{1998Vanloon}. We note that these lines are based on the IRAS 12 $\mu$m passband (which is not identical to $W3$).}
\label{vvv:specgc}
\end{figure}

We obtained near infrared spectra for 37 members of the 816 high amplitude variables from Paper I, all taken with the Magellan 
Baade telescope. All of the 37 stars have red near-infrared colours and SEDs and all are projected within 5\arcmin of an SFR, according to 
our search of the SIMBAD database and the mid-infrared WISE multi-colour images (see Paper I). Some preference was given to sources for 
which a distance estimate to the associated SFR is available in the literature. We note that the VVV disc region contains no large 
SFRs within $\sim$2~kpc, so most of the targets are expected to be more distant. 

We preferentially selected sources with $K_{\rm s}<14.5$ mag at the most recent available epoch in the VVV time series to the 
observations, in order to obtain good data in a relatively brief observation. Some preference was given to sources with signs of a recent eruption in the light 
curve, though the most recent VVV data available were typically taken several months before the observations.
For the first observing run in 2013 we had only 1 night at the telescope, so we observed only sources that either had
$\Delta K_{\rm s} > 2$~mag or a projected location near the massive G305 star forming complex. In the 2014 run we included
more sources with $\Delta K_{\rm s}$ = 1 to 2~mag, in order to better represent the full photometric sample.

The spectroscopic sample covers a fair cross section of the full sample of variables projected 
against SFRs, though it includes a high proportion of systems with eruptive light curves
and relatively few sources in each of the other categories. Of the 37 targets, the light curve 
classifications were: 18 eruptive variables, 5 long-term periodic YSOs, 6 Mira-like 
variables, 2 faders, 3 dippers and 3 short-term variables. The (mean) magnitude range is approximately $11.5 < K_{\rm s} < 14.5$, compared to
$11.5 <K_{\rm s} < 16.25$ for the full photometric sample (see Paper I). The sole exception is VVVv815
(see Appendix A), a somewhat fainter source observed as a target of opportunity.
 
The amplitudes of the 18 candidate eruptive variables in the spectroscopic sample are slightly 
higher than is typical of the 106 eruptive variables in the photometric sample but the difference
is modest: the median $\Delta K_{\rm s} = 1.88$~mag for the former group vs 1.61~mag for the latter 
group. For the full spectroscopic sample there is a stronger bias towards higher amplitude 
systems: the median $\Delta K_{\rm s} = 1.85$~mag, vs 1.44~mag for all the high amplitude variables in 
SFRs (excluding EBs, which are not included in the spectroscopic sample). 

For every object we have 2010--2014 $K_{\rm s}$ photometry from VVV, with the addition of a single 2015 data point arising from the second epoch of contemporaneous $JHK_{\rm s}$ photometry, light curve classification, as well as the information collected in Paper I using SIMBAD \citep{2000Wenger}, Vizier \citep{2000Ochsenbein}, and public near- and mid-infrared surveys from the NASA/IPAC Infrared Science Archive (IRSA).  The latter includes information from 2MASS \citep{2006Skrutskie}, DENIS \citep{1994Epchtein}, {\it Spitzer}/GLIMPSE surveys \citep[see e.g.,][]{2003Benjamin}, WISE Allsky release \citep{2010Wright}, Akari \citep{2007Murakami} and MSX6C \citep{2001Price}. In this work we also use the information from {\it Herschel} Hi-Gal \citep{2010Molinari}, and the recently available data of NEOWISE \citep{2011Mainzer} and DEEP GLIMPSE \citep{2011Whitney}. The 2010--2015 $K_{\rm s}$ light curves of the 37 objects in our sample are presented in Appendix \ref{apen2}. All VVV light curves were derived from v1.3 of the VISTA pipeline developed by the
Cambridge Astronomical Survey Unit (CASU), with the addition of our own saturation correction where necessary (see Paper I). 

Information from the literature allows us to estimate a likely distance to our objects, based on distances to areas of star formation found within 300\arcsec\ in the SIMBAD database. In general these are near kinematic distances estimated from radial velocities of emission lines observed towards HII regions or Infrared Dark Clouds (IRDCS), such as CS (2--1) \citep{1996Bronfman,2008Jackson}, $^{13}$CO(1--0) \citep[e.g.][]{2004Russeil} or OH maser emission \citep[e.g.][]{1998Caswell}. In some other cases distances are assumed from the location of the VVV object near known areas of star formation, such as the G305 star forming complex.

\begin{figure*}
\resizebox{\columnwidth}{!}{\includegraphics{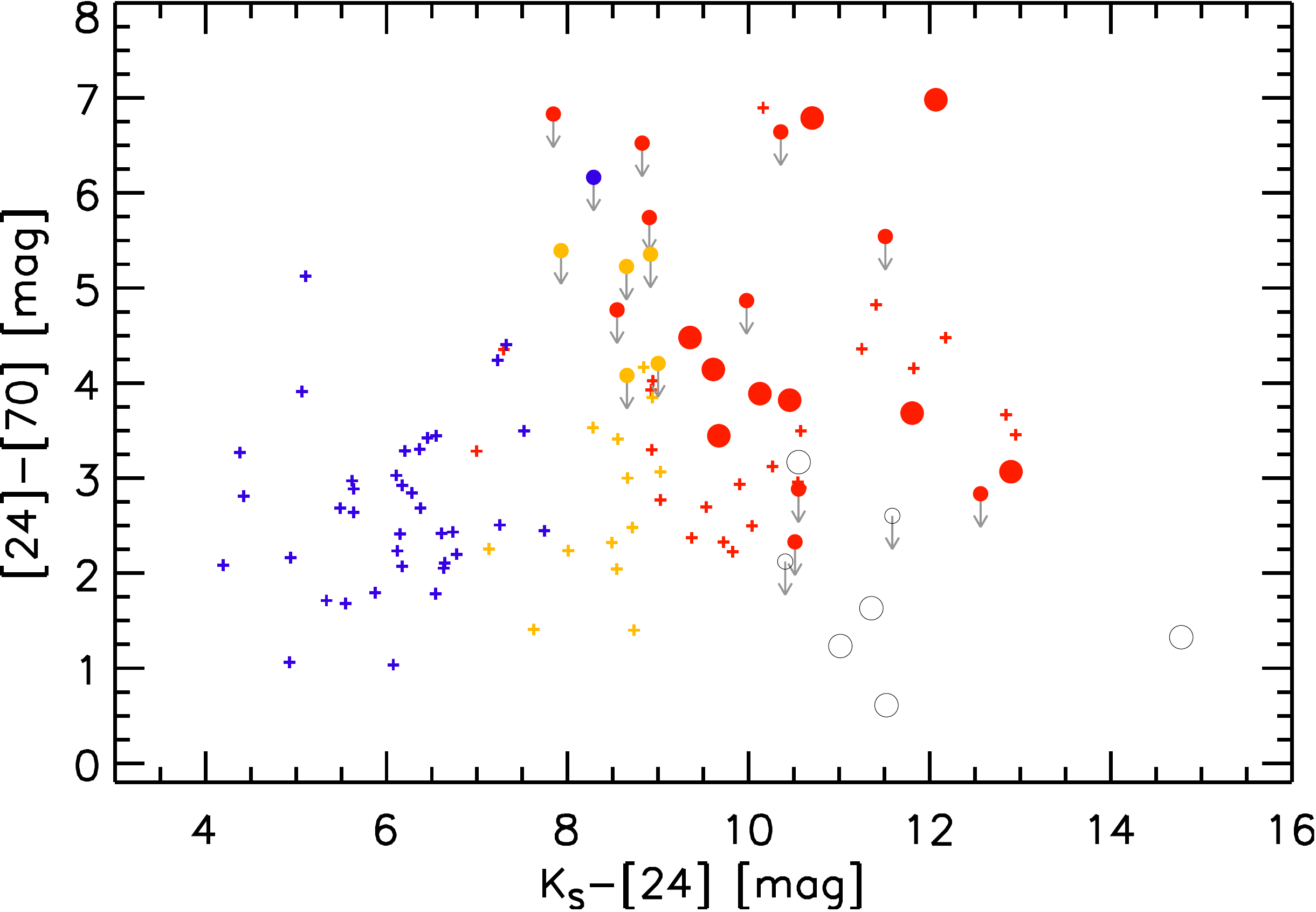}}
\resizebox{\columnwidth}{!}{\includegraphics{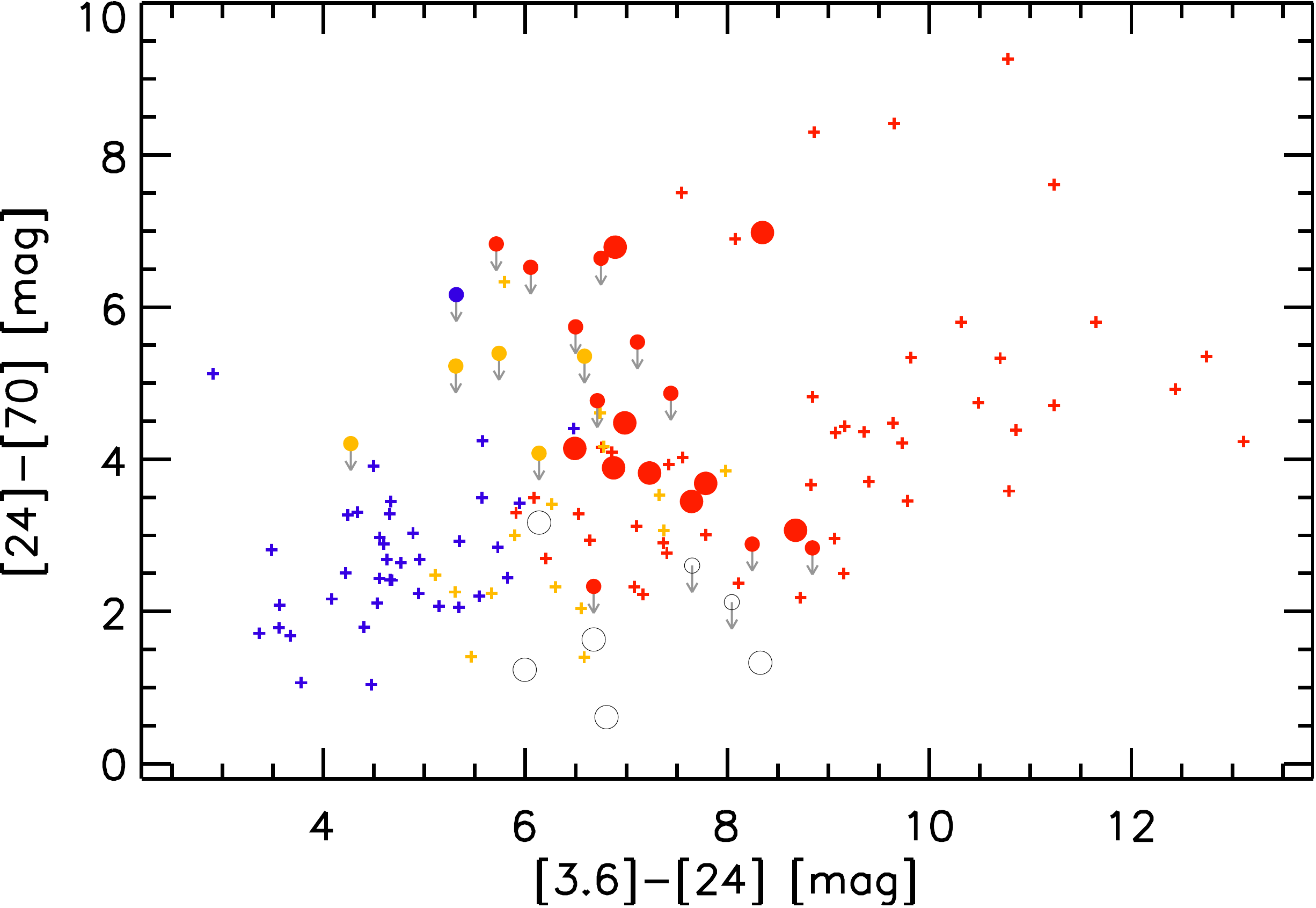}}
\caption{$[24]-[70]$ vs $K_{\rm s}-[24]$ (left) and $[24]-[70]$ vs $[3.6]-[24]$ (right) colour-colour diagrams for our sample. In the plots we show likely YSOs as filled circles and objects which were finally classified as non-YSOs as open circles. Large symbols represent objects with 70 $\mu$m detections, whilst small symbols with grey arrows represent upper limits. In the figure we also show the Perseus YSO sample from \citet{2015Young} as plus symbols. The colours for both VVV objects and YSOs from \citeauthor{2015Young} mark class I (red), flat-spectrum (orange) and class II (blue) YSOs.}
\label{vvv:specgc3}
\end{figure*}

Table \ref{table:vvvspecdample} shows the characteristics of the spectroscopic sample. In column 2, we show the original variable star designation from table 1 in Paper I, columns 3 and 4 show the right ascension and declination of the star, whilst columns 5 to 10 represent the 2010 $J$, $H$ and $K_{\rm s}$ magnitude of the stars along with the respective errors. In column 11 we present $\Delta K_{\rm s}=K_{\rm s, max}-K_{\rm s, min}$, whilst column 12 shows the slope of the SED, $\alpha$, derived in Paper I via a linear fit to data points at wavelengths $2< \lambda < 24~\mu$m. Column 13 shows the likely distance to our objects estimated from SIMBAD (designated as d$_{SFR}$), whilst column 14 shows the light curve classification given in Paper I. Finally column 15 shows the classification given to the objects from the results presented later in this work. Finding charts for these stars are shown in Appendix \ref{vvv:fcharts}.


 
 

\begin{table*}
\begin{center}
\begin{tabular}{l@{\hspace{0.15cm}}l@{\hspace{0.15cm}}c@{\hspace{0.15cm}}c@{\hspace{0.15cm}}c@{\hspace{0.15cm}}c@{\hspace{0.15cm}}c@{\hspace{0.15cm}}c@{\hspace{0.15cm}}c@{\hspace{0.15cm}}c@{\hspace{0.15cm}}c@{\hspace{0.15cm}}c@{\hspace{0.15cm}}c@{\hspace{0.15cm}}l@{\hspace{0.2cm}}l@{\hspace{0.15cm}}}
\hline
No & Variable & $\alpha$  & $\delta$  &  $J$ & $J_{err}$ & $H$ & $H_{err}$ & $K_{\rm s}$ & $K_{\rm s,err}$ & $\Delta K_{\rm s}$ & $\alpha$ & d$_{SFR}$ & Shape & Type\\
 & & (J2000) & (J2000) & (mag) & (mag) & (mag) & (mag) & (mag) & (mag) & (mag) & & (kpc) & (Paper I) & (This work) \\
\hline
1 & VVVv20 & 12:28:27.97 & $-$62:57:13.97  &  17.38 & 0.04 & 14.10 & 0.01 & 11.70 & 0.01 & 1.71 & 0.6 & -- & Eruptive & MNor\\ 
2 & VVVv25 & 12:35:14.37 & $-$62:47:15.63  &  -- & -- & 16.01 & 0.02 & 12.34 & 0.01 & 1.68 & 0.22 & 2.2 & Eruptive & AGB\\ 
3 & VVVv32 & 12:43:57.15 & $-$62:54:45.09  &  16.07 & 0.01 & 14.07 & 0.01 & 12.45 & 0.01 & 2.49 & 0.32 & 2.7 &  LPV-YSO & MNor\\ 
4 & VVVv42 & 13:09:34.64 & $-$62:49:32.52  &  18.45 & 0.10 & 14.26 & 0.01 & 11.94 & 0.01 & 2.16 & 0.98 & 3.7 &  LPV-Mira & AGB?, YSO?\\ 
5 & VVVv45 & 13:11:43.07 & $-$62:48:54.77  &  -- & -- & -- & -- & 13.33 & 0.01 & 2.31 & 1.27 & 3.7 &  LPV-Mira & AGB\\ 
6 & VVVv63 & 13:46:20.48 & $-$62:25:30.81  &  18.53 & 0.09 & 15.79 & 0.02 & 13.81 & 0.01 & 1.44 & 0.9 & 3.5 &  Eruptive & YSO, non-er?\\ 
7 & VVVv65 & 13:47:51.09 & $-$62:42:37.46  &  18.75 & 0.10 & 15.85 & 0.02 & 13.59 & 0.01 & 1.36 & -0.31 & 3.5 &  STV & YSO, non-er\\ 
8 & VVVv94 & 14:22:57.76 & $-$61:05:47.03  &  -- & -- & 16.42 & 0.03 & 13.39 & 0.01 & 1.91 & 0.54 & 2.6 &  Eruptive & MNor\\ 
9 & VVVv118 & 14:51:20.97 & $-$60:00:27.40  &  16.09 & 0.01 & 14.54 & 0.01 & 13.01 & 0.01 & 4.24 & 0.16 & -- &  Eruptive & EXor(MNor?) \\ 
10 & VVVv193 & 15:49:14.34 & $-$54:34:23.66  &  -- & -- & 16.59 & 0.04 & 13.90 & 0.01 & 1.26 & 0.86 & 4.2 & Eruptive & MNor\\ 
11 & VVVv202 & 15:54:26.41 & $-$54:08:29.40  &  17.38 & 0.03 & 13.28 & 0.01 & 11.66 & 0.01 & 1.61 & 1.16 & 4.1 & LPV-YSO & AGB\\ 
12 & VVVv229 & 16:04:24.48 & $-$53:01:14.01  &  -- & -- & -- & -- & 13.01 & 0.01 & 3.7 & 2.41 & 2.3 &  LPV-Mira &  AGB\\ 
13 & VVVv235 & 16:09:35.53 & $-$51:54:14.08  &  -- & -- & 17.95 & 0.22 & 11.70 & 0.01 & 1.26 & 1.15 & 5.3 &  LPV-Mira & AGB\\ 
14 & VVVv240 & 16:13:03.44 & $-$51:41:51.91  &  -- & -- & 16.35 & 0.05 & 12.10 & 0.01 & 1.17 & 0.74 & 3.5 &  LPV-Mira & Nova\\ 
15 & VVVv270 & 16:23:27.14 & $-$49:44:43.96  &  -- & -- & 18.34 & 0.23 & 16.14 & 0.07 & 3.81 & 1.76 & 2.3 &  Eruptive & MNor\\ 
16 & VVVv322 & 16:46:24.57 & $-$45:59:21.04  &  18.48 & 0.13 & 16.82 & 0.08 & 15.25 & 0.04 & 2.63 & 0.91 & 3.1 &  Eruptive & MNor\\ 
17 & VVVv374 & 16:58:33.99 & $-$42:49:55.25  &  17.12 & 0.02 & 13.99 & 0.01 & 11.98 & 0.01 & 2.41 & 0.91 & 2.9 &  Eruptive & MNor\\ 
18 & VVVv405 & 17:09:38.62 & $-$41:38:51.81  &  -- & -- & 17.50 & 0.13 & 13.50 & 0.01 & 2.00 & 1.77 & 2.3 &  Dipper & YSO, non-er?\\ 
19 & VVVv406 & 17:09:57.47 & $-$41:35:48.87  &  17.51 & 0.04 & 14.61 & 0.01 & 12.57 & 0.01 & 2.06 & 1.1 & 2.3 &  Dipper & YSO, non-er?\\ 
20 & VVVv452 & 12:41:58.06 & $-$62:13:42.90  &  18.10 & 0.06 & 15.68 & 0.01 & 13.78 & 0.01 & 2.46 & 0.4 & -- &  Eruptive & MNor\\ 
21 & VVVv473 & 13:10:57.49 & $-$62:35:22.34  &  -- & -- & -- & -- & 14.53 & 0.01 & 1.50 & 1.82 & 3.7 &  LPV-YSO & MNor\\ 
22 & VVVv480 & 13:16:50.32 & $-$62:23:41.61  &  15.81 & 0.01 & 15.08 & 0.01 & 14.58 & 0.01 & 1.70 & -0.28 & 3.7 &  LPV-YSO & YSO, non-er?\\ 
23 & VVVv514 & 14:00:45.37 & $-$61:33:39.95  &  17.40 & 0.04 & 14.80 & 0.01 & 13.26 & 0.01 & 1.73 & -1.46 & 4.3 &  Fader & Nova\\ 
24 & VVVv562 & 14:53:33.59 & $-$59:10:21.73  &  16.68 & 0.02 & 14.20 & 0.01 & 12.46 & 0.01 & 2.79 & -0.01 & 2.9 &  Fader& MNor\\ 
25 & VVVv625 & 15:43:17.95 & $-$54:06:47.29  &  -- & -- & 18.08 & 0.15 & 14.49 & 0.01 & 1.46 & 0.33 & 2.3 &  STV & YSO, non-er\\ 
26 & VVVv628 & 15:44:49.54 & $-$54:07:52.08  &  17.42 & 0.03 & 15.30 & 0.01 & 13.76 & 0.01 & 1.45 & -0.51 & 2.2 &  Dipper & YSO, non-er\\ 
27 & VVVv630 & 15:44:56.13 & $-$54:07:03.18  &  -- & -- & 18.83 & 0.31 & 15.76 & 0.04 & 1.90 & 2.70 & 2.2 &  Eruptive & YSO, non-er\\ 
28 & VVVv631 & 15:45:18.36 & $-$54:10:36.87  &  17.39 & 0.03 & 15.13 & 0.01 & 13.41 & 0.01 & 2.63 & -0.14 & 2.3 &  Eruptive & MNor\\ 
29 & VVVv632 & 15:44:09.80 & $-$53:56:27.78  &  14.35 & 0.01 & 12.95 & 0.01 & 11.87 & 0.01 & 1.51 & 0.29 & 2.5 & STV & YSO, non-er\\ 
30 & VVVv662 & 16:10:26.82 & $-$51:22:34.13  &  -- & -- & 18.55 & 0.24 & 15.30 & 0.03 & 1.83 & 0.81 & 3.1 &  Eruptive & MNor \\ 
31 & VVVv665 & 16:09:57.70 & $-$50:48:09.42  &  -- & -- & 16.57 & 0.04 & 14.09 & 0.01 & 1.63 & 0.95 & 4.3 &  Eruptive & MNor\\ 
32 & VVVv699 & 16:23:44.34 & $-$48:54:55.29  &  -- & -- & -- & -- & 16.19 & 0.09 & 2.28 & 2.78 & 4.2 &  Eruptive & MNor \\ 
33 & VVVv717 & 16:36:05.56 & $-$46:40:40.61  &  -- & -- & -- & -- & 14.37 & 0.02 & 2.47 & 0.81 & -- & LPV-YSO & MNor \\ 
34 & VVVv721 & 16:39:48.77 & $-$45:48:47.96  &  18.49 & 0.16 & 15.96 & 0.05 & 13.98 & 0.01 & 1.86 & 0.87 & 4.2 & Eruptive & FUor \\ 
35 & VVVv796 & 17:12:07.43 & $-$38:41:26.86  &  -- & -- & 16.87 & 0.09 & 12.45 & 0.01 & 2.94 & 1.1 & 1.4 &  LPV-Mira & AGB \\ 
36 & VVVv800 & 17:12:46.04 & $-$38:25:24.63  &  18.89 & 0.20 & 15.80 & 0.04 & 12.89 & 0.01 & 1.65 & 1.44  & 1.4 & Eruptive & MNor \\ 
37 & VVVv815 & 14:26:04.95 & $-$60:41:16.81  &  19.17 & 0.16 & 17.90 & 0.13 & 14.94 & 0.02 & 1.71 & 1.58  & 3.1 & Eruptive & MNor\\ 
\hline
\end{tabular}
\caption{Characteristics of the spectrocopic sample of VVV high amplitude variables. In the last column, ``non-er'' refers to YSOs not classified as eruptive variables.}\label{table:vvvspecdample}
\end{center}
\end{table*}

Fig. \ref{vvv:specgc} (top) illustrates the distributions in $\alpha$ and $K_{\rm s}$ magnitude of the spectroscopic sample (where $K_{\rm s}$ is 
taken from the first epoch of contemporaneous $JHK_{\rm s}$ data from 2010).
The sample has a fairly wide range in $\alpha$, though it is composed mainly of class I and flat spectrum systems. This is fairly 
representative of the photometric YSO sample, especially the eruptive candidates (see Paper I), though the photometric sample
does include a slightly higher proportion of class II YSOs and a few class III YSO candidates. Fig. \ref{vvv:specgc}
also shows the VVV near infrared colour-colour diagram (based on contemporaneous data from 2010) and an $K_{\rm s}-[12]$ vs $H-K_{\rm s}$  diagram.
Sources that we finally conclude are not YSOs are colour coded in red (dusty AGB variables) or blue (cataclysmic variables, more specifically classical novae). In the bottom panel of Fig. \ref{vvv:specgc} we also show fiducial lines for the sequences of Carbon and Oxygen-rich AGB stars \citep[from][]{1998Vanloon}. It is interesting to see that VVV objects that are later classified as AGB stars tend to follow the fiducial line of Oxygen-rich AGB stars. However, we note that not all Oxygen-rich stars in the sample of \citet{1998Vanloon} are observed to follow this line. 

We make use of 70~$\mu$m data from Hi-Gal in order to check whether the evolutionary classes of our sample, as defined by $\alpha$, agree with the expectations for class I, flat-spectrum and class II YSOs. In Fig. \ref{vvv:specgc3} we show the $[24]-[70]$ vs $K_{\rm s}-[24]$ and $[24]-[70]$ vs $[3.6]-[24]$ colour-colour diagrams for our sample, where we compare to the Perseus YSO sample from \citet{2015Young}. The latter use 3.6, 24 and 70 $\mu$m data from {\it Spitzer}, whilst $K_{\rm s}$ photometry arises from a 1\arcsec\ crossmatch of the sample with the 2MASS catalogue. For VVV objects we prefer the use of WISE W1 and W4 filters as it is nearly contemporaneous to Hi-Gal observations which were obtained in 2010. When WISE is not available we make use of GLIMPSE I1 and MIPSGAL 24 $\mu$m photometry. The use of WISE or {\it Spitzer} filters can lead to small differences in the photometry. \citet{2012Cutrirep} shows that WISE/W1 and GLIMPSE I1 only present significant differences for objects with W1$>14$ mag, whilst WISE/W4 tends to be 15 per cent brighter than MIPSGAL 24 $\mu$m photometry. For objects with no 70~$\mu$m detections we use the 90$\%$ completeness limit of 1.75 Jy from \citet{2013Elia}. The comparison of Fig. \ref{vvv:specgc3} shows that our classification agrees with that of \citet{2015Young}. In addition, we can see that the lack of a 70~$\mu$m detection for some VVV sources does not preclude a YSO classification. E.g. for sources with 70~$\mu$m upper limits, the expected location on the 
[24]$-$[70] vs $K_{\rm s}-$[24] diagram based on the \citet{2015Young} data typically corresponds to 70~$\mu$m fluxes that fall below the 
Hi-Gal completeness limit. Finally, we can also observe that objects that are later considered to be more likely non-YSOs tend to fall in a different area than YSOs. This is especially true for VVVv42, VVVv229, VVVv235 and VVVv796. These four objects are classified as having Mira-like light curves in Paper I. For further comparison with the colours of known dusty AGB stars we refer
the reader to \citet{2010Vanloona}, \citet{2010Vanloonb} and \citet{2011Boyer}, though
we note that most of the sources described therein have more modest amplitudes (but these works do include also Mira-like pulsators with $\Delta K > 1$ mag).


\begin{figure}
\resizebox{\columnwidth}{!}{\includegraphics{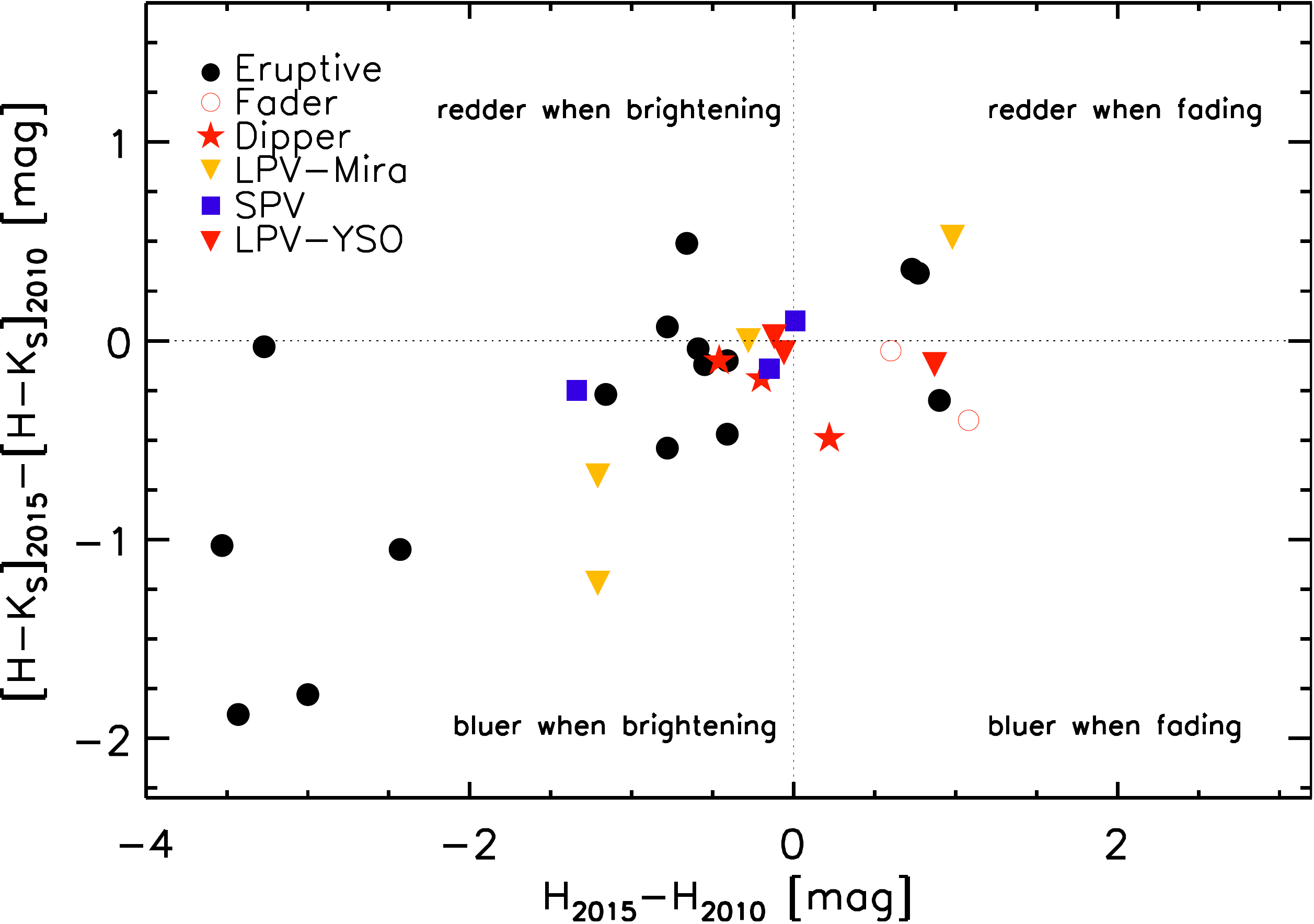}}\\
\resizebox{\columnwidth}{!}{\includegraphics{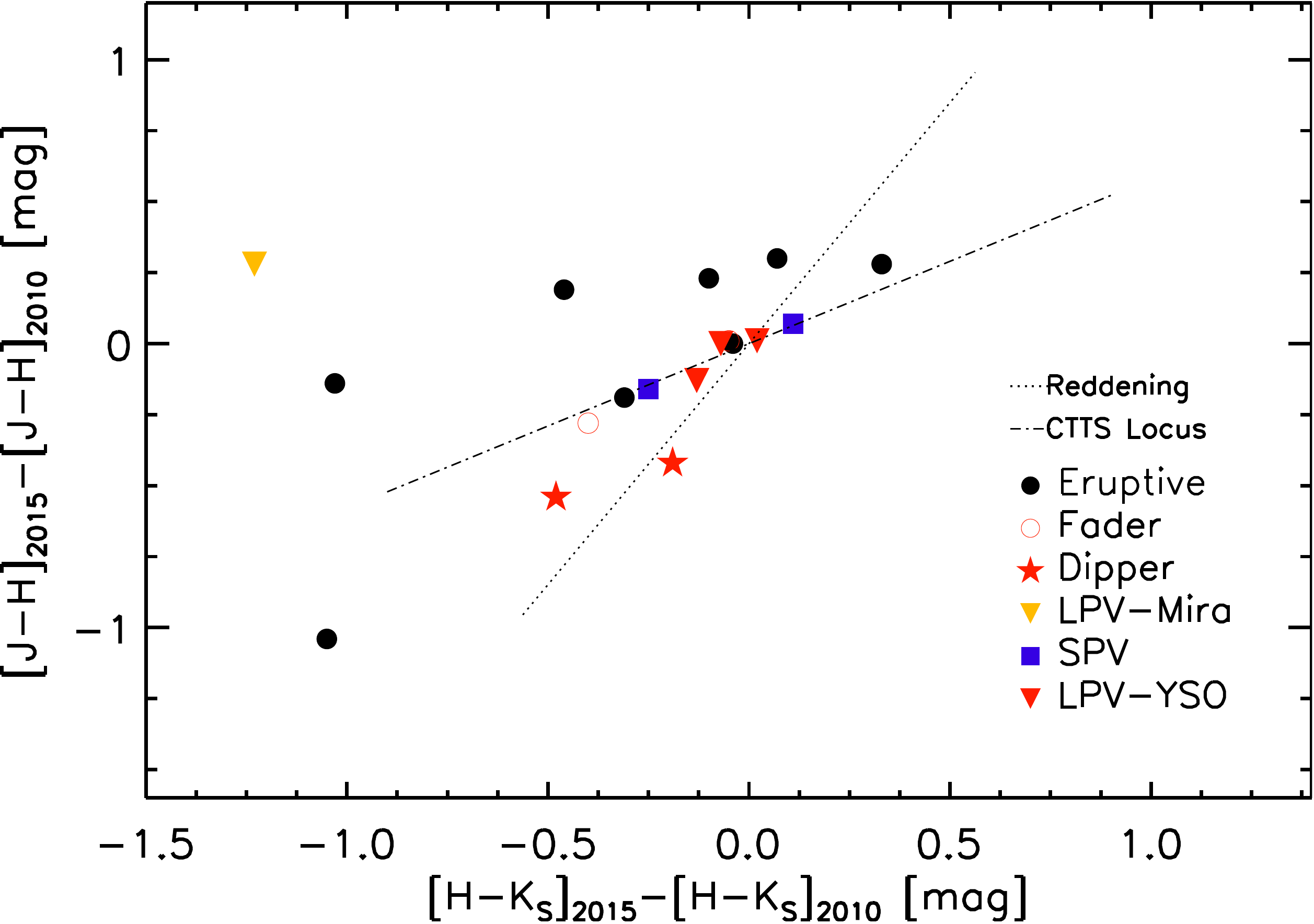}}
\caption{ $\Delta (H-K_{\rm s})$ vs $\Delta H$ (top) and $\Delta (J-H)$ vs $\Delta (H-K_{\rm s})$ (bottom) for objects in the spectroscopic sample with an available second $JHK_{\rm s}$ epoch from VVV. In the plots we mark the different classes from light curve morphology. Different symbols are explained in the plots. In the bottom plot we mark the expected changes which occur parallel to the reddening vector (dotted line) and to the CTTs locus of \citet{1997Meyer} (dot-dashed line).}
\label{vvv:specgc2}
\end{figure}

As we did in Paper I for the full sample, we use the 2 epochs of $JHK_{\rm s}$ photometry from VVV in order to study the mechanism driving variability in our sample. Once again we stress that in most cases the 2 epochs do not sample directly the large $K_{\rm s}$ magnitude changes in the systems. In addition, we note that 6 objects in our sample are undetected in the $J$ and $H$ passbands in at least one epoch, and 20 objects are not detected in the $J$ passband in at least one epoch. Fig. \ref{vvv:specgc2} shows $\Delta (H-K_{\rm s})$ vs $\Delta H$  and $\Delta (J-H)$ vs $\Delta (H-K_{\rm s})$ for objects where this is information is available. In the first plot (upper panel) we see that our sample shows an elliptical distribution with its major axis along the redder when fading/bluer when brightening quadrants, consistent with changes being cause either by variable extinction or accretion \citep{2012Loren}. The second plot shows that for sources with significant changes in colour, these changes are generally not consistent with extinction in the majority of the sample. The 
exception is the objects classified as dippers, for which extinction was already considered to be a more likely explanation for their
variability (see Paper I).

\section{Observations}\label{vvv:sec_datared}
           
\subsection{Magellan Baade/FIRE}

\begin{table}
\begin{center}
\resizebox{\columnwidth}{!}{
\begin{tabular}{@{}l@{\hspace{0.2cm}}c@{\hspace{0.2cm}}c@{\hspace{0.2cm}}c@{\hspace{0.2cm}}c@{\hspace{0.2cm}}c@{\hspace{0.2cm}}}
\hline
Object & $R$ & $T_{exp}$(s) & Airmass & N$_{exp}$ & Date \\
\hline
 VVVv20  & 6000 & 253.6 & 1.4 & 4 & 11 May 2014 \\
 VVVv25  & 6000 & 253.6 & 1.2 & 4 & 27 Apr 2015\\
 VVVv32 & 6000 & 253.6 & 1.2 & 8 & 10 May 2014\\
 VVVv42  & 6000 & 253.6 & 1.4 & 4 &  22 April 2013\\
 VVVv45 & 6000 & 253.6 & 1.2 & 4 &  22 April 2013\\
 VVVv63  & 6000 & 253.6 & 1.4 & 4 & 11 May 2014 \\
 VVVv65   & 6000 & 158.5 & 1.3 & 4 &  11 May 2014\\
 VVVv94 & 6000 & 253.6 & 1.2 & 4 &  10 May 2014\\
 VVVv118   & 6000 & 253.6 & 1.2 & 4 &  22 April 2013\\
 VVVv193  & 6000 & 158.5 & 1.1 & 4 &  11 May 2014\\
 VVVv202  & 6000 & 158.5 & 1.1 & 2 &  11 May 2014 \\
 VVVv229  & 6000 & 158.5 & 1.1 & 4 &  22 April 2013 \\
 VVVv235  & 6000 & 158.5 & 1.1 & 2 &  11 May 2014\\
 VVVv240 & 6000 & 253.6 & 1.2 & 4 &  10 May 2014 \\
 VVVv270  & 6000 & 158.6 & 1.2 & 4 &  11 May 2014 \\
 VVVv322 & 6000 & 158.5 & 1.3 & 4 &  11 May 2014 \\
 VVVv322 & 6000 & 253.6 & 1.1 & 4 &  22 April 2013\\
 VVVv374  & 6000 & 253.6 & 1.1 & 4 &  10 May 2014\\
 VVVv405   & 6000 & 158.5 & 1.3 & 4 &  11 May 2014 \\
 VVVv406   & 6000 & 158.5 & 1.4 & 4 &  11 May 2014\\
 VVVv452   & 6000 & 253.6 & 1.2 & 8 &  10 May 2014\\
 VVVv473   & 6000 & 253.6 & 1.2 & 4 &  22 April 2013\\
 VVVv480   & 6000 & 253.6 & 1.25 & 12 &  22 April 2013\\
 VVVv514   & 6000 & 158.5 & 1.2 & 4 &  11 May 2014\\
 VVVv562  & 6000 & 253.6 & 1.3 & 4 &  11 May 2014\\
 VVVv625   & 6000 & 158.5 & 1.2 & 4 &  11 May 2014\\
 VVVv628   & 6000 & 253.6 & 1.1 & 4 &  11 May 2014\\
 VVVv630   & 6000 & 158.5 & 1.2 & 4 &  11 May 2014\\
 VVVv631   & 6000 & 253.6 & 1.1 & 4 &  10 May 2014\\
 VVVv632   & 6000 & 253.6 & 1.2 & 4 &  10 May 2014\\
 VVVv662   & 6000 & 158.5 & 1.2 & 4 &  11 May 2014\\
 VVVv665   & 6000 & 158.5 & 1.1 & 4 &  11 May 2014\\
 VVVv699   & 6000 & 158.5 & 1.1 & 4 &  11 May 2014\\
 VVVv699   & 6000 & 253.6 & 1.1 & 4 &  22 April 2013\\
 VVVv717   & 6000 & 253.6 & 1.1 & 4 &  22 April 2013\\
 VVVv721   & 6000 & 158.6 & 1.2 & 4 &  11 May 2014\\
 VVVv796   & 6000 & 253.6 & 1.3 & 4 &  10 May 2014\\
 VVVv800   & 6000 & 253.6 & 1.2 & 4 &  10 May 2014\\
 VVVv815 & 6000 & 253.6 & 1.2 & 4 &  22 April 2013\\
 VVVv815 & 300 & 148 & 1.3 & 4 &  8 May 2012\\
\hline
\end{tabular}}
\caption{Spectroscopic observations of VVV objects.}\label{table:vvvobs}
\end{center}
\end{table}

\subsubsection{Low Resolution spectroscopy}

    Follow up of one high amplitude variable VVVv815 was performed on May 8th, 2012 with the FIRE spectrograph mounted on the Magellan Baade Telescope at Las Campanas Observatory, Chile. As detailed in Paper I, VVVv815 was identified via an early search of the VVV time 
series, with slightly different quality criteria to the rest. The observations were carried out in the high-throughput prism mode which provides a continuous coverage from $0.8$--$2.5~\mu$m at low resolving power of $R=\Delta\lambda/\lambda \sim 250$--$350$ with a 0.6\arcsec ~slit. Since the spectrum displayed 
exceptionally strong H$_2$ emission lines (suggesting an accretion-powered outflow) VVVv815 was observed again in echelle mode in a
subsequent run allocated to the topic of this paper, see below. 

    
    The object was observed in the usual ABBA pattern along the slit, with individual exposures of 148 s and at  an airmass of 1.1. Observations of spectroscopic calibrators were carried out for purposes of telluric correction and flux calibration; the calibrators were observed at similar airmass as the objects and with the same instrumental setup. Quartz lamp images were obtained for flat fielding of the data; they consisted of high-voltage (2.2V) images that provide data for the z/J bands but saturate the red end of the spectrum, and low-voltage(1.1V) that generate counts in H/K but are too faint for the blue end. Additionally, ``NeNeAr'' arc lamp images were acquired for wavelength calibration.
    
    The images were reduced using the longslit package of the FIREHOSE software. Using the NeNeAr lamp image FIREHOSE generates an arc solution, which had typical uncertainties of 0.4 pixels or $\sim$ 2.7 \AA. The second step consisted of creating a flat image; this is done by combining the 2.2V (blue) and 1.1V (red) with a smooth weighting function spread over $\sim$ 150 pixels centered on a transition pixel defined by the user. Finally the software traces and extracts the spectrum of the object. Telluric correction and flux calibration were performed in the standard mode using NOAO/TWODSPEC package in IRAF.

\subsubsection{Echelle spectroscopy}    
    
All 37 members of the sample were observed in the echellete mode with the FIRE spectrograph mounted on the Magellan Baade Telescope at Las Campanas Observatory, Chile. The echellete mode provides coverage from $0.8$--$2.5~\mu$m using echelle grating orders 11--32, at a resolving power of $R \sim 6000$ with a 0.6\arcsec ~slit. The spatial scale is 0.18\arcsec ~pixel$^{-1}$. The bulk of the observations were carried out over a total of 3 nights, on April 19th 2013 and May 10th--11th 2014. Thick cloud cover on May 10th restricted
us to bright targets but did not otherwise appear to affect the results. Additional data were also taken on the nights of 26 and 27 
April 2015 in order to investigate the spectroscopic variability of the sample. Data from the 2015 run are only included here for one 
source, VVVv25, for which the 2014 data were of insufficient quality for a clear classification. We defer further discussion of the 
2015 dataset to a future paper that will benefit from additional 2015 VVV time series photometry.
   
   The majority of the variable stars were observed in the typical ABBA pattern along the slit, with individual exposure 
times between 158 and 260~s and a total time on source of $\sim10$--$17$~minutes. Airmasses were in the range 1--1.4. Immediately after observing each object, we observed ThAr lamps for wavelength calibration, and F-type and A0 main sequence stars for relative flux calibration purposes. These were also done in an AB pattern and with exposure times between 10--60 s. Sky and ``Qh'' lamp flats were acquired at the beginning/end of each night to determine the pixel-to-pixel response calibration. Table \ref{table:vvvobs} shows the date, exposure time and number of exposures for all of the VVV objects.  
   
   Data reduction was performed using the echelle mode of the FIREHOSE data reduction software. This determines the boundaries of each order in the image, constructs a master Qh flat, a slit illumination function from the sky twilight flats and determines the tilt of the slit in the detector from a wavelength calibration frame. The software then generates the wavelength solution and traces the object in each order. The wavelength solution has errors between 0.04--0.4 \AA~ depending on the order.
   
   The profile of the object and background sky residuals are fitted using an iterative procedure, where a non-parametric spline function is used for orders with high signal-to-noise, whilst a Gaussian profile is assumed for low signal-to-noise orders. It then performs an optimally-weighted extraction of the object using the profile determined before. 
   
  In the case of the extraction of the telluric calibration stars, the software uses a boxcar extraction as the default, which simply sums the counts of the target within an extraction window of a pre-determined width. We found that the different extraction methods for the object and calibrator yielded odd results for the final spectrum of the object after telluric calibration, i.e. the objects had bluer spectra than expected from their near-infrared colours. This problem was solved by using optimal extraction for the telluric calibration stars as well as the science targets.  
   
  The software removes intrinsic absorption features of HI from the telluric calibration stars following the methods of \citet{2003Vacca}. This procedure is only suitable for A0V stars and it was used for the telluric calibration stars observed in the 2014 run. In the 2013 run we observed F-type stars as calibrators because we were not confident that the software would remove HI features from A0V spectra
with sufficient precision to allow measurement of HI emission in YSOs, such as the Br$\gamma$ line. Subsequent testing showed that the
pipeline does remove HI features from A0 stars very well, so we reverted to A0 stars in 2014, as recommended in the FIRE manual. 
For the 11 science targets observed in 2013, intrinsic features from the F-type calibrator were removed for each order by fitting and 
subtracting them with {\sc SPLOT} in {\sc IRAF}.
  
  The final relative flux calibration and merging of individual orders were performed using the ONEDSPEC tasks TELLURIC and SCOMBINE respectively.
  
  
The reduced and combined FIRE spectra of 37 objects are presented in Appendix \ref{apen2}.

\section{FIRE Spectra}\label{vvv:sec_fire}

\subsection{General properties}

In this subsection and in Section \ref{subsec:EWs} we discuss the general spectroscopic characteristics of the sample, which confirm most of the variables as YSOs
and show that they resemble eruptive systems more than normal class I YSOs.
Then in Section \ref{vvv:sec_varacc} we look at how the spectroscopic characteristics of the confirmed YSOs differ between the different categories of light curves, prior to a discussion on how to classify these stars in Section \ref{vvv:sec_erupclass}. We will present some of the individual spectra throughout the remainder of this work. However, we refer the reader to Appendix \ref{apen2} in order to inspect the whole sample before reading through these sections.

In Table \ref{table:vvvEWs} we show the equivalent widths obtained from the atomic and molecular features. Equivalent widths of individual lines are measured using {\sc SPLOT} in {\sc IRAF} with a Gaussian line profile.

\begin{table*}
\begin{center}
\begin{tabular}{@{}l@{\hspace{0.5cm}}c@{\hspace{0.2cm}}c@{\hspace{0.2cm}}c@{\hspace{0.2cm}}c@{\hspace{0.2cm}}c@{\hspace{0.2cm}}c@{\hspace{0.2cm}}}
\hline
Object & H$_{2} (2.12~\mu$m) &  Br$\gamma (2.16~\mu$m)  & Na I$(2.21~\mu$m) & Ca I$(2.26~\mu$m)  & $^{12}$CO$(2.293~\mu$m)  & Mg I$(2.28~\mu$m) \\[0.1cm]
 & (\AA) & (\AA)  & (\AA) & (\AA)  & (\AA)  & (\AA) \\
\hline
\multicolumn{7}{c}{YSO-Eruptive}\\
\hline
VVVv20 &   $-$9.8$\pm$ 0.4 &   $-$3.8$\pm$ 0.6 &   $-$0.7$\pm$ 0.2 &    -- &   $-$5.3$\pm$ 2.5 &    --\\
VVVv32 &    -- &   $-$5.7$\pm$ 2.2 &    -- &    -- &  $-$19.0$\pm$10.0 &    --\\
VVVv94 &   $-$0.7$\pm$ 0.3 &   $-$7.0$\pm$ 1.5 &    -- &    -- &   $-$5.0$\pm$ 7.0 &    --\\
VVVv118 &    -- &   $-$1.4$\pm$ 1.0 &    -- &    -- &    -- &    --\\
VVVv193 &   $-$1.3$\pm$ 0.6 &   $-$2.2$\pm$ 0.7 &    -- &    -- &   $-$7.0$\pm$ 4.0 &    --\\
VVVv270 &   $-$0.8$\pm$ 0.3 &   $-$2.0$\pm$ 0.7 &    $-$1.7$\pm$0.3 &    -- &   $-$7.0$\pm$ 6.0 &    --\\
VVVv322(13) &   $-$2.0$\pm$ 0.8 &    -- &    -- &    -- &   33.0$\pm$ 7.0 &    --\\
VVVv322(14) &   $-$5.5$\pm$ 2.1 &    -- &    -- &    -- &   25.0$\pm$ 8.0 &    --\\
VVVv374 &   $-$1.9$\pm$ 0.2 &   $-$2.1$\pm$ 0.3 &   $-$0.8$\pm$ 0.2 &    -- &   $-$4.5$\pm$ 1.5 &    --\\
VVVv452 &   $-$1.2$\pm$ 0.6 &   $-$3.6$\pm$ 1.1 &    -- &    -- &   $-$8.0$\pm$ 4.0 &    --\\
VVVv473 &  $-$30.1$\pm$ 3.9 &   $-$1.7$\pm$ 1.2 &    -- &    -- &   $-$1.3$\pm$12.0 &    --\\
VVVv562 &   $-$3.2$\pm$ 0.5 &   $-$3.4$\pm$ 1.1 &    2.0$\pm$ 0.3 &    -- &    -- &    --\\
VVVv631 &    -- &   $-$2.4$\pm$ 0.4 &   $-$1.4$\pm$ 0.2 &    -- &   $-$1.0$\pm$ 1.0 &    --\\
VVVv662 &   $-$0.3$\pm$ 0.1 &   $-$4.1$\pm$ 1.1 &   $-$1.2$\pm$ 0.4 &    -- &   $-$2.0$\pm$ 4.0 &    --\\
VVVv665 &   $-$1.9$\pm$ 0.7 &   $-$2.4$\pm$ 0.4 &   $-$3.4$\pm$ 0.3 &    -- &  $-$52.0$\pm$11.0 &    --\\
VVVv699(13) &  $-$14.1$\pm$ 1.3 &   $-$3.5$\pm$ 0.5 &   $-$5.0$\pm$ 0.4 &    -- &  $-$40.7$\pm$11.1 &   $-$1.4$\pm$ 0.5\\
VVVv699(14) &  $-$15.5$\pm$ 2.4 &   $-$3.1$\pm$ 1.1 &   $-$3.1$\pm$ 0.5 &    -- &  $-$21.3$\pm$10.0 &   $-$0.8$\pm$ 0.3\\
VVVv717 &   $-$0.8$\pm$ 0.3 &   $-$1.1$\pm$ 0.5 &    -- &    -- &   28.0$\pm$ 4.0 &   $-$0.9$\pm$ 0.3\\
VVVv721 &    -- &    -- &    -- &    -- &   14.0$\pm$ 4.0 &    --\\
VVVv800 &   $-$2.0$\pm$ 0.2 &   $-$1.2$\pm$ 0.4 &    -- &    -- &    -- &    --\\
VVVv815(13) & $-$119.0$\pm$36.0 &    -- &    -- &    -- &    -- &    --\\
VVVv815(12) & $-$200.0$\pm$27.0 &    -- &    -- &    -- &    -- &    --\\
\hline
\multicolumn{7}{c}{YSO-Non Eruptive}\\
\hline
VVVv63 &   $-$4.3$\pm$ 0.8 &   $-$2.1$\pm$ 0.5 &    -- &    -- &    -- &    --\\
VVVv65 &    -- &   $-$1.2$\pm$ 0.5 &    -- &    -- &    -- &    --\\
VVVv405 &   $-$7.5$\pm$ 1.8 &   $-$0.9$\pm$ 0.4 &    -- &    -- &    -- &    --\\
VVVv406 &   $-$0.5$\pm$ 0.2 &    -- &    -- &    -- &    -- &    --\\
VVVv480 &   -- &   $-$4.4$\pm$ 2.2 &    -- &    -- &    -- &    --\\
VVVv625 &   $-$0.8$\pm$ 0.2 &   $-$3.6$\pm$ 1.7 &    -- &    -- &    -- &    --\\
VVVv628 &   $-$1.7$\pm$ 0.7 &   $-$1.4$\pm$ 0.5 &    3.1$\pm$ 0.6 &    3.2$\pm$ 0.9 &   13.0$\pm$ 8.0 &    --\\
VVVv630 &    -- &    0.7$\pm$ 0.2 &    1.3$\pm$ 0.2 &    1.8$\pm$ 0.6 &   14.0$\pm$ 5.0 &    --\\
VVVv632 &   $-$1.4$\pm$ 0.2 &   $-$2.1$\pm$ 0.3 &    1.2$\pm$ 0.2 &    1.1$\pm$ 0.2 &    -- &    --\\
\hline
\multicolumn{7}{c}{Non-YSOs}\\
\hline
VVVv25 &    -- &    -- &   $-$0.6$\pm$ 0.3 &    -- &   16.0$\pm$ 7.0 &    --\\
VVVv42 &   $-$0.3$\pm$ 0.2 &    -- &   $-$1.0$\pm$ 0.3 &    -- &    8.0$\pm$ 4.0 &    --\\
VVVv45 &    -- &    -- &    -- &    -- &   33.0$\pm$11.0 &    --\\
VVVv202 &    -- &    -- &    4.7$\pm$ 0.4 &    5.3$\pm$ 0.5 &   37.0$\pm$ 2.0 &    --\\
VVVv229 &    -- &    -- &    -- &    -- &   25.0$\pm$ 7.0 &    --\\
VVVv235 &    -- &    -- &    -- &    -- &   47.0$\pm$12.0 &    --\\
VVVv240 &    -- &   $-$0.7$\pm$ 0.3 &    -- &    -- &    -- &    --\\
VVVv514 &    -- &   $-$7.8$\pm$ 2.3 &    -- &    -- &    -- &    --\\
VVVv796 &    -- &    -- &    -- &    -- &   27.0$\pm$ 7.0 &    --\\
\hline
\end{tabular}
\caption{Equivalent widths of common features found in the near-infrared spectra of YSOs detected in our VVV sample. The EW of CO was measured in {\sc splot} by visually selecting two continuum points around the CO bandhead and estimating the continuum as a linear function between the two points. The EW is estimated by summing (1$-$F$_{i}$/C$_{i}$) over wavelength, where F$_{i}$ and C$_{i}$ are the flux and the continuum at pixel $i$. This inevitably has fairly large errors due to uncertainty in the continuum level, even though the individual transitions are often very clearly detected.}\label{table:vvvEWs}
\end{center}
\end{table*}

We first note that nine variable stars are found to be of a different class, based on consideration of all the data. Two objects are likely novae observed at some time after outburst and seven are likely dust-enshrouded AGB stars. The two novae, VVVv514 and VVVv240, had 
light curves classified as a fader and Mira-like, respectively, in Paper I. We note that novae are known to have very diverse light 
curves (e.g. Gehrz 1988). The light curves of 5/7 of the AGB stars were 
classified as Mira-like in Paper I and they are listed as ``LPV-Mira'' in Table \ref{table:vvvspecdample}. A sixth O-rich AGB star, VVVv202, had an ``LPV-YSO'' 
classification in Table \ref{table:vvvspecdample} and Paper~I since its light curve was not very well fit by a smooth sinusoid. A small number of such 
mis-classifications of LPV-YSO and LPV-Mira types was to be expected, as noted in Paper~I. The light curve of the seventh, VVVv25, was 
classified as eruptive in Paper~I because the sparsely sampled time series does not suggest a periodic nature, except perhaps on a timescale substantially longer than the 4 year baseline. However, the spectrum reveals that it is a C-rich object (which is not consistent 
with YSO status) and our analysis of possible periods in Paper I shows that a period of 400--500 days (typical of a carbon star) is 
amongst the possibilities consistent with the sparsely sampled light curve, most of the troughs being missed in such a case.
In Appendix \ref{vvv:sec_evolved} we analyse these 9 post-main sequence objects and discuss how some of the information from the 
literature and their spectral characteristics can cause such objects to be mis-interpreted as YSOs. These 9 variable stars are not 
included in our discussion of YSO properties, so percentages given below for the incidence of particular features refer to the YSOs 
only.

We find that 28 objects display the typical characteristics of YSOs, such as Br$\gamma$ or H$_{2}$ lines in emission or CO in emission or absorption. In each of these objects the overall trend of the near infrared continuum is either to rise with increasing wavelength 
or to be approximately flat, as expected for class I and flat spectrum YSOs, or other sources suffering high infrared extinction.
Three YSOs, VVVv322, VVVv717 and VVV721, have $\Delta v=2$ CO absorption bands at $2.29$--$2.45~\mu$m that appear to be stronger than expected from the photosphere of a K or M dwarf, and are more in line with that expected from the photosphere of late M-type giant stars. In fact the CO absorption of these three systems is very similar to that of some of our objects which are classified as AGB stars. Such strong CO absorption is characteristic of FUors and FUor-like YSOs \citep{2010Connelley, 2010Reipurth}. In fact the latter authors regard it as the defining spectroscopic characteristic, given the inferred origin in a disc with a high accretion
rate (see Section \ref{vvv:intro}). Two of these three sources also display the broad H$_2$O absorption bands often seen in such systems.

Between 20 and 23 objects show Br$\gamma$ emission (the detections being marginal in 3 cases). This amounts to $71$--$82\%$ of the 28
YSOs. This feature is noticeably broadened ($\sim100$~km~s$^{-1}$ line width) in most cases where it is well 
measured. This broadening is typical of normal YSOs where it is attributed to infalling gas near the stellar photosphere 
\citep[see e.g.][]{1998Muzerolle,2001Davis}. Br$\gamma$ is seen in absorption in only one YSO, VVVv630, which appears not to be an eruptive variable
(see Section \ref{vvv:sec_varacc}).

We find that 22 of the YSO spectra (79\%) show 2.12~$\mu$m H$_{2}$ emission. Of these, 10 objects also show additional lines from H$_{2}$. In every case the emission from this molecule is likely arising from shock-excited emission given that higher excited transitions are usually weak or not present in the spectra. Atomic lines of Na I and Ca I are not common in our sample, with 3 or 4 objects showing absorption at these wavelengths (11--14$\%$) and emission in 5 to 7 objects ($18$--$25\%$). When Na I was observed in emission, CO was usually also in emission (see Table \ref{table:vvvEWs}) following the 
pattern seen in EXors (see Section \ref{vvv:intro}).   

CO emission is observed in 8 to 11 objects ($27$--$37\%$). We find that objects that show this characteristic can be fitted by models (see Appendix \ref{vvv:co}) with temperatures between 2500--3500 K. This temperature range agrees with that expected to give rise to CO emission in gaseous discs of YSOs \citep{1996Najita}. We also find that some objects show evidence of broadened CO profiles. Rotation in Keplerian discs has been shown to produce this feature \citep[see e.g.][]{1996Najita, 2010Davies}. Although this needs a more detailed investigation at higher spectral resolution, the observed broadened profile appears consistent with a similar origin in some of our systems.

Br$\gamma$ is not detected in 4 YSOs. One of these is VVVv815, an object for which only strong H$_{2}$ emission is present (see Appendix \ref{vvv:specvar}). In fact, the 2.12 $\mu$m line has an equivalent width of EW$=-200$~\AA ~in 2012 and EW$=-119$~\AA ~in 2013 which is the largest equivalent width measured for this line across the whole sample. As for VVVv815, also VVVv406 does not show any features at the wavelength of Br$\gamma$, but is also in general a featureless spectrum with weak H$_{2}$~2.12~$\mu$m emission. The remaining 2 objects, VVVv322 and VVVv721, 
are characterized by having CO absorption and broad H$_2$O absorption bands, with a lack of other photospheric absorption features such as Na I and Ca I. These two YSOs are spectroscopically similar to FUors, where CO absorption arises from the accretion disc rather than 
the stellar photosphere. We discuss later whether an FUor classification is appropriate in each case.

CO absorption is also observed in three objects which also show Br$\gamma$ emission or absorption. The most interesting of these is
VVVv717 (see Appendix \ref{vvv:specvar}). It has the FUor-like property of very strong CO absorption, in this case by gas at a 
fairly low temperature (1600 K, see Appendix \ref{vvv:co}). The narrow, relatively weak Br$\gamma$ emission in this object seems to be 
on top of a broader absorption feature. Br$\gamma$ emission is not usually detected in FUors as this accretion-sensitive emission line
is assumed to be quenched by the very high accretion rate in FUors. VVVv717 also shows weak H$_{2}$~2.12~$\mu$m and a marginal 
detection of Mg I 2.28 $\mu$m emission.  No other absorption features from a stellar photosphere are observed. The two other 
objects that show CO absorption are VVVv628 and VVVv630. In both cases the CO absorption seems to arise from a stellar photosphere, given that other absorption features associated with a stellar photosphere, such as Na I and Ca I, are also observed. The temperatures estimated 
from CO models in Appendix \ref{vvv:co} (3100--3200 K) are consistent with those expected for the cool upper layers of the stellar 
photosphere in low-mass YSOs.

Previous spectroscopic studies of YSOs at near-infrared wavelengths report the following frequencies for the atomic and molecular features. In the case of Br$\gamma$ emission, \citet{2013Cooper} finds detection rates of 75$\%$ for high-mass YSOs, whilst \citet{1989Carr} and \citet{2010Connelley} find rates of $71\%$ and $75\%$ respectively for low-mass YSOs.
91$\%$ of intermediate-mass and Herbig Ae/Be stars in the study of \citet{2001Ishii} show this feature in emission. We note that in all of these studies the samples were composed mostly of embedded sources, though not exclusively in every case.

CO emission is observed in 17$\%$ of massive YSOs \citep{2013Cooper}, 20$\%$ and 22$\%$ for low-mass YSOs \citep{1989Carr,2010Connelley}, and 22$\%$ for intermediate-mass and Herbig Ae/Be stars \citep{2001Ishii}. For H$_{2}$ emission, \citeauthor{2013Cooper} finds rates of 56$\%$ (high-mass YSOs), \citeauthor{2001Ishii} report 34$\%$ for intermediate YSOs, whilst low-mass YSOs are found to show this feature in 25$\%$ of the objects of \citeauthor{1989Carr} and 42$\%$ in the sample of \citeauthor{2010Connelley}.

\begin{figure}
\centering
\resizebox{\columnwidth}{!}{\includegraphics{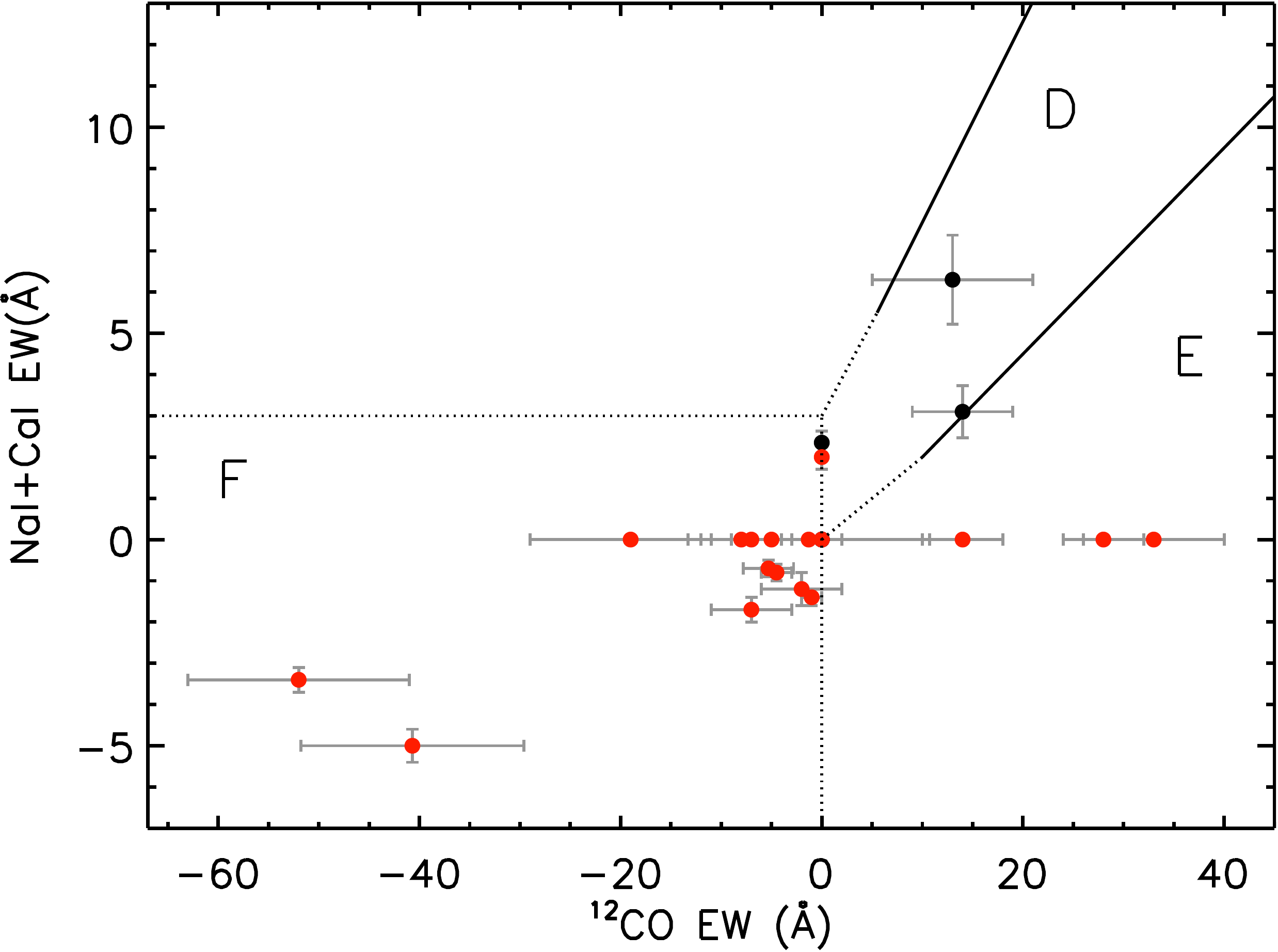}}\label{vvv:ewsa}
\caption{Comparison of the equivalent widths of Na {\sc I}+Ca {\sc I} vs $^{12}$CO for the VVV spectroscopic sample. Regions marked on the plots are based on the work of \citet{2010Connelley}. Objects marked in red represent objects later classified as members of the eruptive variable class, whilst black symbols mark YSOs found not to be part of this variable class.}
\label{vvv:ews}
\end{figure}

The detection frequency for Br$\gamma$ emission in our sample is comparable to the percentages obtained for low- and high-mass YSOs. In the case of CO emission our numbers seem to be higher than those reported in previous studies. For H$_{2}$ emission, the detection rate in our sample appears to be significantly higher. H$_{2}$ and CO emission are commonly associated with outburst events. In addition, the absence of photospheric absorption features in most of the YSOs suggests that these are objects with high veiling from hot circumstellar material. High veiling may be related to a high accretion state \citep{2010Connelley}.

\subsection{Eruptive YSOs vs normal YSOs: equivalent width diagrams}\label{subsec:EWs}



In Fig. \ref{vvv:ews} we plot equivalent widths of CO against those of Na I$+$Ca I for 
the VVV sample. In these plots we mark the distinct regions defined by \citet{2010Connelley} in their study of a large sample of 
class I YSOs (their figure 5). First we describe the results for the flux-limited sample of \citet{2010Connelley}.

\begin{itemize}
\item {\bf Region F} Objects falling in this area were characterized by showing CO, and Na I$+$Ca I emission. Br$\gamma$ emission is also typical in these sources. Objects from the sample of \citeauthor{2010Connelley} with these characteristics had strong veiling from hot circumstellar material, which they suggested is associated with a high accretion rate.

\item {\bf Region D} The majority of class I YSOs in the \citeauthor{2010Connelley} sample fell in this region. Objects in this area had low veiling, showing CO and Na I$+$Ca I absorption. The EWs were consistent with origin in the photosphere of a YSO, falling in between the loci of dwarf and giant photospheres as marked by thick lines in the upper panel of Fig. \ref{vvv:ews}. 


\item {\bf Region E} Objects in this area showed CO absorption in excess of photospheric absorption, as well as a lack of photospheric absorption features. Objects in this area lack Br$\gamma$ emission. The objects in the sample of \citeauthor{2010Connelley} falling in this region were either FUors or ``FUor-like'' YSOs,
a term commonly used to describe sources that have similar spectra to FUors but lack photometric evidence for an eruption at present.
\end{itemize}

The VVV sample is clearly different from that of \citet{2010Connelley}, in that very few of the VVV sample are located in region D,
where most normal class I YSOs were found. Instead, the location of the VVV YSOs is more often in the two regions of the upper panel
expected for highly variable YSOs undergoing episodic accretion:
1) Region F, where objects with high veiling and EXor-like active states of accretion are expected to fall. These objects display Br$\gamma$ emission in addition to shock-excited emission from H$_{2}$. 2) Region E, where FUors are usually found.



 
\subsection{Spectra and light curve categories}\label{vvv:sec_varacc}

We have seen that many of the 28 YSOs in the sample show characteristics that point to high states of accretion.  
In this section we compare their spectroscopic characteristics to their light curve classification from Paper I, and analyse whether 
the observed variability can be explained by variable accretion or another of the known mechanisms in YSOs.

The 28 objects are divided into 4 long-period YSOs, 3 dippers, 1 fader, 3 short-term variable stars and 17 eruptive objects. Unsurprisingly, most of the systems that we regard as bona fide examples of episodic accretion have eruptive light curve classifications. 

\begin{figure*}
\centering
\subfloat{\resizebox{0.45\textwidth}{!}{\includegraphics{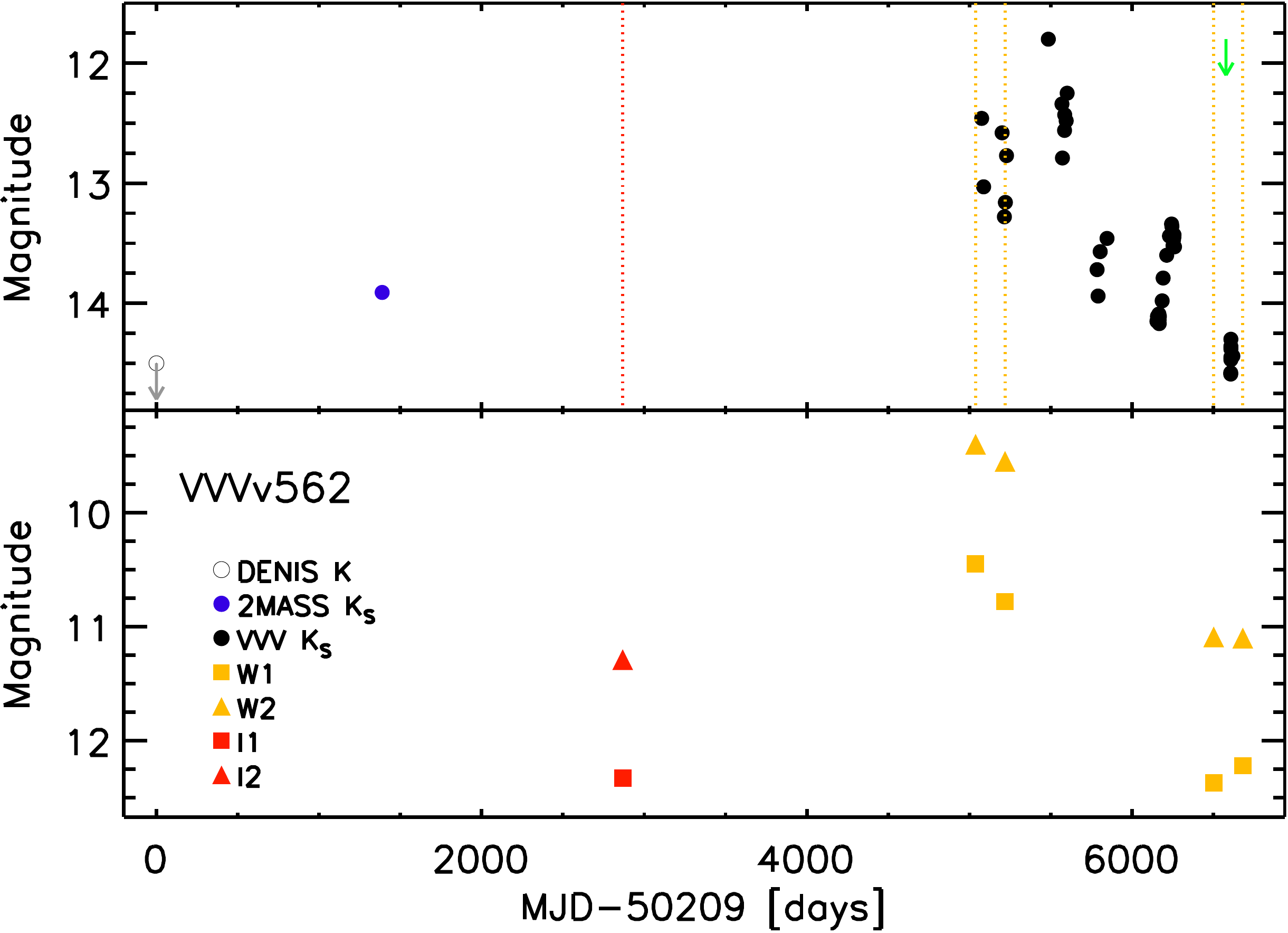}}}
\subfloat{\resizebox{0.45\textwidth}{!}{\includegraphics{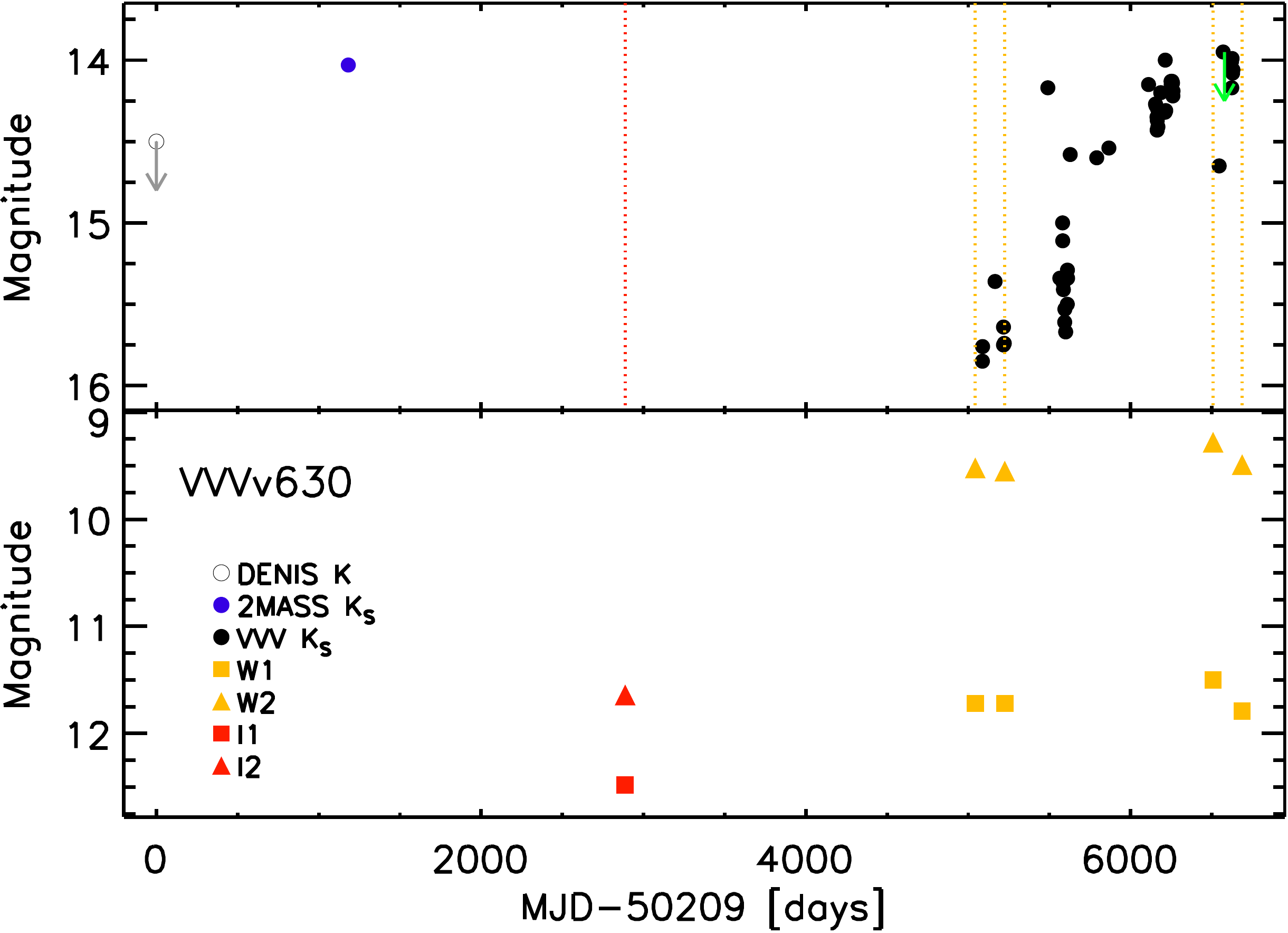}}}\\
\subfloat{\resizebox{0.45\textwidth}{!}{\includegraphics{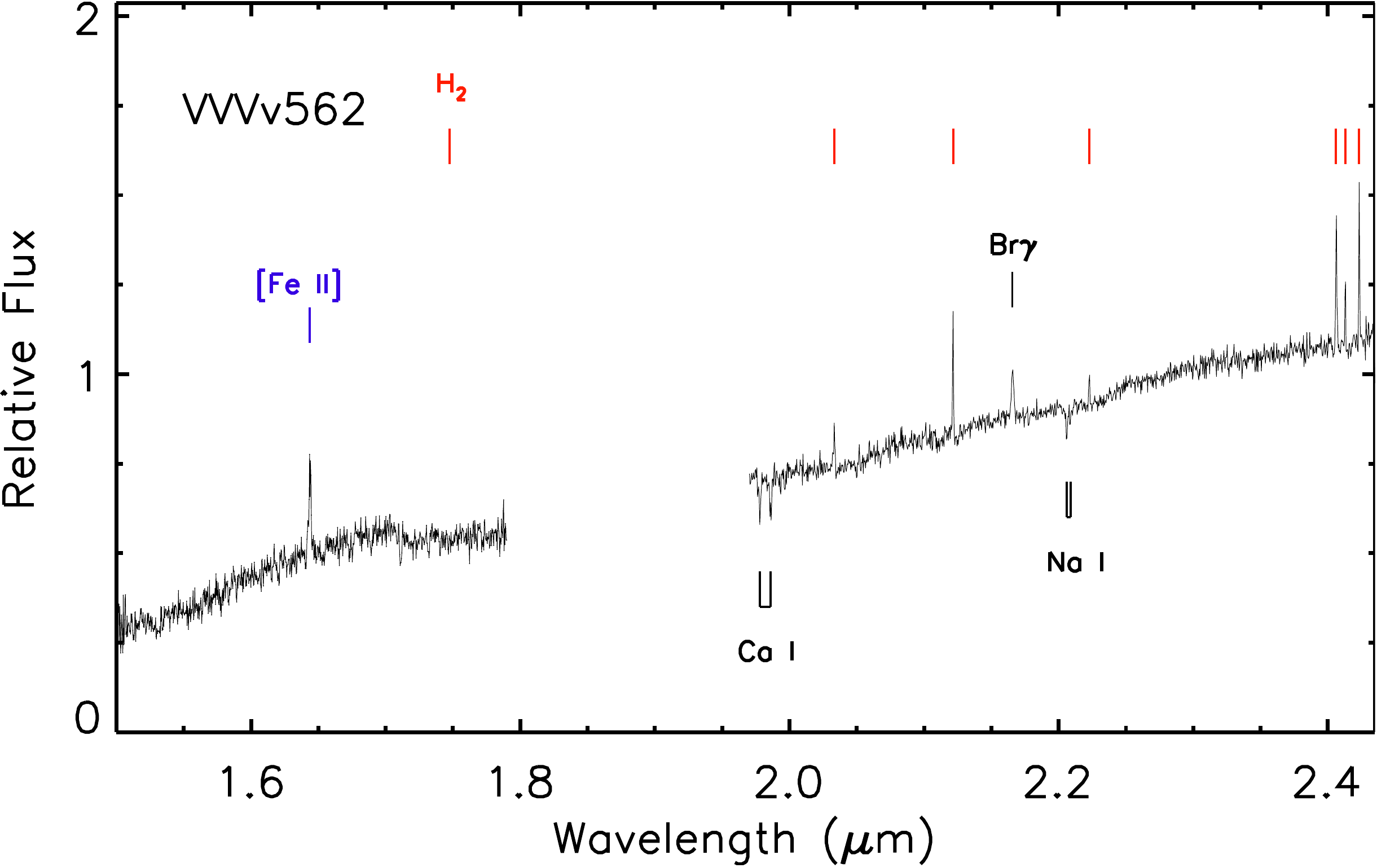}}}
\subfloat{\resizebox{0.45\textwidth}{!}{\includegraphics{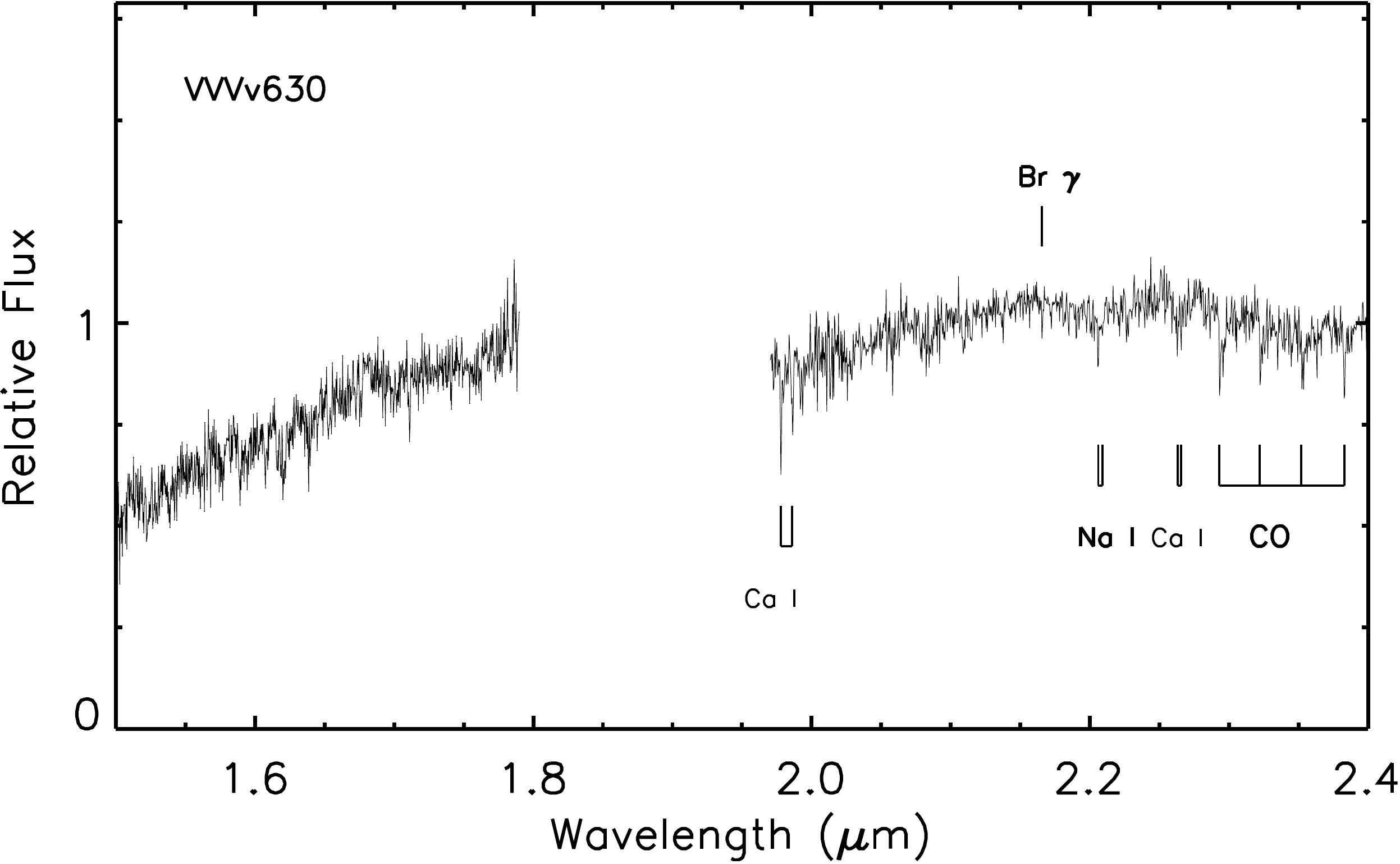}}}
\caption{Near- and mid-infrared photometry (top) and near-infrared spectra (bottom) for objects VVVv562 (left) and VVVv630 (right). Green arrows mark the epoch of spectroscopy. The photometry from different surveys is marked by different symbols that are labelled in each plot. Different emission/absorption lines are also marked in the spectra of the objects. VVVv562 is a fader observed 2~mag below peak brightness that we regard as a likely eruptive variable. VVVv630 has an ``eruptive'' light curve and was observed in a bright state but
we consider that extinction is a more likely cause of the variability given the detection of absorption lines from the stellar photosphere, coupled with colour-magnitude variations along the reddening vector in the 2 epochs of $H$ and $K_{\rm s}$ photometry.}
\label{vvv:varacc_fig}
\end{figure*}

\begin{itemize}

\item {\bf Short-term:} The 3 short-term variable stars, VVVv65, VVVv625, and VVVv632, have lower amplitudes than most of the YSOs in the spectroscopic sample. The spectra of VVVv65 and VVVv625 lack CO emission or absorption features, and show weak Br$\gamma$ emission, with VVVv625 also showing weak H$_2$ emission. The spectrum of VVVv632 resembles that of classical T Tauri stars \citep[e.g. TW Hya, see e.g.][]{2011Vacca} and shows several absorption features (Na {\sc I}, Ca I, Mg I, and Al I) arising from the stellar photosphere. Emission from atomic hydrogen is also present (Br$\gamma$, Pa$\beta$, Pa$\gamma$, Pa$\delta$), as well as from some H$_{2}$ transitions. Finally, the HeI $1.083 \mu$m line is also observed, and it exhibits a P Cygni profile, with absorption extending to $\sim -200$~km s$^{-1}$. This feature suggest the presence of an outflow from the system. Given that clear indicators of a high accretion rate are absent 
and the nature of the light curves is a nearly continuous short-term variability, with relatively low amplitudes ($\le 1.5$~mag) for the 
VVV sample, we consider that the variability in these three objects is likely caused by a different mechanism than variable accretion. 
We note that the VVVv65 spectrum was not taken during a bright state (and the VVVv625 spectrum may or may not have been) so we may be
mistaken in this judgement. However, our main aim here is to identify the YSOs for which evidence of episodic accretion is reasonably
clear.

\item {\bf Dippers:}  VVVv628 has $\Delta K_{\rm s} = 1.45$, Br$\gamma$ and H$_{2}$ emission, and absorption features of Na I, Ca I and CO with typical equivalent widths for a stellar photosphere. Given the lack of veiling despite observation in a bright state, and the nature
of the light curve, it seems likely that another mechanism than accretion may be driving the variability in this object. VVVv405 and VVVv406 have larger amplitude $\Delta K_{\rm s}$~($\geq 2$~magnitudes), with the former showing emission from several transitions of H$_{2}$, as well as Br$\gamma$ emission. 
VVVv406 on the other hand is faint at the epoch of observations and shows a featureless spectrum with weak H$_{2}$ emission. In both cases the two epochs of near-infrared variability are taken at similar $K_{\rm s}$ magnitudes so no information on colour changes could be derived. CO is not detected in these 2 objects. It is difficult to establish whether variable accretion is responsible for the variability in these YSOs, so for the present we classify them as non-eruptive.

\item {\bf Fader:} VVVv562 is the only object in this category. Br$\gamma$, H$_{2}$ and 1.64 $\mu$m [Fe II] lines are seen in
emission. No CO features are detected but we note that this object was observed when $\sim$2~mag fainter than its maximum 
brightness. Inspection of ancillary data from 2MASS, WISE and {\it Spitzer} supports the idea that the object went into an eruption at some time between 2004 and 2010 and has been on a fading trend since then (see Fig. \ref{vvv:varacc_fig}). The colour changes seen between the 2 epochs of $JHK_{\rm s}$ 
photometry (bluer when fainter) disfavour variable extinction. We note that variations in geometry such that absorption increases but so does scattered light can lead to a star becoming bluer at minimum light \citep[an effect sometimes observed in UX Ori stars deep minima due to dust clumps that screen stellar light, see e.g.][]{1999Herbst}. However, in this case the source faded by only 1.1 mag in $H$ between the 2 epochs while becoming 0.35 mag bluer in $H$-$K_{\rm s}$, so an unusually bright source of scattered light would be required. We consider it likely that the 
variability in this object is due to 
changes in the accretion rate. N.B. this is the first of 4 YSOs that we classify as an eruptive variable despite the absence of the
CO emission or strong CO absorption seen in EXors or FUors, respectively. These indicators are likely to be absent during quiescence 
and in any case they may be veiled into invisibility in heavily embedded YSOs, as in OO Ser \citep{1996Hodapp, 2007Kospal} or GM Cha \citep{2003Gomez, 2007Persi}.

\item {\bf Long-term periodic:} In Paper I we discuss the possible mechanisms that can drive periodic variability (P$>$100 days) in YSOs. One of these mechanisms can be variable accretion, which has been invoked to explain the variability in e.g. V371 Ser \citep[where accretion is modulated by a binary companion, see][]{2012Hodapp}. It is very interesting to see that 3/4 YSOs with the long-term periodic classification show characteristics of objects in high states of accretion. VVVv32 and VVVv473 show Br$\gamma$ and CO emission, the latter also showing several transitions of H$_{2}$. VVVv717 shows deep CO absorption with a lack of other absorption features, and additional weak H$_{2}$ and Br$\gamma$ emission (this object is analysed with more detail in Appendix \ref{vvv:specvar}). It is very likely that variable accretion is the driving mechanism for these three objects, which
were all observed in a bright state. Finally, VVVv480 shows only weak Br$\gamma$ emission. It is difficult to establish whether this object was observed at a faint or a bright state, so the absence of evidence for a high accretion rate might be due to observation
in quiescence, but by default we place this YSO in the non-eruptive category. The 2 epochs of $JHK_{\rm s}$ photometry have very similar fluxes
for this source and therefore provide no information on colour changes.

\item{\bf Eruptive:} 
Most objects with an eruptive light curve classification show spectroscopic characteristics indicating a high state of accretion. From 
17 YSOs, 10 show CO emission, 2 show strong CO absorption, broad H$_2$O absorption bands and  a lack of other absorption features (VVVv322 and VVVv721, see Section \ref{vvv:class_ex} and Appendix \ref{vvv:specvar}) and 3 show other characteristics that support variable accretion  (VVVv118, VVVv800 and VVVv815, see Appendix \ref{vvv:specvar}). VVVv63 has a doubtful classification given its somewhat low amplitude, $\Delta K_{\rm s}=1.44$~mag, and a spectrum with weak Br$\gamma$ emission (but H$_{2}$ is clearly present). The accretion luminosity derived for
this source does not support a high state of accretion (see Section \ref{vvv:erupvars}) despite observation at an intermediate state. 
Finally, the variability in VVVv630 is more likely related to extinction events. The near-infrared colour comparison shows that the object becomes bluer when brightening following the reddening line (although this comes from the $H$ vs $(H-K_{\rm s})$ diagram only). The VVV light curve resembles that of an eruptive object, with the spectrum being taken during the suspected bright state. However, the object shows photospheric absorption from CO, Na I and Ca I, with weak Br$\gamma$ emission. This contradicts the expected behaviour from an eruption. 2MASS data shows that the object had a similar $K_{\rm s}$ mag in the bright state to that seen in 1999 (see Fig. \ref{vvv:varacc_fig}). It is possible that VVVv630 went through a dust obscuration event that had cleared at the time of the observations. 

 
\end{itemize}

\section{Analysis and classification}\label{vvv:sec_erupclass}

From 28 likely YSOs, we have seen that 9 are likely not part of the eruptive variable class (3 short-term, 3 dippers, 1 LPV-YSO and 2 eruptive). The remaining 19 VVV objects show characteristics of eruptive variables from their spectra and/or light curves. But to which of the known subclasses of eruptive variables do these objects belong? 

In the classical definition, during outburst, FUors show strong CO absorption at 2.29 $\mu$m with H$_{2}$O absorption sometimes present in their near-infrared spectrum \citep{2010Reipurth}, whilst EXors show strong Br$\gamma$ in emission, with CO bands also in emission and veiling of photospheric features \citep{2009Loren}. By just considering spectroscopic features we could easily classify objects in region F of the equivalent width plots discussed above as EXors, with objects falling in region E as FUors. However, the duration of the outbursts is one of the main characteristics used in the classification of eruptive variables.  FU Orionis stars show outbursts lasting over $>$10 yrs with a minimum timescale of several thousand years for repeat eruptions \citep{2013Scholz}, whilst EXors have recurrent short-lived ($<$1~yr) outbursts and quiescent periods of 5--10 yrs \citep[see e.g.][]{1996Hartmann,2007Fedele,2009Loren}. 

A second problem with the classical definition is that it was defined at optical wavelengths. This has struggled to include some intermediate objects and excludes younger protostars that have higher accretion rates but are too deeply embedded in circumstellar matter to be observed in the optical, such as in the cases of OO Ser, V2775 Ori, HOPS 383 and GM Cha \citep[][]{1996Hodapp,2007Kospal,2007Persi,2011Caratti,2014Audard,2015Safron} or in the GPS sample of \cite{2014Contreras}. Thus, this definition might not apply to our VVV objects given the embedded nature of our sample.

Given the spectroscopic characteristics of the known subclasses of eruptive variables, we could assume that the from the 19 objects with characteristics of eruptive variables, those with strong CO absorption (found in region E of Fig. \ref{vvv:ews}) can be classified as FUors, whilst emission line objects (usually found in region F of Fig. \ref{vvv:ews}) are more likely EXors. In the following we will study whether emission line objects are found at high states of accretion, as expected, by determining accretion luminosities from Br$\gamma$ emission. Finally we will discuss individual examples that demonstrate that the classical definition of eruptive variables fails to classify our VVV sample.



\subsection{Radial velocities}

We estimate radial velocities of objects showing CO emission/absorption by using the cross-correlation technique described in \citet{2015Marocco}, taking the simple CO models that will be explained later in Appendix \ref{vvv:co}, as the radial velocity templates. In every case the  velocities are corrected  to take into account the Earth rotation, its motion around the Sun and the motion with respect to the local standard of rest (LSR). The errors on radial velocity are dominated, in most cases, by the error in the wavelength calibration at the order were the CO region is observed. This error is $\sim0.4$~\AA, which translates into an error of $\sim5.2$ km s$^{-1}$ on radial velocity.

These estimates give us a more reliable measurement of the distance to the objects which can help us determine { accretion luminosities} for some of the emission line stars. It also helps to identify likely AGB variable stars (see Appendix \ref{vvv:sec_evolved}).



Note that we are assuming that the CO emitting region has a similar average velocity to the star. The measurement could be affected if the CO emission/absorption arises from an outflow or wind with a different velocity but in most YSOs the CO arises in the rotating
disc or the stellar photosphere.


\begin{figure}
\centering
\resizebox{\columnwidth}{!}{\includegraphics{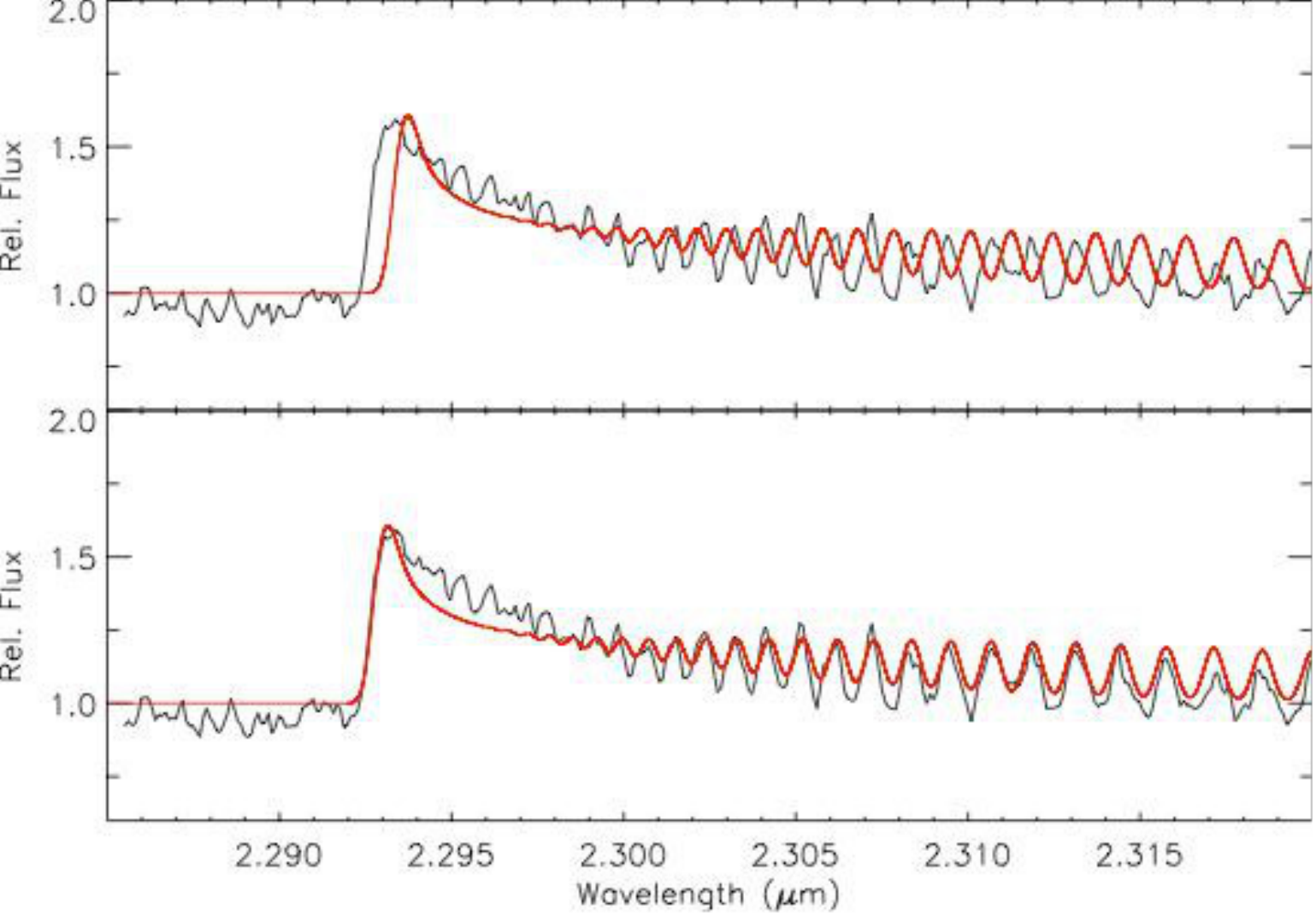}}
\caption{Comparison of the spectrum of VVVv665 (black line) with the CO emission model (red line), without any shifts (top) and after applying the shift derived by the cross-correlation function (bottom). }
\label{vvv:radvel}
\end{figure}


In Fig. \ref{vvv:radvel} we show a comparison of the observed spectrum of VVVv665 and that of the CO model before and after deriving the radial velocities for the object. In Table \ref{table:vvvrvel} we show the estimated values and errors for the objects where we were able to use this method. In Table \ref{table:vvvrvel} we also show the near and far kinematics distances estimated using the Galactic rotation model of \citet{1993Brand}.

\begin{table}
\begin{center}
\begin{tabular}{@{}l@{\hspace{0.4cm}}c@{\hspace{0.4cm}}c@{\hspace{0.4cm}}c@{\hspace{0.4cm}}c@{\hspace{0.2cm}}}
\hline
Object & V$_{LSR}$ & $\Delta$V$_{LSR}$ & d$_{near}$ & d$_{far}$ \\
 & (km s$^{-1}$) & (km s$^{-1}$) & (kpc) & (kpc) \\ 
\hline
VVVv25 & 21.1 & 5.3 & -- & 10.9\\
VVVv32 & $-$28.1 & 13.5 & 1.9 & 7.4\\
VVVv42 & $-$49.0 & 5.5 & 4.0 & 6.1\\
VVVv45 & 16.4 & 5.3 & -- & 11.6 \\
VVVv193 & $-$85.5 & 18.2 & 4.5& 10.2 \\
VVVv202 & $-$84.2 &  5.2 & 4.4 & 10.5 \\
VVVv229 & $-$69.9 &  5.5 & 3.8 & 11.4\\
VVVv235 & $-$92.8 &  5.2 & 4.8 &10.6\\
VVVv270  & $-$87.6 & 12.1 & 4.6 & 11.2 \\
VVVv322 & $-$59.8 &  5.5 & 3.8 & 12.7\\
VVVv628 & $-$41.6 & 5.9 & 2.4 & 12.3\\
VVVv630 & $-$32.9 & 5.9 & 1.9 & 12.8\\  
VVVv665  & $-$73.1 & 5.3 & 4.0 & 11.6\\
VVVv699  & $-$93.5 & 5.3 & 4.9 & 11.0\\
VVVv717 & $-$126.3 &  5.3 & 6.1 & 10.2\\
VVVv721 & $-$89.7 & 5.8 & 4.9 & 11.5\\
VVVv796 & $-$68.7 &  5.3 & 5.1 & 12.1\\
\hline
\end{tabular}
\caption{Radial velocities.}\label{table:vvvrvel}
\end{center}
\end{table}



\subsection{Accretion luminosities}\label{vvv:erupvars}

The majority (16/19) of the eruptive VVV objects show emission line spectra, which is one of the expected characteristics of EXors during outburst. This is not surprising given that only a 2 year time interval of VVV data (2010--2012) was used for our variable search (see Paper I), which 
would favour the detections of EXors over FUors.

\begin{table}
\begin{center}
\begin{tabular}{@{}l@{\hspace{0.4cm}}c@{\hspace{0.4cm}}c@{\hspace{0.4cm}}c@{\hspace{0.2cm}}c@{\hspace{0.2cm}}c@{\hspace{0.1cm}}}
\hline
Object & {\small State} & A$_{K_{\rm s}}$ & $D$ & L$_{acc}^{\dagger}$ & L$_{acc}^{\ddagger}$\\
 & &  mag & kpc & L$_{\odot}$ & L$_{\odot}$ \\ 
\hline
\hline
\multicolumn{6}{l}{Eruptive}\\
\hline
VVVv20 & bright & 2.3 & 2.5$^{d}$ & 3.8 & 15.1 \\
VVVv32 & midpoint & 0.9 & 1.9$^{b}$ & 0.9 & 2.2\\
VVVv94 & bright & 2.5 & 3.5$^{c}$ & 11.1 & 68.2 \\
VVVv118 & faint & 0.5 & 2.2$^{d}$ & 0.1 & 0.1  \\
VVVv193 & bright & 2.5 & 4.5$^{b}$ &  3.2 & 11.9 \\
VVVv270 & bright  & 2.5 & 4.6$^{b}$ & 7.3 & 38.1 \\
VVVv374 & bright & 2.5 & 2.9$^{c}$ & 5.3 & 24  \\
VVVv452 & bright & 1.4 &  2.8$^{d}$ & 2.5 & 8.4 \\
VVVv473 & midpoint? & 2.5 & 3.7$^{c}$ & 0.6 & 1.1 \\
VVVv562 & faint & 1.4 & 2.9$^{c}$ & 0.4 & 0.6\\
VVVv631 & bright & 1.2 & 2.3$^{c}$ & 2.5 & 8.4 \\
VVVv662 & bright & 2.5 & 3.1$^{c}$ & 3.2 & 11.7 \\
VVVv665 & bright & 2.5 & 4.3$^{b}$ & 6.6 & 33.2  \\
VVVv699 & bright? & 32$^{a}$ & 4.9$^{b}$ & 7.4 & 38.9 \\
VVVv699 & bright? & 2.5 & 4.9$^{b}$ & 3.7 & 14.5 \\
VVVv800 & midpoint? & 2.4 & 1.4$^{c}$ & 0.7 &  1.4 \\
\hline
\multicolumn{6}{l}{Non-Eruptive}\\
\hline
VVVv63 & midpoint & 1.7 & 3.5$^{c}$ & 0.6 & 1.1\\
VVVv65 & faint & 1.9 & 3.5$^{c}$ & 0.8 & 1.8 \\
VVVv405 & faint & 2.5 & 2.3$^{c}$ & 0.2 & 0.2 \\
VVVv480 & faint & 0 & 3.7$^{c}$ & 0.2 & 0.3  \\
VVVv625 & bright? & 2.5 & 2.3$^{c}$ & 1.5 & 4.2 \\
VVVv628 & midpoint & 1.1 & 2.3$^{b}$ & 0.1 & 0.1 \\
VVVv632 & bright & 0.3 & 2.5$^{c}$ & 0.6 &  1.0 \\
\hline
\multicolumn{6}{l}{$\dagger$ from \citet{2004Calvet}}\\
\multicolumn{6}{l}{$\ddagger$ from \citet{1998Muzerolle}}\\
\multicolumn{6}{l}{$^{a}$ A$_{V}$ from line ratios, see Appendix \ref{vvv:specvar}}\\
\multicolumn{6}{l}{$^{b}$ Distance from radial velocity}\\
\multicolumn{6}{l}{$^{c}$ Distance from d$_{SFR}$}\\
\multicolumn{6}{l}{$^{d}$ Distance from SED fit}\\
\hline
\end{tabular}
\caption{Accretion luminosity from luminosity of the Br$\gamma$ line. The second column refers to the approximate state (from the light curve) at which the spectrum was taken. }\label{table:vvvaccrate}
\end{center}
\end{table}

From 16 objects with emission line spectra, 15 show Br$\gamma$ emission, the exception being VVVv815. This source was observed well below maximum light and the spectrum shows very strong H$_{2}$ emission, implying the presence of a powerful accretion-driven outflow (see Appendix \ref{vvv:specvar}).
The majority of the emission line objects fall in region F of Fig. \ref{vvv:ews} and are thus characterized by the lack of photospheric absorption lines. CO band-head lines in emission and/or H$_{2}$ lines in emission are also seen in most of these objects. The objects with Br$\gamma$ emission are characterized by having broad emission profiles with FWHM $>100$ km s$^{-1}$ as expected from magnetospheric accretion \citep[see e.g.][]{2001Davis}.


\begin{figure*}
\centering
\resizebox{0.9\textwidth}{!}{\includegraphics{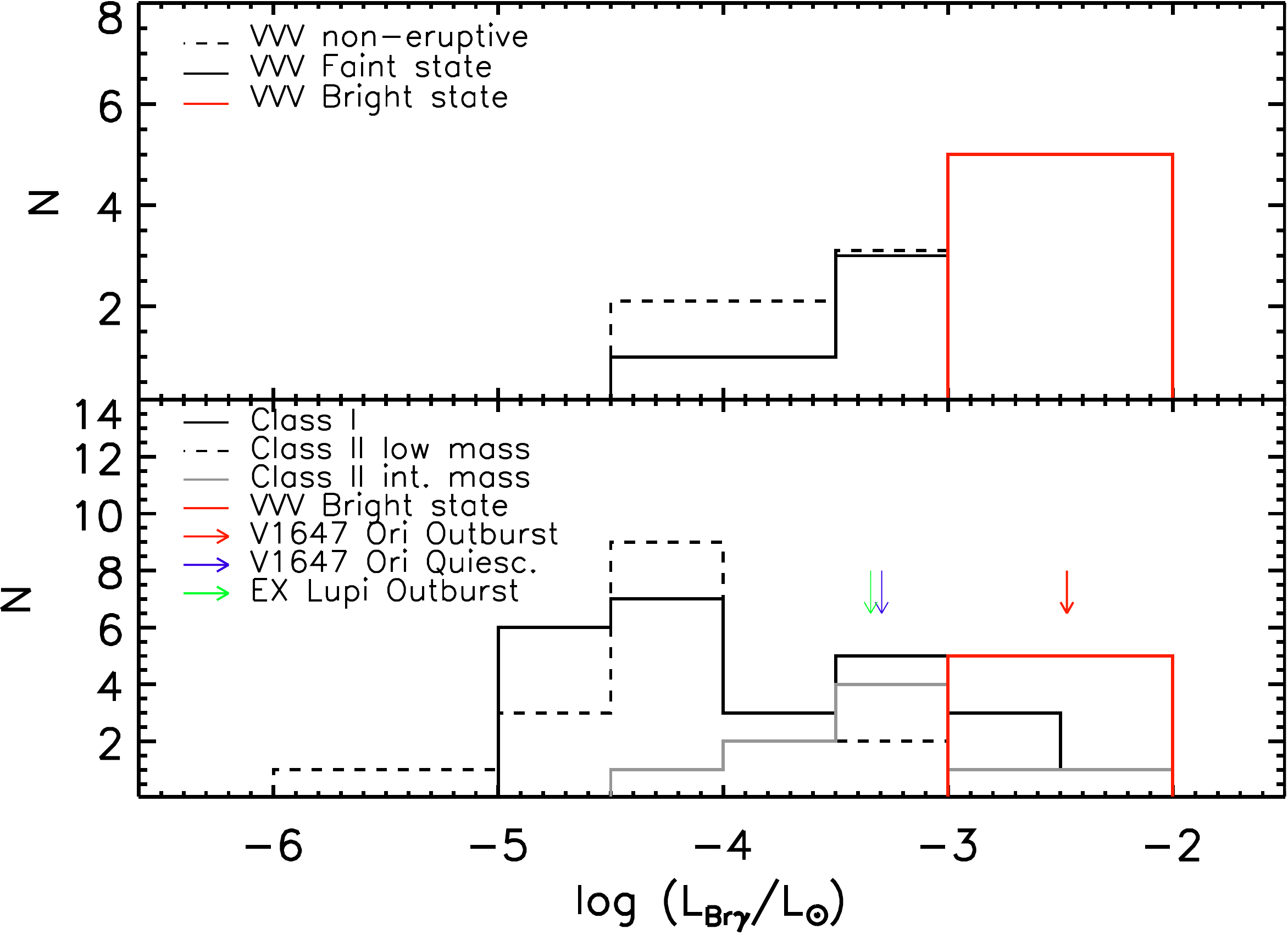}}
\caption{Histogram of Br$\gamma$ luminosities for VVV objects and a sample of YSOs found in the literature (top) Comparison of eruptive VVV objects observed at bright state (red solid line) with eruptive objects observed at faint states (black solid line) and non-eruptive VVV YSOs (dashed black line). (bottom) Comparison of VVV bright objects (solid red line) with the intermediate-mass class II YSO sample of \citet{2004Calvet} (solid grey line) and the low-mass class II YSO sample of \citet{1998Muzerolle} (dashed black line), low-mass class I YSOs found in the literature (solid black line). We also compare with eruptive variables EX Lupi during outburst (green arrow), and V1647 Ori during quiescence (blue arrow) and outburst (red arrow).}
\label{vvv:ysos_acrt}
\end{figure*}


 In order to study whether these objects are in high states of accretion as we expect, we can try to determine the accretion luminosity, $L_{acc}$, of objects in our sample. The latter is obtained from the known correlation between the Br$\gamma$ luminosity and the accretion luminosity in YSOs \citep[see][]{1998Muzerolle,2004Calvet}. Then in principle $\dot{M}$ can be derived from 
 
 \begin{equation}
 L_{acc}=\frac{GM_{\ast}\dot{M}}{R_{\ast}}\left(1-\frac{R_{\ast}}{R_{in}}\right),
 \end{equation} 
 
\noindent \citep[see e.g.][]{1998Gullbring}. However, this last step requires knowledge of stellar (M$_{\ast}$,R$_{\ast}$) and disc parameters (inner disc radius, R$_{in}$). These parameters can be derived from the stellar spectral type and comparison to PMS evolutionary tracks and isochrones in HR diagrams \citep[see e.g][]{2004Calvet,2008Spezzi,2008Aspin}. This method would yield unreliable mass estimates as the majority of our sample have spectral indices of class I and flat-spectrum sources and evolutionary tracks are unreliable (and so rarely published) for ages below 10$^{6}$ yr \citep[see e.g.][]{2000Siess}. In addition to the long-standing uncertainties about the initial conditions used in such evolutionary calculations, \citet{2009Baraffe,2012Baraffe} and others
have pointed out the very concept of a stellar birthline is undermined by episodic accretion because each star is likely to have a different accretion history.

Given the above, $\dot{M}$ can only be estimated by making many assumptions on the star+disc parameters, thus yielding unreliable results. We have shown in Paper I that some of our high-amplitude sample includes some PMS eclipsing binaries. Some of these have spectral indices of flat spectrum sources. Follow up of these systems could provide information on radii and masses for these younger sources, provided the contribution of scattered light can be accounted for.

For the accretion luminosities we first need to determine the flux of the Br$\gamma$ line. We note that several assumptions are made during the derivation of this parameter. Firstly, we do not have $K_{\rm s}$ photometry from the time of the spectroscopic observations. Thus, we assume $K_{\rm s}$ as the magnitude of the closest epoch from the VVV photometry. From this we obtain the continuum level, F$_{\lambda}$ for the Br$\gamma$ line at 2.1659 $\mu $m. In addition, we correct this value for the extinction along the line of sight to the source. This introduces further uncertainties given the difficulties in measuring accurate values of extinction towards class I YSOs. These objects have substantial contributions from scattered light that may cause the reddening, and hence the line luminosity, to be underestimated \citep[see e.g.][]{1998Muzerolle,2010Connelley}. We measure $A_{K{\rm s}}$ by dereddening the sources to the classical T Tauri locus in the $J-H$ vs $H-K$ two colour diagram (see Fig. \ref{vvv:specgc} ) and we present this in Table \ref{table:vvvaccrate} as $A_V$ using the extinction law of \citet{1985Rieke}. For the 9 sources that were not detected in the $J$ filter, we adopt a colour $J-H$=3.5, or A$_V$=23, and deredden them to the red end of the T Tauri locus. This is consistent with the red $H-K$ colours of these systems and the upper limits on $J-H$, which are typically between 2 and 4 magnitudes.

The line flux, $F_{Br\gamma}$, is then estimated from

\begin{equation}
F_{Br\gamma}=F_{\lambda}\times EW\times 10^{0.4A_{\lambda}},
\end{equation}

\noindent with $A_{\lambda}$ from the above values. The line luminosity, L$_{Br\gamma}$, depends on the distance to the object. In our case we have distances from (1) the distance to the nearby SFRs at the location of the VVV objects (d$_{SFR}$), (2) kinematic distances estimated from radial velocities and (3) results from the SED fitting tool (see Appendix \ref{section:vvvsedfits}). For the calculation of L$_{Br\gamma}$ we use the best estimate available for the object: in order of preference we use the kinematic distance (d$_{near}$), then d$_{SFR}$, then the distance from SED fits.

Finally, the accretion luminosity is calculated from the L$_{acc}$--L$_{Br\gamma}$ relations of \citet{2004Calvet} (their equation 3) and of \citet{1998Muzerolle} (their equation 2). \citet{2004Calvet} derive L$_{acc}$ from UV excess for a sample of intermediate mass T Tauri stars (1.5--4 M$_{\odot}$), whilst \citet{1998Muzerolle} uses a similar method for a sample of lower mass classical T Tauri stars. Thus, the former relation is likely to be more suitable for objects in our sample. Table \ref{table:vvvaccrate} shows the values for L$_{acc}$ obtained from both relations.

We can compare our results to the published luminosities for three different sample sets of class I and class II YSOs. (A) The intermediate-mass class II sample of \citet{2004Calvet} and the low-mass class II sample of \citet{1998Muzerolle}. (B) low-mass class I and class II YSOs found in the literature. These are class I objects in \citet{1998Muzerolle}, class I and class II YSOs in the Chamaeleon I and Chamaeleon II clouds \citep{2011Antoniucci} and class I YSOs HH26 IRS, HH34 IRS and HH46 IRS \citep{2008Antoniucci}. (C) Finally, we compare to eruptive variables EX Lupi \citep[2008 outburst, ][]{2010Aspin} and V1647 Ori \citep[in quiescence and outburst, 2006--2008,][]{2007Acosta}.

We note that we could try to compare L$_{acc}$ for these samples directly. However, the results of this comparison would strongly depend on the relation used in estimating L$_{acc}$. Moreover, L$_{acc}$ is derived independently (from UV excess) only in the class II samples of \citet{2004Calvet} and \citet{1998Muzerolle}. The published accretion luminosities of objects in (B) and (C) are calculated from line luminosities using either the \citeauthor{2004Calvet} or the \citeauthor{1998Muzerolle} relations.

Instead we compare Br$\gamma$ luminosities, which given the correlation with L$_{acc}$, gives us a measure of the accretion state of our sample compared with that of typical class I and class II YSOs. The comparison is shown in Fig. \ref{vvv:ysos_acrt}. From the figure we first notice that VVV objects that are classified as eruptive and that are observed during bright states have higher Br$\gamma$ luminosities than the remainder of the sample. Br$\gamma$ luminosities from bright eruptive VVV YSOs are also higher than the majority of intermediate-mass class II YSO as well as low-mass class I and class II YSOs. Unfortunately we are not able to find published Br$\gamma$ luminosities for intermediate-mass class I YSOs.

When comparing to known eruptive variables, we find that VVV objects at bright states have higher L$_{Br\gamma}$ than that of EX Lupi (the prototype of EXors) during its 2008 outburst. The L$_{Br\gamma}$ of VVV sources at bright states is comparable to that of V1647 Ori during outburst. The latter has substantial extinction \citep[A$_{V}\sim10$,][]{2005Muzerolle} and a low to moderate accretion luminosity, L$_{acc}$, during outburst \citep[12 L$_{\odot}$,][determined from the \citeauthor{1998Muzerolle} relation]{2009Aspinb}, which is similar to some sources in our sample (see Table \ref{table:vvvaccrate}). This is further evidence to support our contention that we have identified embedded eruptive variable YSOs similar to the small number of such objects found previously, though a little more luminous on average.

We can estimate a range of values of accretion rates for eruptive objects observed at bright states (open red circles in Fig. \ref{vvv:ysos_acrt}), by adopting plausible values of mass and radius, $M_{\ast}=3$~M$_{\odot}$, R$_{\ast}=5$~R$_{\odot}$ and assuming R$_{in}\gg$~R$_{\ast}$. We stress that this is done for a simple comparison with expected rates in eruptive variables and should not be considered as reliable results. Using the accretion luminosities estimated from the \citet{2004Calvet} relation yields $10^{-6.8}<\dot{M}<10^{-6.2}$~M$_{\odot}$~yr$^{-1}$. Even allowing for a possible order of magnitude under-estimate, these values would be considered at the lower end of accretion rates expected in FUor outbursts ($10^{-3}$--$10^{-6}$~M$_{\odot}$~yr$^{-1}$) and more in line with those seen in nearby low mass EXor outbursts \citep[$10^{-6}$--$10^{-8}$~M$_{\odot}$~yr$^{-1}$, see][]{2014Audard}. This may suggest that the Br$\gamma$ emitters in our sample have lower accretion rates than FUors, which is perhaps to be expected if FUors typically quench Br$\gamma$ emission above a certain accretion rate. (We note that our sample does include three sources with FUor-like spectra and no Br$\gamma$ emission).
Nevertheless, the accretion rates of the Br$\gamma$ emitters are higher than those estimated from SED fitting (section \ref{section:vvvsedfits}). This may be attributed either to the fact that SED fitting used 2010 photometry, an epoch were 
most of the eruptive objects are found at quiescent states, whereas the spectra were taken at epochs close to maximum brightness, thus 
at higher states of accretion. Alternatively, the difference might be due to the fact that the SED fitting tool was not designed to model YSOs undergoing an eruption (see also the discussion about the reliability of the estimated parameters in Appendix \ref{section:vvvsedfits}). 

We note that values of L$_{acc}$ in our sample tend to be lower than the bolometric luminosities (L$_{bol}$) derived from SED fits (by $\sim$1 order of magnitude, specially when comparing the values obtained from the \citeauthor{1998Muzerolle} relation, see Appendix \ref{section:vvvsedfits}). This is somewhat surprising as we might expect that the accretion luminosity dominates the total luminosity in eruptive variables: it is often assumed that in low-mass class I objects L$_{acc}=$L$_{bol}$ \citep[see e.g][]{2009Evans,2009Enoch}. However, \citet{1998Muzerolle} finds that low-mass class I YSOs in their sample show L$_{acc}\sim0.1$~L$_{bol}$. We also note that our sample are most likely intermediate-mass objects and that the central star is expected to contribute an increasing fraction of the total luminosity of the system towards higher masses \citep{1991Calvet}; on average, L$_{acc} \propto M^2$ \citep{2004Calvet}. Although, again this comparison might be affected by the unreliability of the SED fitting tool results (as mentioned above).

In summary, the key result of the accretion luminosity estimates is that eruptive variables in a bright state have higher accretion rates than eruptive variables observed in quiescence.

\subsection{The classification problem}\label{vvv:class_ex}

We are confident that the 16 emission line stars described in Section \ref{vvv:erupvars} are likely to be eruptive variables. However all of them show a mixture of characteristics between those of EXors and FUors.  

Aside from the 16 emission line stars, a further 3 stars (VVVv322, VVVv721 and VVVv717) are classified as eruptive variables. These
show CO 2.29 $\mu$m absorption that appears stronger than expected from a stellar photosphere, with the additional characteristic of not displaying any other features that could be associated with a stellar photosphere. In two cases (VVVv322, VVVv717) we do in fact see emission lines of H$_2$ 
and/or Br$\gamma$ (see Table \ref{table:vvvEWs}, Fig. \ref{d069vc52im2} and Appendix \ref{vvv:specvar})
 In two sources (VVVv322, VVVv721) we also observe absorption from H$_{2}$O. However, only VVVv721 shows an outburst duration that could place it under the FUor category, i.e. longer than the 5 year time baseline of VVV. For these 3 objects we cannot measure an accretion rate but we are confident that they are eruptive variables. However, two of the three show a mixture of characteristics between those of EXors and FUors.  



To illustrate the difficulties found in classifying these likely eruptive variable stars, we will discuss two objects from the previous sample, one of the emission line objects and one object showing strong CO absorption. We present similar discussions of several other objects with mixed characteristics in Appendix \ref{vvv:specvar}. They include one source with CO in emission but H$_2$O in absorption (VVVv631), one with a
FUor-like spectrum but a quasi-periodic light curve (VVVv717), one with a somewhat EXor-like spectrum but a longer outburst duration and H$_2$ emission (VVVv699) and one (VVVv118) with FUor-like H$_2$O absorption but very brief, EXor-like outbursts. Appendix \ref{vvv:specvar} also includes descriptions of two eruptive YSOs with neither EXor-like nor FUor-like spectra (VVVv800 and VVVv815).




\subsubsection{VVVv270}\label{vvv:v270}

\begin{figure*}
\centering
\subfloat{\resizebox{0.48\textwidth}{!}{\includegraphics{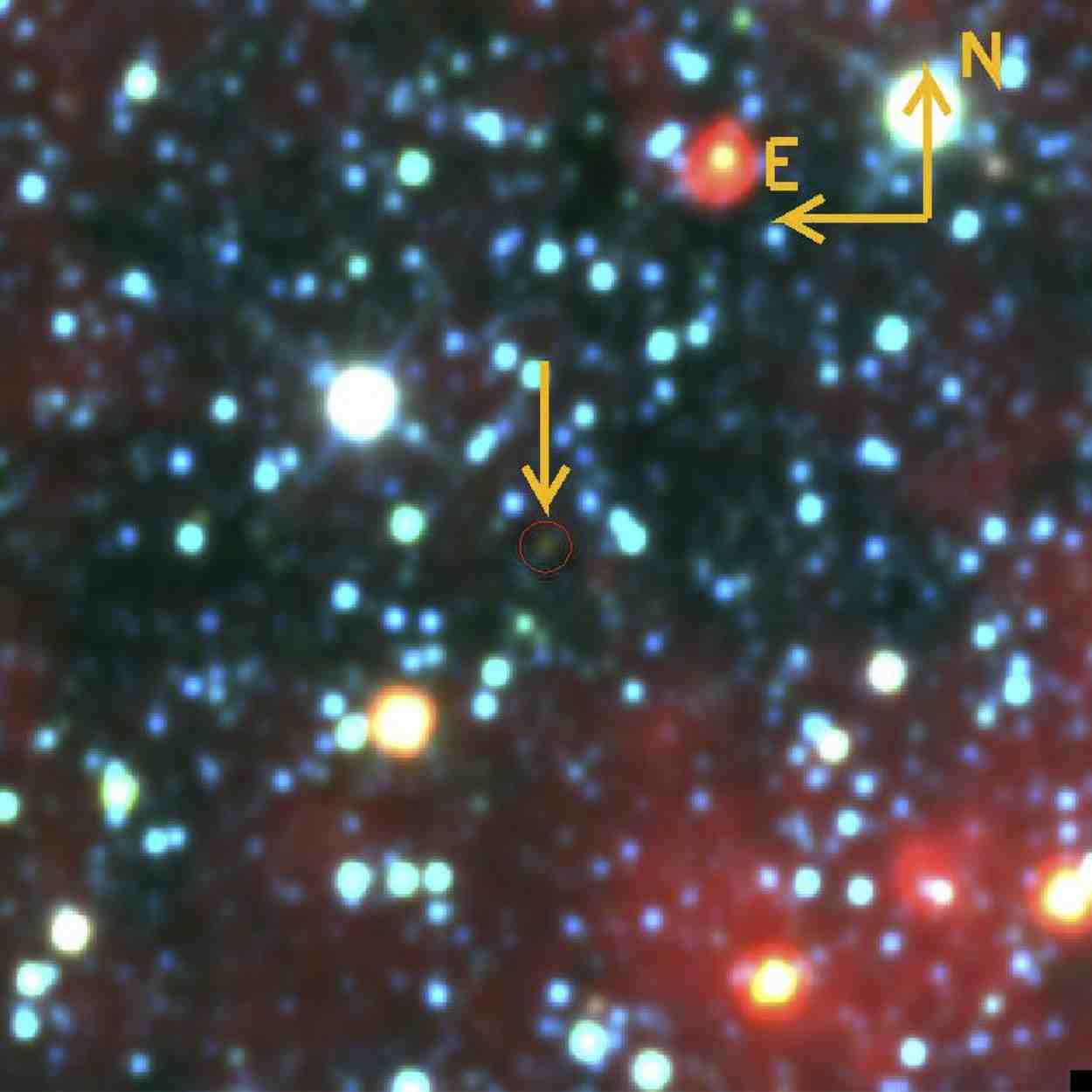}}}
\subfloat{\resizebox{0.48\textwidth}{!}{\includegraphics{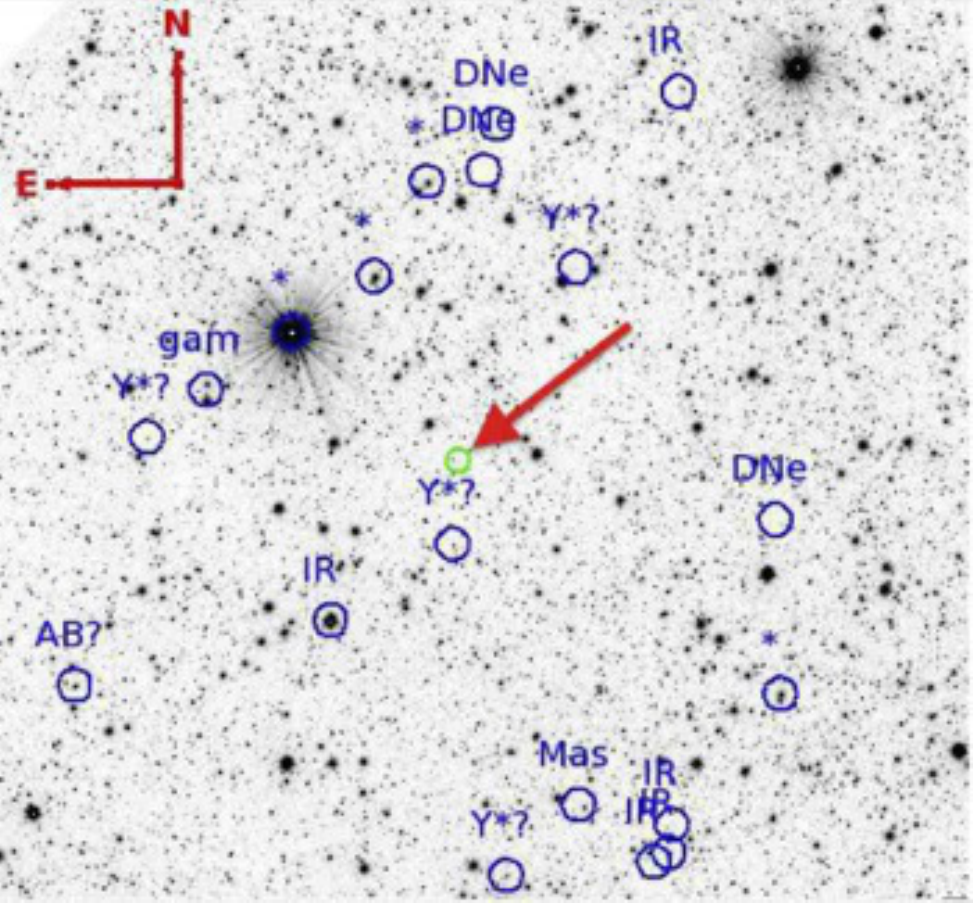}}}\\
\subfloat{\resizebox{0.48\textwidth}{!}{\includegraphics{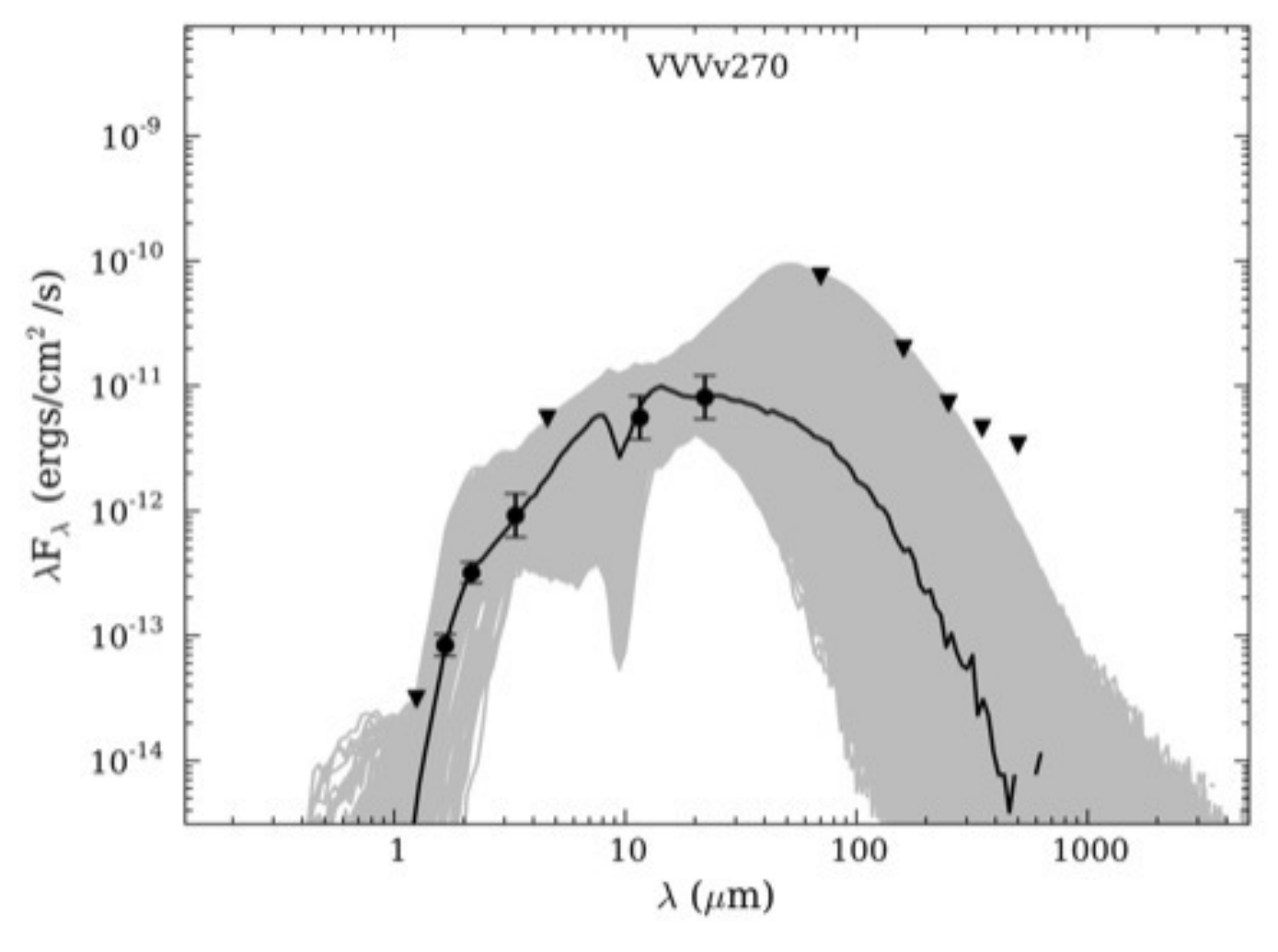}}}
\subfloat{\resizebox{0.48\textwidth}{!}{\includegraphics{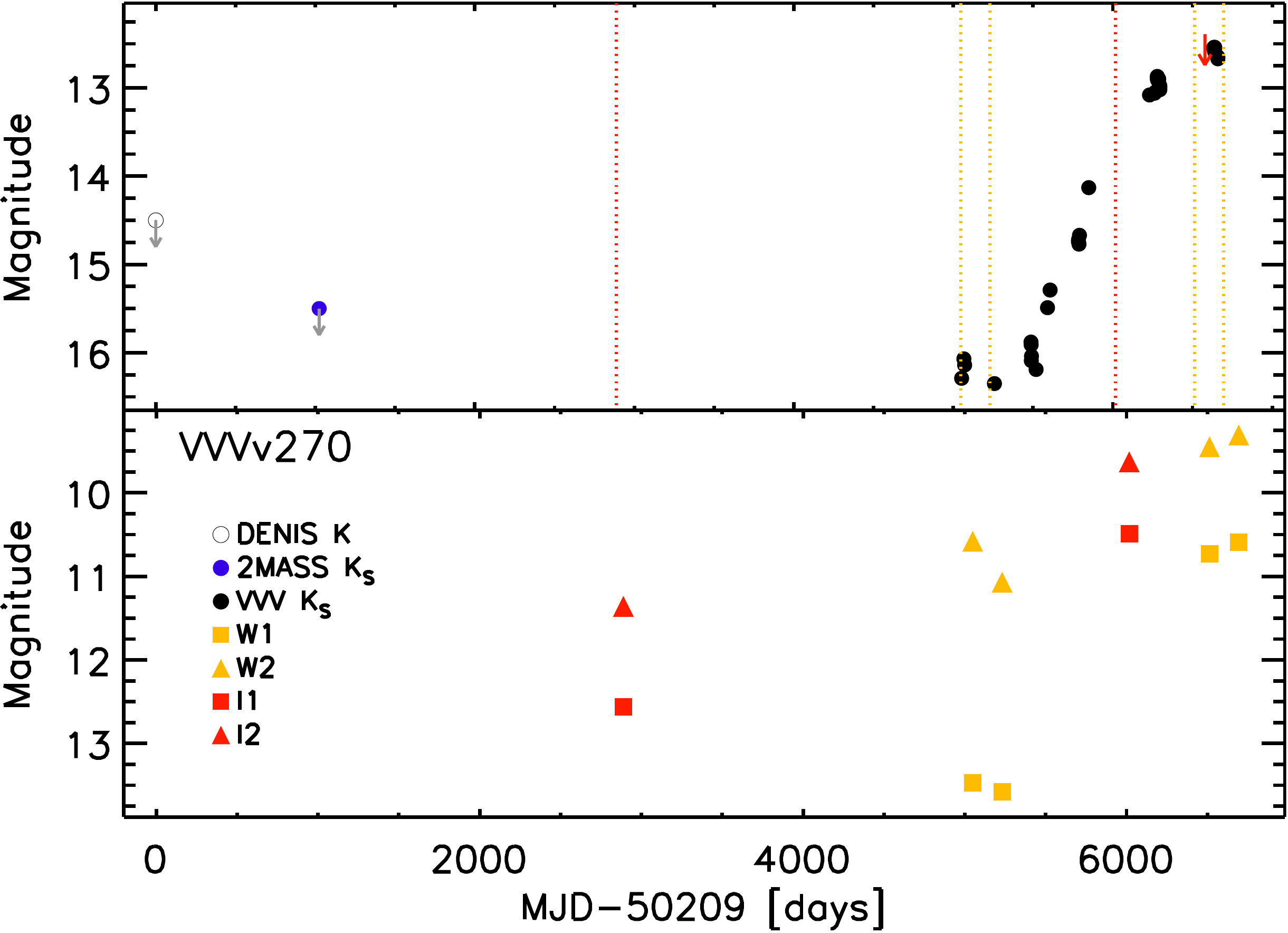}}}\\
\subfloat{\resizebox{0.6\textwidth}{!}{\includegraphics{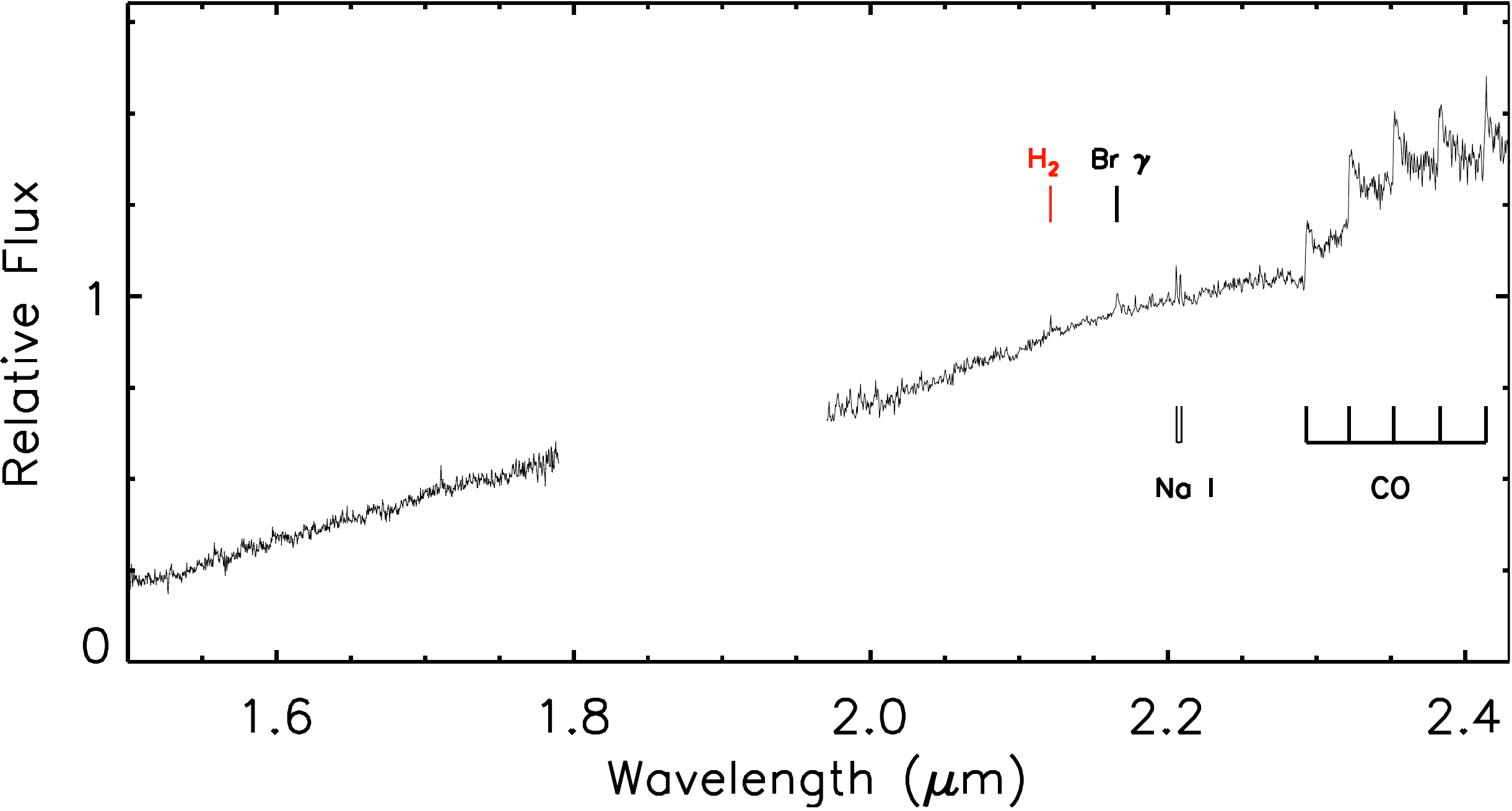}}}
\caption{(top left) False colour WISE image (blue=3.5 $\mu$m, green=4.6 $\mu$m, red=12 $\mu$m) of a 10\arcmin$\times10$\arcmin\ area centred on VVVv270. The location of the object is marked by the arrow. (top right) $K_{\rm s}$ image of a 10\arcmin$\times$10\arcmin ~area centred on VVVv270. The location of the object is marked by the red arrow and green circle. In addition, blue circles and labels mark objects found in a SIMBAD query with a 5\arcmin ~radius. (medium left) SED of VVVv270 along with \citet{2007Robitaille} YSO models that fulfil the criteria of $\chi^{2} - \chi^{2}_{best} < 3N $, with N the number of data points used to generate the fits. This image is generated by the fitting tool. (medium right) Near- and mid-infrared photometry of VVVv270. Data arising from different surveys are marked by different symbols which are labelled in the figure. The red arrow marks the approximate date of the spectroscopic follow-up. (bottom) FIRE spectrum of VVV270. The spectroscopic features found for this object are marked in the graph.}
\label{vvv:erupfour1}
\end{figure*} 


This object began an outburst in late 2011 and remains bright in the most recent data from 2015. The amplitude is $\Delta K_{\rm s}\sim3.8$~magnitudes (see Fig. \ref{vvv:erupfour1}). This star has not been detected in 2MASS nor DENIS, which is to be expected for the magnitude of the star prior to this current outburst ($K_{\rm s}\sim16$ magnitudes). The source brightened smoothly from late 2011 to 2014, remaining at similar brightness in 2015.
This slow rise resembles what has been observed in the classical FUor V1515 Cyg and the likely FUor object VVVv721 (Appendix \ref{vvv:specvar}). However, its spectrum shows strong emission from Br$\gamma$, CO, Na I and H$_{2}$, which is more expected for EXors in outburst rather than FUors.  The accretion luminosity of VVVv270 is amongst the highest in the sample (see Table \ref{table:vvvaccrate}), which supports the object being in a high state of accretion.


The object is associated with several star formation tracers within 5\arcmin, such as IR sources, \citet{2008Robitaille} intrinsically red objects and IRDCs identified with {\it Spitzer} from \citet{2009Peretto}. The only distance information from SIMBAD sources within 5\arcmin ~arises from the methanol maser Caswell CH3OH 333.931$-$00.135, with a radial velocity of V$_{LSR}=-36.8$~km s$^{-1}$ \citep{1996Ellingsen}. However this object is located 222\arcsec ~from VVVv270 and the radial velocity of VVVv270 ($-87.6\pm$10.9 km s$^{-1}$) is too different for there to be a physical
association. Our estimate from the VVV radial velocity places the object at d$\sim$4.6 kpc.

\subsubsection{VVVv322}\label{vvv:d069}

\begin{figure*}
\centering
\subfloat{\resizebox{0.5\textwidth}{!}{\includegraphics{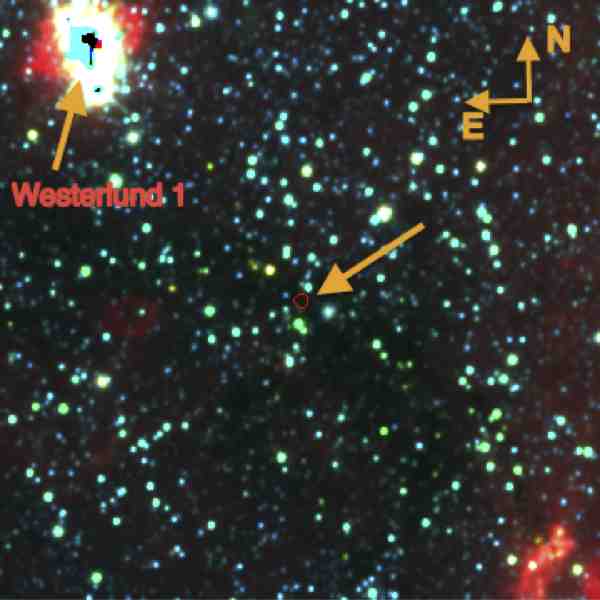}}}
\subfloat{\resizebox{0.5\textwidth}{!}{\includegraphics{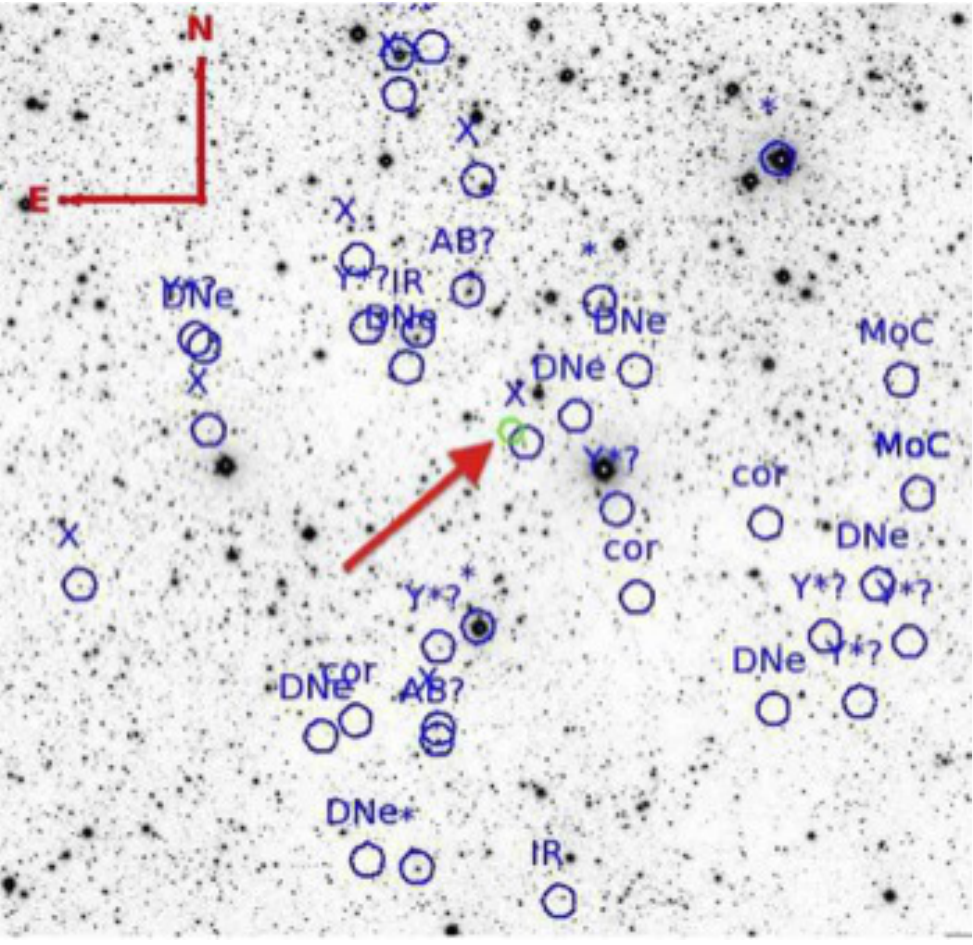}}}\\
\subfloat{\resizebox{0.5\textwidth}{!}{\includegraphics{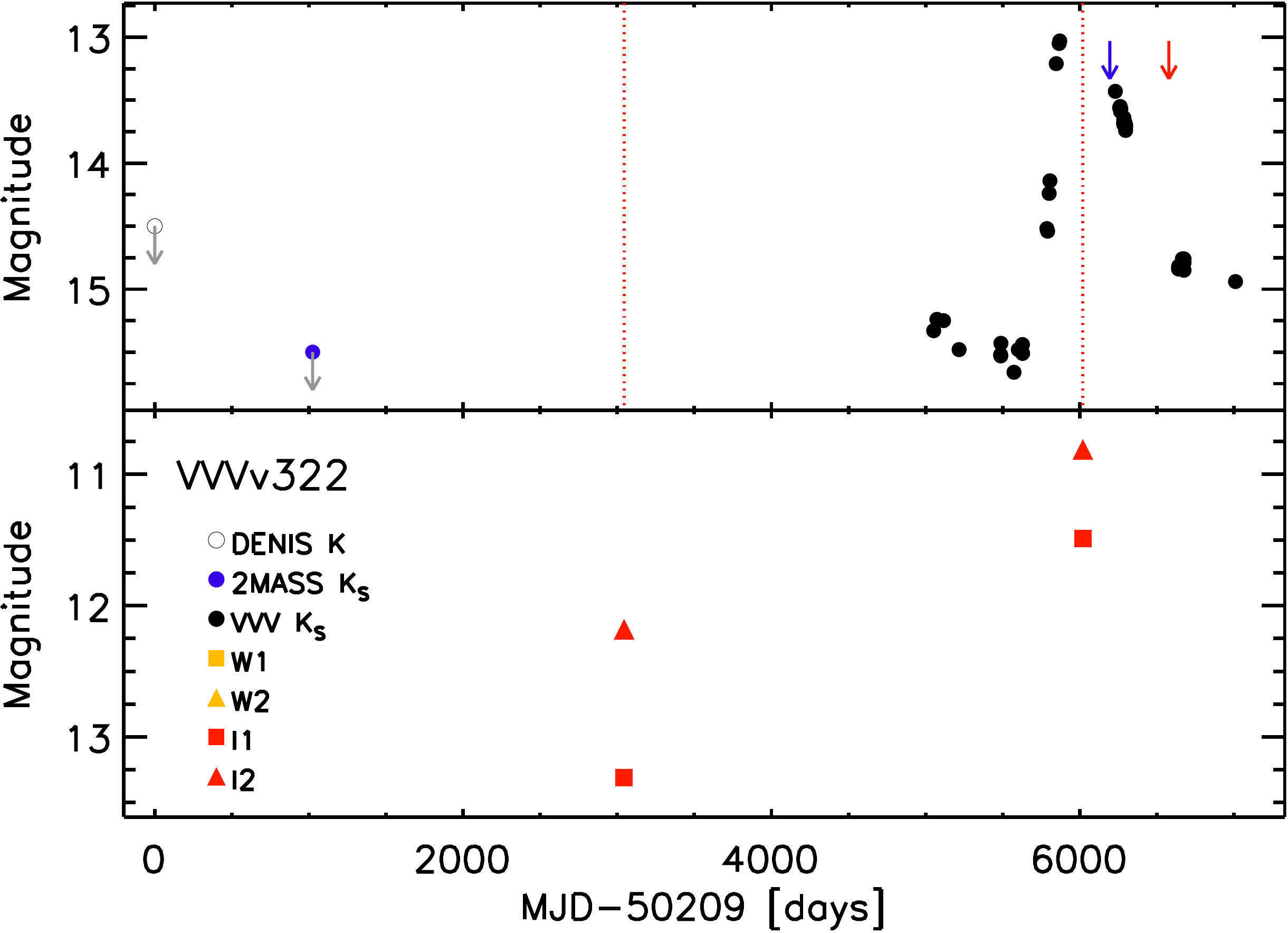}}}
\subfloat{\resizebox{0.5\textwidth}{!}{\includegraphics{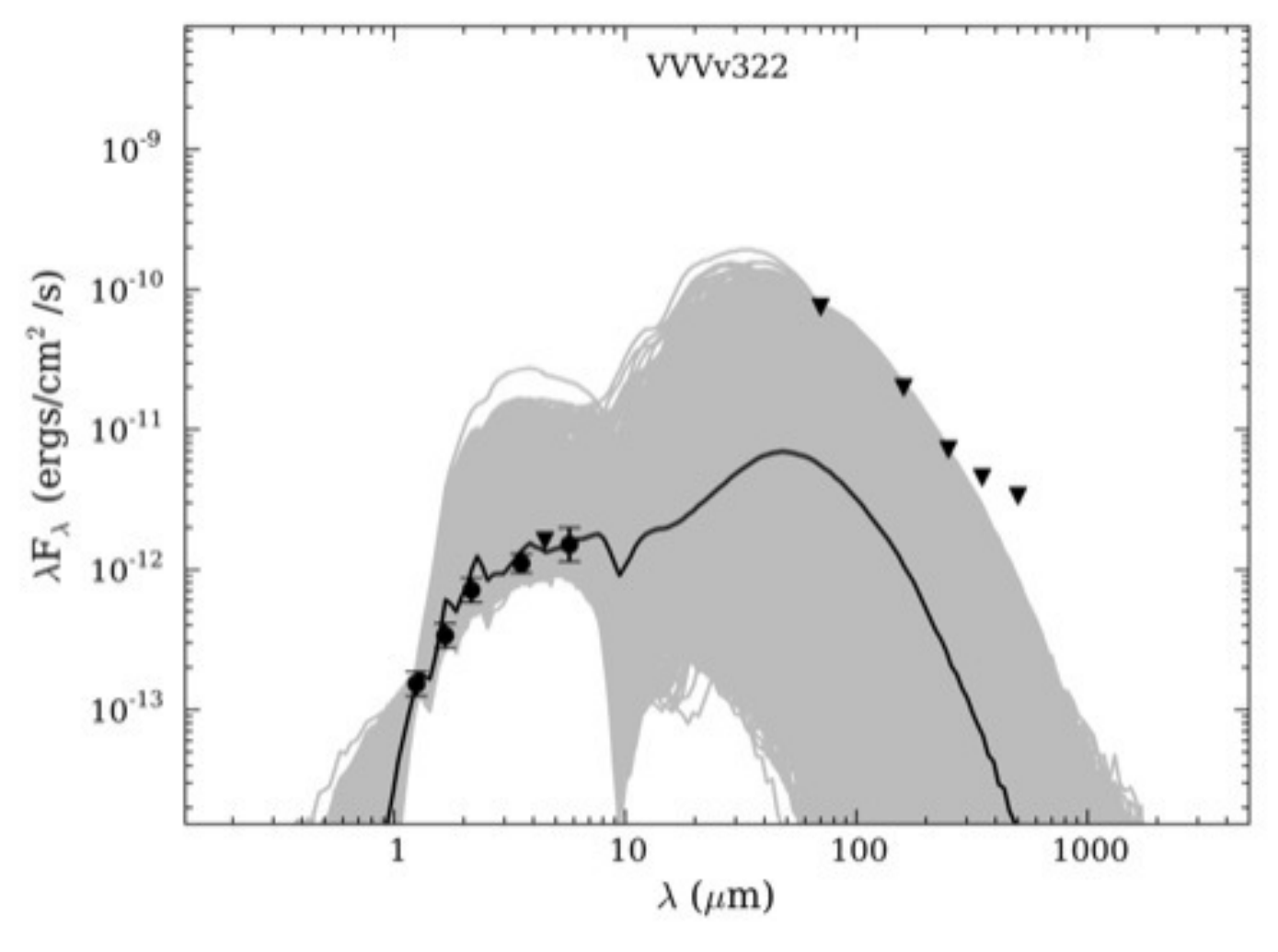}}}
\caption{(top left) False colour WISE image (blue=3.5 $\mu$m, green=4.6 $\mu$m, red=12 $\mu$m) of a 20\arcmin$\times$20\arcmin ~area centred on VVVv322. The location of the object is marked by the arrow. The SFR Westerlund 1 is also shown in the image. (top right) $K_{\rm s}$ image of a 10\arcmin$\times$10\arcmin ~area centred on VVVv322. The location of the object is marked by the red arrow and the green circle. In addition, blue circles and labels mark objects found in a SIMBAD query with a 5\arcmin ~radius. (botton, left) Near- and mid-infrared photometry of VVVv322. The approximate dates of spectroscopic follow up are marked by blue (2013) and red(2014) arrows. (bottom, right) SED of VVVv322 along with \citet{2007Robitaille} YSO models that fulfil the criteria of $\chi^{2} - \chi^{2}_{best} < 3N $, with N the number of data points used to generate the fits. This image is generated by the fitting tool.}
\label{d069vc52im1}
\end{figure*}

The projected location of this object is found close to several indicators of star formation as revealed by SIMBAD (see Fig. \ref{d069vc52im1}). These include molecular clouds, IRDCs, likely YSOs \citep[as determined by their mid-infrared colours by][]{2008Robitaille} and most interestingly several X-ray sources that could be associated with the young massive cluster Westerlund 1, which is located at $d=3.55\pm0.17$~kpc and has an age of $4$--$5$~Myr \citep{2005Clark}. The projected location of VVVv322 is 651\arcsec ~from the centre of the massive cluster. The object is classified as being associated with SFRs in Paper I, although whether this association is with Westerlund 1 is not clear, owing to the separation on the sky. However, we note that the near kinematic distance to the object ($d=3.8$~kpc, see Table \ref{table:vvvrvel}), is remarkably similar to the distance of the cluster. Therefore the association seems likely.

The near-infrared colours of the star show an $(H-K)$ excess that can be attributed to warm circumstellar material, placing it to the right of the 
reddened T Tauri locus in Fig. \ref{vvv:specgc}. The SED of the star is constructed using {\it Spitzer} photometry (due to non-detection in WISE) 
and the spectral index is $\alpha=0.92$, corresponding to a class I YSO. The {\it Spitzer} and VVV measurements are not contemporaneous, but the star appears to have been at quiescence prior to the 2011 outburst (see below), giving some confidence that $\alpha$ is estimated correctly. 

The light curve of VVVv322 (Fig. \ref{d069vc52im1}) shows the object was in a quiescent state at $K_{\rm s} \sim$15.5 mag before going into outburst between the end of the 2011 coverage and the beginning of the 2012 campaign. The object had apparently been at this faint state for $>10$ years given that is not detected by DENIS nor 2MASS. The brightness of the object is already declining by 2013. Due to the lack of complete coverage we cannot be certain of the amplitude of the outburst but it is likely to be $\Delta K_{\rm s} > 2.5$~mag. Mid-infrared photometry from the GLIMPSE I \citep{2003Benjamin} and DEEP GLIMPSE \citep{2009Churchwell} surveys also shows a large change with 1.8 and 1.4 magnitudes at 3.6 and 4.5 $\mu$m respectively. The DEEP GLIMPSE data was taken in October 2012, which would place it close to the maximum brightness of VVVv322. Given that the object appears to have been at a quiescent state for a long time prior to 2011, we can then consider the GLIMPSE I measurements as being pre-outburst data. The mid-infrared change is in agreement with the expected change from variable accretion in protostars \citep[see e.g.,][]{2012Fischer, 2013Scholz} 

The observed $K_{\rm s}$ light curve during 2013 implies a decline rate of $\sim$ 1.57 mag yr$^{-1}$. This rate implies that the duration of the outburst would not be longer than approximately 3 years. VVV photometry from July/August 2014 shows that this trend continued, returning the object 
to a brightness close to the quiescent state of 2010. However, the 2015 photometry shows that the rate of decline has slowed in the most recent 
data.


\begin{figure*}
\centering
\subfloat{\resizebox{0.8\textwidth}{!}{\includegraphics{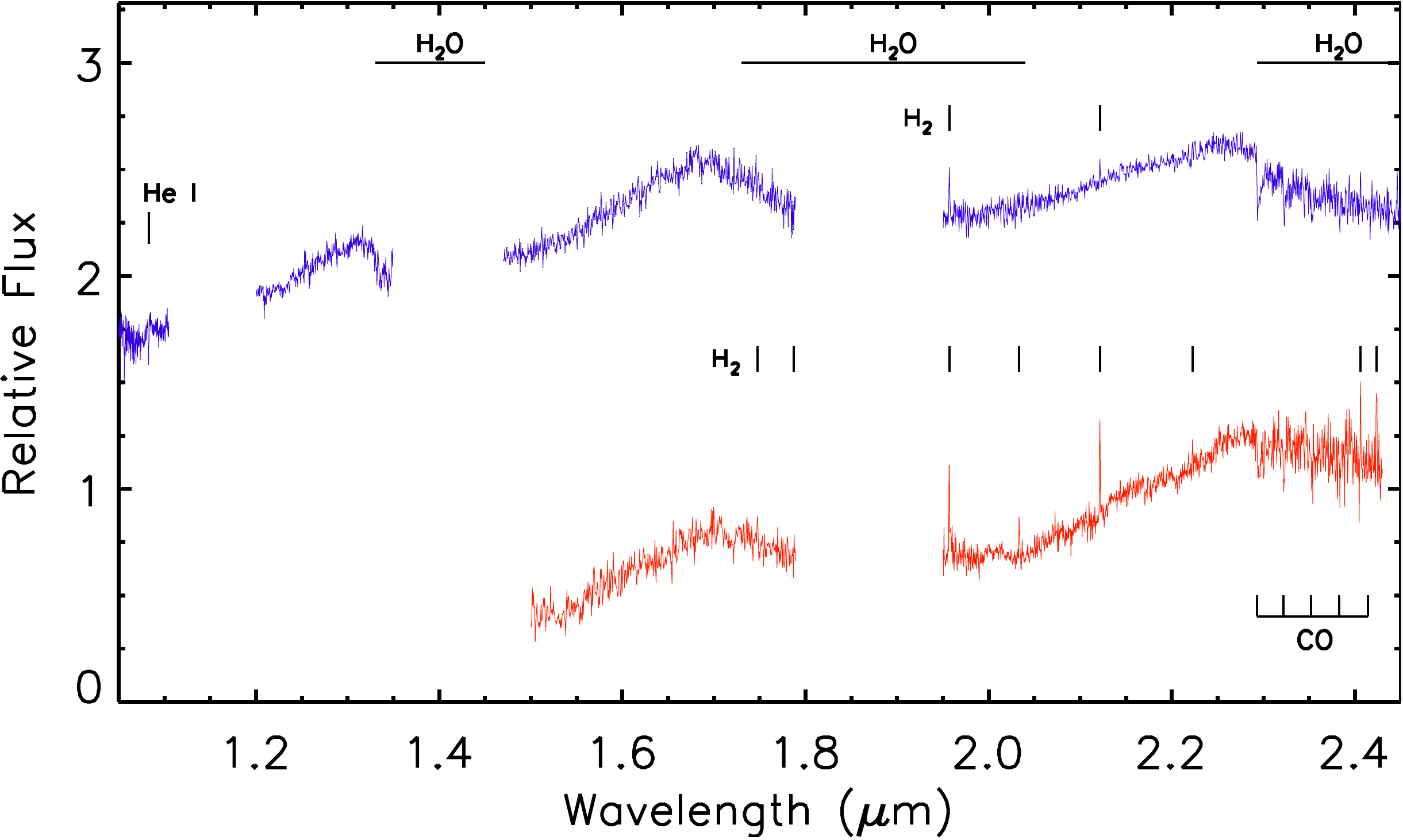}}}\\
\subfloat{\resizebox{0.8\textwidth}{!}{\includegraphics{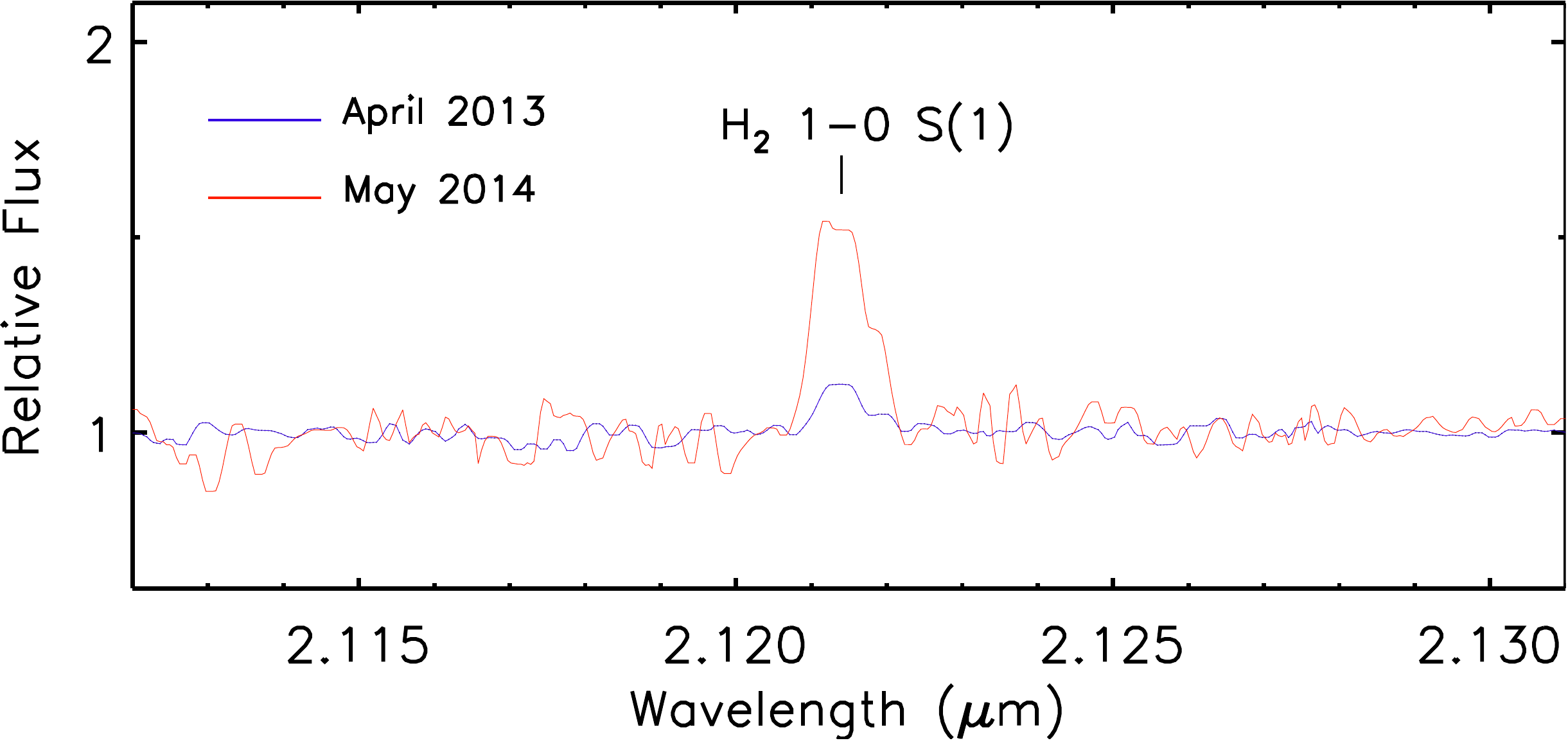}}}
\caption{(top) Graph comparing the 2013 (blue) and 2014 (red) FIRE spectra of VVVv322, where we mark the observed absorption features of H$_{2}$O and CO, along with emission lines from H$_{2}$. (bottom) 2013 and 2014 observations in the region 2.10--2.13 $\mu$m comparing the 2.12 $\mu$m 1--0 S(1) H$_{2}$ emission line.}
\label{d069vc52im2}
\end{figure*}

The spectrum for this object was obtained in April 2013, when it was slightly below maximum brightness, based on 
inspection of the light curve. The spectrum shows the presence of strong H$_{2}$O absorption bands at 1.3--1.4, 1.7--2, 2.3--2.5 $\mu$m  and CO absorption at $2.29$~$\mu$m. In addition the 1--0 S(1) H$_{2}$ line at 2.12~$\mu$m is seen weakly in emission. The spectrum lacks any other absorption features that could be associated with a stellar photosphere (see Fig. \ref{d069vc52im2}). 

The evidence presented above supports the scenario of VVVv322 being an eruptive YSO. However, the classification of this object within the known classes of eruptive variables seems to be uncertain. The 2013 spectrum resembles that of FU Orionis stars during periods of high accretion, \citep[see e.g. V900 Mon, ][]{2012Reipurth} and it would agree with the fact that the star was observed at a point close to maximum brightness. However, the duration of the outburst would seem to be much shorter than expected for FU Orionis stars, but longer than for EXors. In this sense this object resembles more recently discovered objects which display mixed characteristics of eruptive variable classes \citep[see e.g. V1647 Ori, V2775 Ori,][]{2007Fedele, 2011Caratti}.

A second spectrum was taken in 2014\footnote{Two other objects, VVVv815 and VVVv699, have repeated spectroscopy. Their characteristics are presented in Appendix \ref{vvv:specvar}.}, which allows us to note some interesting points about this source. The May 2014 spectrum was taken at a time where the object has apparently faded by about 1.3 magnitudes. By this time the H$_{2}$O and CO features seem to have weakened (see Fig. \ref{d069vc52im2}). However, the H$_{2}$ line at 2.12 $\mu$m has become stronger. The spectrum, as in 2013, does not show any apparent photospheric features

We note that the weakening of the H$_{2}$O and CO features by 2014 agrees with the fact that the luminosity of the star had faded by this epoch.  The lack of Br$\gamma$ emission in FU Orionis objects has been attributed to the fact that for high accretion rates ($\dot{M} > 10^{-5}$~M$_{\odot}$~yr$^{-1}$) there is a breakdown in the magnetospheric accretion model \citep[see e.g.][]{2012Fischer}. It would be very interesting to see that with VVVv322 returning to a quiescent state, the star would show absorption features arising from a stellar photosphere rather than the disc and Br$\gamma$ returning to emission in the spectrum of the object.

The H$_2$ 2.12 $\mu$m emission feature has become stronger between 2013 and 2014 (see Fig. \ref{d069vc52im2}), with the equivalent width increasing by a factor of 3--4. The emission does not appear to be spatially extended at either epoch. The presence of this feature could be associated with shocks from molecular outflows linked to this particular outburst event. While this is not certain, 
continuous monitoring and high resolution imaging could help clarify this picture. A direct link between an outburst and an outflow has been 
observed in V2494 Cyg \citep{2013Magakian}.

\begin{table*}
\begin{center}
\begin{tabular}{@{}l@{\hspace{0.35cm}}l@{\hspace{0.25cm}}l@{\hspace{0.2cm}}c@{\hspace{0.2cm}}c@{\hspace{0.2cm}}c@{\hspace{0.2cm}}}
\hline
Object & $\Delta K_{\rm s}$ & Variability & Spectrum & Classification\\
\hline
\multicolumn{5}{c}{YSO-Eruptive}\\
\hline
VVVv20 & 1.71 & $\Delta t_{out} \sim 2.5$~yr & Br$\gamma$, CO, Na I, H$_{2}$ emission & Eruptive, MNor\\
VVVv32 & 2.50 &  $\Delta t_{out} \sim 3.3$~yr, periodic? & Br$\gamma$, CO emission & Eruptive, MNor\\
VVVv94 & 1.91 &  $\Delta t_{out}> 3$~yr & Br$\gamma$, CO, H$_{2}$ emission & Eruptive, MNor\\
VVVv118  & 4.24 &  $\Delta t_{out}< 150$~days, repetitive & Br$\gamma$ emission, H$_{2}$O absorption(?) & Eruptive, EXor(MNor?)\\
VVVv193 & 1.26 &  $\Delta t_{out}> 2$~yr? & Br$\gamma$, CO, H$_{2}$ emission & Eruptive, MNor\\
VVVv270  & 3.81 & $\Delta t_{out} >3.5$~yr, ongoing & Br$\gamma$, CO, Na I, H$_{2}$ emission & Eruptive, MNor\\
VVVv322 & 2.63 &  $\Delta t_{out} \sim 3.5$~yr & CO, H$_{2}$O absorption & Eruptive, MNor\\
VVVv374 & 2.41 &  $\Delta t_{out}> 3.5$~yr, ongoing? & Br$\gamma$, CO, Na I, H$_{2}$ emission & Eruptive, MNor \\
VVVv452  & 2.46 &  $\Delta t_{out} \sim 2.2$~yr & Br$\gamma$, CO, H$_{2}$ emission & Eruptive, MNor\\
VVVv473  & 1.50 &  $\Delta t_{out} \sim 275$~days, periodic & Br$\gamma$, CO, H$_{2}$ emission & Eruptive, MNor\\
VVVv562  & 2.79 &  $\Delta t_{out}> 2$~yr? & Br$\gamma$, H$_{2}$ emission. Na I absorption & Eruptive, MNor\\
VVVv631  & 2.64 &  $\Delta t_{out}>4$~yr, ongoing? & Br$\gamma$, CO, Na I emission. H$_{2}$O absorption & Eruptive, MNor\\
VVVv662  & 1.83 & $\Delta t_{out}> 2$~yr & Br$\gamma$, CO, Na I, H$_{2}$ emission & Eruptive, MNor\\
VVVv665  & 1.63 &  $\Delta t_{out}> 2$~yr & Br$\gamma$, CO, Na I, H$_{2}$ emission & Eruptive, MNor\\
VVVv699  & 2.28 & $\Delta t_{out}> 4$~yr, ongoing & Br$\gamma$, CO, Na I, H$_{2}$ emission & Eruptive, MNor\\
VVVv717 & 2.47 &  $\Delta t_{out}> 2$~yr, periodic? & CO absorption & Eruptive, MNor\\
VVVv721 & 1.86 &  $\Delta t_{out}> 5$~yr, ongoing & CO absorption & Eruptive, FUor\\
VVVv800  & 1.65 & $\Delta t_{out}> 3$~yr & Br$\gamma$, H$_{2}$ emission & Eruptive, MNor\\
VVVv815 & 1.71 &  $\Delta t_{out} \sim 1.8$~yr & H$_{2}$ emission & Eruptive, MNor\\
\hline
\multicolumn{5}{c}{YSO-Non Eruptive}\\
\hline

VVVv63  & 1.44 & $\Delta t \sim 1$~yr? & Br$\gamma$, H$_{2}$ emission & non-eruptive? \\
VVVv405  & 2.00 &  $\Delta t> 1$~yr & Br$\gamma$, H$_{2}$ emission & non-eruptive?\\
VVVv406  & 2.06 &  $\Delta t> 1$~yr & H$_{2}$ emission & non-eruptive?\\
VVVv480  & 1.70 &  $\Delta t \sim 1$~yr & Br$\gamma$, H$_{2}$ emission & non-eruptive?\\
VVVv630  & 1.90 &  $\Delta t> 1$~yr & Br$\gamma$, H$_{2}$, CO, Na I, Ca I absorption & non-eruptive\\
VVVv625  & 1.46 &  $\Delta t< 100$~days & Br$\gamma$, H$_{2}$ emission & non-eruptive\\
VVVv628  & 1.45 &  $\Delta t< 1$~yr & Br$\gamma$ (?), CO, Na I, Ca I absorption & non-eruptive\\
VVVv632  & 1.51 &  $\Delta t< 100$~days & Br$\gamma$, H$_{2}$ emission. Na I, Ca I absorption & non-eruptive\\
VVVv65  & 1.36 &  $\Delta t< 100$~days & Br$\gamma$ emission  & non-eruptive.\\
\hline
\multicolumn{5}{c}{\large{Non-YSOs}}\\
\hline
VVVv25 & 1.68 &  $P \sim 2000$ ~days? & CO absorption & Evolved star\\
VVVv42 & 2.16 &  $P \sim 812$~days, periodic & cold CO absorption & Evolved star?, YSO?\\
VVVv45 & 2.32 &  $P \sim 560$~days, periodic & CO absorption & Evolved star\\
VVVv229 & 3.70 &  $P \sim 978$~days, periodic & CO absorption & Evolved star\\
VVVv235 & 1.26 &  $P \sim 769$~days, periodic & CO absorption & Evolved star\\
VVVv796 & 2.94 &  $P \sim 796$~days, periodic & CO absorption & Evolved star\\
VVVv202 &  1.61    &  $P > 1000$~days, periodic & $^{12}$CO, $^{13}$ CO, Na I, Ca I absorption & Evolved star\\
VVVv514  & 1.73 & $\Delta t> 500$~days? & Br$\gamma$, HeI (?) emission & Nova\\
VVVv240  & 1.17     &  $\Delta t> 500$~days? & Br$\gamma$, CI emission & Nova\\
\hline
\end{tabular}
\caption{$\Delta K_{\rm s}$, timescale of the variability and main features observed in the spectra of VVV objects from our sample. The last column provides a likely classification as eruptive, non-eruptive or non-YSO. For eruptive YSOs the timescale relates to the approximate duration of the outburst (marked as $\Delta t_{out}$). In non-eruptive YSOs the timescale corresponds to the duration of the observed large amplitude changes, in some cases these could correspond to outburst durations such as in objects marked as non-eruptive?. Finally, for non-YSOs we provide the periods for AGB stars and the approximate duration of what could be eruptions in novae.}\label{table:vvvlcerup}
\end{center}
\end{table*}




\subsection{Final remarks on classification} 

 In the previous sections, we have shown that most spectra of our young eruptive variables fail the classical FUor/EXor classification. In fact, only one object in our sample would fall into the classical FUor definition, VVVv721, which shows the characteristic CO absorption as well as a long-lasting outburst, at least extending beyond the coverage from VVV, of these objects. 

The majority of the remaining YSOs show characteristics that are more commonly found in EXors, such as an emission line spectrum. However, H$_{2}$ emission is common in our objects, and in some cases [Fe II] emission is also present in their spectra. These features were not observed in the sample of EXors studied by \citet{2009Loren}. In addition, the duration of the outburst is longer than expected from this subclass, but also shorter than the long-lasting outburst of classical FUors. Both VVVv322 and VVVv717, two of the three objects that show strong, FUor-like, CO absorption, have light curve properties that are different to classical FUors.

 The different characteristics of our sample to that of the known classes of eruptive variables could be explained by the fact that the majority of our sample is optically invisible and found at early evolutionary stages. This is certainly true when we compare to classical EXor events (usually associated with instabilities in discs of class II YSOs). In addition, the classical FUors are often described as being at the boundary between T Tauri stars and protostars with an envelope \citep[see e.g.][]{2014Audard}, comparable to flat-spectrum YSOs. Inspection of their SEDs reveals that none of the classical, optically visible FUors have steeply rising class I SEDs \citep[see][]{2014Gramajo}. 
 
Considering the arguments presented above, we are confident that these 19 objects are new additions to the eruptive variable class, with 4 more objects where it is unclear if variable accretion is responsible  for the observed variability. Of the 19 eruptive
sources, 16 are classified as class I and 3 as flat spectrum sources, from their value of $\alpha$. 17/19 are optically very faint, 
defined here by non-detection in the VVV Z band images. 
 
The characteristics of our sample are in line with what has been observed in a small but increasing number of new members of the eruptive variable subclass. For example, V1647 Ori went into outburst in 2003 and showed a near-IR spectrum with strong Br$\gamma$ and CO emission \citep[see e.g.][]{2013Ninan}. This object returned to quiescence after 2 yrs and then underwent a second outburst in 2008 \citep{2009Aspinb}, but now showing CO weakly in absorption at 2.29 $\mu$m. OO ser \citep{2012Hodapp} is a deeply embedded object that shows a featureless near-IR  spectrum and an outburst duration of less than 10 yr. V2775 Ori \citep{2011Caratti}, went into outburst in 2011 and its near-IR spectrum shows the characteristic H$_{2}$O and CO absorption of FUors, but it is unclear if the  outburst duration is comparable to that for FUor objects. V723 Car is a highly embedded class I object that went into outburst between 1999 and 2003 reaching peak magnitude of $K=12.9$~mag. The light curve of the object after the outburst shows an erratic behaviour with $\Delta K=2$~mag. The near-infrared spectrum of V723 Car shows H$_{2}$, Br$\gamma$ and CO emission. \citet{2015Tapia} argue that these characteristics make V723 Car different from EXors and FUors. V1180 Cas is a flat-spectrum YSO that shows an outburst duration that is between that of both EXors and FUors \citep{2011Kun} and shows an emission line spectrum during outburst \citep{2014Antoniuccib}. V512 Per is a class I object that went into outburst in 1990 \citep{1991Eisloeffel} and has remained close to the maximum brightness ever since \citep{2014Hodapp}, resembling the duration of FUor outbursts. However, the object shows strong CO emission during outburst \citep{1991Eisloeffel} and drives an H$_{2}$ molecular outflow \citep{2014Hodapp}. V2492 Cyg is a class I YSO that went into outburst in 2010 and shows an emission line spectrum with an outburst duration of $>3$~yr \citep{2011Covey,2013Kospal}. \citet{2011Covey} point out that this object differs from FUors and EXors and resembles the characteristics of V1647 Ori.

A label of some sort is likely to be useful to categorise these eruptive YSOs that show a mixture of characteristics of FUors and EXors. 
We propose ``MNors'' (pronounced ``emnors'') given the resemblance with V1647 Ori, the illuminating star of the McNeil's Nebula 
(McNeil's Nebular Object or MNO). Given the characteristics of our sample and their similarities with some of the recent discoveries from the literature, MNors could be considered as objects that: 1) have SEDs of class I or flat-spectrum sources, 2) show outburst durations $>$1.5 yr but shorter than those usually associated with FUors, 3) show spectroscopic characteristics of eruptive variables, i.e. CO emission, absorption, and 4) usually have 2.12 $\mu$m H$_{2}$ emission, most likely due to a molecular outflow. As exceptions to item (3) sources lacking CO emission or absorption may still be MNors, e.g. if the spectrum is taken in
quiescence or there is strong veiling of the photosphere and the innermost parts of the disc (e.g. see the individual discussions of VVVv800 and VVVv815 in Appendix \ref{vvv:specvar}).

Our definition of the MNor subclass includes  sources with diverse spectroscopic properties. The
classification is therefore provisional, pending further study. The most likely scenario is one in
which the known FUor, EXor and MNor subclasses are not different types of objects but are instead different representations of the same phenomena, i.e. episodic accretion. For example, \citet{2015Vorobyov} show that their disc fragmentation models, one of the mechanisms that is invoked to explain episodic accretion,  can produce isolated luminosity bursts, similar to FUors, or clustered bursts, i.e. a succession of closely packed bursts, similar to EXors and the majority of the objects in our sample.



The 19 new episodic variables approximately double the number of the spectroscopically-verified eruptive variable YSOs known so far. These new additions increase the number of embedded members of the class by a large factor.

Table \ref{table:vvvlcerup} shows the classification of the objects according to their light curves, as well as their spectroscopic characteristics.

\section{Summary and Conclusions}\label{vvv:sec_summary}

We have presented intermediate resolution infrared spectra of 37 VVV high amplitude variable stars, 18 of which have eruptive light curve classification from Paper I. The observed systems generally have
class I or flat spectrum SEDs, typical of the full VVV photometric sample of YSOs. In the following we summarize the main points of our work.

\begin{itemize}
\item From the sample, nine stars show 
characteristics that indicate they are AGB stars or novae rather than YSOs (see Appendix \ref{vvv:sec_evolved}). Amongst the 28 YSOs, the incidence
of Br$\gamma$ emission is high, as in normal YSOs, but the incidences of CO emission and especially H$_{2}$ are higher than in 
normal embedded YSOs. The locations of the YSOs in the equivalent width diagrams of \citet{2010Connelley} are in general very different
from those of normal YSOs, especially for VVV sources followed up during a bright state.


\item Three YSOs show the characteristic strong CO absorption and lack of photospheric features that are associated with FUors, though
only one of these has an outburst of sufficient duration to be considered a likely classical FUor.

\item Of the 25 other YSOs, we classify 16 as likely eruptive variables. These show emission line spectra with Br$\gamma$, CO 
and/or H$_{2}$ almost always present. The H$_{2}$ emission line ratios are consistent with shock-excited emission from molecular 
outflows. The latter is not observed in classical EXors \citep[see e.g.][]{2009Loren}. They have light curves which are characterized by large variations ($\Delta K_{\rm s} > 1.5$~magnitudes), 
with apparent durations of the bright state of 1 to 5 years, which is longer than the characteristic timescale of EXors \citep{2012Loren}. Of the 16 systems, 12 
have eruptive light curve classifications, 1 is a fader, 
and 3 are long-term periodic variables. We note that the periodicity in these 3 is not perfect: such objects show short term scatter
and the sparse sampling of the light curves prevents precise period determination. As noted in Paper I, some of the eruptive 
variables show repetitive bursts (a quasi periodic behaviour) so the distinction between these provisional light curve 
classifications is not clear cut.

\item Three objects show outburst durations that extend over the 5 year period covered by VVV, which would be expected if these are FUor-like outbursts. However, for two of these, the near-infrared spectra do not show the usual CO and H$_{2}$O absorption characteristic of the FUor class \citep{1996HK}. Instead, they are dominated by emission from CO, H$_{2}$ and Br$\gamma$, which is typical in EXors. 

\item The predominance of outburst durations longer than EXors but shorter than FUors, coupled with a mixture of EXor and FUor 
spectroscopic characteristics, leads us to propose a new classification, MNors, for these embedded eruptive variable YSOs. 
The name follows V1647 Ori, the illuminating source of the McNeil nebula, which is one of a small but growing number of such
eruptive variables discovered in recent years \citep[see e.g.][]{2007Fedele,2009Aspin}. While these might have been considered rare exceptions in the past, they now appear to
represent the majority of eruptive YSOs at earlier evolutionary stages. This designation is of course provisional, given that these objects are heterogeneous and so 
further distinctions and sub-classes might be needed in future. The paucity of EXor-like systems with short duration outbursts appears to be a genuine difference between optically visible class II eruptive YSOs and their embedded counterparts. The small number of FUor candidates is likely a selection effect given the short (2 year) duration of the dataset used for the initial search. The fact that FUor outbursts are seen in nearby SFRs amongst optically visible YSOs with ”flat spectrum” SEDs but MNor outbursts were not until
recently may simply be due to small number statistics.

\item Overall we have classified 19/28 YSOs as eruptive variables. Two or three of the remainder might also also be eruptive systems
observed in quiescence. The results validate our contention in Paper I that members of the VVV photometric sample with eruptive light
curves are mainly bona fide YSOs undergoing episodic accretion, along with some of the LPV-YSOs and faders also.
The short term variables and dippers generally show weaker evidence for episodic accretion, as was expected. The classification of 1 or 2 of the eruptive systems
could be disputed but that would not affect our broader conclusion about the properties of the eruptive 
population. The 18 spectroscopically observed candidates with eruptive light curves are fairly
representative of the 106 systems with this classification in the full photometric sample, with
only a moderate bias towards higher amplitude (and brighter mean magnitude). Much work remains 
to be done to explore the full VVV sample of high amplitude variables, particularly sources in 
the other light curve categories.

\item The accretion luminosities of YSOs estimated from Br$\gamma$ emission show that those objects classified as eruptive and observed at a bright state have higher L$_{acc}$ than the remainder of the YSO sample. These also show higher accretion luminosities than normal low mass class I YSOs and low and intermediate mass class II YSOs, comparable to those of known Br$\gamma$-emitting eruptive variables. The accretion rates of the Br$\gamma$ emitters are found to be lower than those of FUors, which is unsurprising if we expect this emission line to be quenched at the high accretion rates of FUors \citep{2014Audard}. The variability amplitudes are also slightly lower on average than those of classical FUors, the data nonetheless support the view that episodic accretion-events lasting a few years
are common amongst class I and flat spectrum YSOs. In this context, FUors may represent the more extreme examples of a common phenomenon, and there may be no clear distinction between YSOs with relatively small 
($\sim$1 mag) eruptions and the lower level accretion variability commonly observed in long duration studies
of normal YSOs. A variability-based census of normal YSOs in VVV will be invaluable to quantify the incidence of episodic accretion.

       
       
\item For three objects we have repeat observations that allows us to study the spectroscopic variability of eruptive variables. Object VVVv322 is of special interest. The light curve of the object shows a large outburst and its near-infrared spectrum closely resembles those of classical FUors. However, the duration of the outburst appears to be $\sim$ 3 years, much shorter than expected for classical FUors. The spectrum of the object also shows variations 
as absorption features from CO and H$_{2}$O become weaker as the star goes back into quiescence.

       
\end{itemize}
 

\section*{Acknowledgments}

This work was supported by the UK's Science and Technology Facilities Council, grant numbers ST/J001333/1, ST/M001008/1 and ST/L001403/1.

We gratefully acknowledge the use of data from the ESO
Public Survey program 179.B-2002 taken with the VISTA 4.1m telescope and
data products from the Cambridge Astronomical Survey Unit. Support for
DM, and CC  is provided by the Ministry of Economy, Development, and 
Tourism’s Millennium Science Initiative through grant IC120009, awarded to
the Millennium Institute of Astrophysics, MAS. DM is also supported by the Center for 
Astrophysics and Associated Technologies PFB-06, and Fondecyt Project No. 1130196.
This research has made use of the SIMBAD database, operated at CDS, Strasbourg, France; 
also the SAO/NASA Astrophysics data (ADS). A.C.G. was supported by the Science Foundation of Ireland, grant 13/ERC/I2907. We also acknowledge the support of CONICYT REDES project No. 140042 ``Young variables and proper motion in the Galactic plane. Valparaiso-Hertfordshire collaboration''

C. Contreras Pe\~{n}a was supported by a University of Hertfordshire PhD studentship in the earlier stages of this research.

We thank Janet Drew for her helpful comments on the structure of the paper.

\bibliographystyle{mn2e}
\bibliography{ref.bib}

\clearpage

\appendix

\section{Individual Objects.}\label{vvv:specvar}

\begin{itemize}
\item {\bf VVVv721} Previously identified as a likely YSO in the \citet{2008Robitaille} catalogue of intrinsic red objects, the SED and near-infrared colours of the star are consistent with a class I object. The star is found within 244\arcsec ~of the dark cloud SDC G338.734+0.584 \citep{2009Peretto}, molecular clouds [RC2004] G338.7+0.6$-$72.1 and [RC2004] G338.7+0.6$-$62.2 \citep{2004Russeil}, and HII region GAL 338.74+00.64 \citep{2014Brown}, which have measured radial velocities between $-$62 and $-$72.1 km s$^{-1}$ . The use of CO model templates yield V$_{LSR}=-89.7\pm2.6$ km s$^{-1}$ for VVVv721. Other star formation tracers are found within 300\arcsec ~of the source (HII region, YSOs) and the WISE false colour image reveals evidence of active star formation in the area (Fig. \ref{d107vc13}). 

The light curve shows a slow rise with $\Delta K_{\rm s}=1.9$~magnitudes between 2010 and 2014. The classical FUor V1515 Cyg is known for displaying a slow rise in its light curve, and taking approximately 25 years to reach its maximum brightness \citep{1991Kenyon}. This slow rise has been explained as arising from thermal instabilities that spread from the inner regions towards the outer parts of the accretion disc \citep[see e.g][]{2014Audard}. However, VVVv721 is detected in 2MASS with $K_{\rm s}=13.23\pm0.03$ magnitudes, thus being at least 0.7 magnitudes brighter than the first VVV epoch. The light curve in Fig. \ref{d107vc13} also shows the variability observed at 3.4 and 4.6 $\mu$m using the photometry from the WISE \citep{2010Wright} and NEOWISE \citep{2011Mainzer} surveys. The mid-infrared data points also show large variability with $\Delta$W1$=1.53$ and $\Delta$W2$=1.02$ magnitudes, values which agree with the expected change from variable accretion in YSOs \citep{2013Scholz}. These changes do not appear to be related to variable extinction along the line of sight. 

VVVv721 is an optically invisible star, as it is not detected in $Z$ nor $Y$ bands in VVV. The spectrum of the star (Fig. \ref{d107vc13}) rises longwards of 1.3 $\mu$m , and it is nearly featureless, apart from the deep CO feature characteristic of FUors. Absorption from H$_2$O is also present in the spectrum.  

Some doubt will remain as to whether the duration of the outburst is in line with the expected duration in FUor objects. Nevertheless this variable star is very likely a new addition to the FUor eruptive variable class.
 






 
\item {\bf VVVv717} This object has been previously classified as a likely YSO by \citet{2008Robitaille}, based on {\it Spitzer} and MIPS photometry. The SED and near-infrared colours are those of a class I object. It is found within 5\arcmin ~of six IRDCs from \citet{2009Peretto}, whilst two other likely YSOs from the \citeauthor{2008Robitaille} catalogue are found within 30\arcsec ~from this object (see Fig. \ref{d106vc30}). There are no distance estimates to objects around this source. 

VVVv717 is an optically invisible star, not being detected in $J$ nor $H$ bands  in VVV. Fig. \ref{d106vc30} shows a red rising, nearly featureless spectrum, with no flux at wavelengths shorter than 1.4 $\mu$m. The object shows strong and broad CO absorption and although it does not show any absorption from H$_{2}$O, we note that other deeply embedded FUor objects have been classified as such even when lacking this feature \citep[e.g. OO Ser, PP13S,][]{1996Hodapp,2001Aspin}. Br$\gamma$ is also apparent in this object, with relatively weak emission that seems to be 
on top of a broader absorption feature. This emission line is usually not observed in FUors as this accretion-sensitive emission line
is assumed to be quenched by the very high accretion rate of these systems. VVVv717 also shows weak H$_{2}$~2.12~$\mu$m and a marginal detection of Mg I 2.28 $\mu$m emission. Further follow up of the system (Contreras Pe\~{n}a et al., in prep.) shows that the H$_{2}$~2.12~$\mu$m has become much more evident.

The light curve shows some possible periodicity with $P>400$~days. It was classified as a YSO with a long-term period in Paper I. It rose in brightness during 2010, remained bright in 2011, showed a drop in brightness in 2012, recovered in 2013, and then showed a deeper drop in 2014. The object shows $K_{s,max}-K_{s,min}=2.47$~magnitudes. The 2013 data points reveal that this object was probably close to maximum brightness at the time of the spectroscopic follow up. The mid-infrared data points from WISE and NEOWISE seem to follow the behaviour of the VVV photometry. The change in $W1$ of $\sim$1.1 magnitudes agrees with that expected from large changes in the accretion rate of a YSO \citep{2013Scholz}. The changes in colour from WISE and NEOWISE are also not consistent with variable extinction. The fact that the object is highly variable over a somewhat short period time (that could point to periodicity) is confirmed by the GLIMPSE I detection of the object. The star shows [3.6]=10.12 and [4.5]=8.81, which  are close to the maximum brightness observed during the 2010--2014 period. In addition VVVv717 is not detected in 2MASS nor DENIS, which makes the object fainter than at least K=15.3 at the epoch of the surveys ($\sim1999$). This is fainter than any of the data points measured during the coverage of VVV.




While the observed emission lines should be sufficient to rule out an
AGB star classification, we also test that possibility by estimating
the distance to such an object using the Akari method, which is further explained in Appendix \ref{vvv:sec_evolved}. VVVv717 was not detected at 9~$\mu$m by the Akari Survey, so we use the 8~$\mu$m measurement from 
{\it Spitzer}/GLIMPSE I instead. This should give similar results because the two passbands are fairly similar (6.0--11.0~$\mu$m for Akari vs 6.5--9.3~$\mu$m for {\it Spitzer}).
VVVv717 shows [8.0]=6.8 mag or 0.12 Jy. We use the $K_{\rm s}-[8]$ colour in equation A4 of \citet{2011Ishihara} to derive the mass loss rate of an OH/IR or carbon star with similar colours to VVVv717.

This yields $\dot{M}=10^{-4.9}$~M$_{\odot}$~yr$^{-1}$. According to equation A3 of \citet{2011Ishihara} an OH/IR or carbon star with this mass loss rate and 8 $\mu$m flux, would be located at $70^{+41}_{-27}$ kpc from the Sun, where the large uncertainties reflect the 
large uncertainty in the $K_{\rm s}-$[8] colour based on non-contemporaneous data as well as the scatter in the distance vs colour 
relation. This distance would place VVVv717 well outside the Galactic disc. We have checked if extinction could explain the red colour of VVVv717; \citet{1998Schlegel} maps give A$_{Ks}$=2.16 magnitudes at the coordinates of the object, which implies E(K$_{\rm s}$-[9])$\sim$1 mag. This value is within the uncertainties assumed in the $K_{\rm s}-$[8] colour, and it is not large enough to put the object within the Galactic disc. We conclude that an AGB star interpretation is very unlikely.

The near kinematic distance derived from the radial velocity of VVVv717, of $d=6.1$~ kpc, also contradicts the large distances derived from the AGB scenario.


We conclude that this star is very likely an eruptive YSO. However, it is also impossible to place it within the original classification for this type of stars.

\item {\bf VVVv631}, also in an area rich in star formation, went into outburst during 2010--2011 and has remained in outburst 
since that time, though perhaps beginning to fade in 2015. It has $\Delta K_{\rm s}\sim2.5$~magnitudes (Fig. \ref{vvv:erupfour2a}). This star was detected in 2MASS with $K_{\rm s}=14.11\pm0.12$, suggesting the object had been in a quiescent state prior to the VVV coverage. The 2014 data show that the object was at a bright state during our follow-up. The spectrum shows Br$\gamma$ (2.16 $\mu$m) and Pa $\beta$ (1.28 $\mu$m) emission as well as Na I, CO  and possible Mg I (1.502 $\mu$m) emission. The $\Delta v$=3--1 CO bandhead is more prominent than the $\Delta v$=2--0 bandhead (this is confirmed in the recent spectrum from the 2015 run, Contreras Pe\~{n}a et al., in prep). This object also shows blueshifted 1.08 $\mu$m He I absorption, probably from a wind. This feature is blueshifted by $\sim2.5$~\AA~ ($-70$~km s$^{-1}$), but with a velocity profile extending to $\sim-220$~km s$^{-1}$. This feature has been observed in FUors and EXors, but on its own is not a discriminant of FUor-like activity \citep{2012Fischer}. 

It is very interesting to see that the object displays absorption from H$_{2}$O. The object IRAS 06297+1021(W) in the sample of \citet{2010Connelley} shows a remarkably similar spectrum. In that case the authors had mixed feelings about classifying it as an FUor-like object due to its rich emission spectrum. However, the observed photometric behaviour of our object makes it a very strong candidate for being part of the eruptive variable class.

The fact that we are seeing both CO emission (assumed to arise from the innermost parts of the disc) and H$_{2}$O absorption would suggest that two separate regions are contributing to the near-infrared spectrum. H$_{2}$O is never seen in emission in the near-infrared spectra of YSOs, presumably because the innermost part of the disc is too hot for the molecule to remain bound. 


\item {\bf VVVv118} classified as a likely YSO by \citet{2008Robitaille}, VVVv118 is found 124\arcsec~ from IRAS 14477$-$5947, a high-mass YSO candidate \citep[see e.g.][]{2012Guzman}, and 94\arcsec~ from infrared bubble [SPK2012] MWP1G317420$-$005582 \citep{2012Simpson}. The star is detected in 2MASS with $K_{\rm s}=12.49\pm0.04$. Fig. \ref{vvv:erupfour1a}  shows that the object displays large variations ($>2$~mag) over short timescales (with changes occurring on timescales less than 100 days). This short timescale resembles that of objects in the STV category of Paper I. However, both the short term changes and the total amplitude ($\Delta K_{\rm s}=4$~mag)
are much larger than those of all other objects in that category, allowing us to definitely discard rotational modulation from hot/cold spots as the driving physical mechanism for the variability of VVVv118. In addition colour changes (bluer when fainter) seen in the two epochs of near-IR photometry disfavours the possibility of variable extinction, so the light curve was classified as eruptive in Paper I. This large $\Delta K_{\rm s}$ over short timescales is consistent with an eruptive variable of the EXor type. However, the spectrum (Fig. \ref{vvv:erupfour1a}) displays not only weak Br$\gamma$ emission but also a probable detection of
weak H$_{2}$O absorption, the latter being commonly found in FUor variables rather than EXors. The presence of CO absorption at 2.29~$\mu$m cannot be confirmed. Assuming that water absorption is present, this represents yet another example of eruptive variables in our sample showing a mixture of characteristics between those of EXors and FUors. 

\item {\bf VVVv800} is located in the vicinity of HII region RCW120 \citep[d$\sim$1.4 kpc,][]{2009Deharveng} and has already been classified as a YSO by \citet{2010Martins} (source 76 of YSOs around RCW120). The latter have obtained a near-infrared spectrum of the object, which shows a lack of absorption features and only has H$_{2}$ emission at 2.03, 2.12, 2.22 $\mu$m. Our observation of this object (Fig. \ref{vvv:erupfour2b}) shows a similar behaviour, with several transitions from molecular hydrogen present, as well as Br$\gamma$ emission. This object also shows strong emission at 1.64 $\mu$m from [Fe II]. The variable star seems to be driving a powerful outflow. VVV 2014 photometry shows that our observations were taken close to the maximum brightness of the object (see Fig. \ref{vvv:erupfour2b}). We note also that this object is detected in both 2MASS and DENIS with $K_{\rm s}=10.82\pm0.03$ and $K=11.02\pm0.08$ respectively, suggesting that this object had gone into a bright state prior to our VVV coverage. 
 The accretion luminosity for this object is at the low end of L$_{acc}$ for eruptive VVV objects (see Table \ref{table:vvvaccrate}), which is not expected given that this object was observed close to the brightest point in the light curve. The absence of CO emission may cast some doubt on its eruptive nature but this could be explained by high extinction towards the inner disc region, e.g. due to an edge-on disc. Such a geometry would also be expected to veil the Br$\gamma$ emission 
line: as noted earlier we should not expect the individual estimates from
this line to be accurate in every case. We note that the spectrum of VVVv800 resembles that of the eruptive variable GM Cha, an object with a red-rising featureless continuum dominated by H$_{2}$ emission \citep[see e.g.][]{2007Persi}.
The long duration of the bright state appears to rule out rotational modulation by bright or dark spots and the colour changes (bluer when fainter) seen in the two epochs of $JHK_s$ colour data disfavours extinction as the cause of variability, leaving only
accretion as the likely cause of variability. Given that the light curve is well described as a $\sim 1$~mag outburst with additional short timescale scatter, this object may be regarded at the low end of what might be termed an eruptive variable. \cite{2015Rice} observed lower level accretion variations in some YSOs on timescales of years, as well as shorter timescales. Detection of objects such as VVVv800 perhaps indicates that there is no clear distinction in amplitude between normal low level variations in accretion rate and episodic 
accretion. The other eruptive variables in our sample that were observed in
a bright state have higher derived accretion luminosities and amplitudes.

\item {\bf VVVv815} This object was discovered as a high amplitude variable in an early data release of VVV 2010 data and is not selected in the 2010--2012 analysis (see Paper I). It was found to be an intrinsically red object from mid-IR colours and classified as a likely YSO by \citet{2008Robitaille}. The star is likely associated with an SFR (see Fig. \ref{g314im1}), and its projected location is 79\arcsec ~from HII region [WHR97] 14222$-$6026 \citep[V$_{LSR}$=$-$44 km s$^{-1}$, d$\sim$2.5 kpc][]{1997Walsh}. Its SED is consistent with a class I object ($\alpha=1.58$).

  VVVv815 was observed in low-resolution with FIRE in May 2012. The spectrum shows no flux in the $J$ or $H$ bands, and is dominated by strong emission of $\nu=1$--$0~ \Delta J=-2$ rovibrational transitions from molecular hydrogen , with the 1--0 S(1) 2.12 $\mu$m being the most prominent feature (EW$=-200$\AA). The lack of higher excited lines in the spectrum of this object points to the emission arising from molecular shocks. 
  
  The 2013 medium resolution echelle spectrum from FIRE shows very similar characteristics (see Fig. \ref{g314im2}), although the equivalent width of the 1--0 S(1) 2.12 $\mu$m decreased to $-$119~\AA. The higher quality of this spectrum allows us to derive a ratio 1--0 S(1)/2--1 S(1) of $\sim6$, which agrees with models of emission arising from molecular shocks \citep[see e.g.][]{1995Smith}.

\begin{figure}
\centering
\subfloat{\resizebox{\columnwidth}{!}{\includegraphics{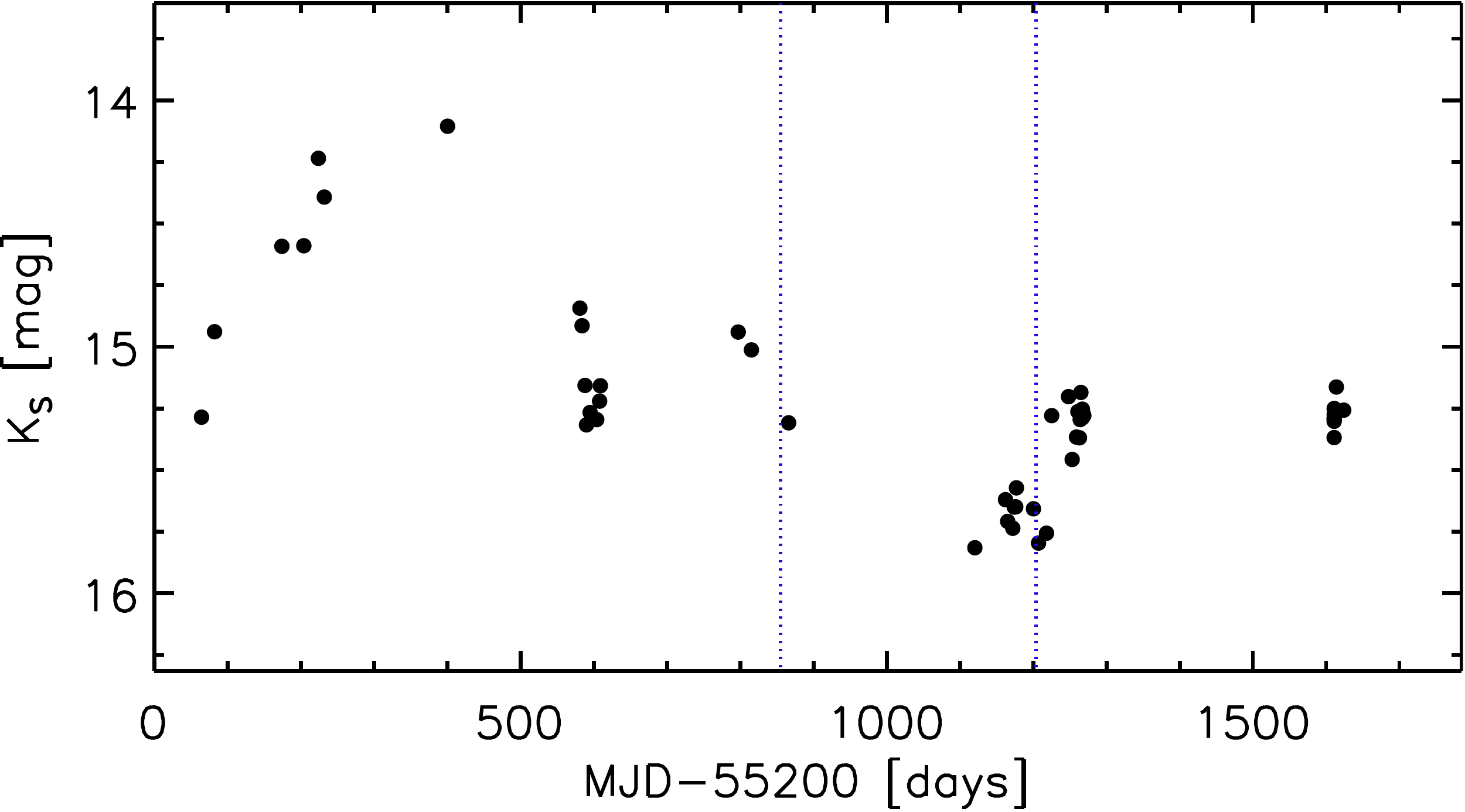}}}\\
\subfloat{\resizebox{\columnwidth}{!}{\includegraphics{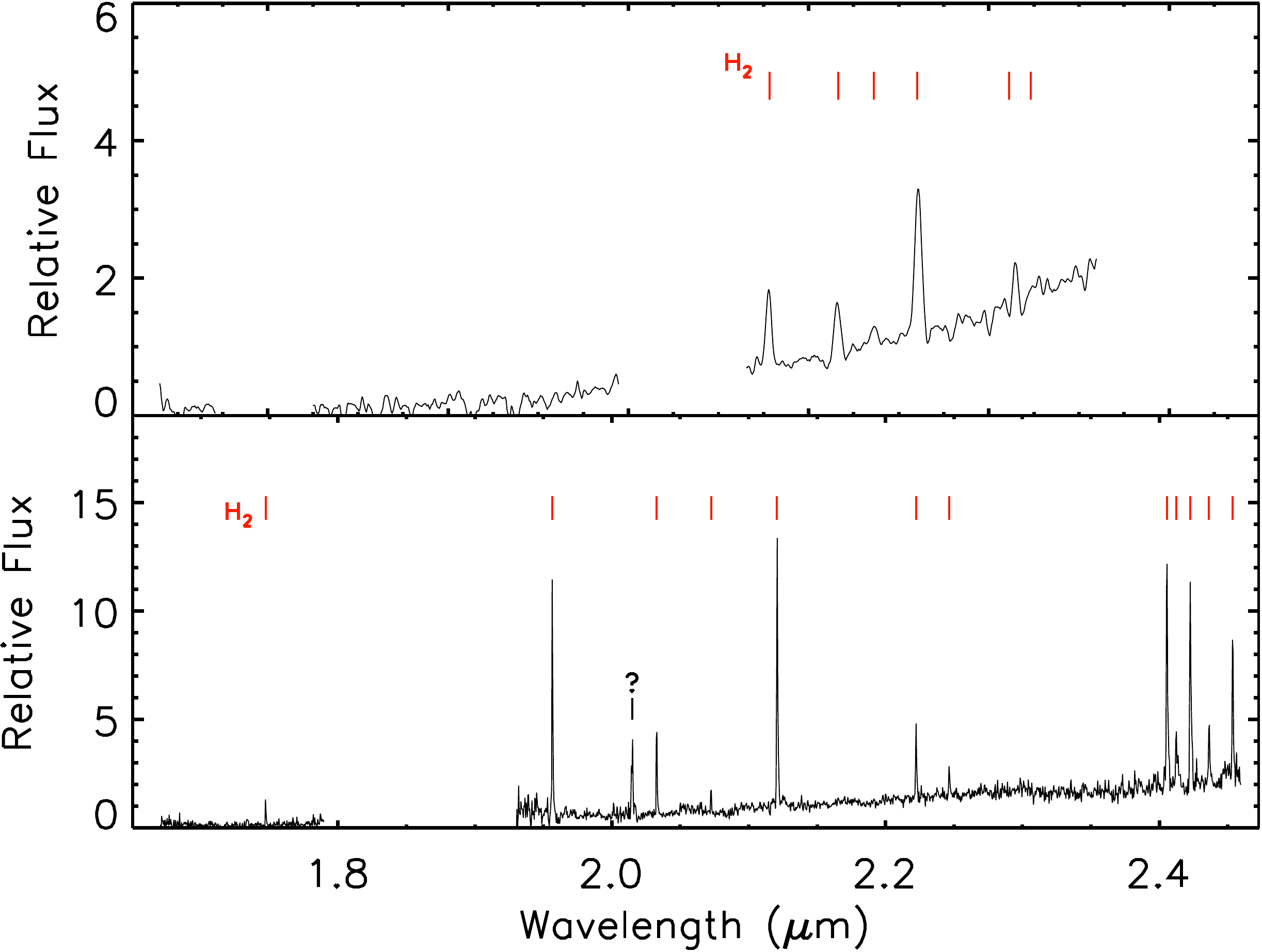}}}
\caption{(top) $K_{\rm s}$ light curve of VVVv815, where the arrows mark the dates of the 2012 and 2013 spectroscopic follow-up. (middle) 2012 FIRE low resolution spectrum of VVVv815. (bottom) 2013 FIRE high resolution spectrum of VVVv815. In both spectra the observed rovibrational transitions of H$_{2}$ are marked at the top of each graph. }
\label{g314im2}
\end{figure}


\item {\bf VVVv699} \citet{2008Robitaille} finds the {\it Spitzer} IRAC colours of this object to be consistent with those of a YSO, and it is thus classified as such in their study. The automatic search for mid-infrared detections within 1\arcsec~ of the source only yields a GLIMPSE I \citep{2003Benjamin} detection. Inspection of the images arising from WISE show the possible presence of VVVv699 (Fig. \ref{d104vc12im1}) and there is one detection at 1.3\arcsec~ from the VVV source. Comparison of the magnitudes from WISE $W1,W2$ with {\it Spitzer} I1,I2 shows an agreement between the two measurements, which suggests that VVVv699 had been in quiescence previous to the 2011 outburst. This is also supported by the fact that the source falls below the detection limit in both 2MASS and DENIS surveys (see Fig. \ref{d104vc12im1} ). Thus use of {\it Spitzer} photometry to build the SED of VVVv699 is preferred given that the centroid of the detection is closer to the VVV position. Our analysis shows the object as a class I YSO, with $\alpha=2.08$.

The light curve (Fig. \ref{d104vc12im2}) shows an outburst between the 2010 and 2011 campaigns of VVV, with $\Delta K \sim 2.2$~mag. The fact that the object had been on a quiescent state prior to 2010 seems to be supported by mid-IR information (see above). In addition, the object seems to be displaying some shorter-term, lower-amplitude variability ($\Delta K < 1$~mag) during 2012--2013.

Fig. \ref{d104vc12im2} shows that VVVv699 is located within 300\arcsec ~of several indicators of active star formation, such as likely YSOs from \citet{2008Robitaille}, 6 IRDCs, and is within the broken GLIMPSE infrared bubble [CPA2006] S43 \citep{2006Churchwell}. These infrared bubbles are expected to be formed around hot young stars in massive SFRs \citep{2006Churchwell}. Unfortunately none of the indicators have available distance information. The radial velocity of the object (see Table \ref{table:vvvrvel}) suggests that VVVv699 is located at  a Heliocentric distance of 4.9 kpc.


The 2013 follow-up (Fig. \ref{d104vc12im1}) reveals a rich emission line spectrum, with strong H$_{2}$ lines from several transitions and Br$\gamma$ emission. CO and Na I are also in emission. CO and Na I emission are usually attributed to a hot inner disc at a distance of a few au from the central star. The observations from 2014 (Fig. \ref{d104vc12im2}) show a decrease in the strength of several lines, most noticeable in the CO and Na I lines, with also an apparent decrease in the intensity of the Br$\gamma$ emission. The H$_2$ emission on the other hand seems to increase, an effect that is apparent in all of the observed transitions. We do note that the changes in Br$\gamma$ and H$_{2}$ are within the measurement errors and do not appear to be larger than $10\%$. 

\begin{figure}
\centering
\resizebox{\columnwidth}{!}{\includegraphics{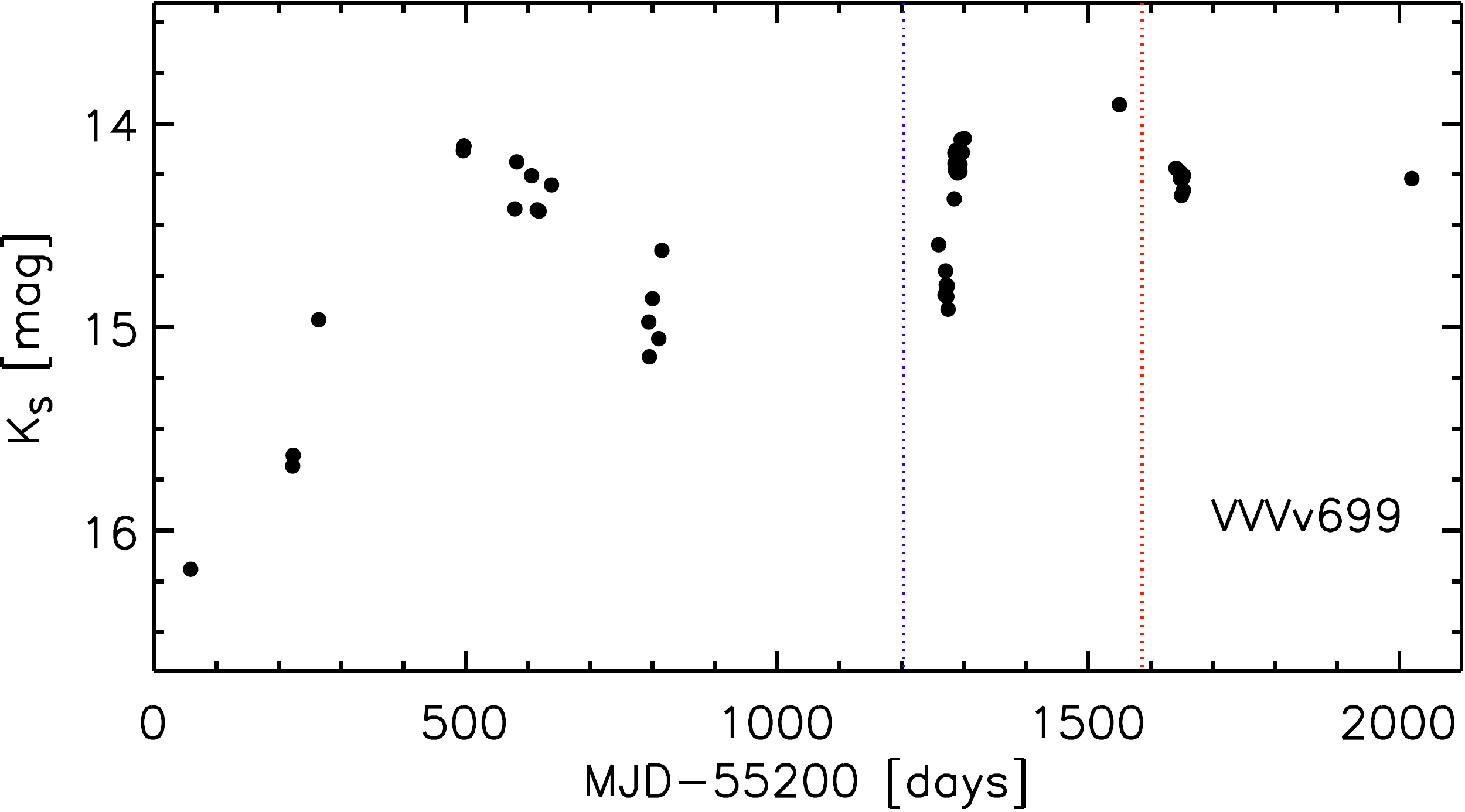}}\\
\resizebox{\columnwidth}{!}{\includegraphics{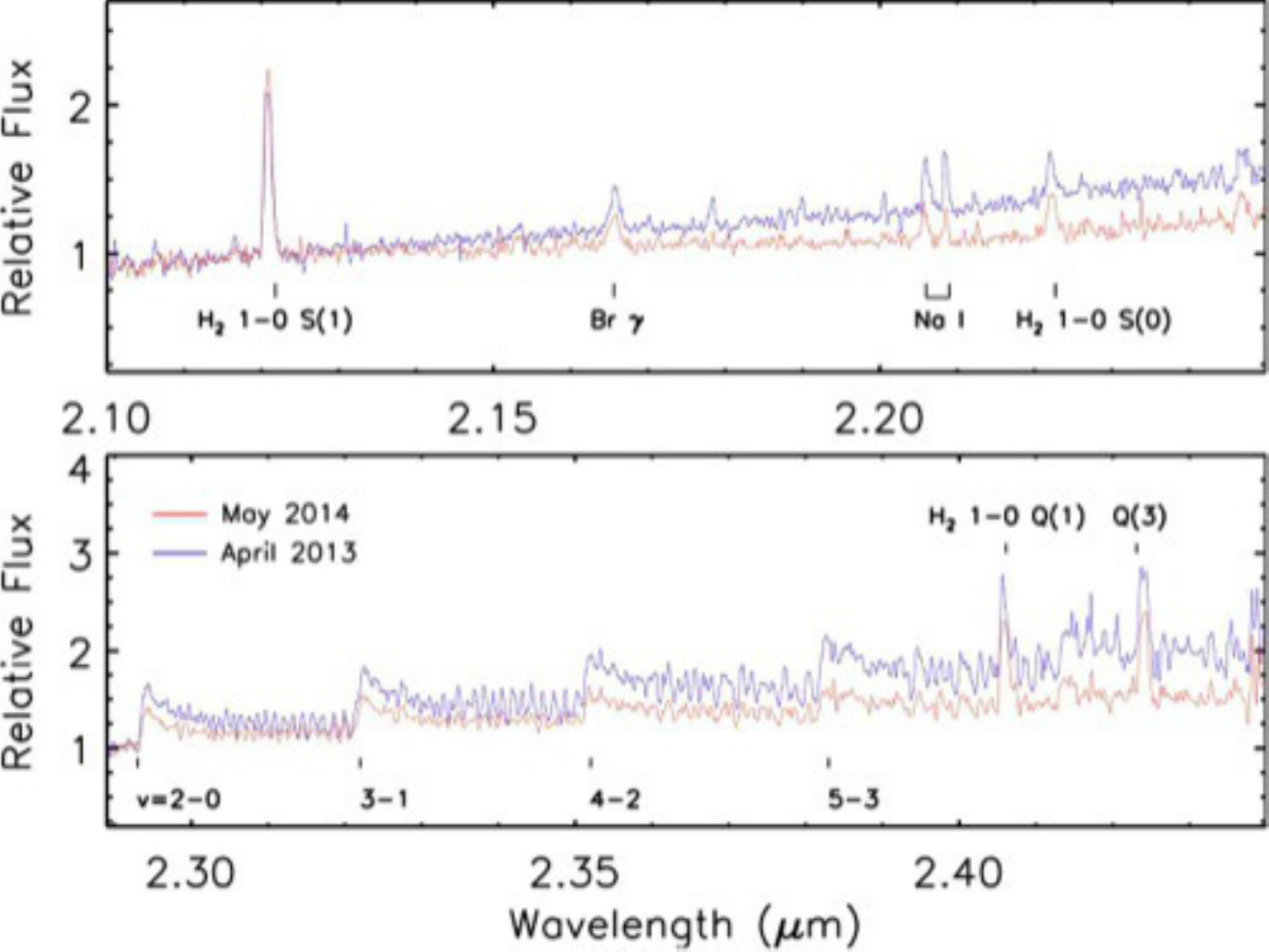}}
\caption{(top) $K_{\rm s}$ light curve of VVVv699. The dashed lines in the graph mark the dates of the April 2013 (blue) and May 2014(red) FIRE observations. (middle) Graph comparing the 2013 (red) and 2014 (blue) FIRE spectra of VVVv699 in the region 2.1--2.25 $\mu$m. Emission lines observed in this object are marked in the graph. (bottom) Similar comparison to the previous graph, but in the region  2.29--2.42 $\mu$m.}
\label{d104vc12im2}
\end{figure}

\citet{2014Connelley} have studied the spectroscopic variability of a sample of class I YSOs. They have found that tracers of mass accretion, Br$\gamma$ and CO emission, are variable at all timescales, with CO being most variable (up to a factor of 3) on a timescale of 1--3 yr which corresponds to a spatial scale of within 1--2 au in a Keplerian disc \citep{2014Connelley}. These authors also find that changes in mass flow onto the central star, as traced by Br$\gamma$, are coupled to changes in the disc surface temperature (which give rise to the CO emission), whenever there is a large change in the CO emission and on long timescales. The observed variability in VVVv699 over a period of 1 year, seems to support the latter. We observed a large variation of up to a factor of 2 in CO emission, accompanied by an apparent change in Br$\gamma$ emission. However, we cannot rule out the possibility that changes in EWs are affected by changes in 
veiling.

Unfortunately, we are not able to estimate whether this relates to a change in the accretion rate between the two epochs. We do not have contemporaneous photometry to the spectroscopic measurements, and as we saw before there seems to be a short-term variability with $\Delta K \sim 0.5$~magnitudes. Thus, trying to assign a K band brightness to the dates of the spectroscopic observations from the light curve of VVVv699 would yield unreliable results. Most importantly, we have already stated the difficulties in deriving this parameter in Section \ref{vvv:erupvars}. 


We can estimate a range of values for the accretion luminosity from the 2013 observations. Following Section \ref{vvv:erupvars} we can derive the flux of the Br$\gamma$ emission, F$_{br\gamma}$, using the measured equivalent widths, the extinction $A_{V}$ to the system and the continuum flux from K band photometry.


In Section \ref{vvv:erupvars} we estimated the visual extinction to the source by adopting a colour $J-H$=3.5, which yields A$_V$=23. This can be considered as a lower limit of A$_{V}$. In this section we attempt to derive this parameter with a different method.  

The visual extinction to the source can be estimated using the H$_{2}$ 1--0 Q(3)/S(1) ratio, $r_{obs}$, since these lines originate in the same upper state. We compare the observed ratio to the intrinsic one \citep[0.7, see ][and references therein]{2007Beck}, and derive $A_{V}$ assuming the \citet{1989Cardelli} extinction law. Then,

\begin{equation}
0.404A_{V}\left[\left(\frac{1}{\lambda_{Q(3)}}\right)^{1.61}-\left(\frac{1}{\lambda_{S(1)}}\right)^{1.61}\right]=\log\left(\frac{0.7}{r_{obs}}\right)
\end{equation}

\noindent where we estimate A$_{V}=42$ and A$_{V}=23$ magnitudes for 2013 and 2014, respectively. This change in A$_{V}$ would imply $\Delta K_{\rm s}\sim2.2$ magnitudes, which seems highly unlikely from the light curve of this object. We note that \citet{2010Connelley} argue against this ratio as a reliable measurement of the visual extinction, given the large discrepancies that they found between the values using this method and those found through either photometry or modelling of the continuum. They suspect that the closeness of the Q(3) line to a telluric absorption feature is likely to affect the observed ratio. A$_{V}$ is also found to be variable in their sample. The \citet{1998Schlegel} maps yield an upper limit of the visual extinction of A$_{V}=35.3$ magnitudes \citep[30.8 magnitudes in the revised values of][]{2011Schlafly}. 


\begin{figure}
\centering
\resizebox{\columnwidth}{!}{\includegraphics{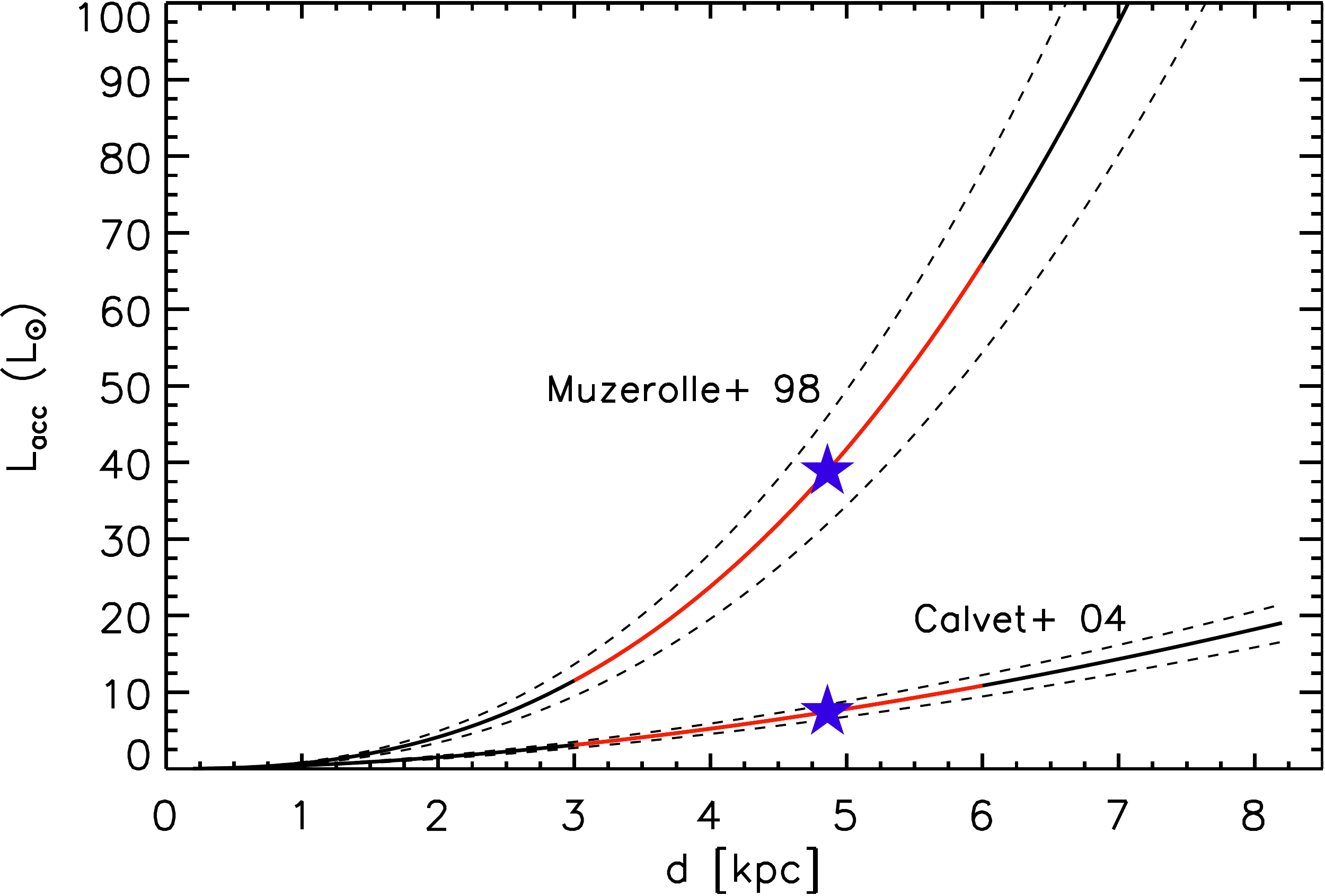}}
\caption{Accretion luminosity of VVVv699, estimated from the observed flux of the Br$\gamma$ line in the 2013 observations, and for the distance range of infrared bubbles from \citet{2006Churchwell}. L$_{acc}$ is derived from the \citet{2004Calvet} and the \citet{1998Muzerolle} relations. The red line marks the estimated L$_{acc}$ for $3<d<6$ kpc. Errors on L$_{acc}$ are shown as dashed lines. The accretion luminosity estimated using the distance from radial velocities is marked by a blue star.}
\label{d104vc12im3}
\end{figure}


We then assume $K=14.1$, which corresponds to the approximate mean brightness in the 2013 epochs from VVV. From this we obtain the continuum level, F$_{\lambda}$ for the Br$\gamma$ line at 2.1659 $\mu $m. In addition, we correct this value for the extinction along the line of sight to the source, assuming $A_{V}=32$ ( a simple mean of the values obtained using line ratios). The line flux is then estimated following Section \ref{vvv:erupvars}. We finally obtain $F_{Br\gamma}=(7.54\pm1.08)\times10^{-15}$ erg s$^{-1}$ cm$^{-2}$.






In order to estimate the line luminosity we take a range of values for the distance of VVVv699 between 0.2--8.2 kpc, which corresponds to the distribution of near kinematic distances of Galactic bubbles in \citet{2006Churchwell}. The accretion luminosity is derived  from the \citet{2004Calvet} and \citet{1998Muzerolle} relations. We obtain a large range in L$_{acc}$, up to 20 L$_{\odot}$ and $>100$~L$_{\odot}$ from the \citeauthor{2004Calvet} and \citeauthor{1998Muzerolle} relations, respectively. The peak distribution of the \citeauthor{2006Churchwell} sample with known distances is found to be at $d=4.2$~kpc, a similar distance to that estimated from the radial velocity of the object, with the majority of the sample being located between 3--6 kpc. If we assume the latter range for the distance of [CPA2006] S43, then the maximum luminosities derived from the two relations reduce to 10 and 60 L$_{\odot}$ respectively. Using the near kinematic distance from radial velocities yields L$_{acc, Calvet}=7.4$~L$_{\odot}$ and L$_{acc, Muzerolle}=38.9$~L$_{\odot}$. These values are also higher than the ones obtained in section \ref{vvv:erupvars}. This is due to the lower value of A$_{V}$ estimated from dereddening the $J-H$ colour of VVVv699 to the T Tauri locus in section \ref{vvv:erupvars}. However, this particular value might not represent the true A$_{V}$ to the system as VVVv699 is not detected in $J$ nor $H$. Nevertheless, these values of L$_{acc}$ support the object being in a high state of accretion.

\end{itemize}

\clearpage

\section{Plots for Individual Objects}

\begin{figure*}
\centering
\subfloat{\resizebox{0.48\textwidth}{!}{\includegraphics{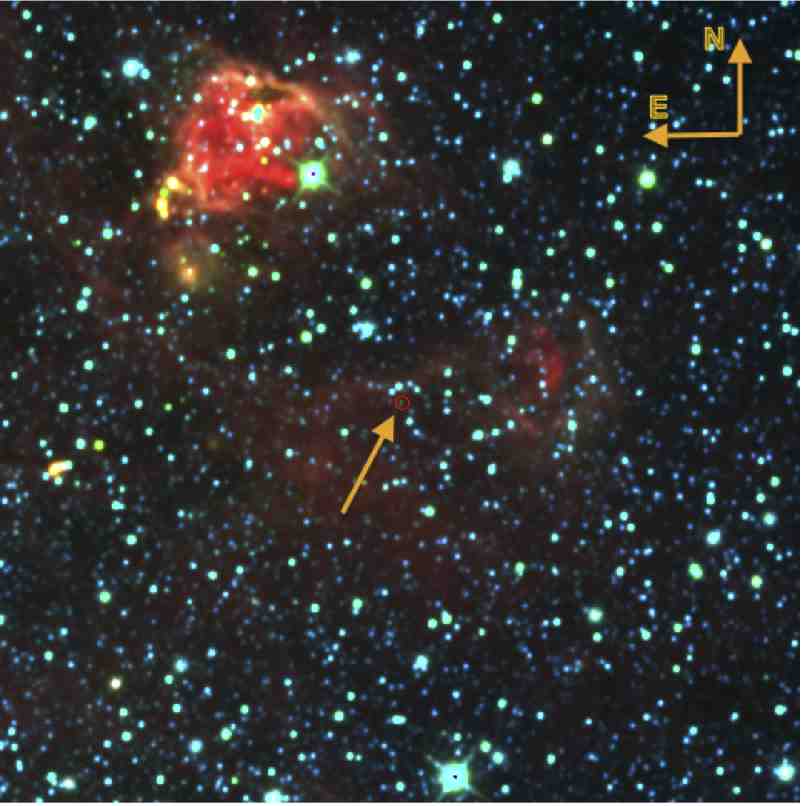}}}
\subfloat{\resizebox{0.48\textwidth}{!}{\includegraphics{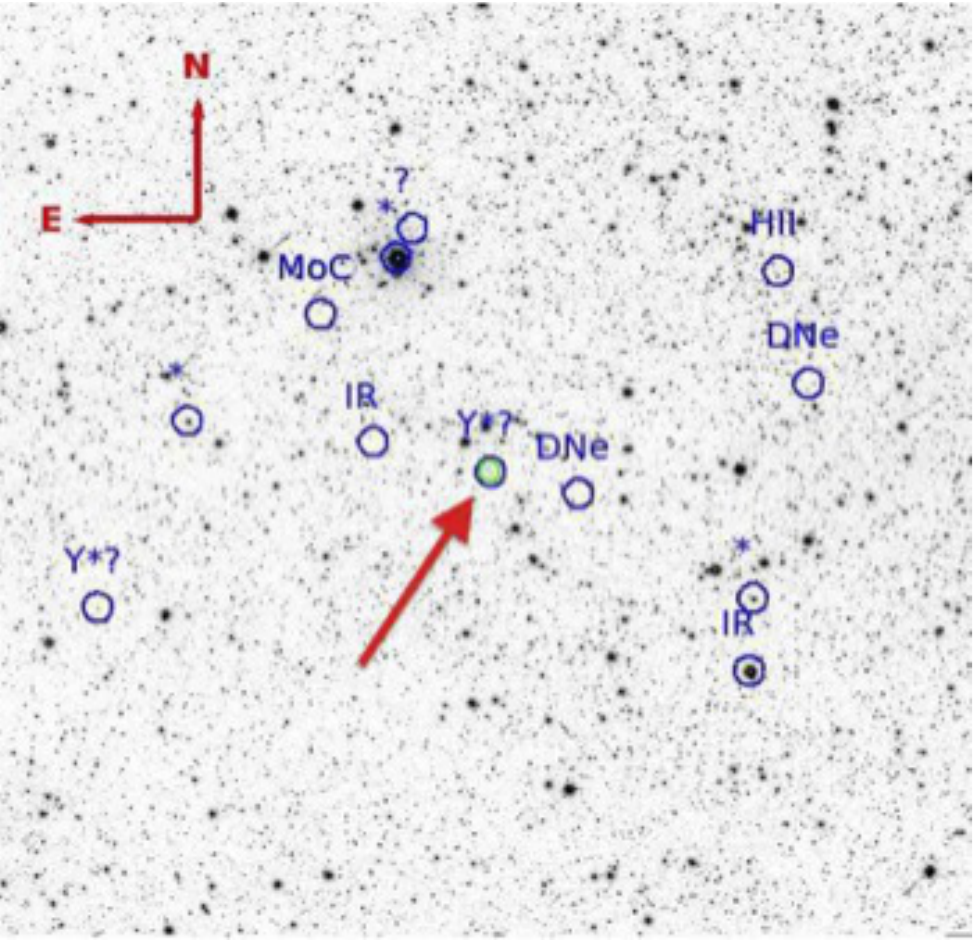}}}\\
\subfloat{\resizebox{0.48\textwidth}{!}{\includegraphics{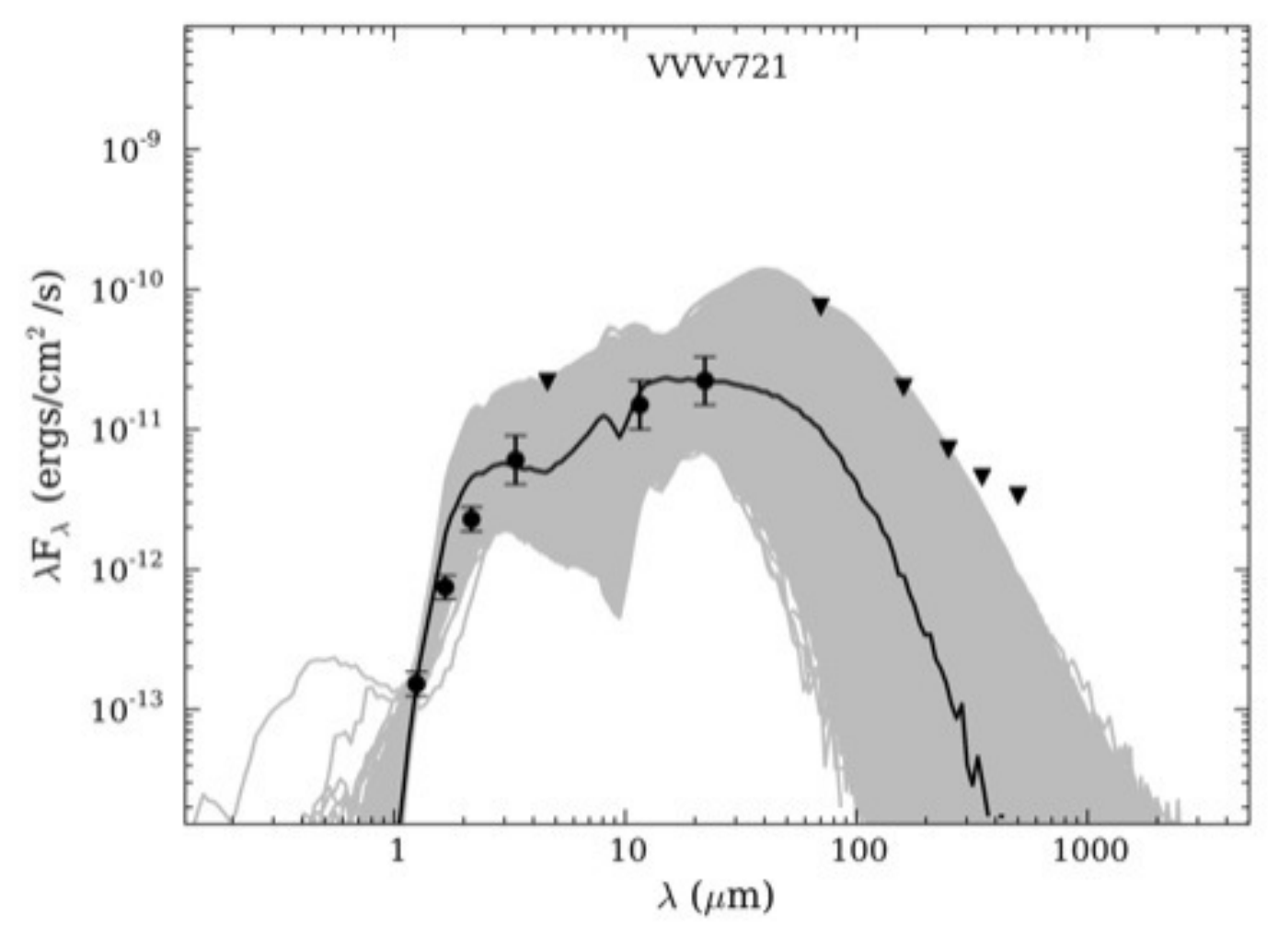}}}
\subfloat{\resizebox{0.48\textwidth}{!}{\includegraphics{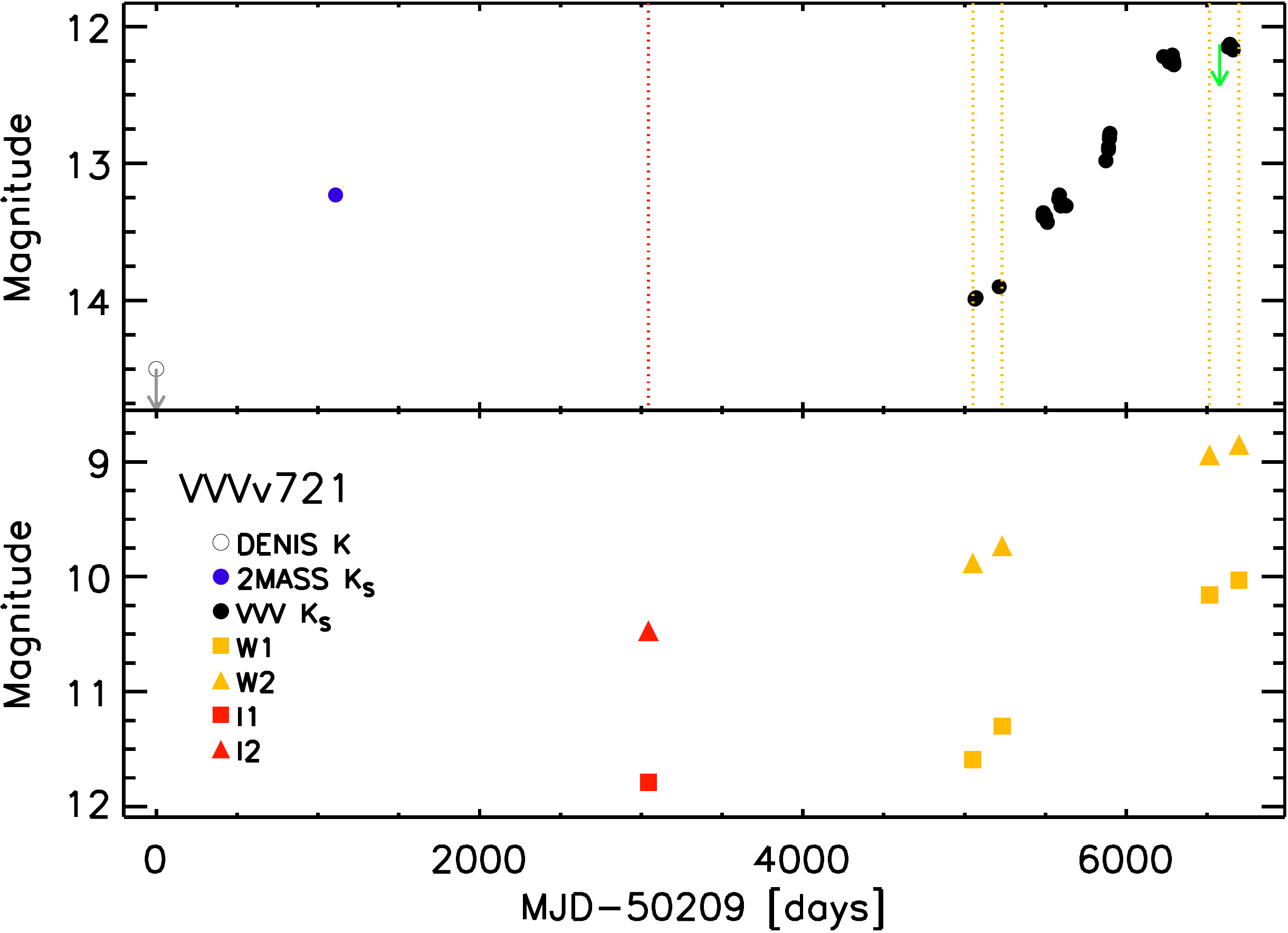}}}\\
\subfloat{\resizebox{0.65\textwidth}{!}{\includegraphics{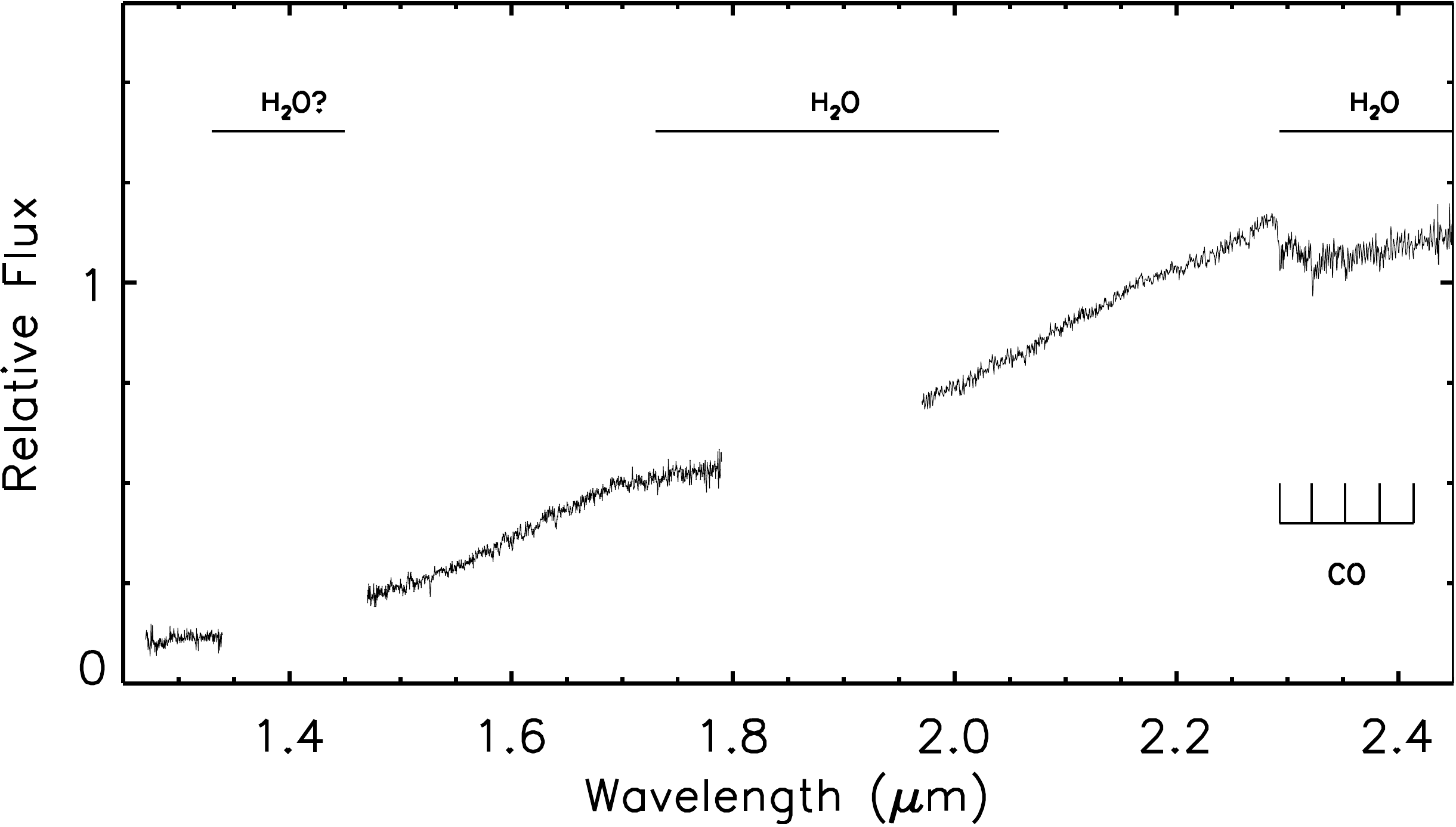}}}
\caption{(top left) False colour WISE image (blue=3.5 $\mu$m, green=4.6 $\mu$m, red=12 $\mu$m) of a 0.44$^{\circ}\times0.44^{\circ}$ area centred on VVVv721. The location of the object is marked by the arrow. (top right) $K_{\rm s}$ image of a 10\arcmin$\times$10\arcmin ~area centred on VVVv721. The location of the object is marked by the red arrow and the green circle. In addition, blue circles and labels mark objects found in a SIMBAD query with a 5\arcmin ~radius. (middle left) SED of VVVv721 along with \citet{2007Robitaille} YSO models that fulfil the criteria of $\chi^{2} - \chi^{2}_{best} < 3N $, with N the number of data points used to generate the fits. This image is generated by the fitting tool. (middle right) Near-infrared ($K_{\rm s}$) and mid-infrared photometry of VVVv721. Data arising from different surveys are marked by different symbols which are labelled in the figure. The green arrow marks the time when the spectrum was taken. (bottom) FIRE spectrum of VVVv721. The spectroscopic features found for this object are marked in the graph.}
\label{d107vc13}
\end{figure*}

\begin{figure*}
\centering
\subfloat{\resizebox{0.48\textwidth}{!}{\includegraphics{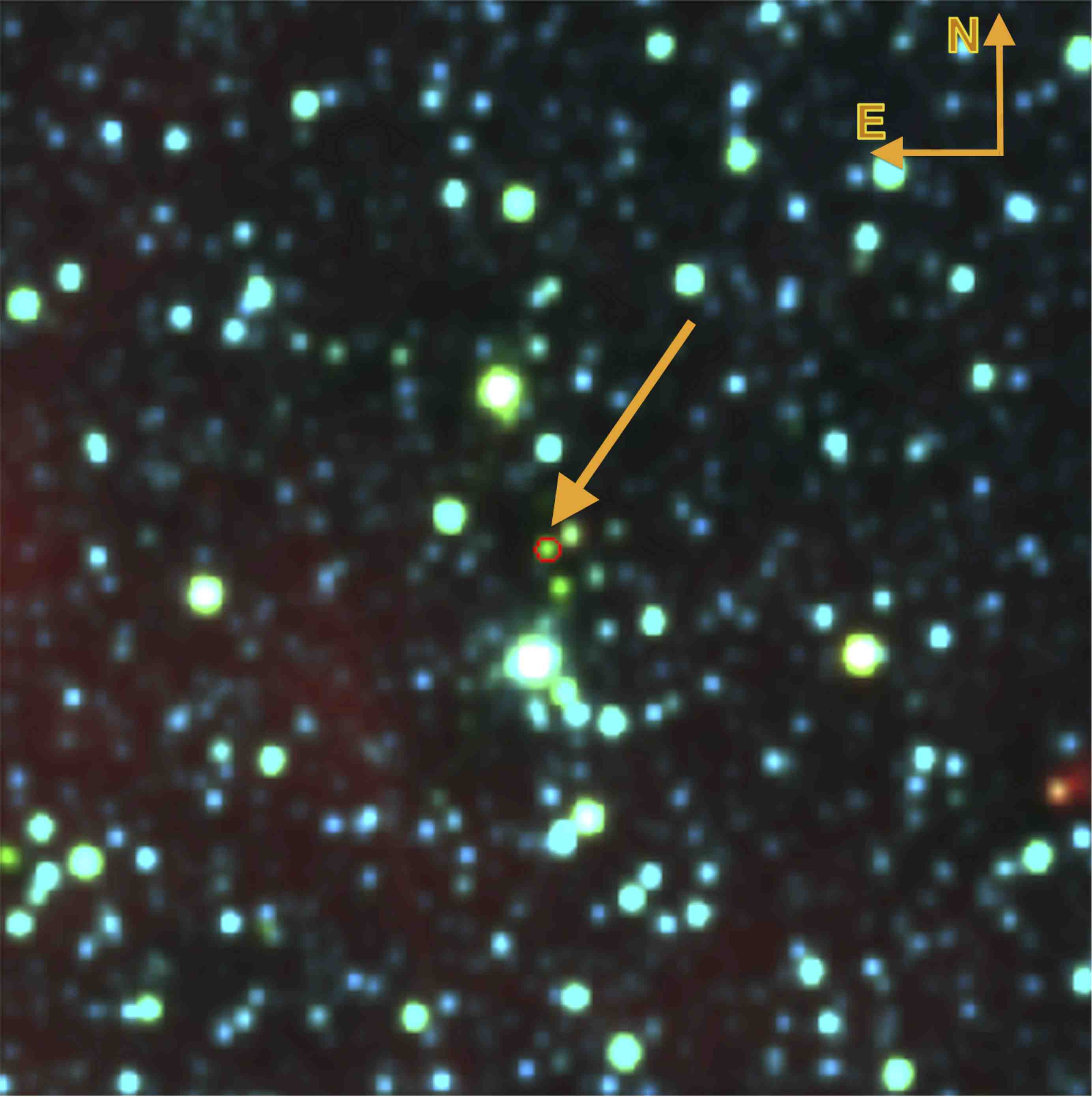}}}
\subfloat{\resizebox{0.48\textwidth}{!}{\includegraphics{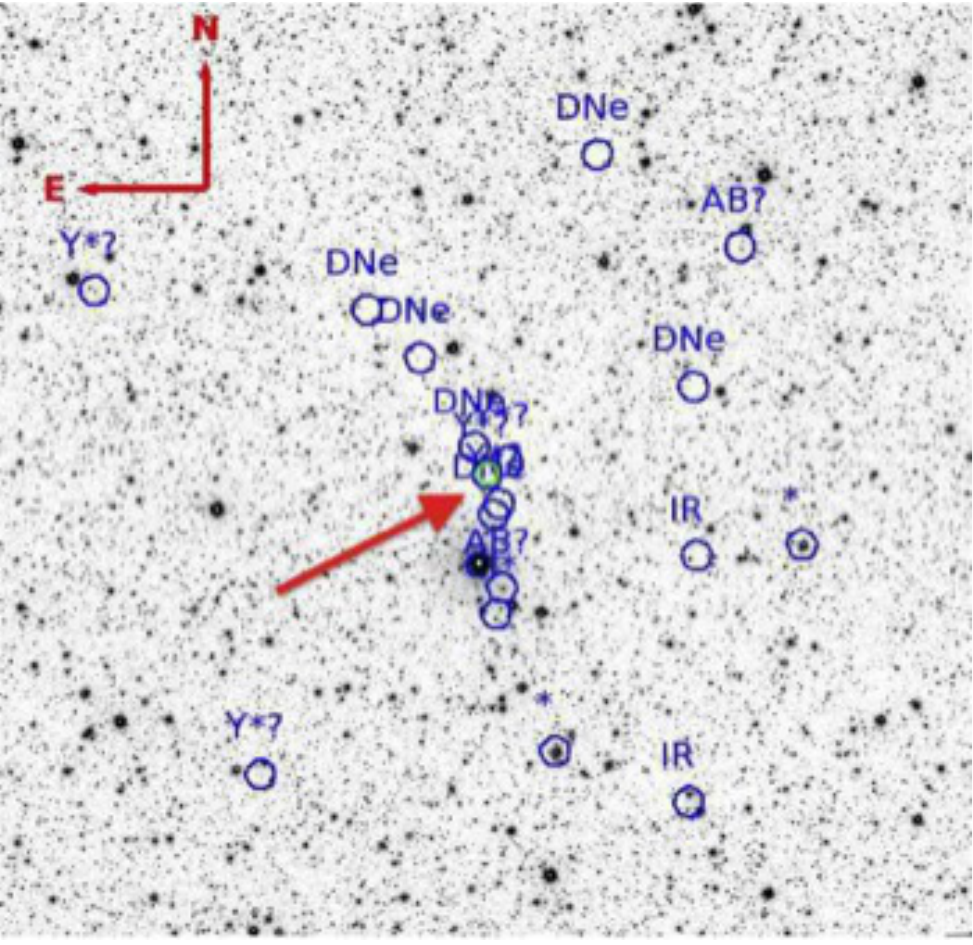}}}\\
\subfloat{\resizebox{0.48\textwidth}{!}{\includegraphics{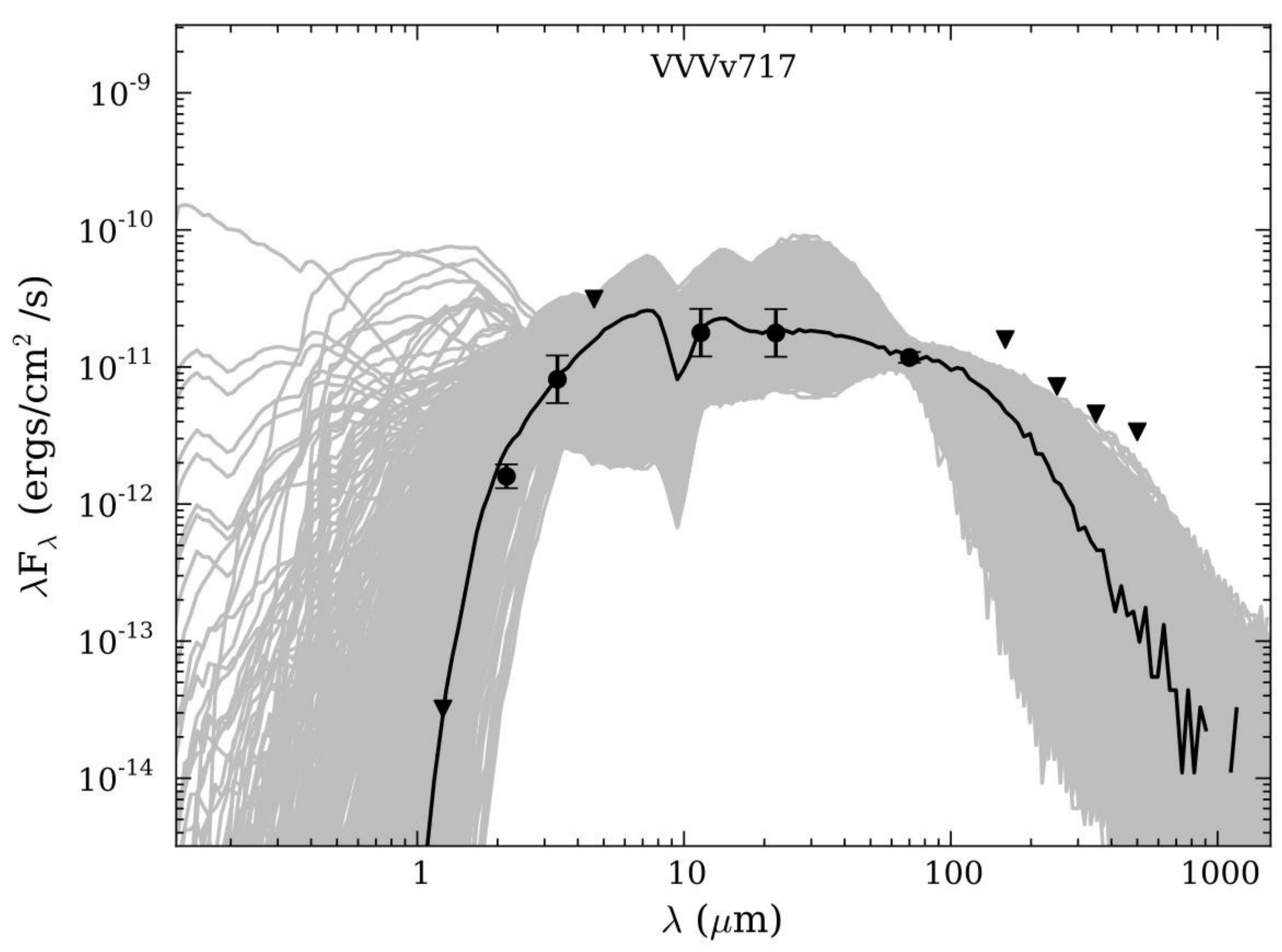}}}
\subfloat{\resizebox{0.48\textwidth}{!}{\includegraphics{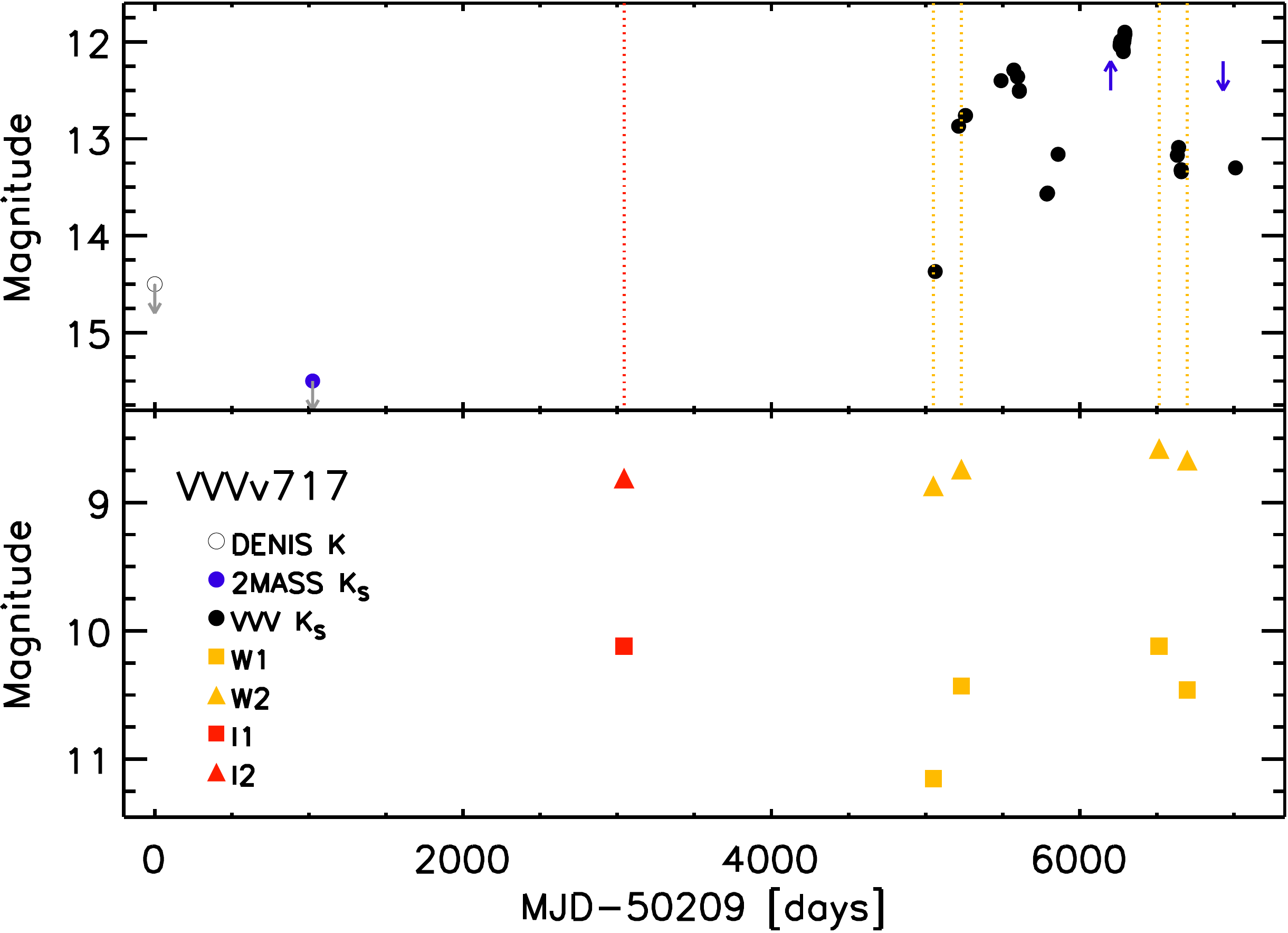}}}\\
\subfloat{\resizebox{0.65\textwidth}{!}{\includegraphics{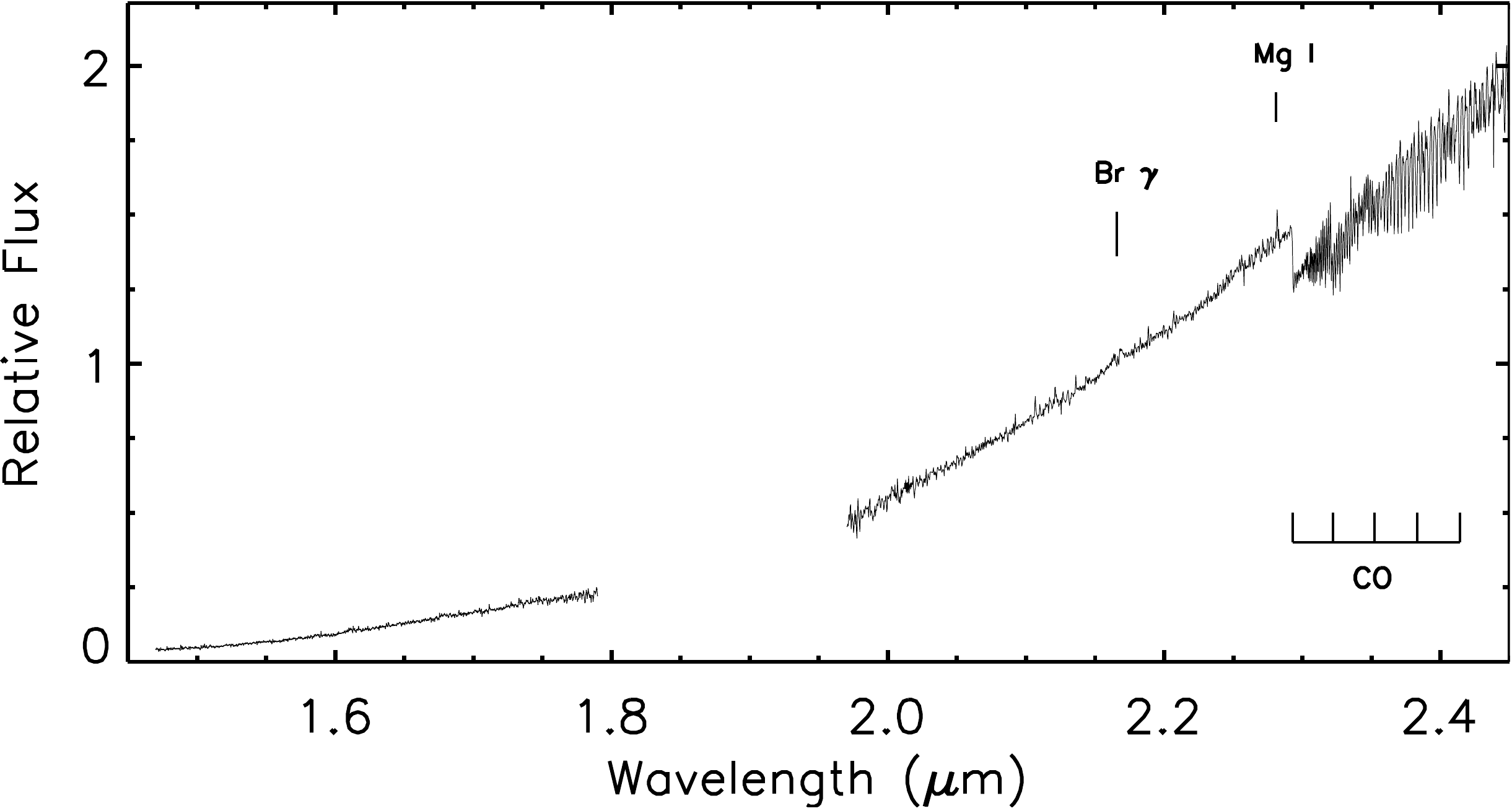}}}
\caption{Same as Fig. \ref{d107vc13}. In the top left we present a 10\arcmin$\times$10\arcmin ~WISE false colour image of the area near VVVv717.}
\label{d106vc30}
\end{figure*} 

\begin{figure*}
\centering
\subfloat{\resizebox{0.48\textwidth}{!}{\includegraphics{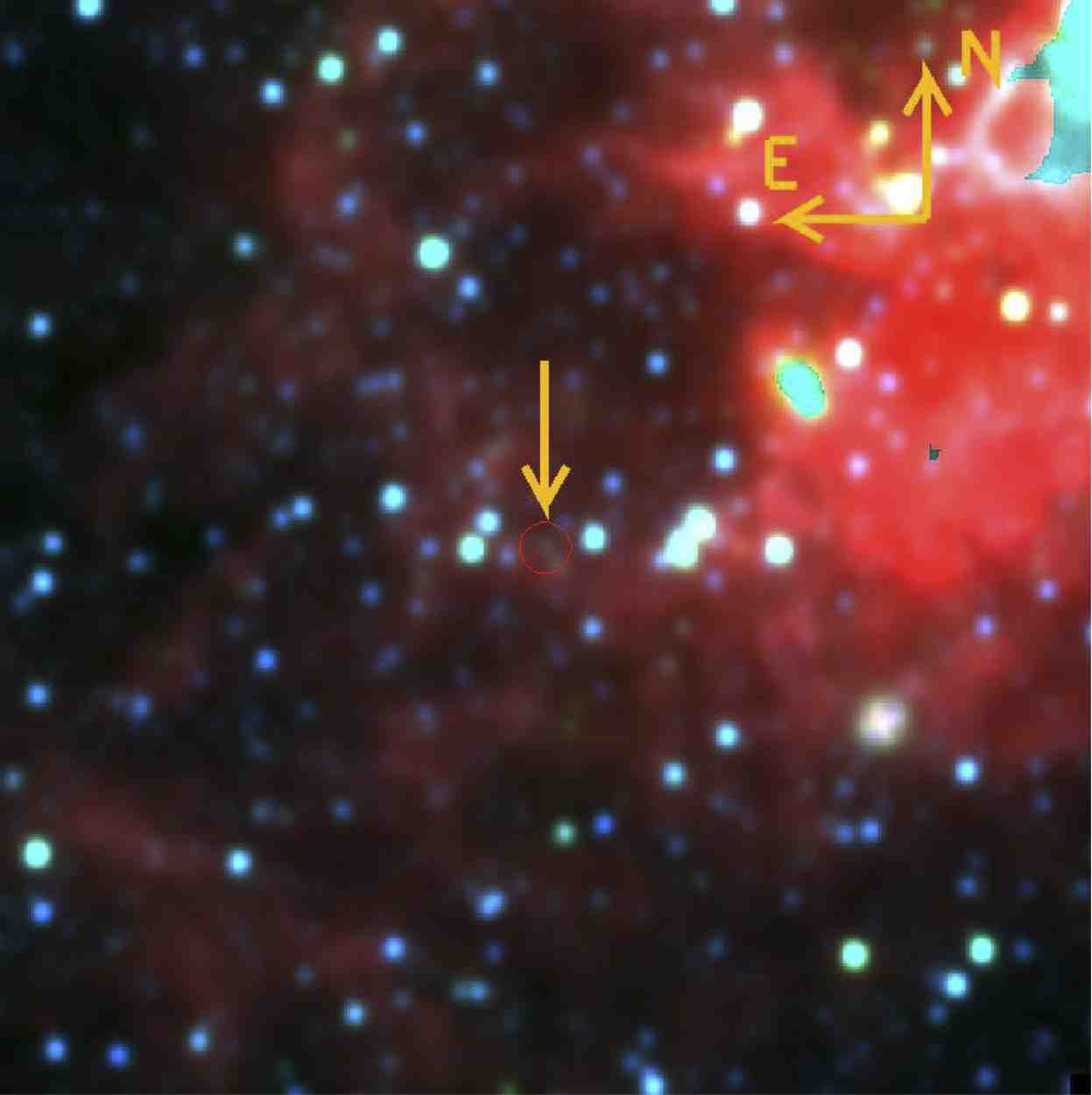}}}
\subfloat{\resizebox{0.48\textwidth}{!}{\includegraphics{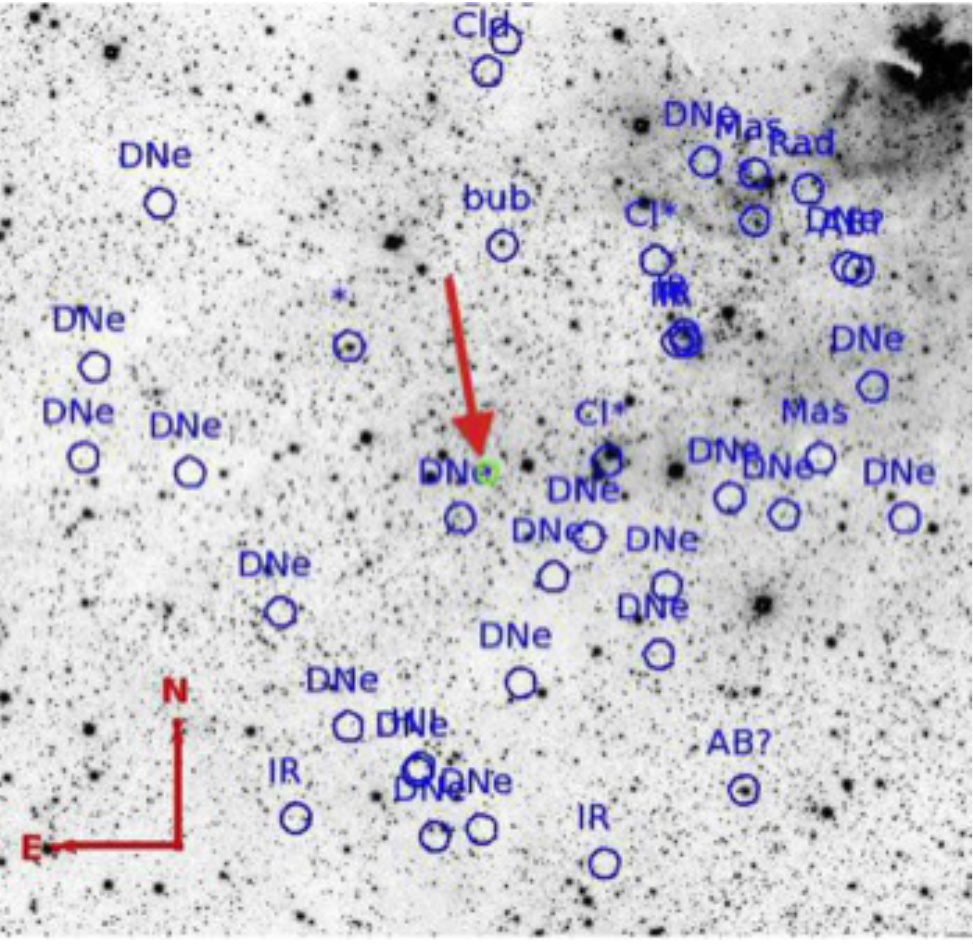}}}\\
\subfloat{\resizebox{0.48\textwidth}{!}{\includegraphics{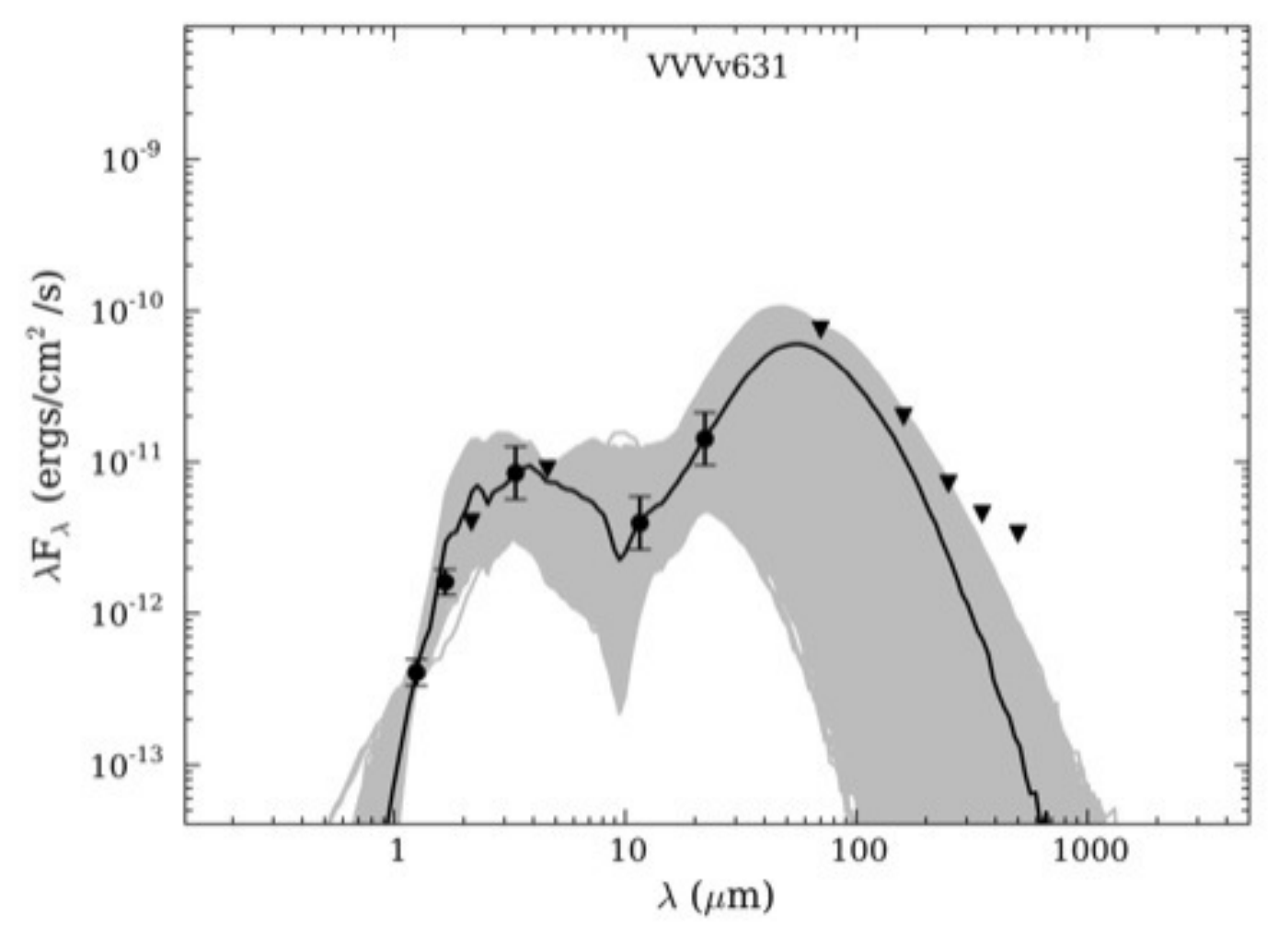}}}
\subfloat{\resizebox{0.48\textwidth}{!}{\includegraphics{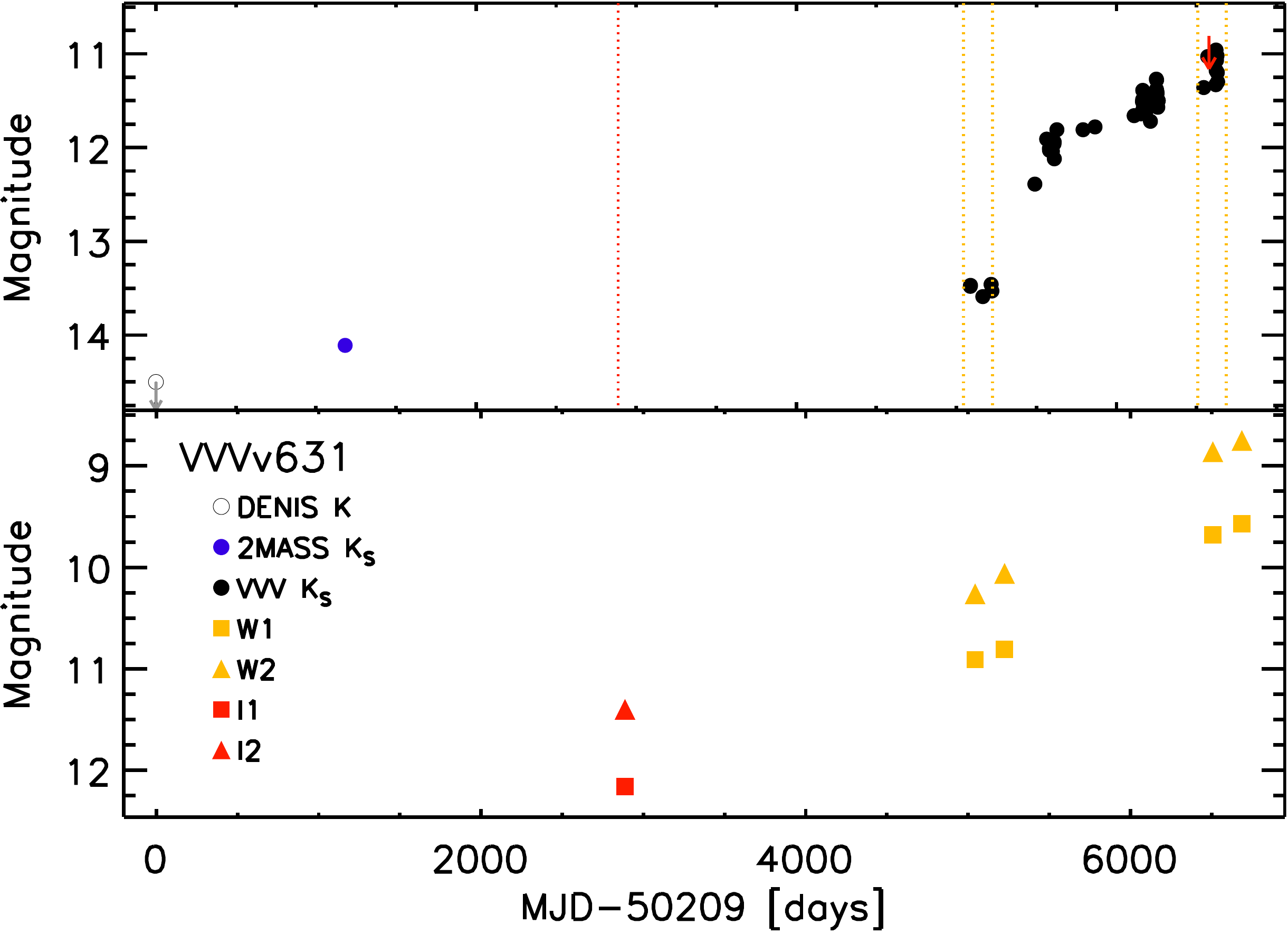}}}\\
\subfloat{\resizebox{0.65\textwidth}{!}{\includegraphics{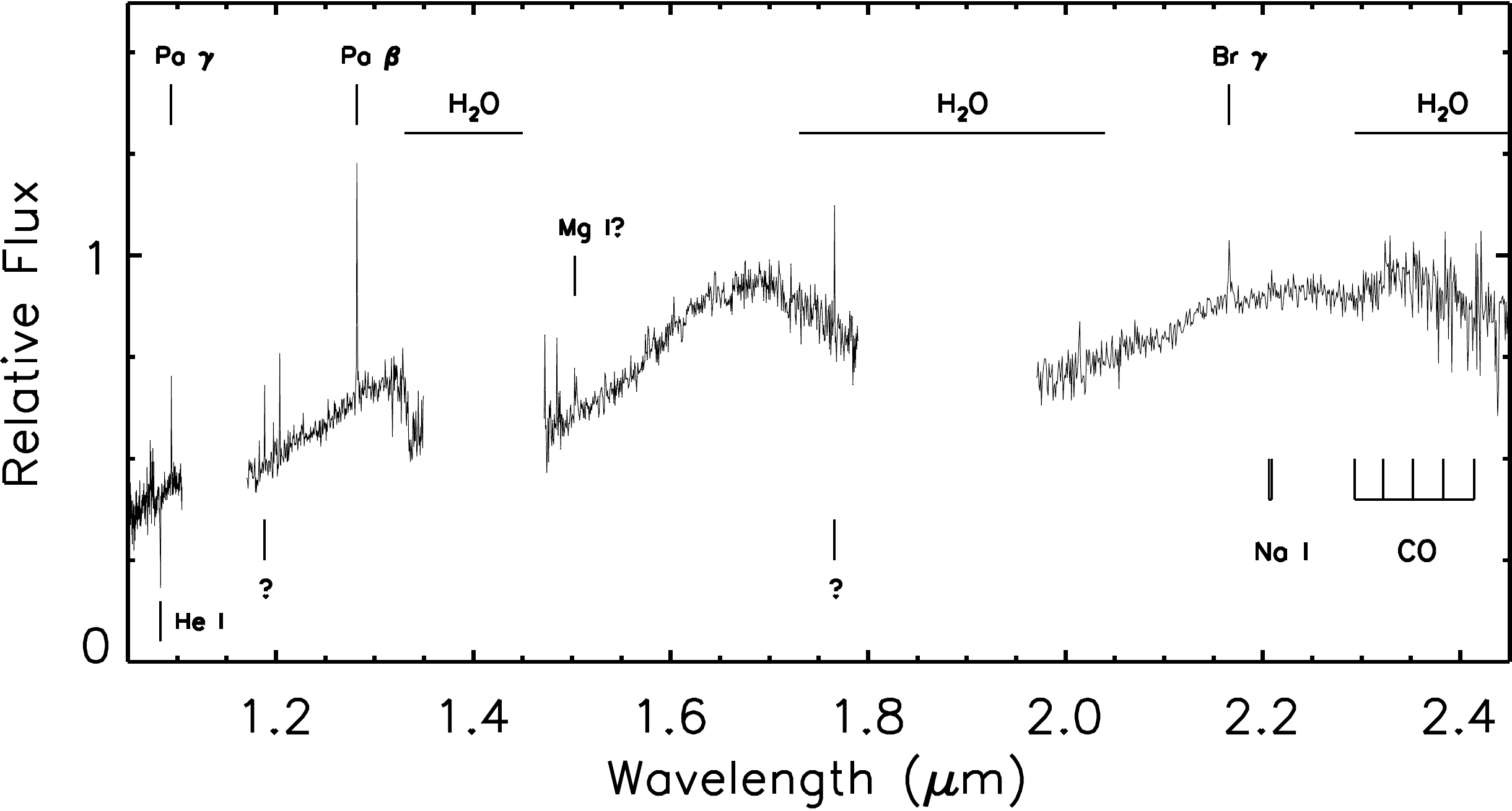}}}
\caption{Same as Fig. \ref{d107vc13}. In the top left we present a 10\arcmin$\times$10\arcmin ~WISE false colour image of the area near VVVv631.}
\label{vvv:erupfour2a}
\end{figure*}

\begin{figure*}
\centering
\subfloat{\resizebox{0.48\textwidth}{!}{\includegraphics{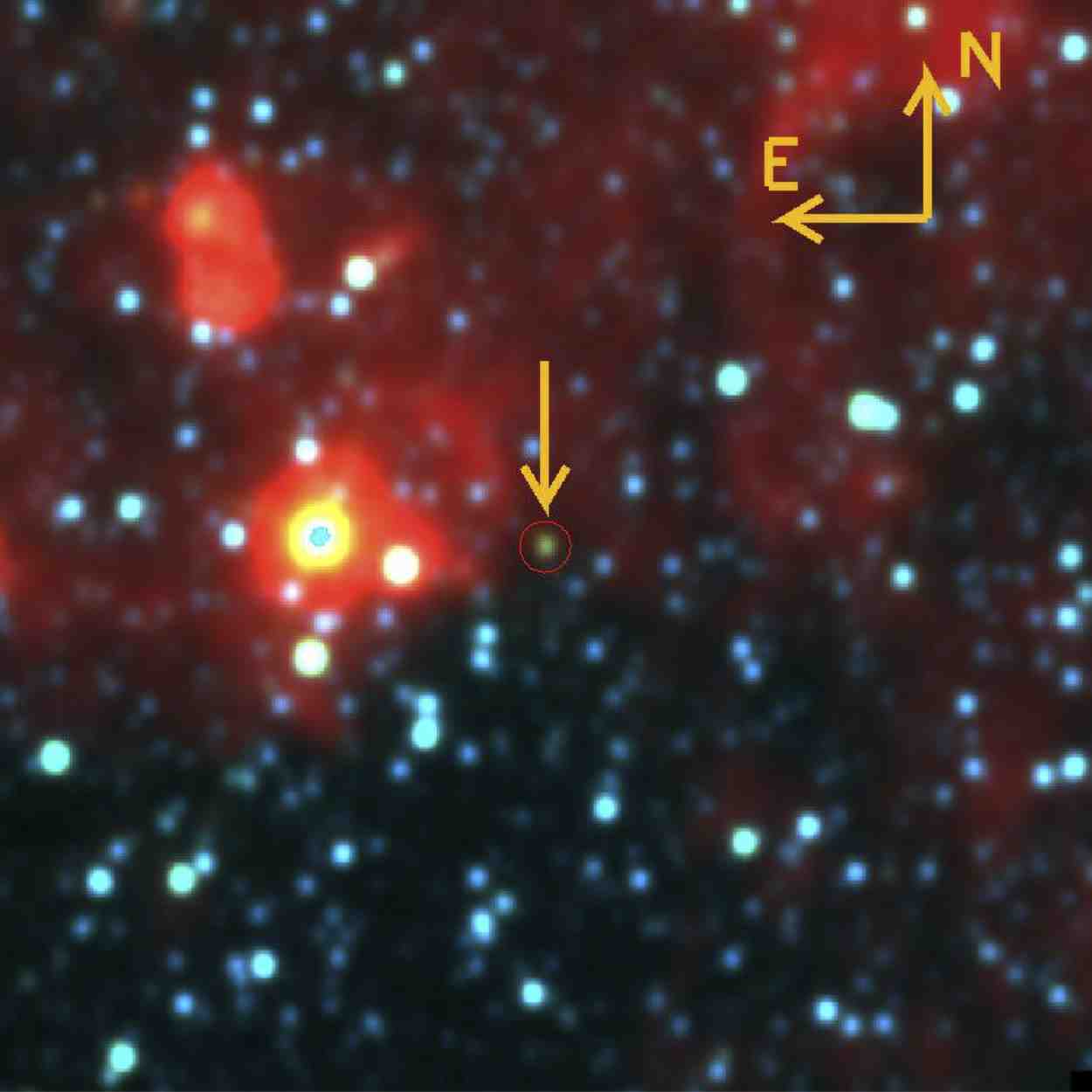}}}
\subfloat{\resizebox{0.48\textwidth}{!}{\includegraphics{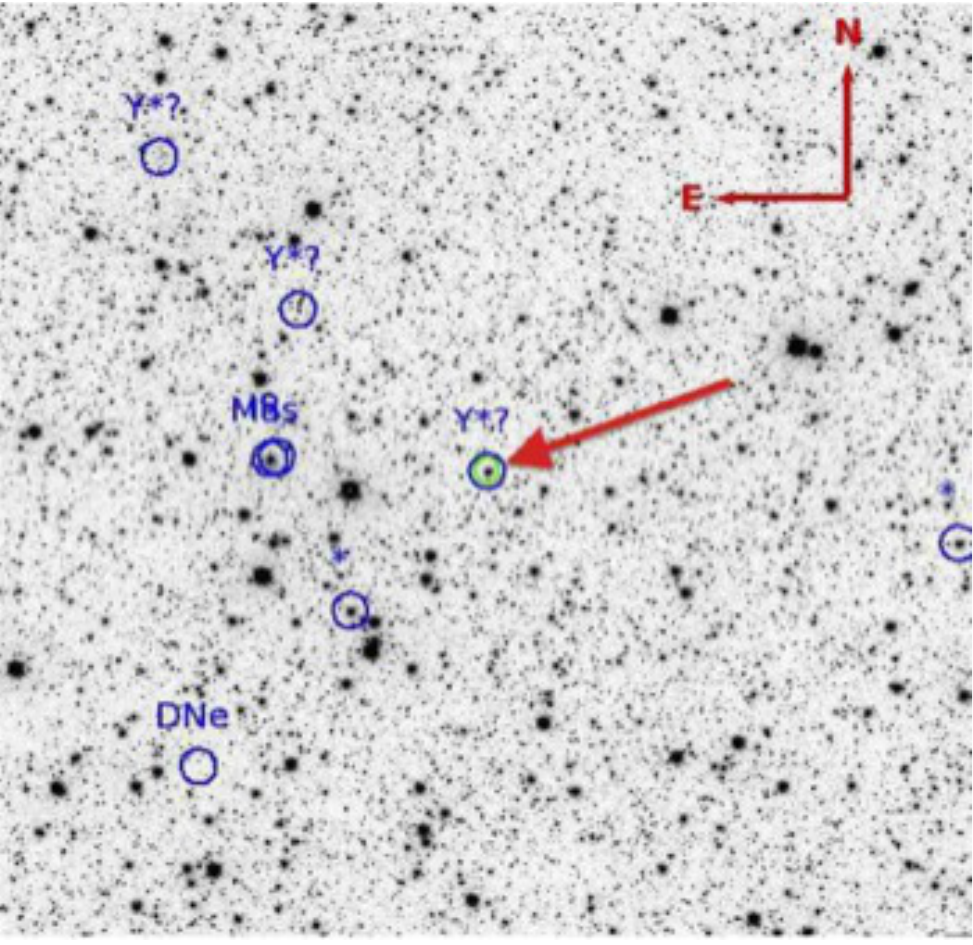}}}\\
\subfloat{\resizebox{0.48\textwidth}{!}{\includegraphics{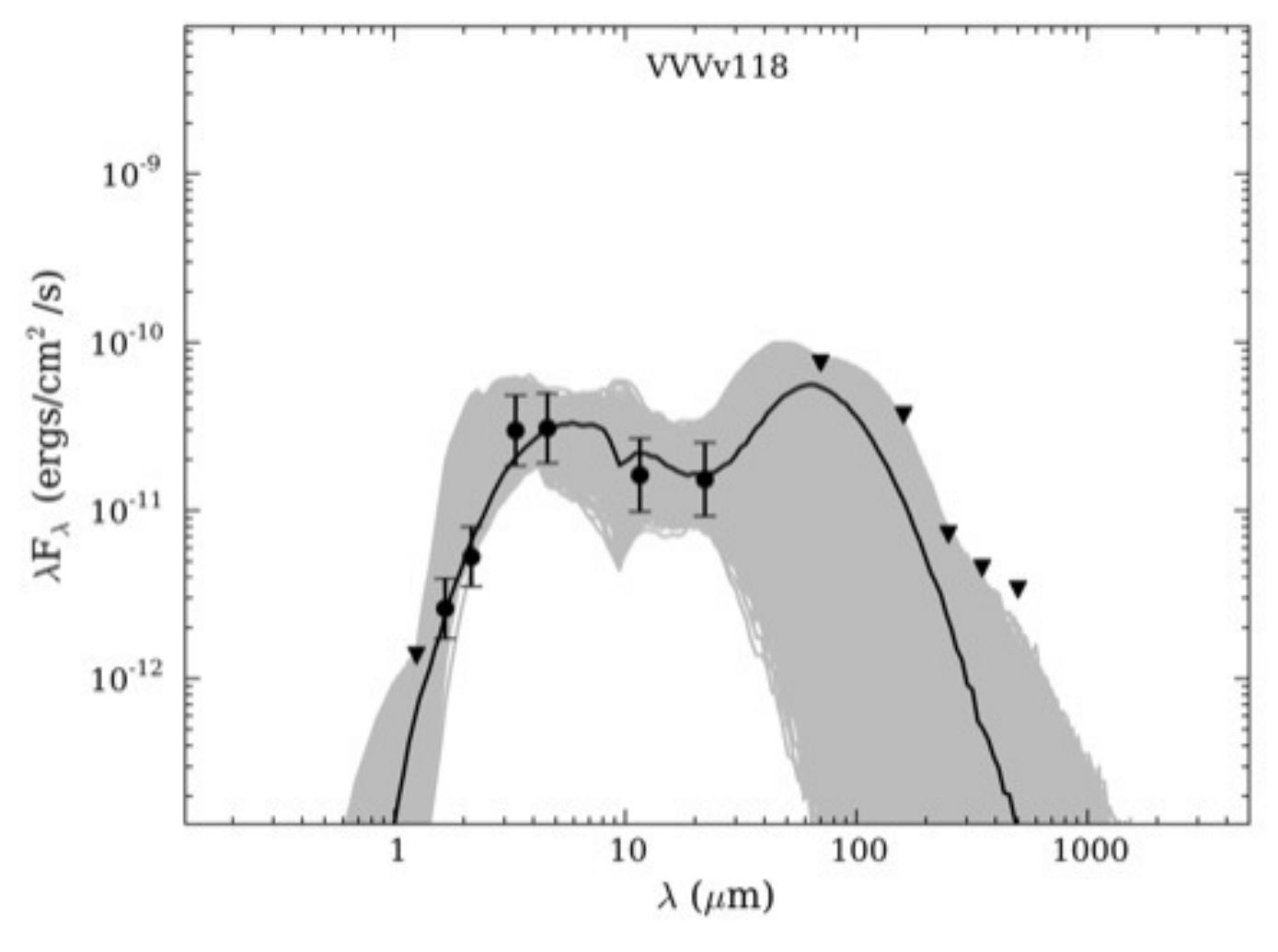}}}
\subfloat{\resizebox{0.48\textwidth}{!}{\includegraphics{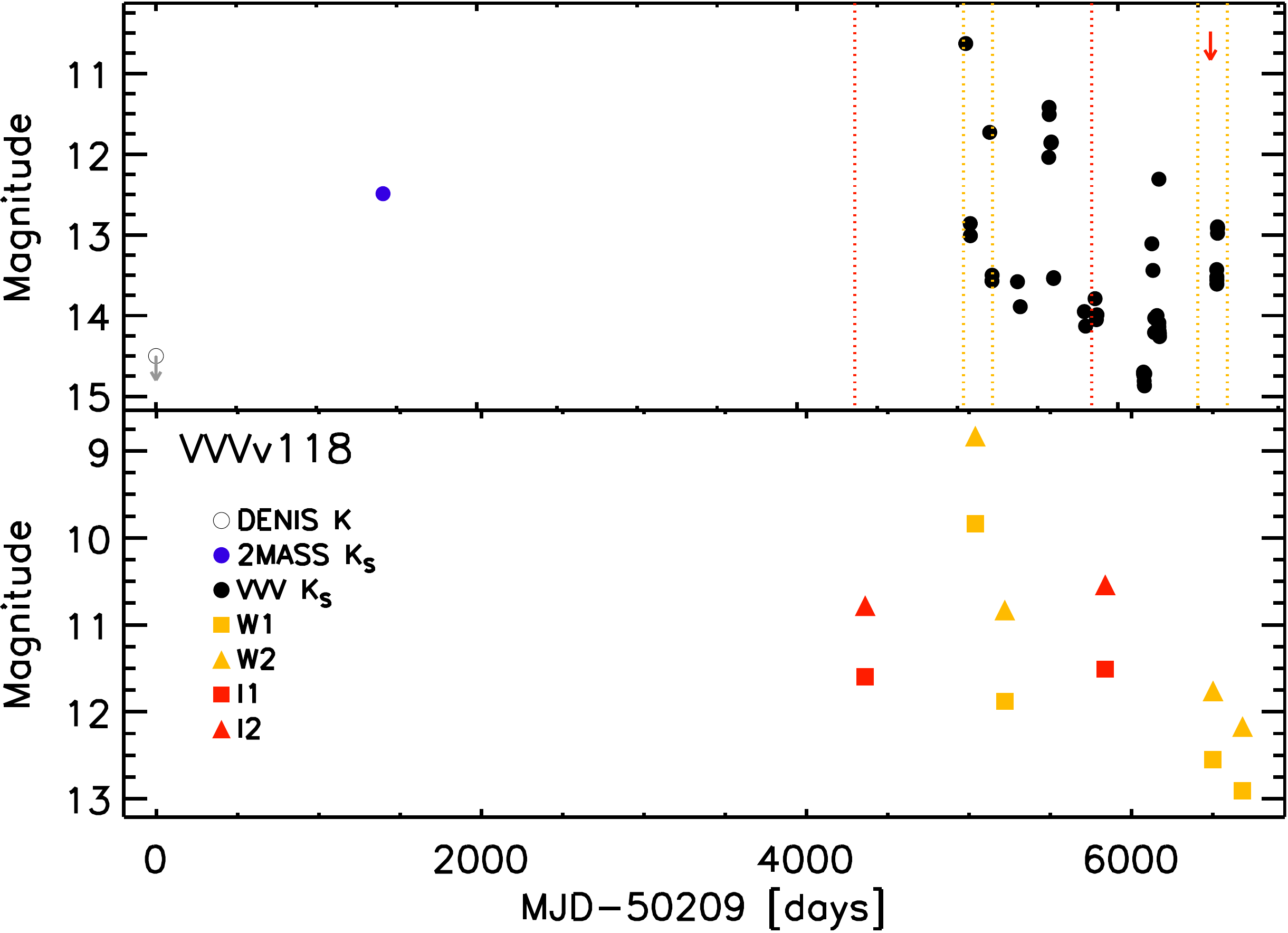}}}\\
\subfloat{\resizebox{0.65\textwidth}{!}{\includegraphics{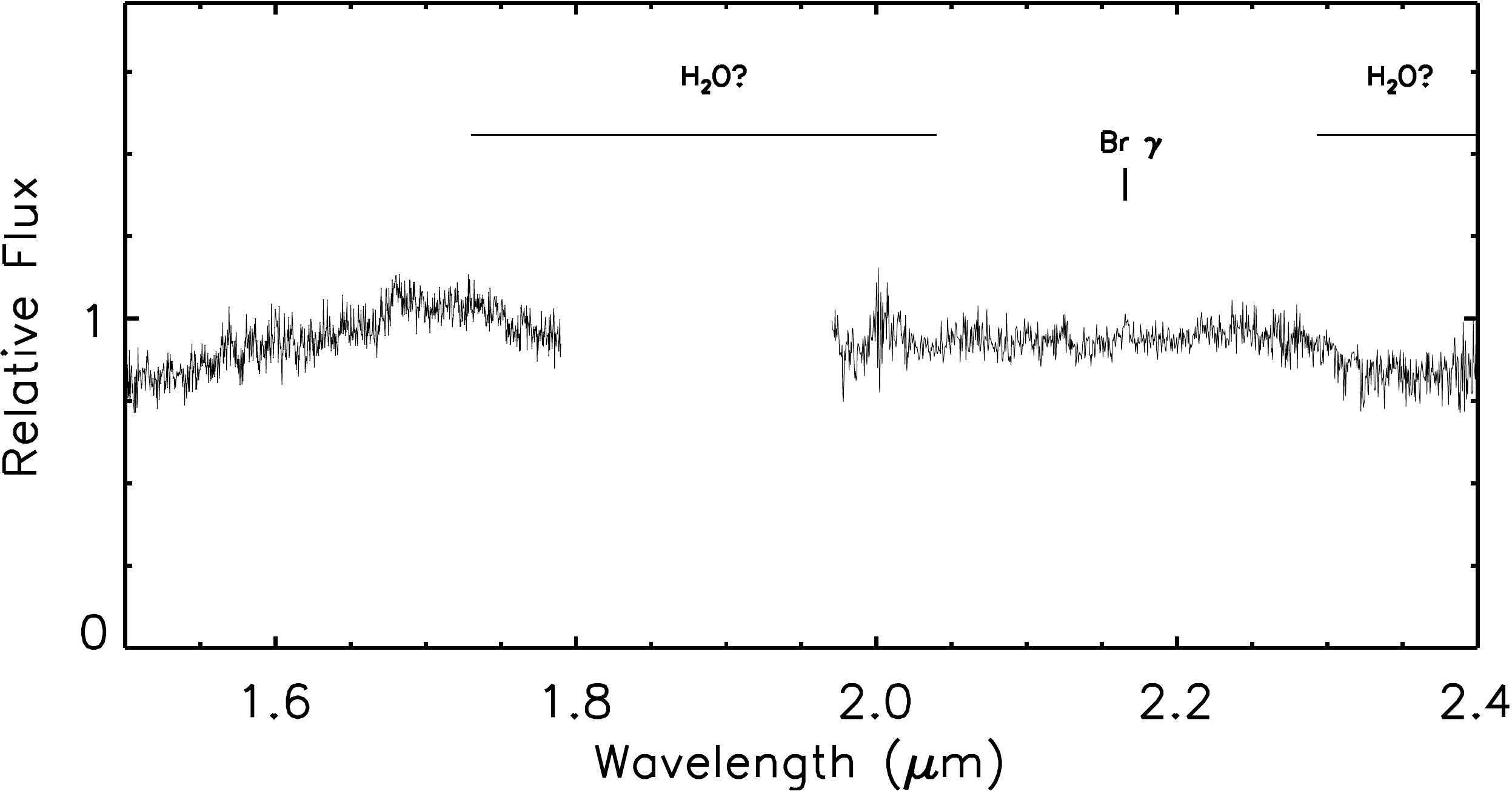}}}
\caption{Same as Fig. \ref{d107vc13}. In the top left we present a 10\arcmin$\times$10\arcmin ~WISE false colour image of the area near VVVv118.}
\label{vvv:erupfour1a}
\end{figure*}

\begin{figure*}
\centering
\subfloat{\resizebox{0.48\textwidth}{!}{\includegraphics{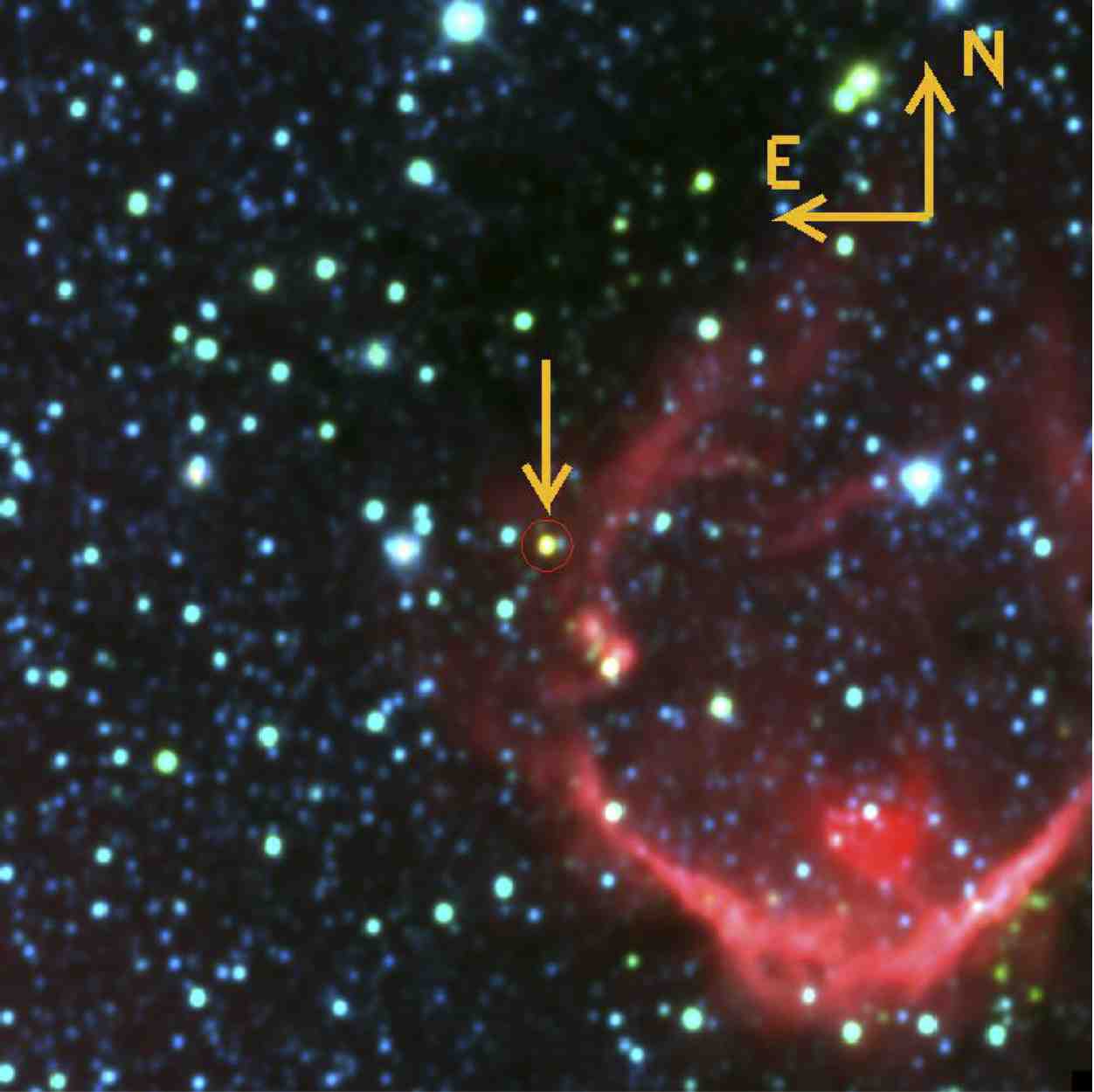}}}
\subfloat{\resizebox{0.48\textwidth}{!}{\includegraphics{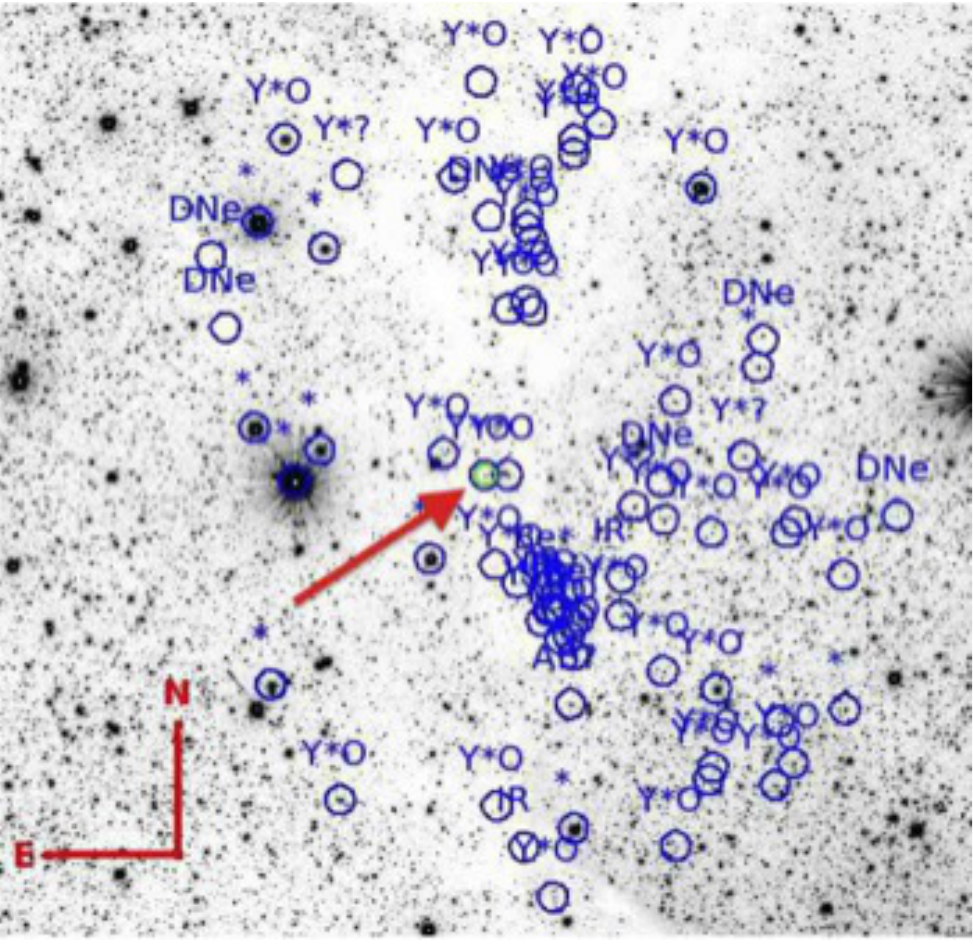}}}\\
\subfloat{\resizebox{0.48\textwidth}{!}{\includegraphics{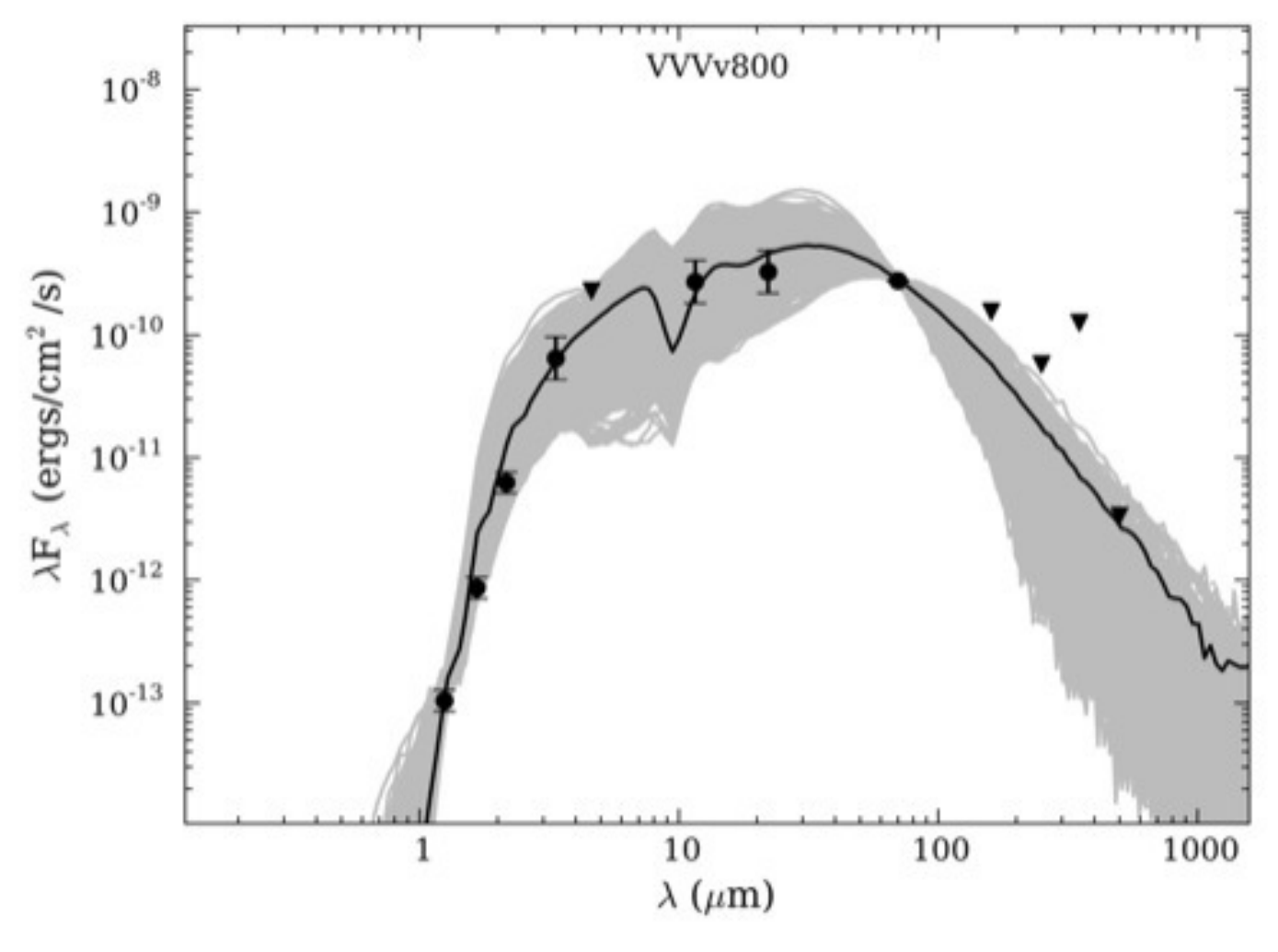}}}
\subfloat{\resizebox{0.48\textwidth}{!}{\includegraphics{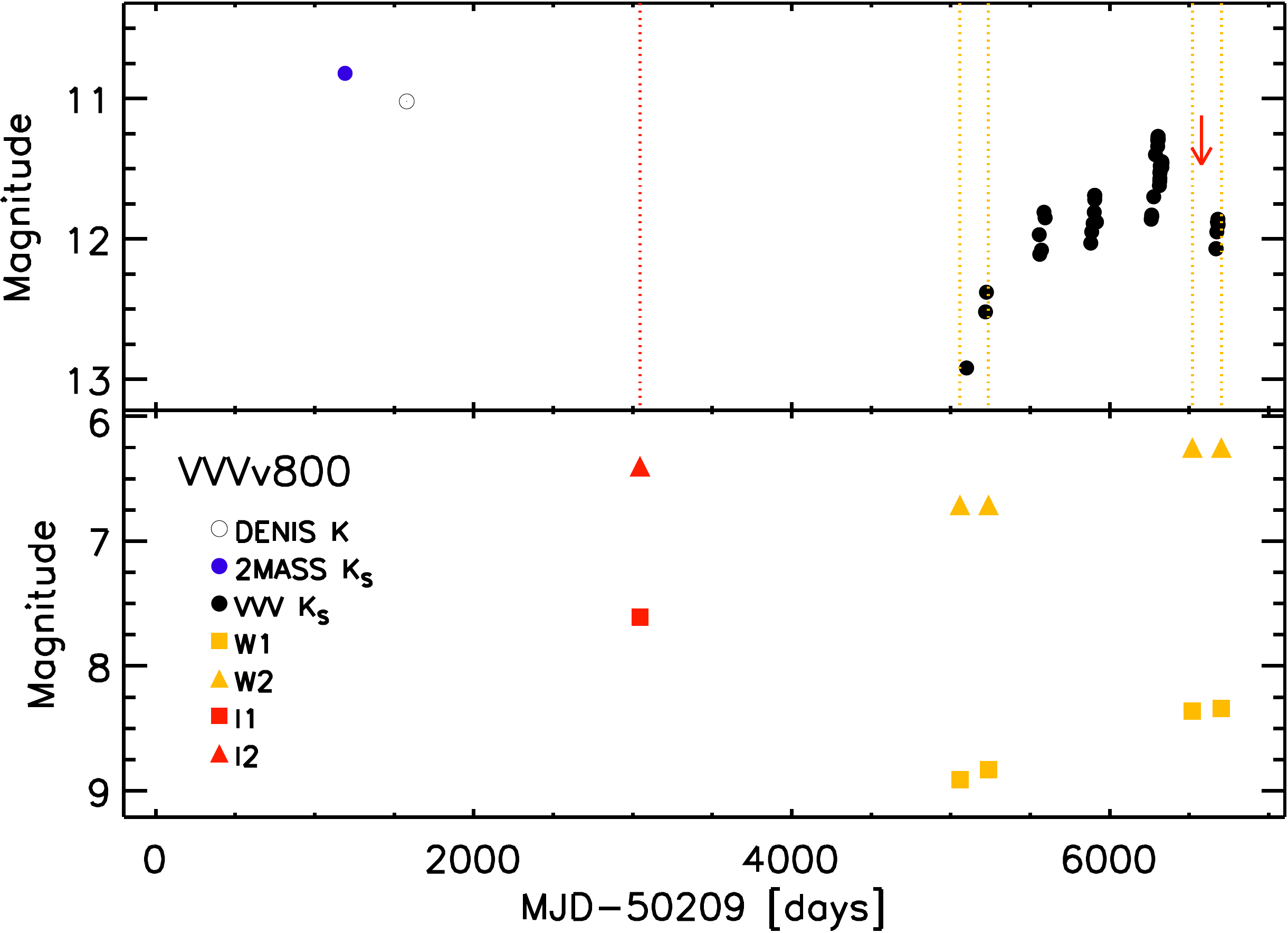}}}\\
\subfloat{\resizebox{0.65\textwidth}{!}{\includegraphics{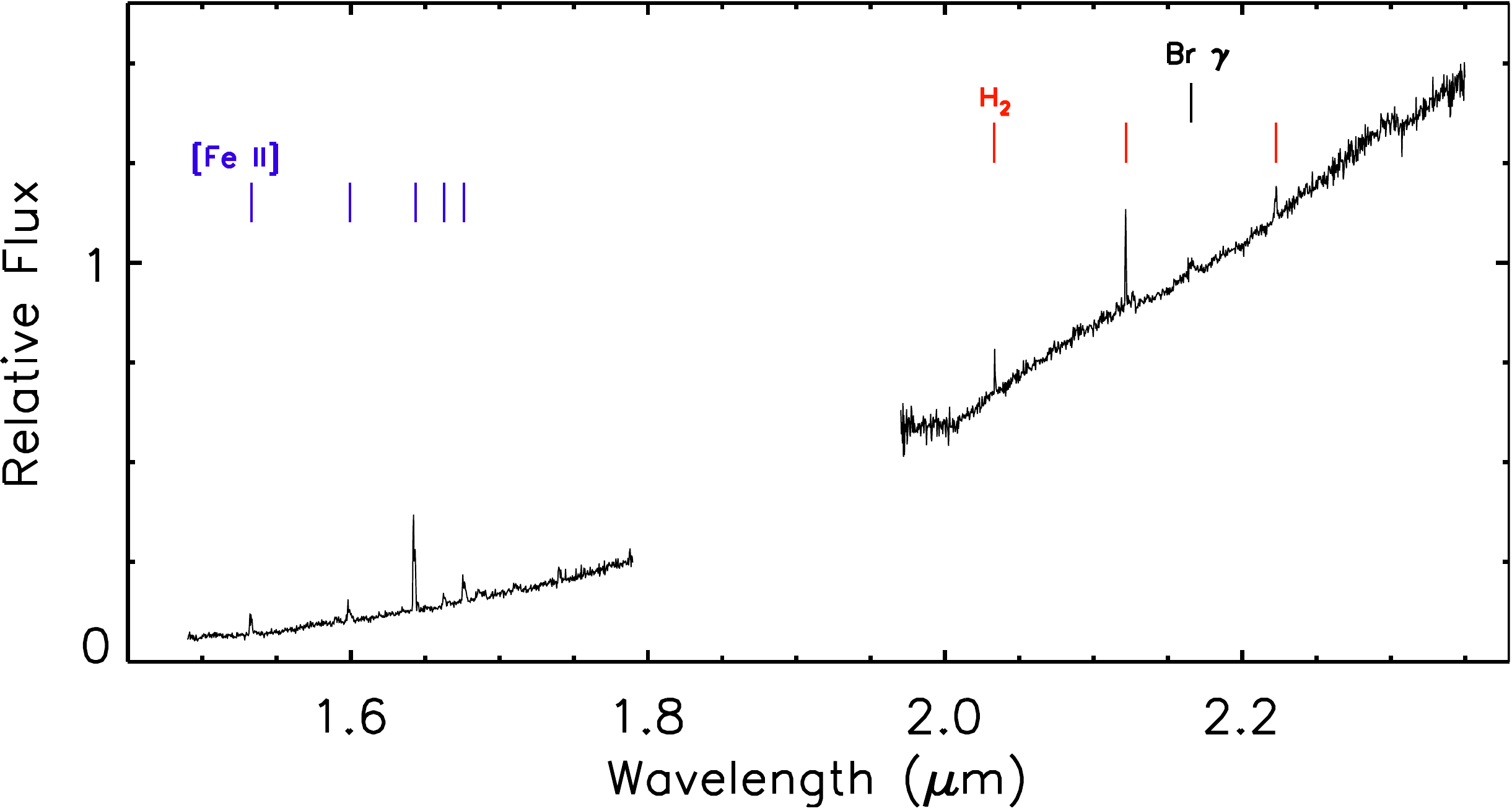}}}
\caption{Same as Fig. \ref{d107vc13}. In the top left we present a 10\arcmin$\times$10\arcmin ~WISE false colour image of the area near VVVv800.}
\label{vvv:erupfour2b}
\end{figure*}

\begin{figure*}
\centering
\subfloat{\resizebox{0.48\textwidth}{!}{\includegraphics{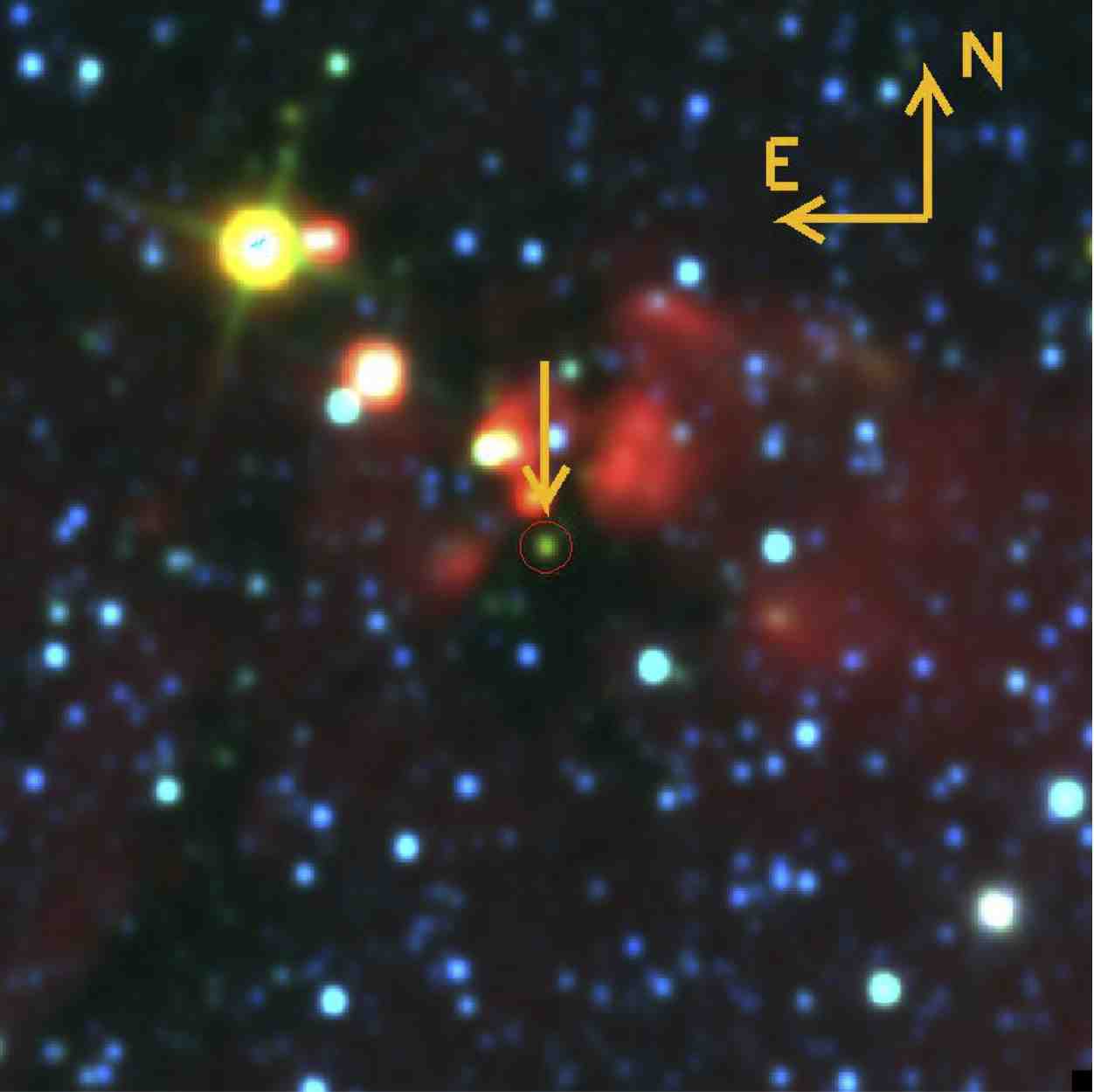}}}
\subfloat{\resizebox{0.48\textwidth}{!}{\includegraphics{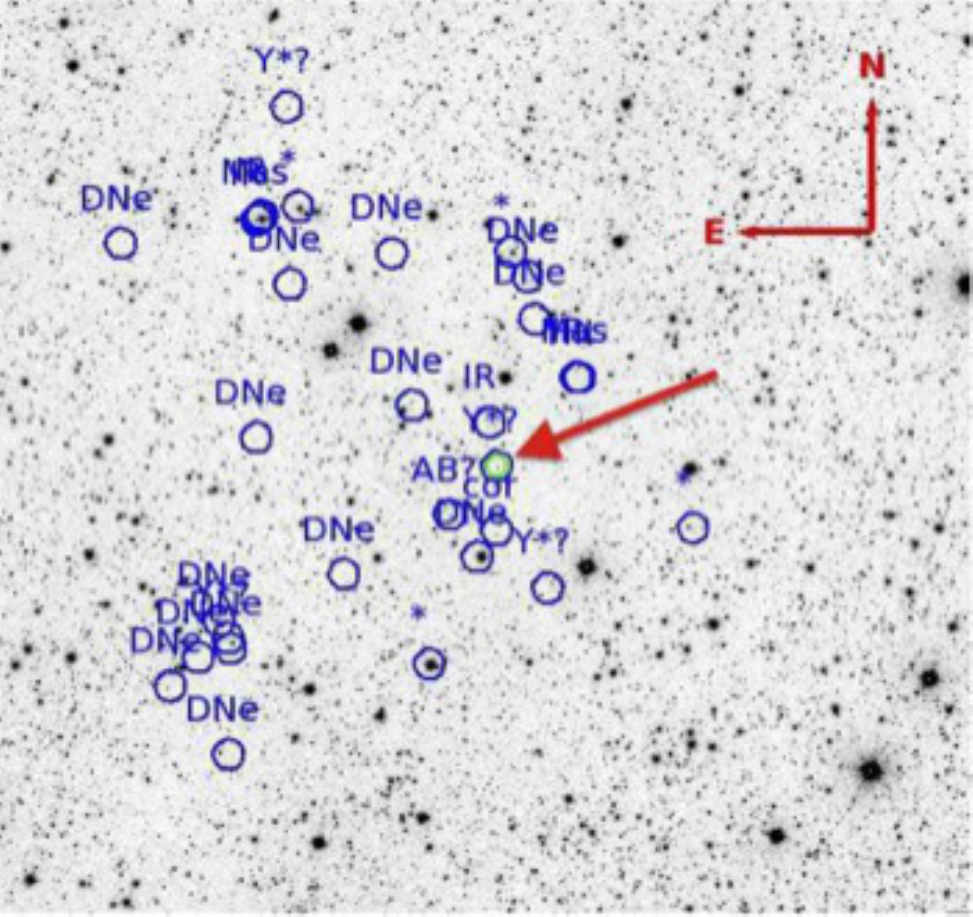}}}\\
\subfloat{\resizebox{0.48\textwidth}{!}{\includegraphics{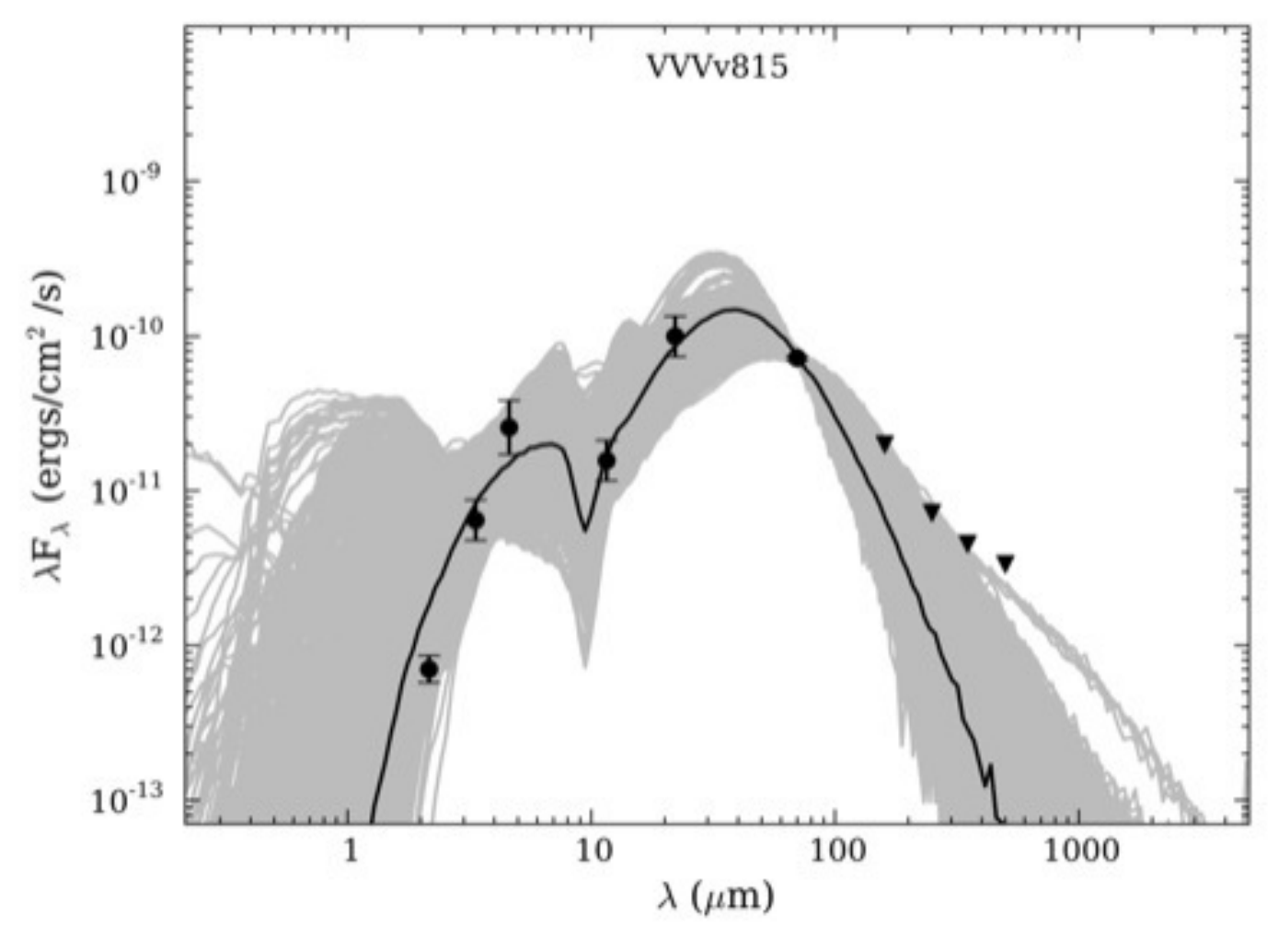}}}
\subfloat{\resizebox{0.48\textwidth}{!}{\includegraphics{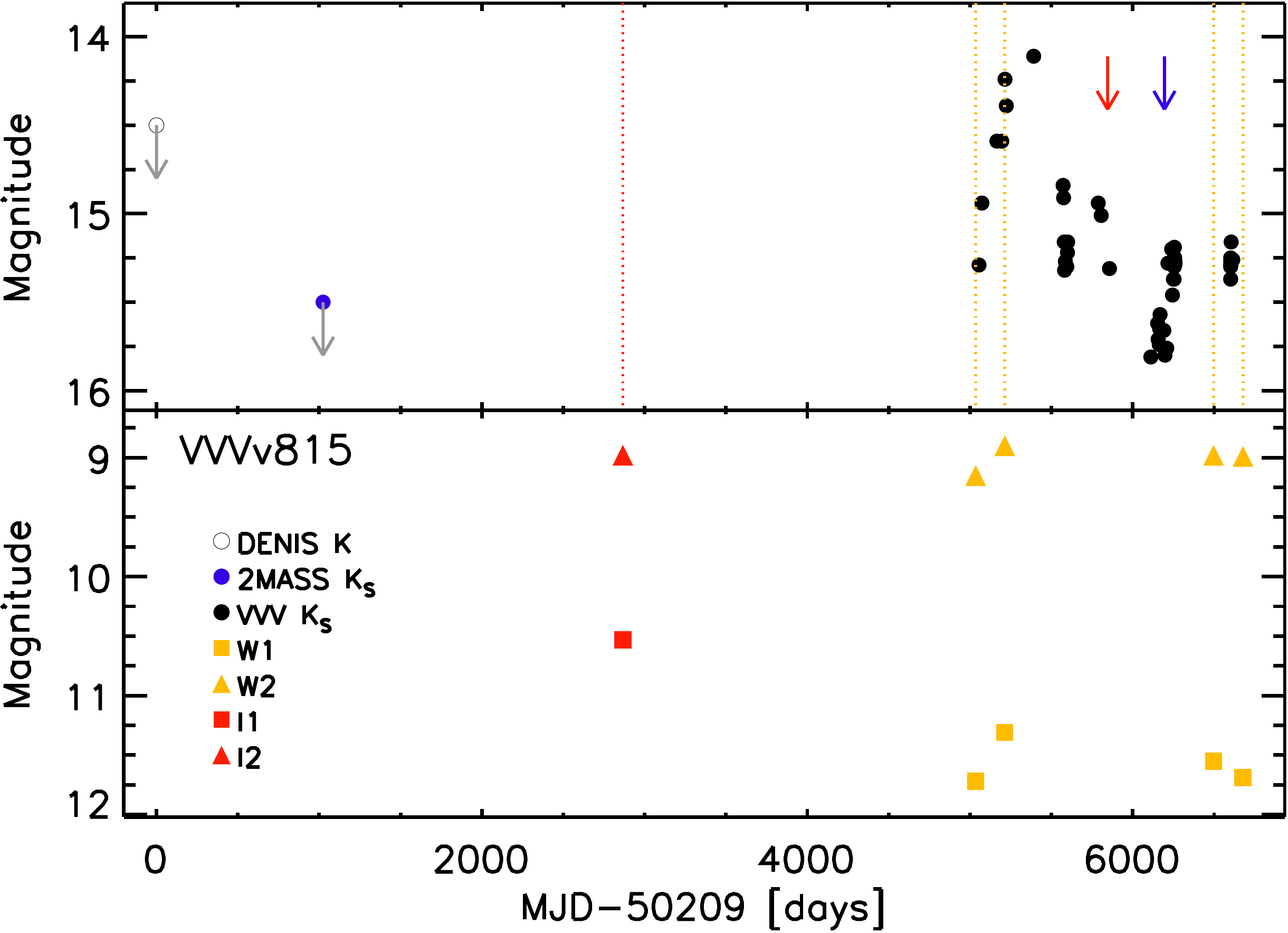}}}\\
\subfloat{\resizebox{0.65\textwidth}{!}{\includegraphics{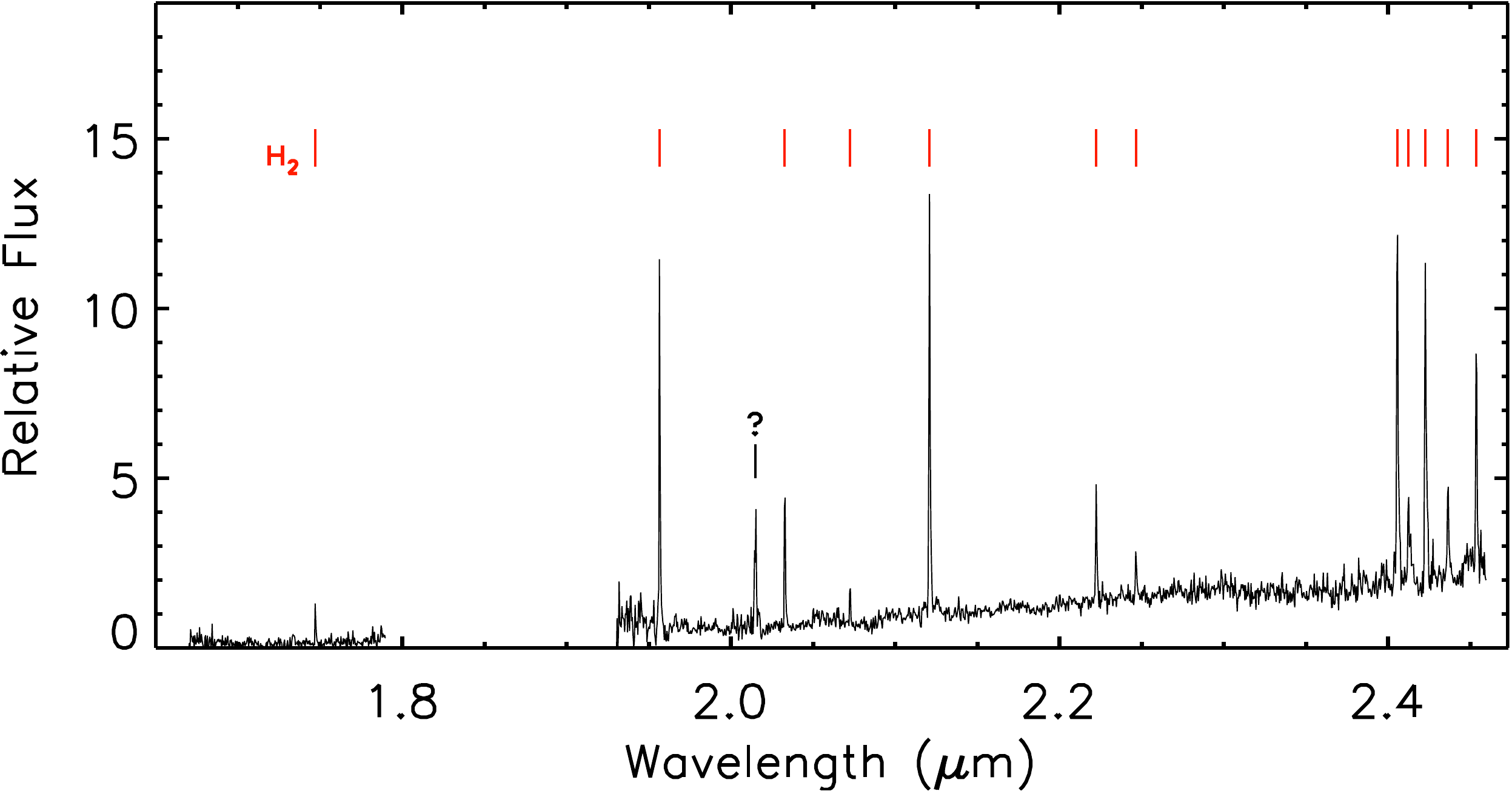}}}
\caption{Same as Fig. \ref{d107vc13}. In the top left we present a 10\arcmin$\times$10\arcmin ~WISE false colour image of the area near VVVv815.}
\label{g314im1}
\end{figure*}

\begin{figure*}
\centering
\subfloat{\resizebox{0.48\textwidth}{!}{\includegraphics{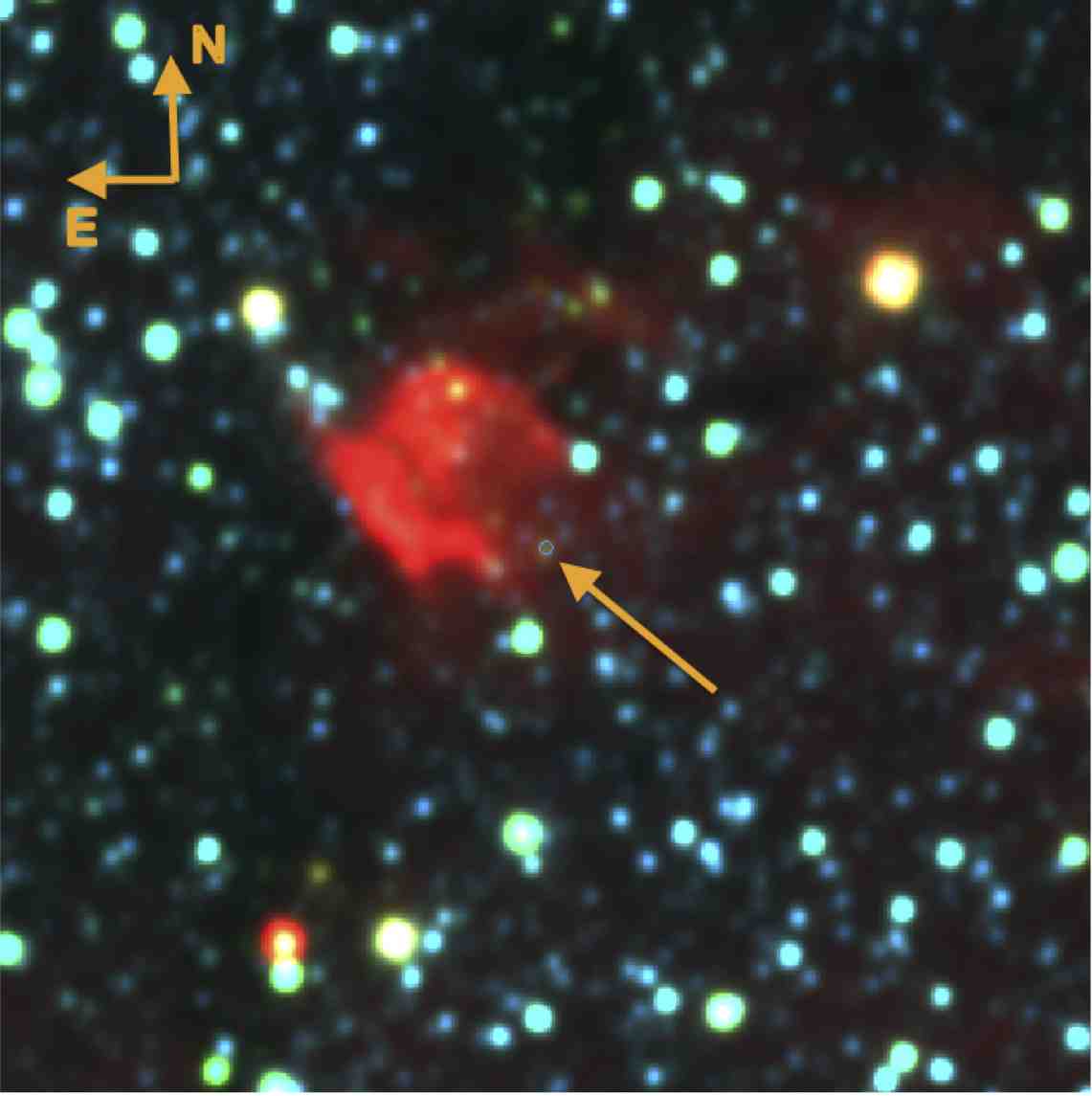}}}
\subfloat{\resizebox{0.48\textwidth}{!}{\includegraphics{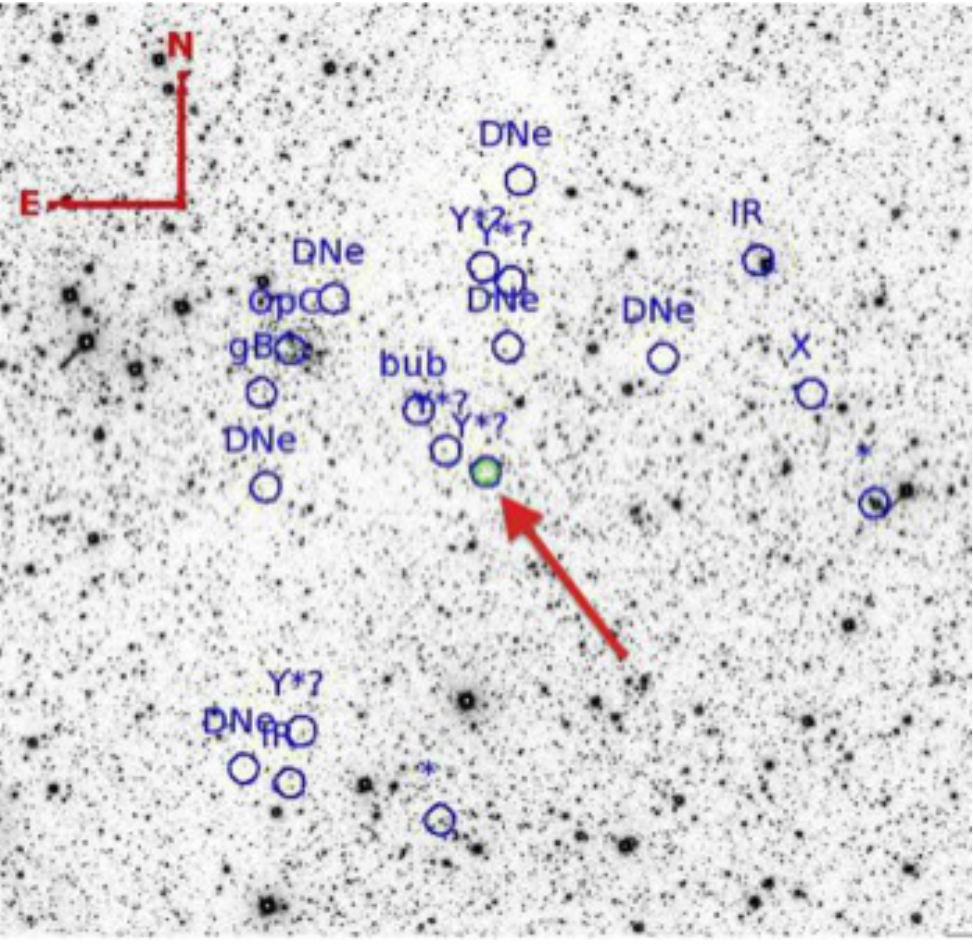}}}\\
\subfloat{\resizebox{0.48\textwidth}{!}{\includegraphics{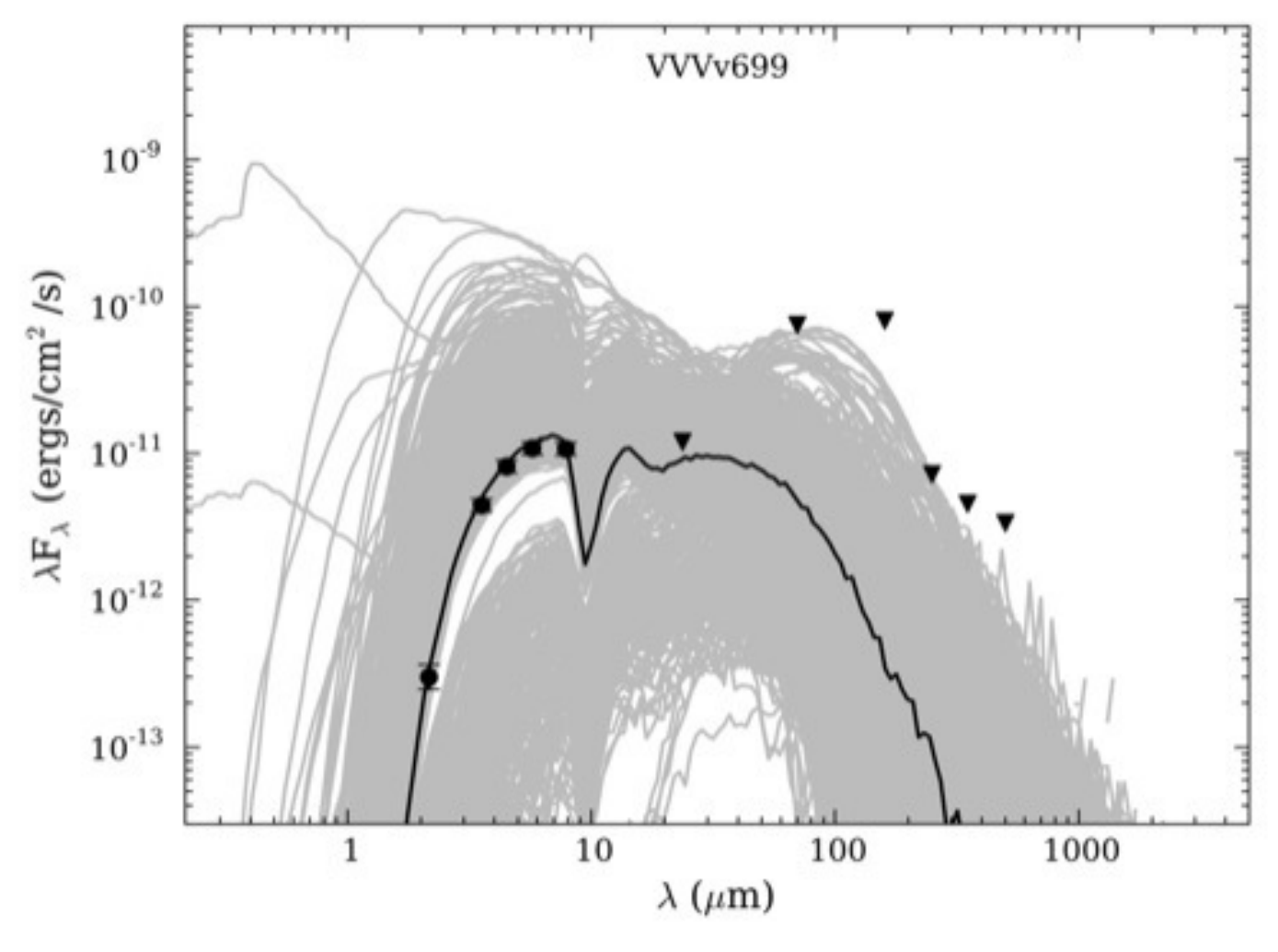}}}
\subfloat{\resizebox{0.48\textwidth}{!}{\includegraphics{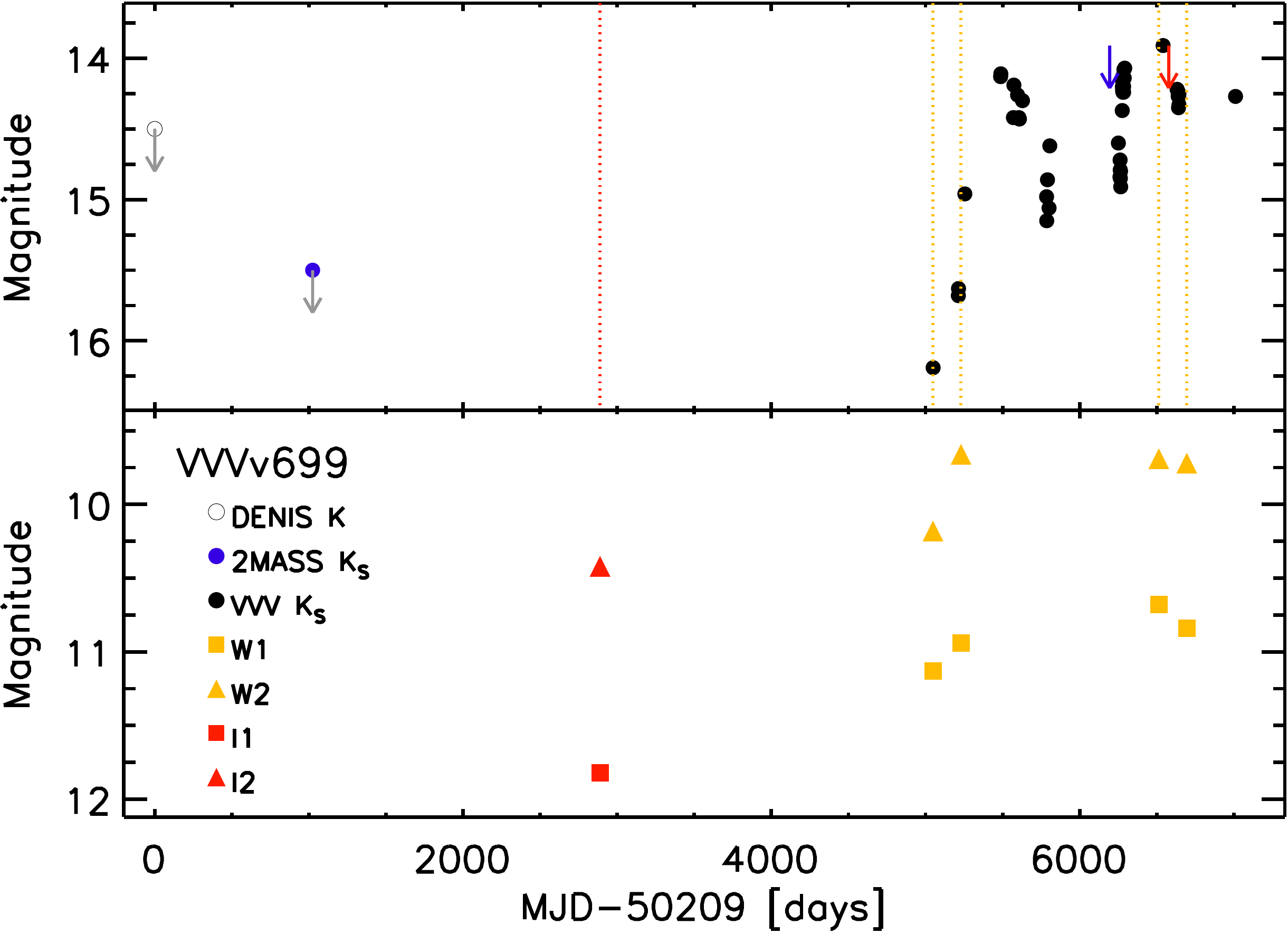}}}\\
\subfloat{\resizebox{0.65\textwidth}{!}{\includegraphics{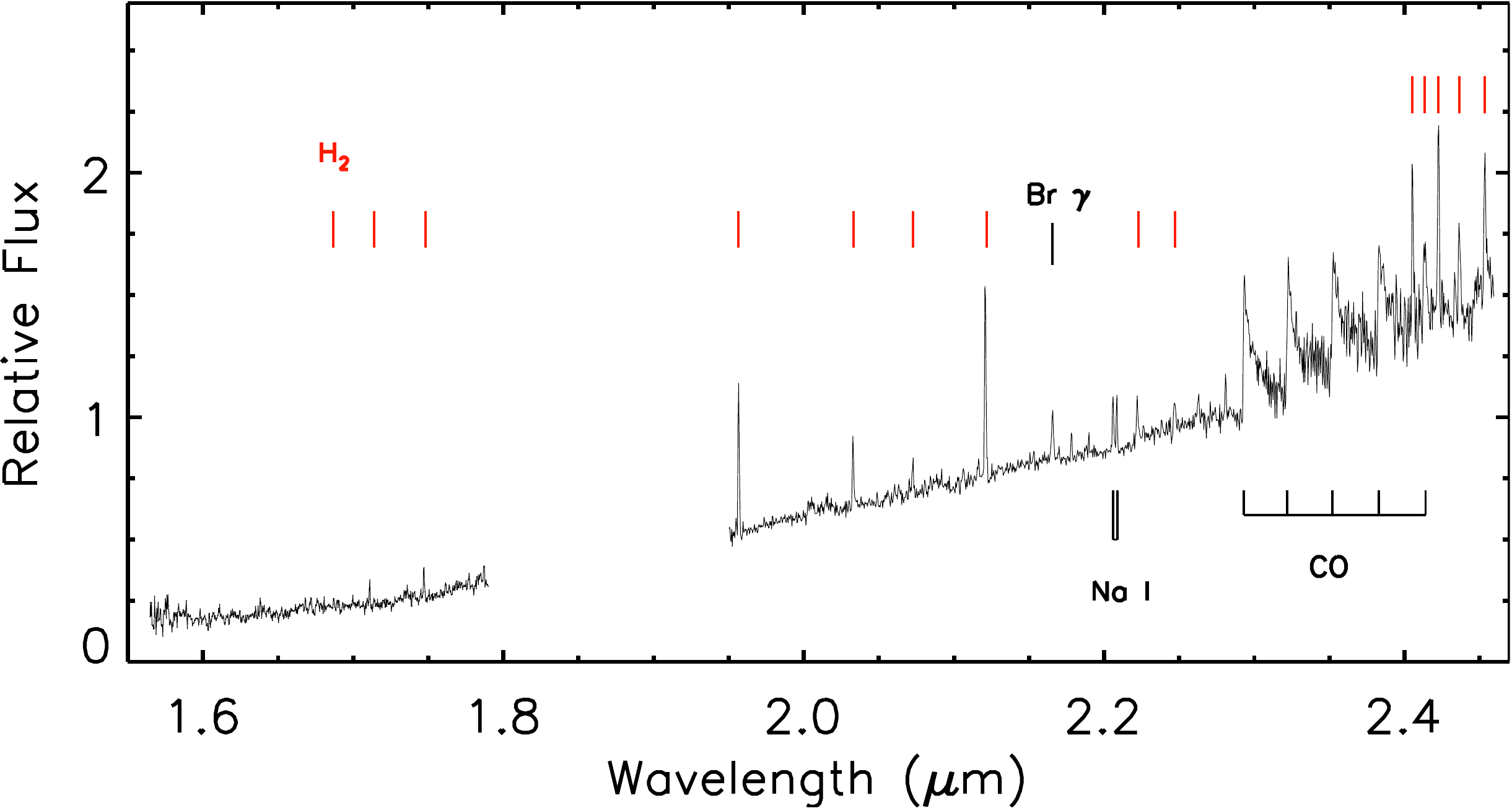}}}
\caption{Same as Fig. \ref{d107vc13}. In the top left we present a 10\arcmin$\times$10\arcmin ~WISE false colour image of the area near VVVv699.}
\label{d104vc12im1}
\end{figure*}

\clearpage

\section{Evolved Objects}\label{vvv:sec_evolved}

In this section we analyse the objects that show characteritics of being from a different variable class, not related to YSOs. These are found to be likely contaminants of our YSO sample and were not included in the analysis in Section \ref{vvv:sec_fire}. 

\subsection{Novae}

\begin{figure*}
\centering
\resizebox{0.45\textwidth}{!}{\includegraphics{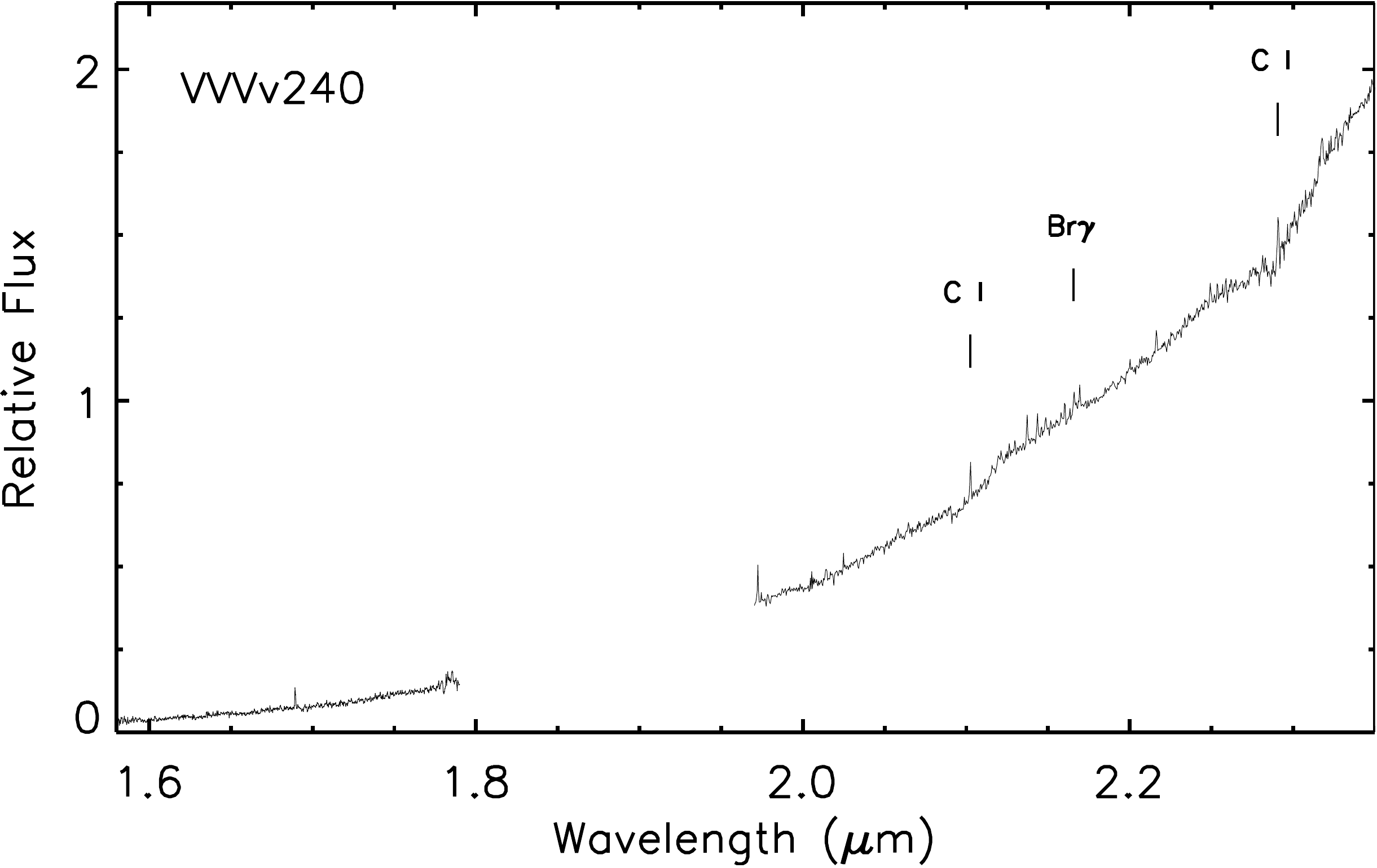}}
\resizebox{0.45\textwidth}{!}{\includegraphics{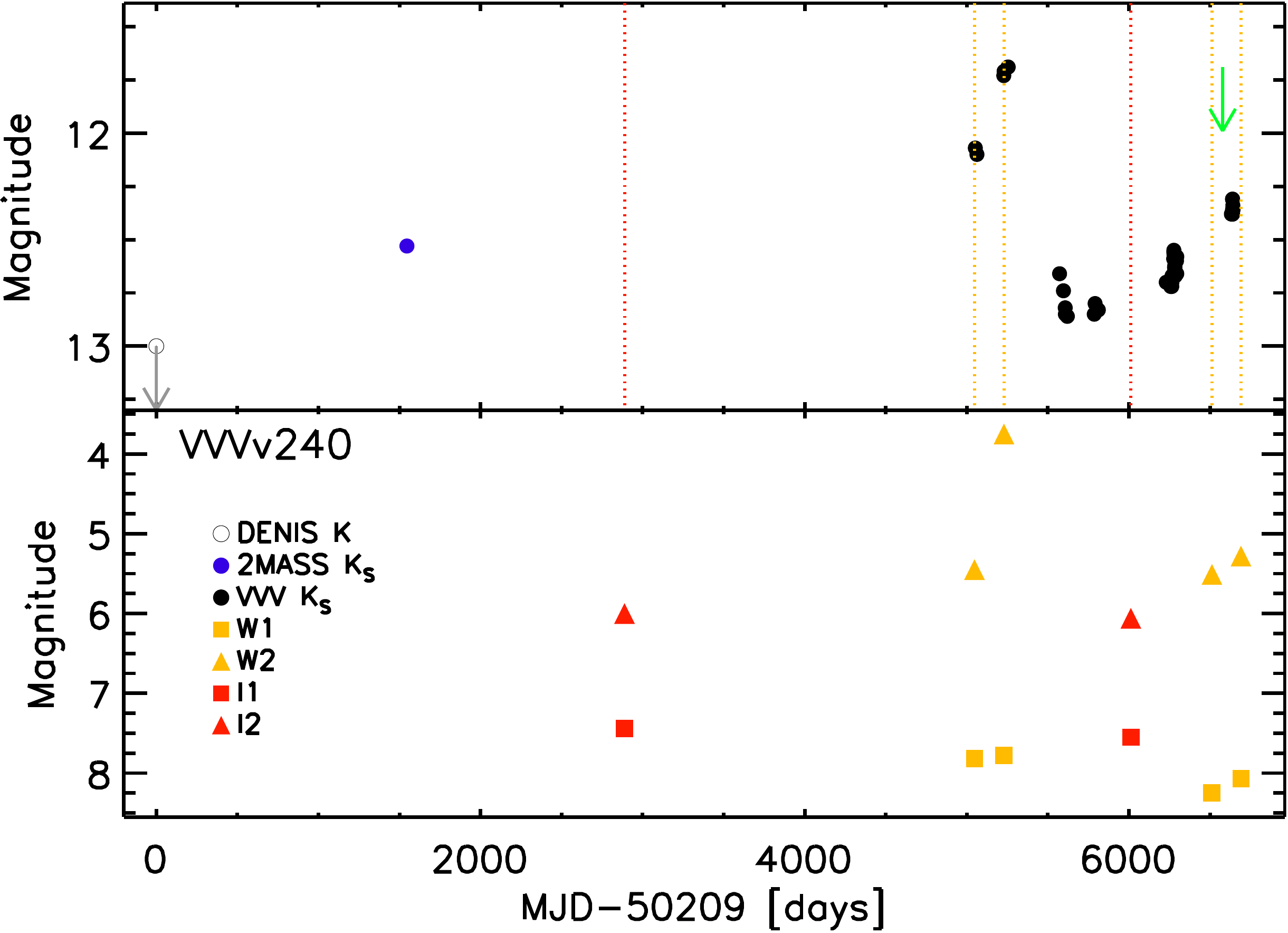}}\\
\resizebox{0.45\textwidth}{!}{\includegraphics{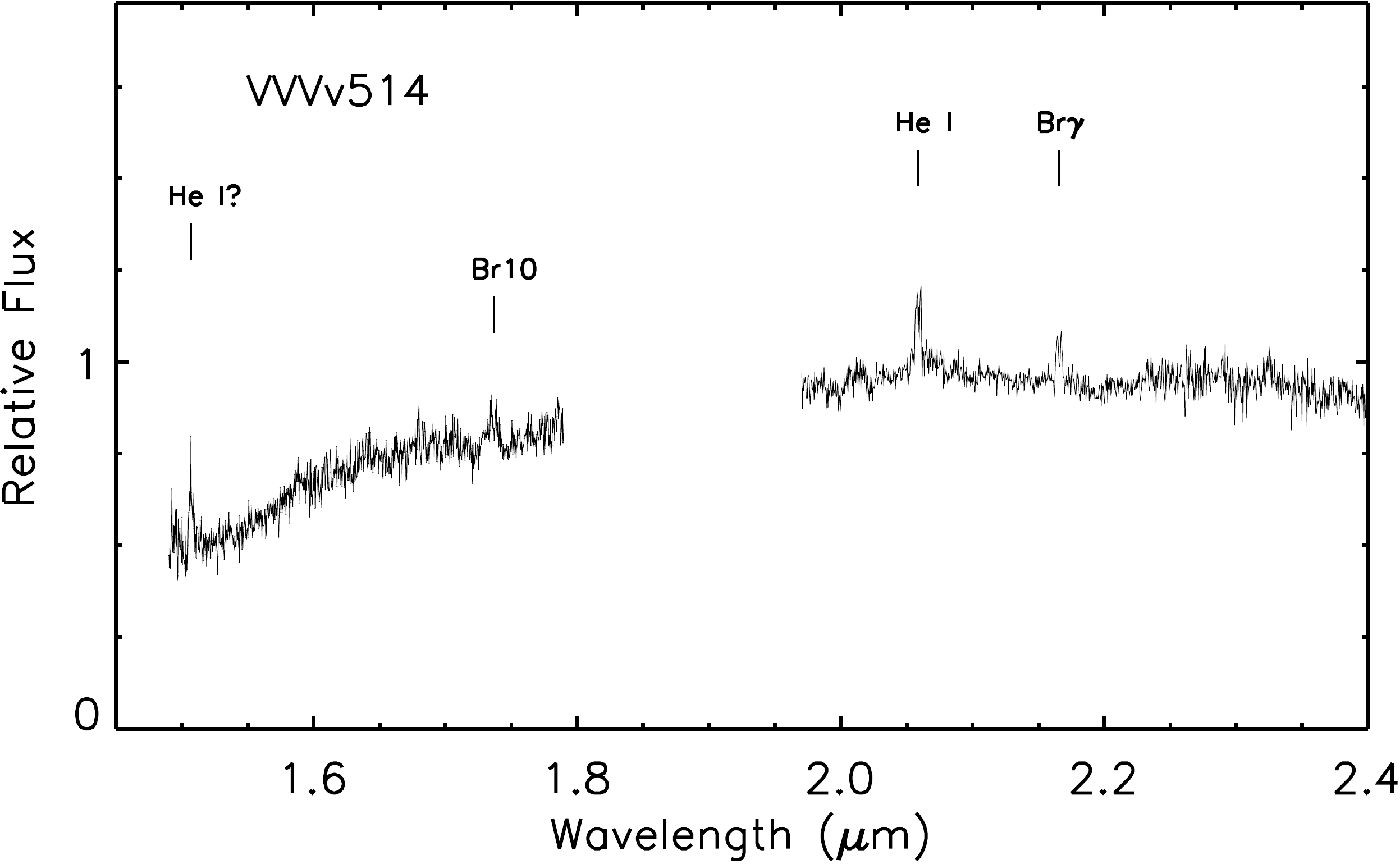}}
\resizebox{0.45\textwidth}{!}{\includegraphics{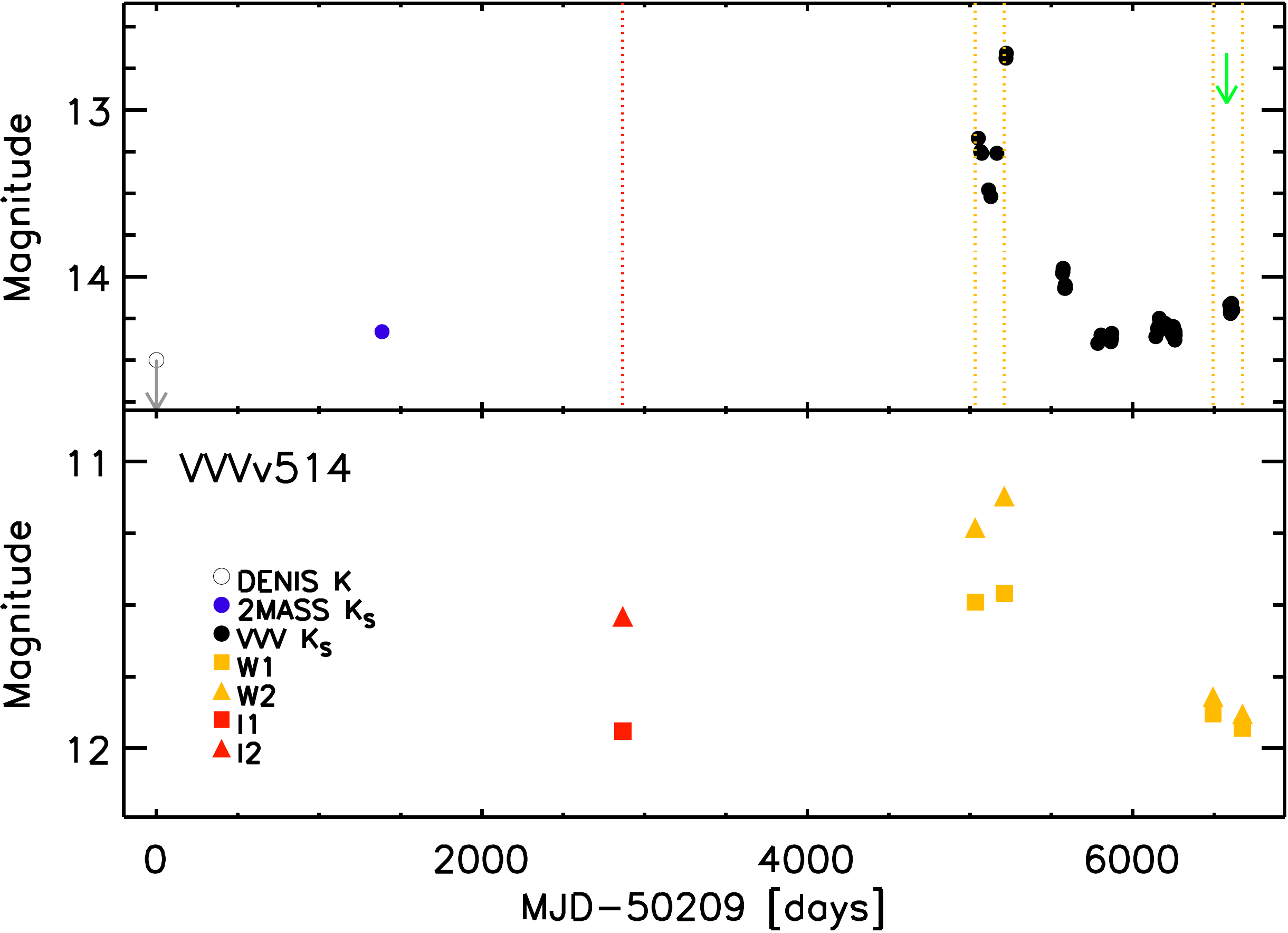}}
\caption{FIRE spectra (left), near- and mid-infrared photometry (right) of likely novae VVVv240 (top) and VVVv514 (bottom).}
\label{vvv:notyso}
\end{figure*}

Two objects are classified as likely novae. VVVv514 shows Br$\gamma$, Br10 (1.73 $\mu$m) and He I (2.058 $\mu$m) emission. This object shows a blue SED ($\alpha=-1.41$) and was classified as a fader from its light curve in Paper I. Inspection of its light curve in Fig. \ref{vvv:notyso} shows that the object went into outburst prior to 2010 and a second peak is apparent at the end of the 2010 campaign. These spectroscopic and photometric characteristics are similar to nova V1493 Aql \citep{2004Venturini}.

The second object, VVVv240, is classified as a long-term periodic variable (Mira-like) in Paper I. The poor sampling of the data probably led to this classification. The object has a rich emission line spectrum (Fig. \ref{vvv:notyso}), which includes weak Br$\gamma$ and C I lines (2.10, 2.29 $\mu$m). The presence of these carbon lines are characteristics of Fe II novae \citep[see e.g. V476 Scuti,][]{2013Das}. 

We note that this classification is based entirely on some similarities of our objects with novae. We do not discard the possibility of these objects belonging to a different class of variable stars.

\subsection{AGB stars}

\begin{figure*}
\centering
\subfloat{\resizebox{0.47\textwidth}{!}{\includegraphics{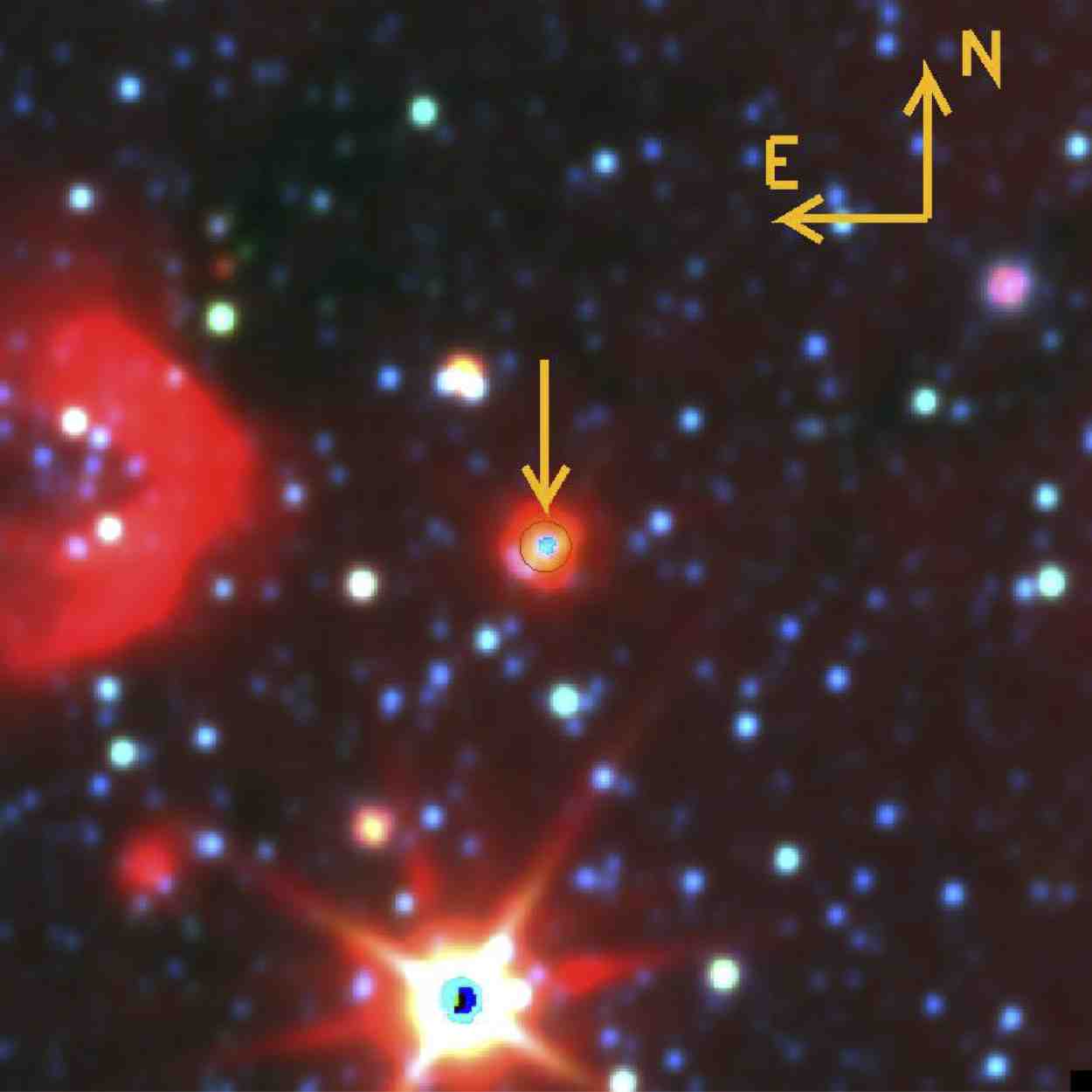}}}
\subfloat{\resizebox{0.47\textwidth}{!}{\includegraphics{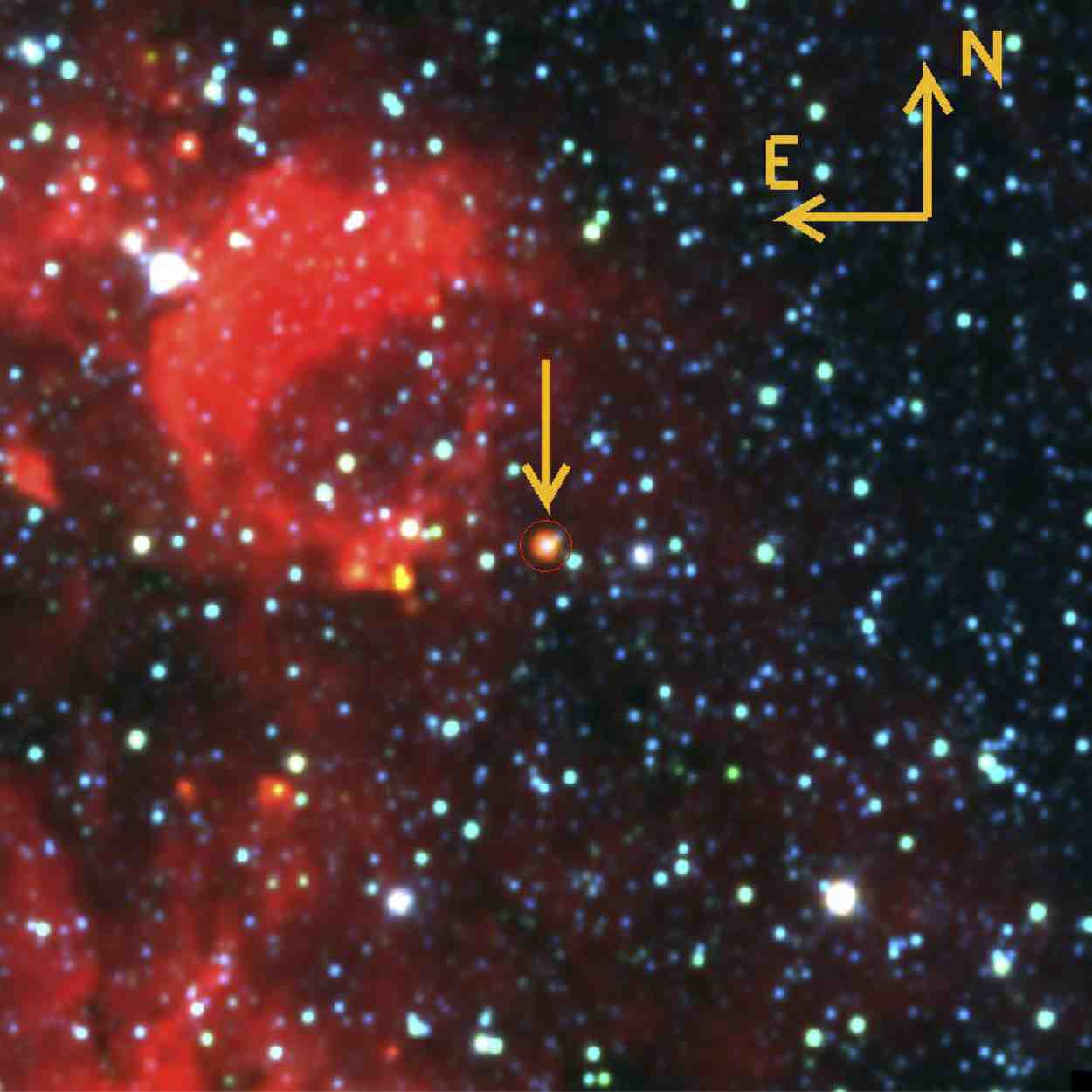}}}\\
\subfloat{\resizebox{0.5\textwidth}{!}{\includegraphics{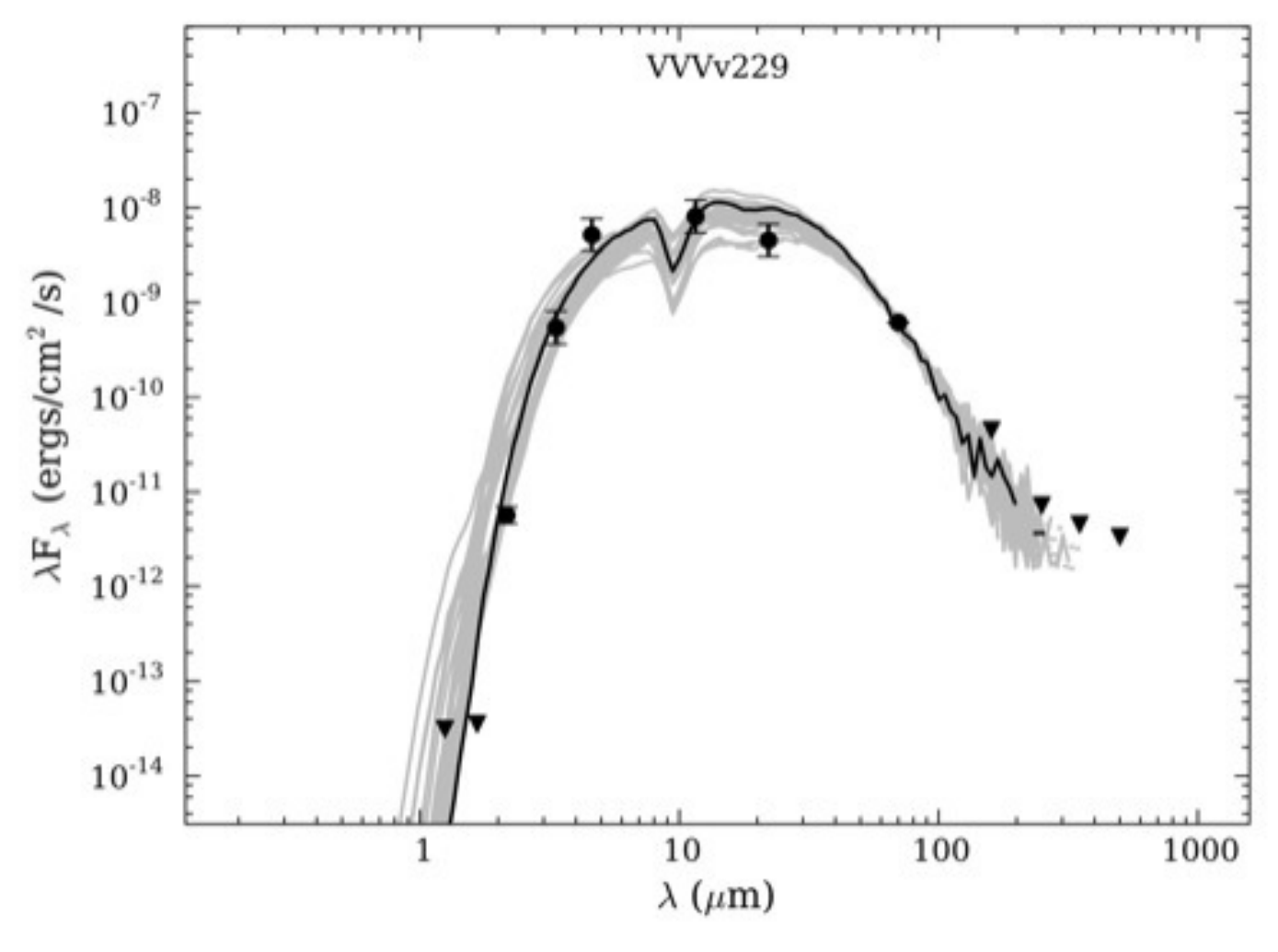}}}
\subfloat{\resizebox{0.5\textwidth}{!}{\includegraphics{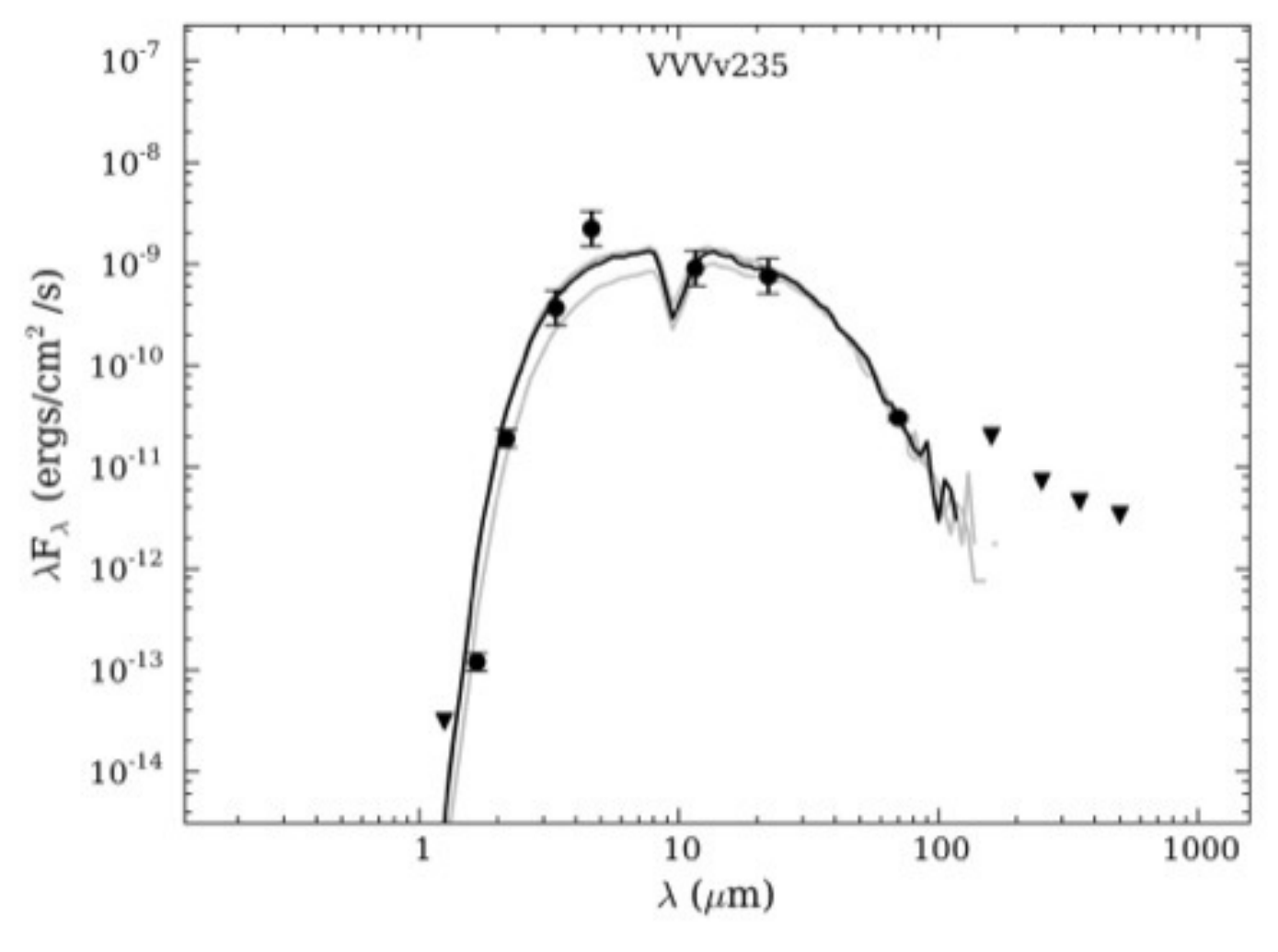}}}
\caption{(top) WISE false colour images of 10\arcmin$\times$10\arcmin areas centred on objects VVVv229 (left) and VVVv235 (right). (bottom) \citet{2007Robitaille} SED fits for the same objects. }
\label{vvv:agbex}
\end{figure*}

We find that 7 VVV objects are more likely AGB stars. Most of these display long-term periodic variability in Paper I, where 5 are classified as displaying Mira-like light curves, whilst VVVv202 is classified as YSO-like. VVVv25 was originally given an eruptive light curve classification. The spectra of these objects are dominated by strong CO absorption, and in many cases a lack of other photospheric absorption features, similar to FUors. These objects also show an apparent association with SFRs, e.g. VVVv42 and VVVv45 are located within the G305 star forming complex (see Fig. \ref{vvv:agbex} for other examples). These characteristics lead to an early idea of these objects being (eruptive) YSOs because the light curves are in some cases not truly periodic. However the recognition that the light curves of dusty-AGB stars 
can have additional trends due to variable extinction on timescales of 
years, superimposed on the periodic light curve, revived our view that 
these sources can be dusty AGB stars. Also, sources with Mira-like light curves tend to have SEDs that are
less well fit by the YSO SED fitting tool (see Fig. \ref{vvv:agbex} and Appendix \ref{section:vvvsedfits})
which cast doubt on a YSO classification.


We study the AGB scenario following \citet{2014Contreras}, estimating whether an AGB star at the magnitudes of the VVV sources can be found within the Galactic disc. In \citet{2014Contreras} we estimated this parameter by 1) Assuming the source is  a Mira variable star with M$_{K}=-$7.25 \citep[][]{2008Whitelock} and 2) From the observed $K_{\rm s}-[9]$ colour of the star and following appendix A of \citet{2011Ishihara}.


We have shown in Paper I that Mira variable stars are very likely to saturate in VVV, even when considering large distances and extinctions towards the source. Then, evolved long period variables in the sample are very likely AGB stars with thick circumstellar envelopes, which makes them fainter and thus able to appear at the magnitudes covered in VVV. Given this, method 1 above would not be applicable to our sub-sample of periodic variable stars. This is confirmed by the fact that the periods observed in this sub-sample are not found within 100--400 days, the typical period of Mira variable stars.

The distance estimate from method 2 is subject to several sources of error. The distance equation \citep[A4 in][]{2011Ishihara} depends on the observed 9 $\mu$m flux as well as the mass loss rate of the object. The latter is derived from the $K-[9]$ colour of the star. Given the high variability on our sample this value can have errors of about $\pm1$ magnitudes. In addition the $\log \dot{\mathrm{M}}$--$(K-[9])$ relation shows large scatter in fig. A2 of \citeauthor{2011Ishihara} which is on the order of $\Delta [\log \dot{\mathrm{M}}$(M$_{\odot}$~yr$^{-1}$)$]=\pm0.5$~dex. These errors propagate into the distance estimate, which for typical values of mass loss rates and F$_{9 \mu m}$ fluxes of our sample, yield $\Delta (\log d)\sim0.2$~dex.

When discussing the AGB scenario in the following, we will only take into account the distance derived from $K_{\rm s}-[9]$~colours and appendix A in \citet{2011Ishihara}.

\begin{itemize}

\item {\bf VVVv202} The light curve of this object was classified as long-term periodic variable, but YSO-like in Paper I. We note that the latter classification might have been due to the star being saturated in some of the VVV epochs. The projected location of VVVv202 is found in an area of active star formation (see Fig. \ref{vvv:agbex}), being 96\arcsec\ from HII region [CH87] 327.759$-$0.351 \citep[V$_{LSR}=-75$ km s$^{-1}$,][]{2014Brown}, and 202\arcsec\ from dense core [PLW2012] G327.721$-$00.383$-$075.1 \citep[V$_{LSR}=-95$ to $-55$ km s$^{-1}$,][]{2012Purcell}. The radial velocity of VVVv202 (V$_{LSR}=-84.2$ km s$^{-1}$, see Table \ref{table:vvvrvel}) agrees with those of the star forming regions. We are unable to compare its distance from radial velocity to that of an AGB star given that it is not detected in Akari nor GLIMPSE.

The spectrum of VVVv202 shows strong CO absorption, with Na I and Ca I also in absorption. In addition this object shows absorption features from $^{12}$CO~$\nu=3$--0, 4--1 bandheads in its H-band spectrum, as well as strong absorption from $^{13}$CO $\nu=2$--0, 3--1 at 2.3448 and 2.3739 $\mu$m. Absorption from $^{13}$CO is usually observed in the spectrum of K--M giant and supergiant stars \citep[see e.g.][]{1997Wallace}. The $^{13}$C/$^{12}$C ratio increases as a consequence of the first dredge up during the star's ascent through the giant branch. The appearance of this feature (in emission) has also been proposed to help distinguish evolved Be from Herbig Ae/Be stars by \citet{2009Kraus}. This is very likely an evolved star.

\item {\bf VVVv25} This object is optically invisible and shows a very similar SED and spectrum to object GPSV15 \citep{2014Contreras}. WISE and VVV images with SIMBAD sources overlaid reveal possible association with SFRs. 

It is unclear whether the light curve is periodic: if so then the period indicated by the Lomb-Scargle fit from Paper I is very long at 
$\sim$5100~days.  However, and as noted in section \ref{vvv:sec_fire}, our analysis of possible periods in Paper I shows that a period of 400--500 days (typical of a carbon star) is 
amongst the possibilities consistent with the sparsely sampled light curve. VVVv25 brightened from 2010 to 2012 and remained stable in 2013. The 2014 photometry shows a small decline, which appears to have continued into 2015, based on the $K_{\rm s}$ datum from the 2nd epoch of VVV multi-colour data.
In addition the DENIS ($K=10.99\pm0.06$) and 2MASS ($K_{\rm s}=10.59\pm0.02$) photometry indicates that this object had been in a bright state previously. 
However, the light curve of the object (until 2014) is consistent with that of an eruptive YSO.

The potential classification as an AGB star is discussed based on similar arguments used for the previous object. A carbon- or oxygen-rich AGB star losing mass at a rate of $\sim 10^{-4.8}$~M$_{\odot}$yr$^{-1}$ (estimated from the $K-[9]$ colour of VVVv25), would be located at a Heliocentric distance of $29^{+17}_{-11}$~kpc. The radial velocity of VVVv25 indicates $d\sim10.9$~kpc.

 

The resemblance of the spectrum with likely eruptive variables from the GPS and its apparent strong association with a SFR, would favour a YSO interpretation. However, the radial velocity distance is not far off from the lower end of the distance to an AGB star. In addition the 2015 spectrum of the object, reveals the presence of an absorption bandhead at 1.76 $\mu$m, which is more likely from C$_{2}$. This is generally observed in carbon stars, making VVVv25 more likely an evolved object.

\item {\bf VVVv42} The star is 1.5\arcsec ~from IRAS source IRAS 13063$-$6233, which is classified as a stellar object in SIMBAD. The object is located within 300\arcsec ~of 5 IRDCs and is in the vicinity of the G305 star-forming complex. The spectrum of the object is very red and does not show any features that could be associated with a stellar photosphere. The source resembles GPS variable GPSV3 as it is dominated by absorption of low temperature rotational transitions from CO (see Appendix \ref{vvv:co}). However, doubts regarding a YSO classification arise given its periodic light curve (see Appendix \ref{apen2}) which led to a "Mira-like" light curve 
classification in Paper I.



The object is detected at 9 $\mu$m in the Akari mission with F$_{9\mu m}=7.53$~Jy. The $K-[9]$ colour would correspond to a dust-enshrouded AGB star losing mass at a rate of 10$^{-4.6}$~M$_{\odot}$yr$^{-1}$ in the work of \citet{2011Ishihara}. A star with that observed flux and mass loss rate would be located at a distance of $\sim$ 14$^{+7}_{-4}$ kpc according to equation A3 of \citeauthor{2011Ishihara}, which would place VVVv42 within the Galactic disc. The distances estimated from the radial velocity of VVVv42 are $d_{near}=4.0$ kpc and $d_{far}=6.1$ kpc. We see that $d_{near}$ places this object within the G305 SFC, but $d_{far}$ does not completely disagree with the distance from Akari.

This object is very likely an evolved star. However, some doubt remains given the near-infrared spectrum of the object. To our knowledge, dust-enshrouded AGB stars have not previously been seen with the cold CO signatures observed in VVVv42. However, this feature could be explained if we are observing the circumstellar dust rather than the stellar photosphere, with CO arising within the circumstellar envelope at T$\sim1000$~K. AGB stars show acetylene absorption in the L-band that can be explained by a similar effect \citep[see e.g.][]{2008Vanloon}.

 
\item {\bf VVVv235} The variable star is likely associated with OH maser OH 330.93$-$0.14 \citep{1997Sevenster} which is located 1.4\arcsec ~from the VVV object. This source is also located 164\arcsec ~from several masers (see Fig. \ref{vvv:agbex}) that form the maser cluster OH 330.953$-$0.182, which is associated with massive star formation \citep{2010Caswell}. 

The $K_{\rm s}$ light curve of VVVv235 is also classified as Mira-like in Paper I, which is confirmed by the mid-infrared information from WISE and NEOWISE. We used the phase dispersion minimization \citep[{\sc pdm}, ][]{1978Stellingwerf} in IRAF to estimate a period in between 760--790 days. Given this, we have to consider the possibility that the maser emission is not related to star formation but rather to an evolved star. Galactic OH/IR stars are characterized by the presence of this type of maser emission. In addition the SED of the object does not appear to be fit by the YSO models, having the largest $\chi^{2}$ in the sample (Table \ref{table:vvvsedfits}). 


We find that a dust-enshrouded AGB star with the measured 9 $\mu$m flux from the Akari survey could be located at d$\sim24.7^{+14}_{-9}$ kpc, which is within the Galactic disc edge (located at $\sim$20.8 kpc at this longitude). However, this distance is much larger than the one estimated from radial velocities. The radial velocity of the object is found to be V$_{LSR}=-92.8\pm0.3$ km s$^{-1}$ ($d_{near}=4.8$~kpc, $d_{far}=10.6$~kpc). The radial velocity agrees with the velocities found for masers in  the maser cluster OH 330.953$-$0.182 \citep[$-$80 to $-$100 km s$^{-1}$][]{2010Caswell}. However, the separation of VVVv235 with respect to the maser cluster is large (164\arcsec). In addition the far kinematic distance is not too dissimilar to the lower limit of the distance from the Akari flux. \citet{1998Schlegel} maps show a large value of the extinction towards VVVv235, with A$_{V}=178$ magnitudes. This large value would have a strong effect even at 9 $\mu$m, bringing the distance into agreement with the kinematic distance, and thus making an AGB classification more likely.

The classification of this object is very likely that of an evolved object.

\item {\bf VVVv796} This star has been previously identified as a YSO in the area of HII region RCW 120 ($d\sim$1.4 kpc) in the work of \citet{2009Deharveng}. These authors identify it based on near- and mid-infrared colour-colour selection and it is estimated to be a class I or class II object; our value of $\alpha$ identifies this star as a class I object. Images from VVV (with SIMBAD sources overlaid) and WISE show a concentration of YSOs around VVVv796. We note that these are likely YSOs associated with RCW120, classified as such based only on near- and mid-infrared colours in the works of \citet{2007Zavagno} and \citet{2009Deharveng}.

This is a deeply embedded, optically invisible star, which is only detected at $H$ and $K_{\rm s}$ in VVV. Accordingly, the spectrum of the object lacks any flux at $J$ band. Some flux can be observed at wavelengths greater than 1.5 $\mu$m and some H$_2$O absorption could be present. The spectrum steeply rises beyond 2 $\mu$m showing deep $^{12}$CO absorption. We note that $^{13}$CO could also be present in the spectrum of the object.

The light curve shows $\Delta K_{\rm s}=2.6$~magnitudes, and it is also a Mira-like long-term periodic variable in Paper I. Near- and mid-infrared photometry from other surveys proves the high variability of this object. VVVv796 is detected in 2MASS with $K_{\rm s}=9.69\pm0.02$ and in DENIS with $K=11.47\pm0.11$. The WISE and NEOWISE photometry is contemporary to VVV. The data shows $\Delta$W$1=1.5$ and $\Delta$W$1=1.3$ magnitudes, where changes in colours show that extinction might be involved. The variability data from WISE also seem to support the periodic variations. 

We use {\sc pdm} in IRAF to determine a period of $P\sim780$--820 days, with some difficulties in phase folding the light curve. The object is detected in the Akari survey and has $F_{9\mu m}=3.1$~Jy. Using similar arguments as before, we find that a dust-enshrouded AGB star with that flux would be located at a distance of $18^{+10}_{-7}$ kpc, which puts it within the Galactic disc. The radial velocity yields $d_{near}=5.1$~kpc, $d_{far}=12.1$~ kpc. 

First, we can see that the radial velocity of the object shows that VVVv796 is not associated with RCW120.  This was to be expected given the large separation from RCW120 ($\sim$ 30\arcmin). Inspection of the Hi-gal 70 $\mu$m image shows some diffuse emission around the object, but does not conclusively prove an active area of star formation.

The far kinematic distance agrees within the errors with the distance estimated to a dust-enshrouded AGB star. The light curve of VVVv796 and the possible $^{13}$CO absorption also supports the latter interpretation. 

The object seems more likely and evolved object.

\item {\bf VVVv45} This object is located within several star formation indicators associated with the G305 Star Forming Complex \citep[see e.g.][]{2012Faimali}. VVVv45 is source 33941 in table 2 of  the study of YSOs towards the G305 star-forming complex of \citet{2009Baume}. This object is selected based on its near-infrared colours, but no information can be found whether the authors consider this as a likely YSO and/or a member of the complex.

The SED of this star is consistent with a class I object, with $\alpha=1.26$. This is an optically invisible object which is not detected at $J$ nor $H$ bands in VVV photometry. The spectrum of the object resembles that of deeply embedded FUor stars, with no flux at wavelengths shorter than 1.9 $\mu$m and steeply rising in the K band, with deep CO absorption at 2.29~$\mu$m. This object does not show any absorption from H$_2$O. We note that $^{13}$CO at 2.34 $\mu$m might  be present in the spectrum of the object.

The light curve of VVVv45 looks clearly to be periodic resembling the light curves of AGB variable stars, and with an amplitude of $\Delta K_{\rm s}=2.2$~ magnitudes. This object had also been detected in 2MASS with $K_{\rm s}=13.53\pm0.04$ magnitudes. The photometry from the WISE surveys also shows the object to be highly variable at mid-infrared wavelengths, with $\Delta W1=1.4$ and $\Delta W2=1.6$ magnitudes. The colour change does not seem consistent with variable extinction. The WISE data also supports the periodic variability. We use {\sc pdm} within IRAF to search for a period in the $K_{\rm s}$ light curve of the object. We find P$\sim520$--530 days, although with some difficulty in phase folding the data. This object has an LPV-Mira classification from paper I. 



VVVv45 is detected at 9 $\mu$m in the Akari mission with F$_{9\mu m}=0.52$~Jy. The $K-[9]$ colour would correspond to a dust-enshrouded AGB star losing mass at a rate of $\sim10^{-4.6}$~M$_{\odot}$yr$^{-1}$ in the work of \citet{2011Ishihara}. A star with that observed flux and mass loss rate would be located at a distance of $44.6^{+26}_{-16}$ kpc according to equation A3 of \citeauthor{2011Ishihara} The estimated distances to a dust-enshrouded AGB star are much larger than the distance to the Galactic disc edge. However, we note that the large extinction towards this source, with A$_{V}\sim$137 mag from \citet{1998Schlegel} maps, could put the object within the Galactic disc. The radial velocity of VVVv45 (see Table \ref{vvv:radvel}) places the object at $d\sim11.6$~kpc

The spectrum of this object points to an embedded YSO classification. However, the distance derived from its radial velocity places it outside the G305 star-forming complex and if the $^{13}$CO is present in the spectrum, then this object is more likely to be an evolved object. The periodicity of the light curve also supports an AGB interpretation. This object is more likely an evolved star.

 
\item {\bf VVVv229} The projected location of this star is within 300\arcsec ~from several indicators of active star formation. The star is not detected in both $J$ and $H$ bands in VVV and its spectrum shows the lack of flux at these wavelengths, rapidly rising towards 2 $\mu$m and only showing CO absorption at 2.29 $\mu$m. This is all in agreement with a FUor classification. However, there are several characteristics that point towards a different classification of the star. The light curve is clearly periodic with an amplitude of $\Delta K_{\rm s}=3.7$~magnitude, making it the most extreme in the spectroscopic sample and the second most extreme in the whole VVV sample, with the first being VVVv360, an OH/IR star. Its $K-[12]$ colour (using WISE filter $W3$) is 14 magnitudes (see Fig. \ref{vvv:specgc}), which is typical of dust-enshrouded AGB stars \citep[see the discussion on these type of variable stars in][]{2014Contreras} and larger than any other object in the spectroscopic sample. The star is clearly not well fitted by the YSO models of \citet{2007Robitaille}, having the second largest $\chi^{2}$ of the spectroscopic sample (see Appendix \ref{section:vvvsedfits}). 

The observed 9 $\mu$m flux is also the largest of the sample, with F$_{9\mu m}=37.8$~Jy. An OH/IR star with this flux can easily be found within the Galactic disc at a distance of $8.6^{+4}_{-3}$ kpc and show up in VVV. Distance which somewhat agrees with $d_{far}$ of $11.4$ kpc from the radial velocity of the source.

Thus, this object is much more likely an evolved star rather than a YSO.

\end{itemize}

\clearpage 

\section{SED fits}\label{section:vvvsedfits}

\begin{table*}
\begin{center}
\begin{tabular}{@{}l@{\hspace{0.5cm}}c@{\hspace{0.2cm}}c@{\hspace{0.2cm}}c@{\hspace{0.2cm}}c@{\hspace{0.2cm}}c@{\hspace{0.2cm}}c@{\hspace{0.2cm}}c@{\hspace{0.2cm}}c@{\hspace{0.2cm}}c@{}}
\hline
Object & $M_{\ast}$ &  $\log \dot{M}_{env}$  & $\log \dot{M}_{disc}$  & $\log Age$ & $ d $ & $\log L_{bol}$  & $\chi^{2}_{best}/N$ & $N_{fits}$ & $N_{fits2}$ \\
 & \footnotesize{(M$_{\odot}$)} &\footnotesize{(M$_{\odot}$~yr$^{-1})$} & \footnotesize{(M$_{\odot}$~yr$^{-1})$} & \footnotesize{(yr)} & \footnotesize{(kpc)} & \footnotesize{$(L_{\odot})$} &  &  & \\
\hline
VVVv20 &  3.0$\pm$2.7 &  $-$5.7$\pm$ 0.7 &   $-$8.1$\pm$1.5 &  5.2$\pm$0.6 & 2.5$^{+3.6}_{-2.4}$ &  1.5$\pm$1.2 &  0.09 & 10000 &  9715\\[0.09cm]
VVVv32 &  2.5$\pm$2.2 &  $-$5.7$\pm$ 0.8 &   $-$8.3$\pm$1.4 &  5.3$\pm$0.7 & 3.1$^{+3.6}_{-3.0}$ &  1.3$\pm$1.1 &  0.01 & 10000 &  8169\\[0.09cm]
VVVv63 &  2.2$\pm$1.6 &  $-$6.6$\pm$ 1.0 &   $-$8.7$\pm$1.6 &  6.3$\pm$0.5 & 3.4$\pm$3.3 &  1.0$\pm$1.2 &  0.01 & 10000 &  4572\\[0.09cm]
VVVv65 &  1.8$\pm$1.7 &  $-$6.3$\pm$ 1.0 &   $-$9.0$\pm$1.4 &  6.0$\pm$0.6 & 2.1$^{+2.9}_{-2.0}$ &  0.7$\pm$1.2 &  0.01 & 10000 &  5152\\[0.09cm]
VVVv94 &  2.4$\pm$1.5 &  $-$7.3$\pm$ 0.8 &   $-$8.4$\pm$1.4 &  6.3$\pm$0.4 & 0.9$^{+1.3}_{-0.8}$ &  1.1$\pm$0.9 &  0.01 &  6200 &  3303\\[0.09cm]
VVVv118 &  2.5$\pm$2.1 &  $-$7.2$\pm$ 0.9 &   $-$9.0$\pm$1.6 &  6.2$\pm$0.4 & 1.5$^{+2.7}_{-1.4}$ &  1.1$\pm$1.2 &  0.04 & 10000 &  3647\\[0.09cm]
VVVv193 &  1.7$\pm$1.5 &  $-$6.7$\pm$ 0.9 &   $-$9.1$\pm$1.3 &  6.0$\pm$0.5 & 1.1$^{+1.9}_{-1.0}$ &  0.5$\pm$0.9 &  0.01 & 10000 &  6563\\[0.09cm]
VVVv270 &  2.3$\pm$1.8 &  $-$5.4$\pm$ 0.9 &   $-$8.0$\pm$1.4 &  5.3$\pm$1.0 & 3.5$\pm$3.3 &  1.3$\pm$0.8 &  0.01 & 10000 &  8190\\[0.09cm]
VVVv322 & 1.5$^{+1.6}_{-1.4}$ &  $-$5.2$\pm$ 0.7 &   $-$8.1$\pm$1.4 &  5.0$\pm$0.8 & 3.6$\pm$3.2 &  0.9$\pm$0.9 &  0.01 & 10000 &  8613\\[0.09cm]
VVVv374 &  4.2$\pm$2.1 &  $-$7.7$\pm$ 0.4 &   $-$8.6$\pm$1.4 &  6.6$\pm$0.3 & 1.2$^{+1.5}_{-1.1}$ &  2.3$\pm$0.6 &  0.35 &   610 &    26\\[0.09cm]
VVVv405 &  4.8$\pm$4.1 &  $-$5.3$\pm$ 0.9 &   $-$6.9$\pm$1.6 &  4.7$\pm$1.3 & 3.7$^{+4.4}_{-3.6}$ &  2.3$\pm$1.2 &  0.05 &  2154 &  1846\\[0.09cm]
VVVv406 &  2.8$\pm$2.6 &  $-$5.2$\pm$ 0.6 &   $-$8.0$\pm$1.7 &  5.0$\pm$0.7 & 2.2$^{+3.0}_{-2.1}$ &  1.6$\pm$1.0 &  0.01 & 10000 &  9300\\[0.09cm]
VVVv452 &  2.1$\pm$1.8 &  $-$5.6$\pm$ 0.9 &   $-$8.5$\pm$1.5 &  5.5$\pm$0.9 & 2.8$\pm$2.7 &  1.2$\pm$1.0 &  0.02 & 10000 &  8171\\[0.09cm]
VVVv473 &  1.5$^{+2.0}_{-1.4}$ &  $-$5.8$\pm$ 0.6 &   $-$8.7$\pm$1.3 &  5.2$\pm$1.0 & 0.5$^{+0.7}_{-0.4}$ &  0.7$\pm$1.1 &  1.17 &  4222 &  3053\\[0.09cm]
VVVv480 &  1.4$\pm$1.3 &  $-$6.5$\pm$ 1.1 &   $-$9.0$\pm$1.5 &  6.1$\pm$0.7 & 4.2$\pm$3.9 &  0.5$\pm$1.1 &  0.04 & 10000 &  5324\\[0.09cm]
VVVv562 &  2.8$\pm$2.0 &  $-$6.3$\pm$ 1.0 &   $-$8.7$\pm$1.4 &  6.1$\pm$0.7 & 2.8$^{+3.1}_{-2.7}$ &  1.4$\pm$1.2 &  0.01 & 10000 &  5946\\[0.09cm]
VVVv625 &  0.8$^{+1.0}_{-0.7}$ &  $-$7.4$\pm$ 0.9 &  $-$10.2$\pm$1.4 &  6.1$\pm$0.4 & 0.3$^{+0.9}_{-0.2}$ &  0.0$\pm$0.7 &  0.08 & 10000 &  4348\\[0.09cm]
VVVv628 &  0.7$^{+1.1}_{-0.6}$ &  $-$7.3$\pm$ 1.0 &  $-$10.3$\pm$1.7 &  6.2$\pm$0.5 & 1.0$^{+1.8}_{-0.9}$ & $-$0.4$\pm$1.0 &  0.11 & 10000 &  3456\\[0.09cm]
VVVv630 &  0.7$^{+0.9}_{-0.6}$ &  $-$7.3$\pm$ 1.1 &  $-$11.1$\pm$2.2 &  6.2$\pm$0.5 & 0.4$^{+1.1}_{-0.3}$ & $-$0.4$\pm$0.8 &  0.01 & 10000 &  2742\\[0.09cm]
VVVv631 &  1.8$\pm$1.7 &  $-$5.4$\pm$ 0.6 &   $-$8.3$\pm$1.4 &  5.1$\pm$0.6 & 2.5$^{+2.8}_{-2.4}$ &  1.0$\pm$0.9 &  0.01 & 10000 &  8986\\[0.09cm]
VVVv632 &  2.7$\pm$2.1 &  $-$5.6$\pm$ 0.8 &   $-$8.1$\pm$1.5 &  5.4$\pm$1.0 & 2.5$\pm$2.4 &  1.5$\pm$0.9 &  0.22 &  4769 &  3667\\[0.09cm]
VVVv662 &  0.3$^{+0.5}_{-0.2}$ &  $-$5.9$\pm$ 0.4 &   $-$9.1$\pm$1.3 &  5.1$\pm$0.4 & 0.3$\pm$0.2 & $-$0.2$\pm$0.4 &  0.03 &  2934 &  2890\\[0.09cm]
VVVv665 &  1.8$\pm$1.1 &  $-$7.6$\pm$ 0.7 &   $-$8.8$\pm$1.4 &  6.2$\pm$0.5 & 0.4$\pm$0.3 &  0.8$\pm$0.7 &  0.20 &  1230 &   539\\[0.09cm]
VVVv699 &  5.6$\pm$2.4 &  $-$7.8$\pm$ 0.7 &   $-$7.3$\pm$1.1 &  6.5$\pm$0.3 & 5.1$\pm$3.5 &  2.7$\pm$0.7 &  0.02 &  1960 &    50\\[0.09cm]
VVVv717 &  2.9$\pm$2.2 &  $-$7.6$\pm$ 0.8 &   $-$8.4$\pm$1.5 &  6.3$\pm$0.4 & 1.8$^{+2.6}_{-1.7}$ &  1.4$\pm$1.2 &  0.03 &  9474 &  2988\\[0.09cm]
VVVv721 &  2.4$\pm$2.1 &  $-$5.4$\pm$ 0.8 &   $-$8.0$\pm$1.5 &  5.2$\pm$0.9 & 2.9$\pm$2.8 &  1.4$\pm$1.0 &  0.03 & 10000 &  8186\\[0.09cm]
VVVv800 &  4.5$\pm$3.9 &  $-$5.4$\pm$ 0.9 &   $-$7.0$\pm$1.5 &  4.7$\pm$1.1 & 3.8$^{+4.4}_{-3.7}$ &  2.2$\pm$1.2 &  0.18 &  1573 &  1389\\[0.09cm]
VVVv815 &  2.8$\pm$2.5 &  $-$5.4$\pm$ 0.9 &   $-$7.1$\pm$1.5 &  4.5$\pm$1.1 & 3.0$^{+3.1}_{-3.0}$ &  1.6$\pm$1.0 & 0.48 &  4336 &  4011\\[0.09cm]
\hline
\multicolumn{10}{c}{Non-YSOs}\\
\hline
VVVv25 &  7.7$\pm$4.9 &  $-$7.5$\pm$ 0.8 &   $-$7.2$\pm$1.5 &  6.4$\pm$0.3 & 2.7$^{+3.1}_{-2.6}$ &  3.1$\pm$0.9 &  0.60 &  1214 &     2\\[0.09cm]
VVVv42 &  8.4$\pm$5.1 &  -- &   $-$8.0$\pm$1.9 &  6.4$\pm$0.3 & 2.0$\pm$1.9 &  3.2$\pm$0.9 &  1.59 &   139 &     0\\[0.09cm]
VVVv45 &  6.6$\pm$3.7 &  $-$7.5$\pm$ 1.1 &   $-$7.9$\pm$1.4 &  6.5$\pm$0.3 & 2.4$\pm$2.3 &  2.9$\pm$0.8 &  2.48 &  3464 &    39\\[0.09cm]
VVVv202 &  2.8$\pm$2.3 &  $-$6.8$\pm$ 1.0 &   $-$9.4$\pm$1.5 &  6.2$\pm$0.4 & 1.0$^{+2.2}_{-0.9}$ &  1.2$\pm$1.1 &  0.38 &  6090 &  2879\\[0.09cm]
VVVv229 &  7.5$\pm$2.4 &  -- &   $-$8.1$\pm$0.9 &  6.4$\pm$0.2 & 0.5$\pm$0.4 &  3.3$\pm$0.4 &  3.01 &    62 &     0\\[0.09cm]
VVVv235 & 10.0$\pm$1.8 &  -- &   $-$8.4$\pm$0.3 &  6.4$\pm$0.1 & 1.3$\pm$0.5 &  3.7$\pm$0.3 &  3.61 &     3 &     0\\[0.09cm]
VVVv240 &  6.8$\pm$3.2 &  -- &   $-$7.4$\pm$1.2 &  6.4$\pm$0.3 & 1.6$\pm$1.5 &  3.0$\pm$0.7 &  0.40 &  1180 &     0\\[0.09cm]
VVVv514 &  0.4$^{+0.5}_{-0.3}$ &  $-$7.7$\pm$ 0.9 &  $-$12.7$\pm$1.9 &  6.4$\pm$0.4 & 0.3$\pm$0.2 & $-$0.7$\pm$0.6 &  0.01 & 10000 &  2319\\[0.09cm]
VVVv796 &  9.5$\pm$4.1 &  -- &   $-$7.5$\pm$1.6 &  6.3$\pm$0.2 & 2.7$\pm$2.0 &  3.5$\pm$0.7 &  0.15 &   430 &     0\\
\hline
\end{tabular}
\caption{Parameters derived from the \citeauthor{2007Robitaille} models SED fitting as explained in the text.}\label{table:vvvsedfits}
\end{center}
\end{table*}

We use the command-line version of the SED fitting tool of \citet{2007Robitaille} to provide a rough estimate of the physical properties of objects in the sample. We note that in our analysis, this tool is used as a measurement of how well the SED of our objects can be fitted by that of a YSO, and we do not take the values of physical parameters as reliable estimates for our variable stars. Nevertheless, these results do allow us to make some general remarks about the likely characteristics of the sample. 

For the SED fits, we use photometry arising from VVV ($JHK_{\rm s}$, Paper I), WISE $W1-W4$ \citep{2010Wright}, or (when necessary) {\it Spitzer} IRAC $I1-I4$ and MIPS 24 $\mu$m data from the GLIMPSE and MIPSGAL surveys \citep{2003Benjamin, 2009Carey}. We note that this process is unreliable given that the \citeauthor{2007Robitaille} fits might not be able to describe these highly variable sources. In addition, the near and mid-infrared photometric measurements are not contemporaneous. To minimise the latter problem, we combine the contemporaneous VVV $JHK_{\rm s}$ data from 2010 with {\it WISE} data taken in the same year, in preference to {\it Spitzer} data from the earlier 
GLIMPSE survey. If a {\it WISE} detection is lacking due to source confusion, which occurred for 7 sources, 
{\it Spitzer} GLIMPSE photometry is fed to the fitting tool. The effect of using {\it Spitzer} data taken in 2003--4 is discussed below. 
A 2\arcsec cross match radius was used for {\it WISE} and MIPSGAL matches to each VVV source, whereas a 1\arcsec radius was used 
for GLIMPSE data. In an additional attempt to minimise the effect of using non-contemporanoues data, we also feed the tool with larger uncertainties for the data. Typically we used 0.2 mag for near-infrared ($JHK_{\rm s}$) observations and 0.4 mag for mid-infrared data from {\it WISE} or {\it Spitzer}.
Far-infrared fluxes are estimated from Hi-GAL PACS and/or SPIRE photometry \citep{2010Molinari}. Detections at 70 $\mu$m are given as data points to the tool, whilst detections at longer wavelengths are given as upper limits. If the objects are not detected at these wavelengths, then we provided the fitting tool with upper limit fluxes from \citet{2013Elia}. The interstellar extinction and distance to the objects where left as free parameters, with A$_{V}=$1--50 magnitudes and $d=0.1$--13 kpc.


Table \ref{table:vvvsedfits} shows the weighted mean values for different parameters of the spectroscopic sample. The weights are taken as the inverse of $\chi^{2}$, so results from the best fits are given greater weights when estimating the mean value. The values are determined using the models for which $\chi^{2}-\chi_{best}^{2} < 3N$ \citep[as suggested in][]{2007Robitaille}, where $N$ is the number of data points used for the fits. These physical parameters include the stellar age, mass, $M_{\ast}$, envelope infall rate, $\dot{M}_{env}$, disc mass accretion rate, $\dot{M}_{disc}$, luminosity of the system, $L_{tot}$, and distance to the system, $d$. We also provide information regarding the $\chi^{2}$ per data point of the best-fitting model, $\chi^{2}_{best}/N$ and the number of fits used to derive the weighted mean values, $N_{fits}$. We note that the envelope infall rate is derived only using those models which had a non-zero value for this parameter. The number of models which fulfil this condition are given by $N_{fits2}$ in the table. In this table we have divided objects according to the final classification given in this work, i.e. between objects with characteristics of YSOs and those that are more likely evolved objects, such as AGB stars and Novae.

The values of $\chi^2/N$ provided in Table \ref{table:vvvsedfits} are small in many cases. This might suggest that the error bars are over-estimated (which is a result of our attempt to minimise the effect of using non-contemporaneous data), or alternatively that the fitting tool has so many free parameters and such a
large library of model results that it can often find a perfect fit between YSO data and a given YSO model, with suitable reddening and inclination of the system. Given this, and as mentioned before, the values obtained from SED fits are only used to make general remarks regarding the likely properties of the spectroscopic sample.

We studied the effect that using photometry from a different survey or giving a narrow range in distance, had on the final values of physical parameters presented in Table \ref{table:vvvsedfits}. In both cases we find that, although the weighted means suffer a change, these variations are always smaller than the errors on the estimation of the physical parameter. Thus, we always prefer the use of WISE photometry given that these observations overlap with VVV (or at least occur a few months apart). Due to the possibility of mis-association with an SFR due to chance projection, we prefer the use of a range 0.1--13 kpc for all the fits.



\begin{figure}
\centering
\resizebox{\columnwidth}{!}{\includegraphics{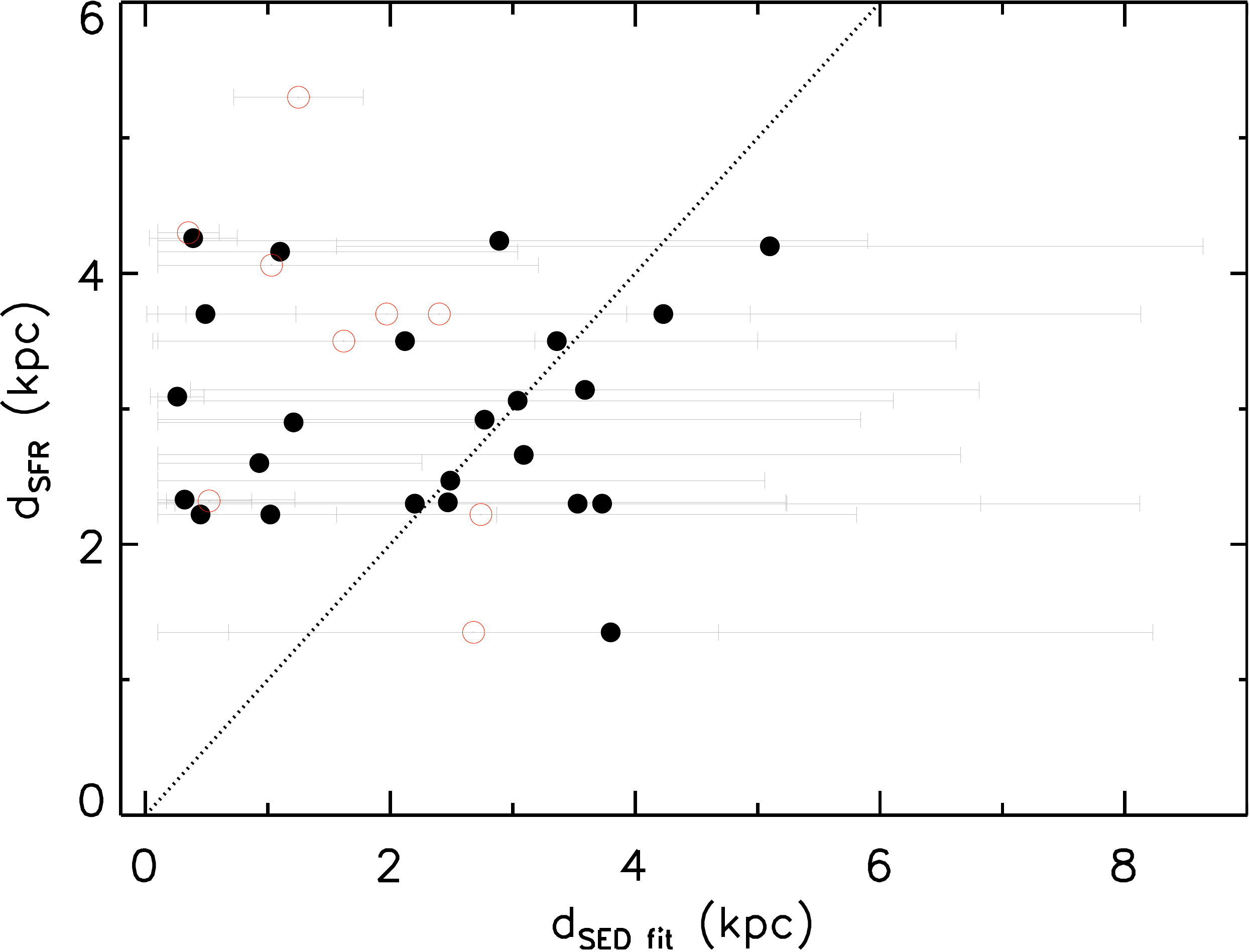}}
\caption{Comparison of the distances from SFRs taken from literature sources, d$_{SFR}$ and those provided by the results from 
SED fits. Objects later classified as more likely non-YSOs are marked by red circles. The dotted line marks the identity line. For the sake of simplicity we do not give errors on d$_{SFR}$ 
because this is difficult to do accurately.}
\label{vvv:seddistances}
\end{figure}

From Table \ref{table:vvvsedfits}, we first see that objects which are classified as Mira-like from their light curves in Paper I, and that also have other characteristics that support an AGB classification as discussed later on this work, tend to have poor fits to the model grid. In fact, we see that 26/28 YSOs show $\chi^{2}/N <0.4$, whilst the majority of non-YSOs (6/9) show $\chi^{2}/N>$0.4. The three non-YSOs with $\chi^{2}/N <0.4$, VVVv202, VVVv514 and VVVv796 are classified as non-YSOs based on their spectroscopic and photometric properties (see Appendix \ref{vvv:sec_evolved}).  

The use of a large range in distance allows to check whether the distance provided by the fitting tool, $d_{fit}$, agrees with that of the SFR to which the object appears to be associated with. Fig. \ref{vvv:seddistances} shows the comparison of $d_{fit}$ with
the literature based estimate $d_{SFR}$, for objects where $d_{SFR}$ is available. It is hard to see a correlation between the two measurements,
even considering only the likely YSOs (black circles in Fig. \ref{vvv:seddistances}). However,
in most cases the distances agree within the (large) errors estimated from
the SED fits, so the apparent lack of correlation may well be due to the
large uncertainties. The d$_{SFR}$ values will also be uncertain, or simply
wrong in cases of chance projection. Fortunately, both methods agree
that distances for most of the YSOs are larger than 2 kpc. The stellar masses and luminosities of most of the objects are also found to be larger than the ones observed in the UKIDSS GPS sample of \citet{2014Contreras}. This agrees well with our suggestion in Paper I that the VVV is detecting a larger number of more distant, higher luminosity YSOs than UGPS 
due to the mid-plane location of the VVV sample.
The fitted masses typically correspond to intermediate mass YSOs but they span a wide range. The possibility that some of the highly variable YSOs in the full VVV photometric sample may be high mass objects is the subject of a separate paper (Kumar et al., in prep).


The disc accretion rates indicated by the SED fits are typically a few $\times 10^{-9}$~M$_{\odot}$~yr$^{-1}$. This is lower than the expected
level for low mass EXors in nearby star formation regions during outburst \cite[e.g.][]{2014Audard} and lower than the 
$10^{-3}$ to $10^{-6}$~M$_{\odot}$~yr$^{-1}$ seen in the accretion events of FUors. As we have shown, the sample is dominated by emission line 
objects with EXor-like spectra rather than FUors, but these accretion rates still appear low for relatively luminous YSOs at distances
of a few kpc. However, we use 2010 photometry for the SED fits, an epoch in which the majority of the objects, from inspection of their light 
curves, appear to be at quiescent states (with the exception of some objects that show fading light curves such as VVVv562, see Section \ref{vvv:sec_varacc}). In Section \ref{vvv:erupvars} we estimate accretion luminosities, L$_{acc}$, using the Br$\gamma$ emission line observed in most of the YSOs. The accretion rates derived from L$_{acc}$ for these objects show higher values than those from SED fits, and more in line with the values expected in eruptive variables. The spectra were typically taken when the sources were brighter.

\clearpage
 
\section{CO Model}\label{vvv:co}

In this section we present a first attempt to determine the temperature of the region responsible for the CO emission/absorption in some of our objects. This is based on the modelling of the CO band emission from MCW 349  (a peculiar B[e] star) done by \citet{2000Kraus}. A brief explanation of the basics of the model is presented below.

The energy of the rovibrational states ($v,J$) of a diatomic molecule are expanded in the following way \citep[from][]{1932Dunham} 

\begin{equation}
E(v,J)=\sum_{k,l} Y_{k,l}\left(v+\frac{1}{2}\right)^{k}\left(J^{2}+J\right)^{l},
\end{equation}  

\noindent with $Y_{k,l}$ the Dunham coefficients, which for CO are taken from \citet{1991Farrenq}. Now, assuming that the gas is in local thermodynamical equilibrium (LTE) then the levels are populated according to
  
\begin{equation}
n_{v,J}=\frac{n}{Z}(2J+1)e^{\frac{-E(v,J)}{kT}},
\end{equation}

\noindent where $n$ and $T$ are the total number density and temperature of CO molecules and $Z$ corresponds to the partition function. The latter has been derived by summing  over vibrational levels up to $v=15$ and rotational levels up to $J=110$, where the states with larger quantum number contribute little to the partition function due to their low level populations.

The absorption coefficient is calculated from 

\begin{equation}
\kappa_{\nu}=\frac{c^{2}n_{v,J}A_{vJ;v'J'}}{8\pi\nu^{2}}\left(\frac{2J+1}{2J'+1}\frac{n_{v',J'}}{n_{v,J}}-1\right)\phi(\nu),
\end{equation}

\noindent with $v$ and $J$ the indices for the upper energy level, whilst $v'$ and $J'$ are for the lower energy level. $\phi(\nu)$ is the line profile function and $A_{vJ;v'J'}$ the Einstein A coefficient, which are taken from \citet{1996Chandra}.

The absorption coefficient is assumed to be constant along the line of sight; in this way the optical depth is simply the product of the absorption coefficient per CO molecule times the CO column density, i.e.

\begin{equation}
\tau_{\nu}=N_{CO}\times\sum_{lines} \frac{\kappa_{\nu}}{n}
\end{equation}

The line profile function is assumed to be Gaussian (neglecting disc rotation) of the form

\begin{equation}
\phi(\nu)=\frac{c}{\sqrt{\pi}\nu_{0}v_{g}}\exp\left[-\frac{c^{2}}{v_{g}^{2}}\left(\frac{\nu-\nu_{0}}{\nu_{0}}\right)^{2}\right],
\end{equation}

\noindent with $v_{g}$ the width of the Gaussian profile, where the dominant broadening mechanism corresponds to the instrumental resolution. In these preliminary calculations we assume $v_{g}=50$ km s$^{-1}$, the instrumental resolution of the FIRE spectrograph. 

Finally the flux of the absorbing CO gas will be given by

\begin{equation}
I_{\nu}=B(\lambda,T)e^{-\tau_{\nu}},
\end{equation}

\noindent and 

\begin{equation}
I_{\nu}=B(\lambda,T)(1-e^{-\tau_{\nu}}),
\end{equation}

\noindent for the emitting gas.

\begin{table}
\begin{center}
\begin{tabular}{@{}l@{\hspace{0.5cm}}c@{\hspace{0.2cm}}c@{\hspace{0.2cm}}}
\hline
Object & T (K) & N$_{\mathrm{CO}}$ (cm$^{-2}$)  \\
\hline
\multicolumn{3}{l}{Emission} \\
\hline
VVVv32 & 2500 & $3\times10^{20}$ \\
VVVv193 & 2500 & $1\times10^{20}$ \\
VVVv270  & 2700 &   $1\times10^{20}$\\
VVVv665  & 2900 &  $6\times10^{20}$\\
VVVv699  & 2900 & $5\times10^{20}$ \\
\hline
\multicolumn{3}{l}{Absorption}\\
\hline
VVVv25 & 1800 & $2\times10^{20}$ \\
VVVv42 & 1100 & $9\times10^{19}$ \\
VVVv45 & 2100 & $5\times10^{20}$\\
VVVv229 & 1600 & $3\times10^{20}$ \\
VVVv235 & 2100 & $5\times10^{20}$ \\
VVVv322 & 2700 & $2\times10^{20}$ \\
VVVv628 & 3200 & $1\times10^{20}$\\
VVVv630 & 3100 & $1\times10^{20}$\\
VVVv717 & 1600 & $2.5\times10^{20}$ \\
VVVv721 & 2200 & $1\times10^{20}$\\
VVVv796 & 2800 & $2\times10^{20}$ \\
\hline
\end{tabular}
\caption{Temperature, T, and column density, N$_{\mathrm{CO}}$, estimated from fitting the CO model to the observed spectrum of VVV objects.}\label{table:vvvcomodel}
\end{center}
\end{table}

In these preliminary calculations we assume that this is the emission/absorption arising from a non-moving cloud of gas. We intend to include the effect of rotation  and inclination of an accretion disc as well as the varying of the temperature and density profiles with distance to the central star \citep[following e.g.][]{1996Najita}. In addition, in these preliminary attempts the best models are obtained by manually varying the temperature and column density of the gas until obtaining a model that best described the observations. 

\begin{figure}
\begin{center}
\resizebox{0.8\columnwidth}{!}{\includegraphics[]{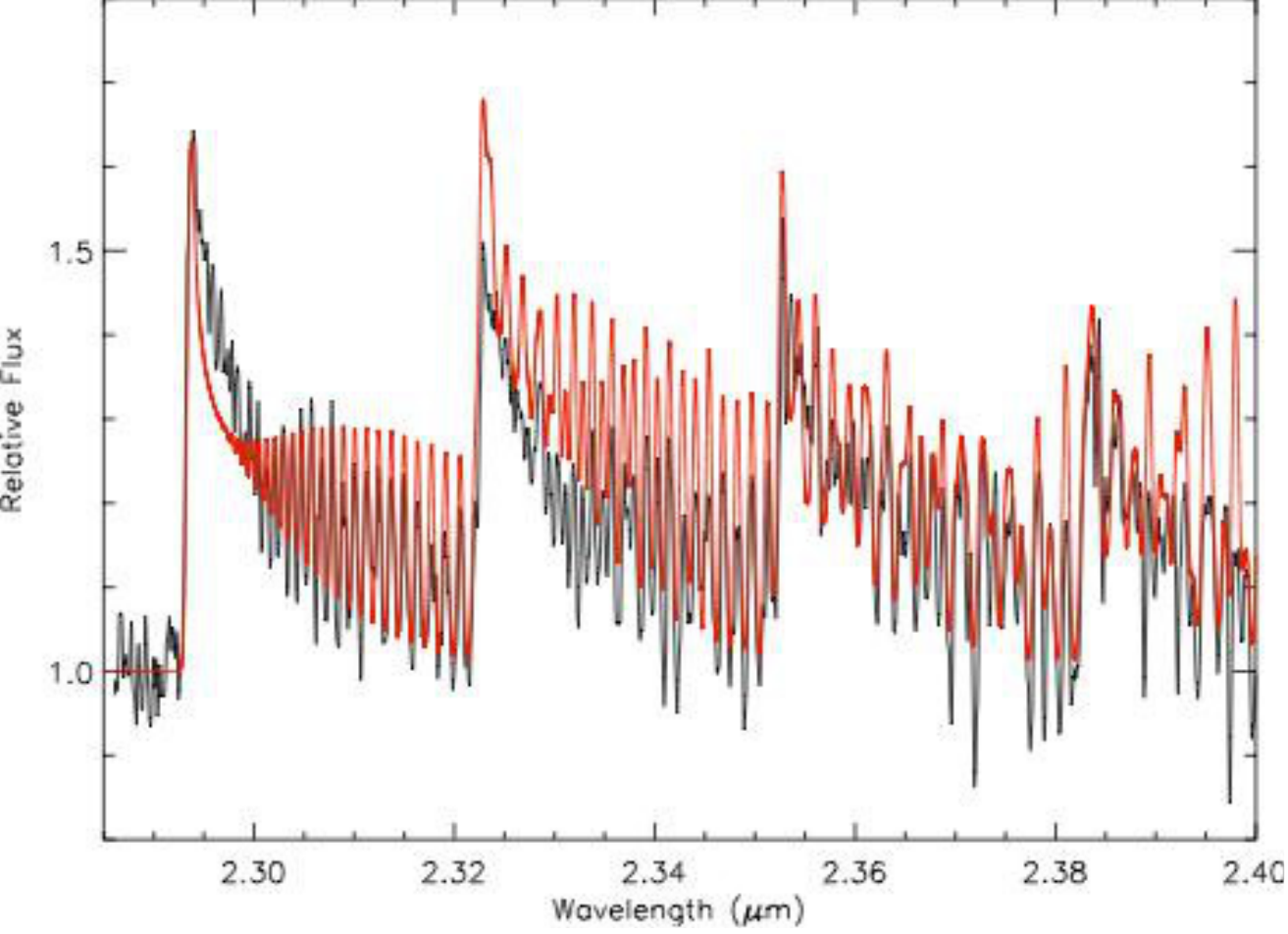}}
\caption{Observed spectrum (black line) and CO model (red line) for VVVv665. The model is derived using T=2500 K and N$_{\mathrm{CO}}$=$6\times10^{20}$~cm$^{-2}$.}
\label{fig2}
\end{center}
\end{figure}  

In the case of CO emission we find that objects that show this characteristic can be fitted by models with temperatures between 2500-3500 K and column densities in between $9\times10^{19}$--$6\times10^{20}$~cm$^{-2}$ (see Table \ref{table:vvvcomodel}). The range in temperatures agrees with those expected to give rise to CO emission in gaseous discs of YSOs.  Fig. \ref{fig2} shows the results of applying the model (T=2900 K, N$_{\mathrm{CO}}=6\times10^{20}$~cm$^{-2}$) to VVVv665. Here we observe that the observed profile of the $\nu=2$--0 bandhead seems to be broader than the one produced by the model. Increasing the values of the column densities produces a broadening of the bandhead profile \citep{2000Kraus}. However, even with high values of N$_{\mathrm{CO}}$ the broad CO profile of VVVv665 cannot be reproduced. Rotation in Keplerian discs have been shown to produce broadened CO bandhead profiles  \citep[see e.g.][]{1996Najita, 2010Davies}. Although this needs a more detailed investigation, the observed broadened profile might give evidence of the presence of an accretion disc in VVVv665.

\begin{figure}
\begin{center}
\resizebox{0.8\columnwidth}{!}{\includegraphics[]{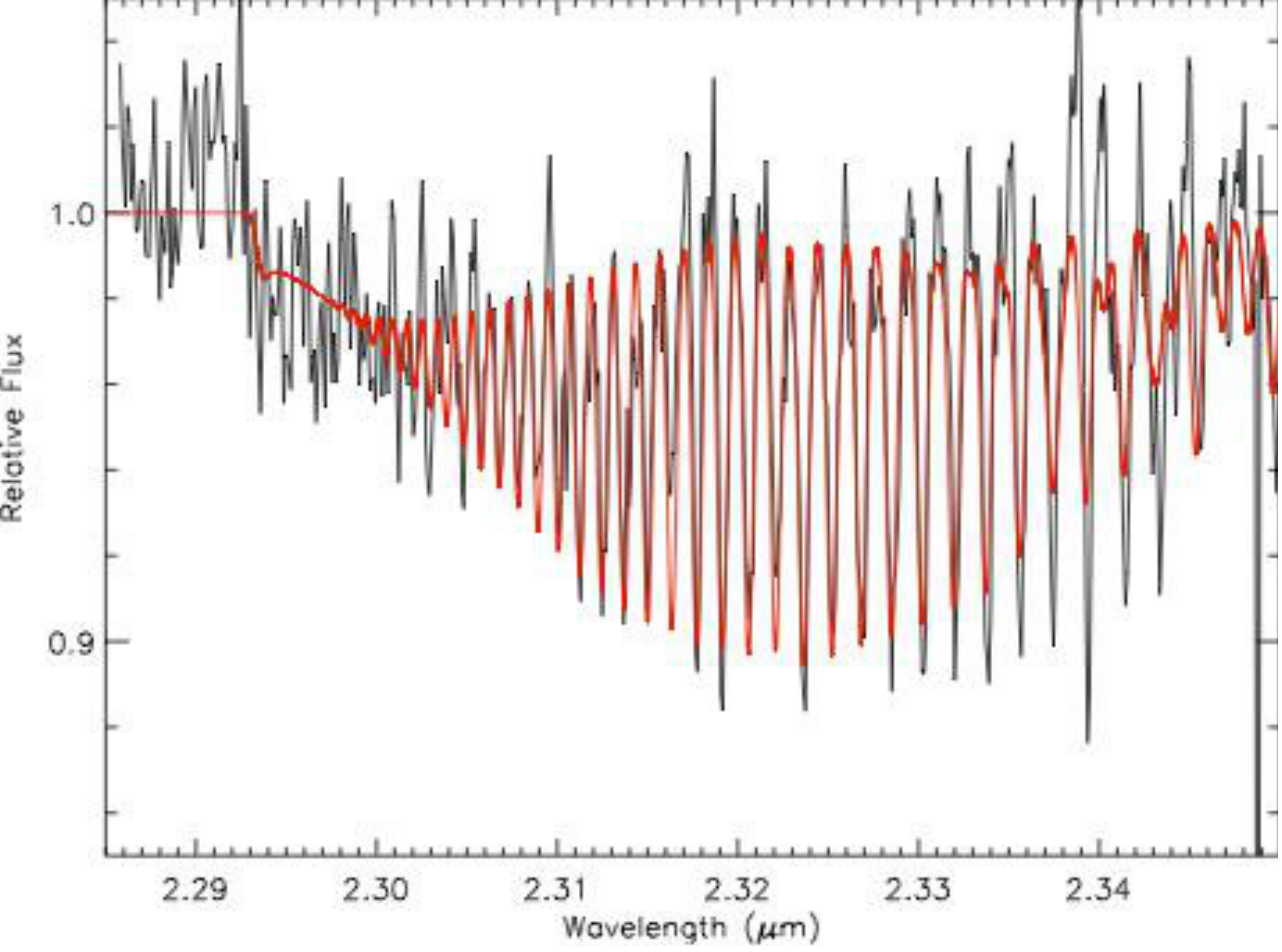}}
\caption{Observed spectrum (black line) and CO model (red line) for VVVv42. The model is derived using T=1100 K and N$_{\mathrm{CO}}$=$9\times10^{19}$~cm$^{-2}$.}
\label{fig1}
\end{center}
\end{figure}

We also estimate temperatures for objects showing CO absorption, a sample mainly composed of the FUor candidates that were shown in previous sections. One particularity interesting case is VVVv42. In here the absorption seems to arise from cooler CO gas given the lack of prominent bandhead absorption and stronger absorption from low-$J$ transitions of the CO $\nu=2$--0 state. The observations are best fitted by models with $T\sim$1100 K. The object resembles likely eruptive variables GPSV3 and GPSV15 in \citet{2014Contreras}. Applying the model to these objects yield similar results ($T\sim1100$--$1800$ K). \citet{2014Contreras} have argued that absorption in these objects might be arising from an ejecta (as in GPSV15) or from outer parts of the disc. \citet{2010Davies} shows that the CO absorption observed in W33A is due to cool gas ($\sim$30 K) in the envelope of the massive YSO. It is possible then that we are observing a similar effect in our VVV object. Fig. \ref{fig1} shows the results of applying the model to VVVv42.

\clearpage

\section{Spectra and light curves of VVV high amplitude variables}\label{apen2}

Here we show the 37 Magellan/FIRE spectra and the 2010--2015 VVV light curves for all the high amplitude 
variables discussed in this paper. We mark some of the most prominent features observed in the spectra.
The preliminary light curve (LC) classifications from Paper I are indicated for each source. Note that
these were based on the 2010--2014 photometry only, not including the final datum from 2015, and a few of 
these initial classifications were found to be incorrect after consideration of the spectra and other 
data; see main text. The final classifications (FC) for the objects are also indicated in each figure. In the light curve plots, the date of spectroscopic follow-up is marked with a blue dashed line.

\begin{figure*}
\resizebox{0.75\textwidth}{!}{\includegraphics{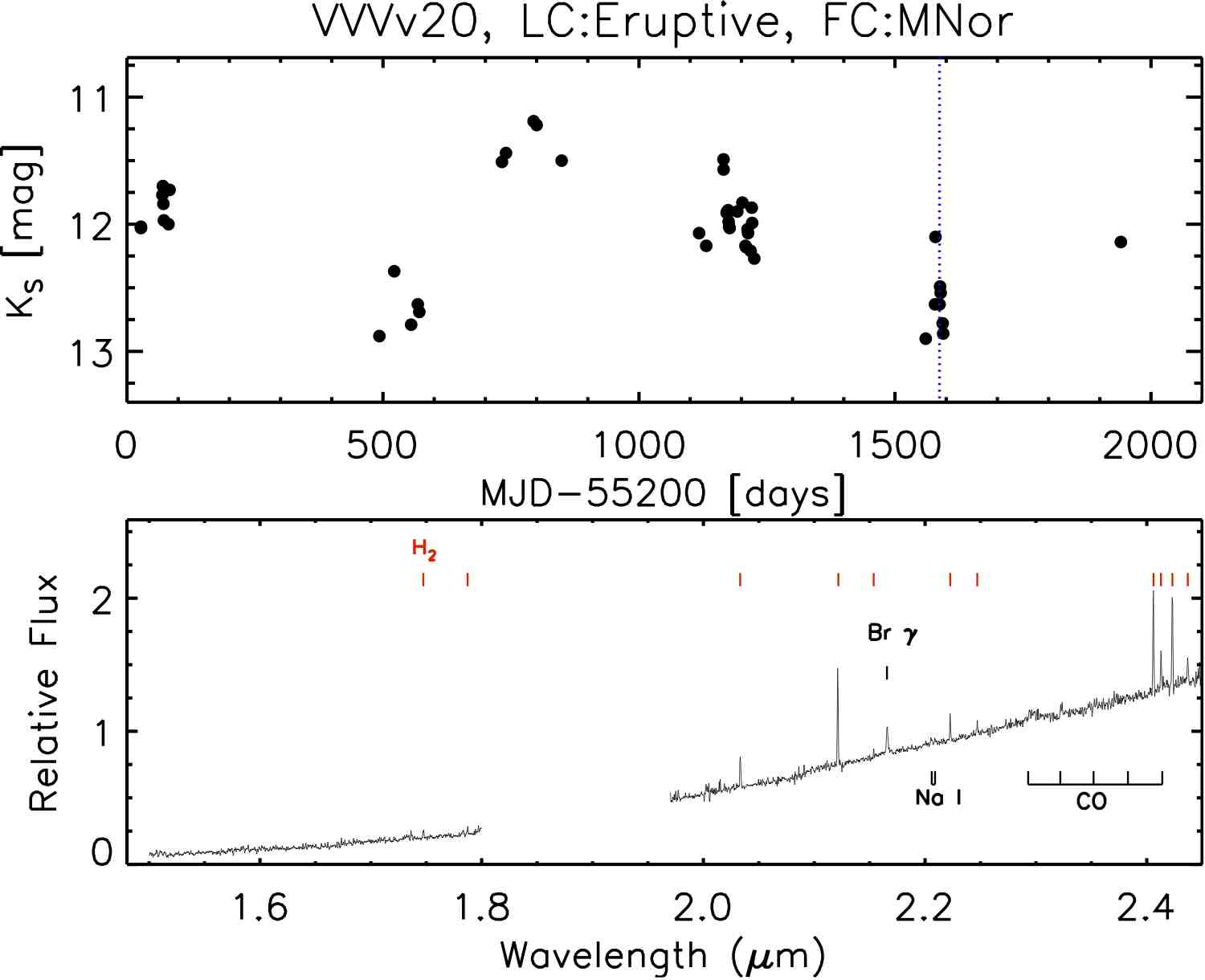}}\\
\resizebox{0.75\textwidth}{!}{\includegraphics{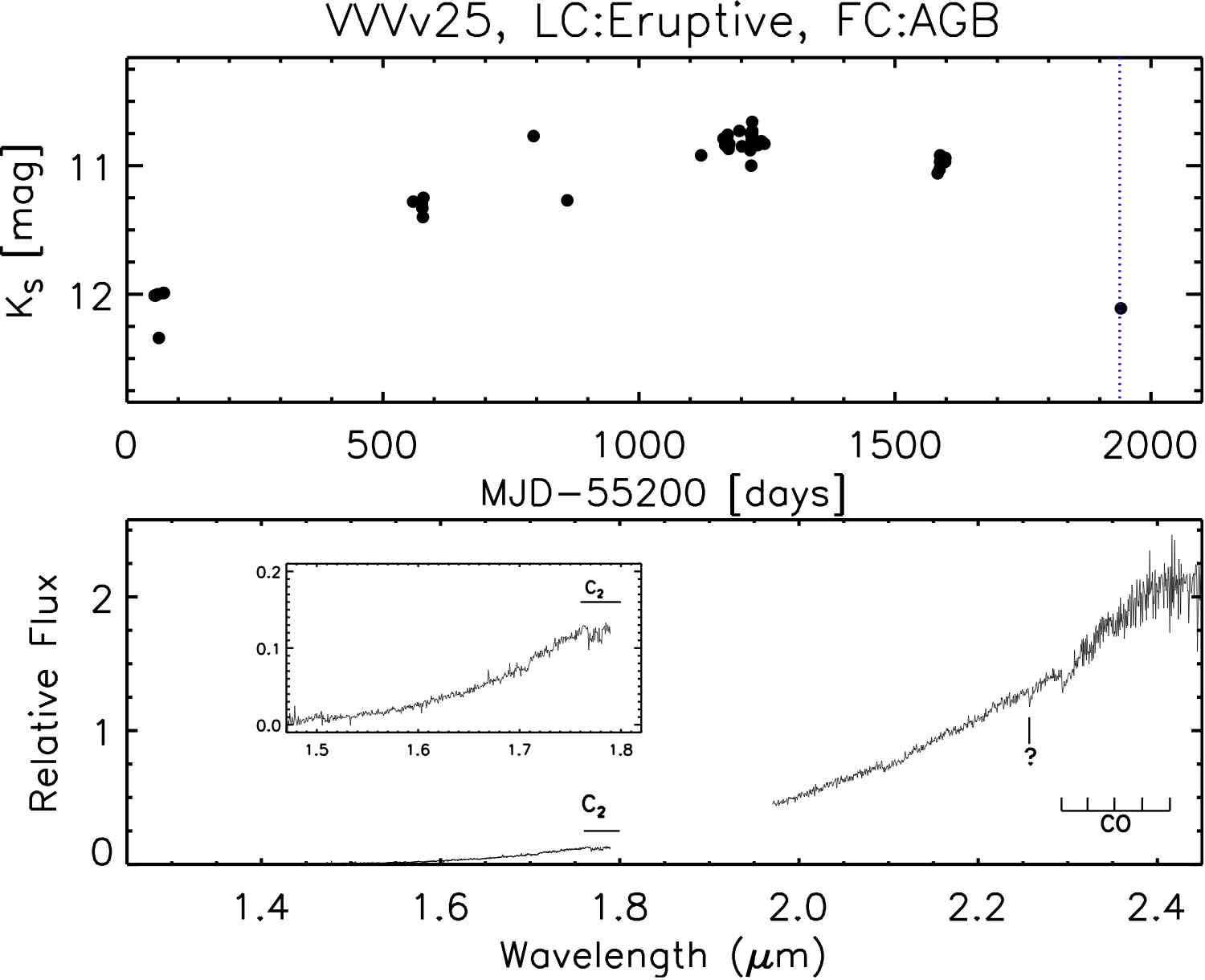}}
\caption{FIRE spectra and $K_{\rm s}$ light curves.}
\label{apen:fig1}
\end{figure*}

\begin{figure*}
\resizebox{0.75\textwidth}{!}{\includegraphics{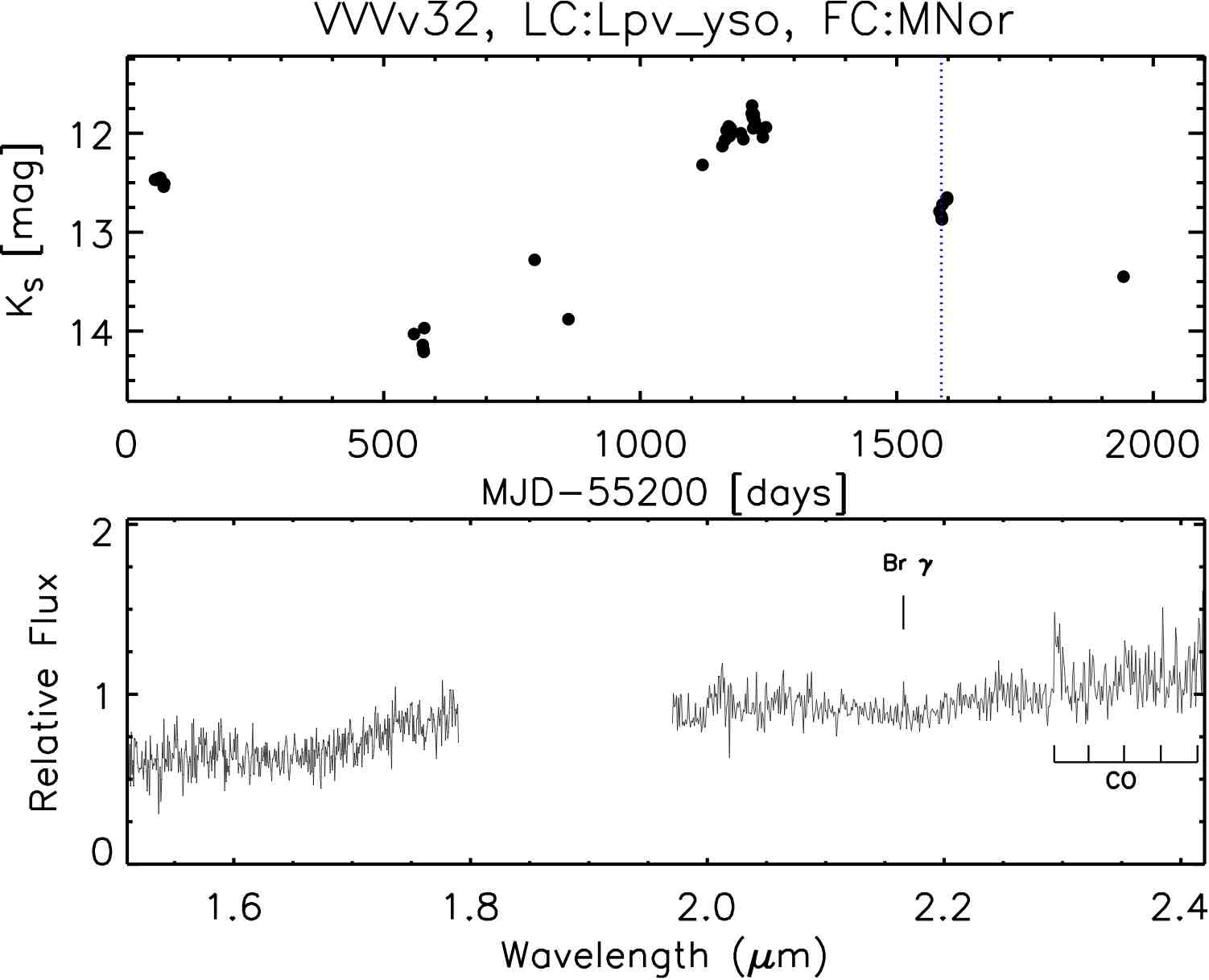}}\\
\resizebox{0.75\textwidth}{!}{\includegraphics{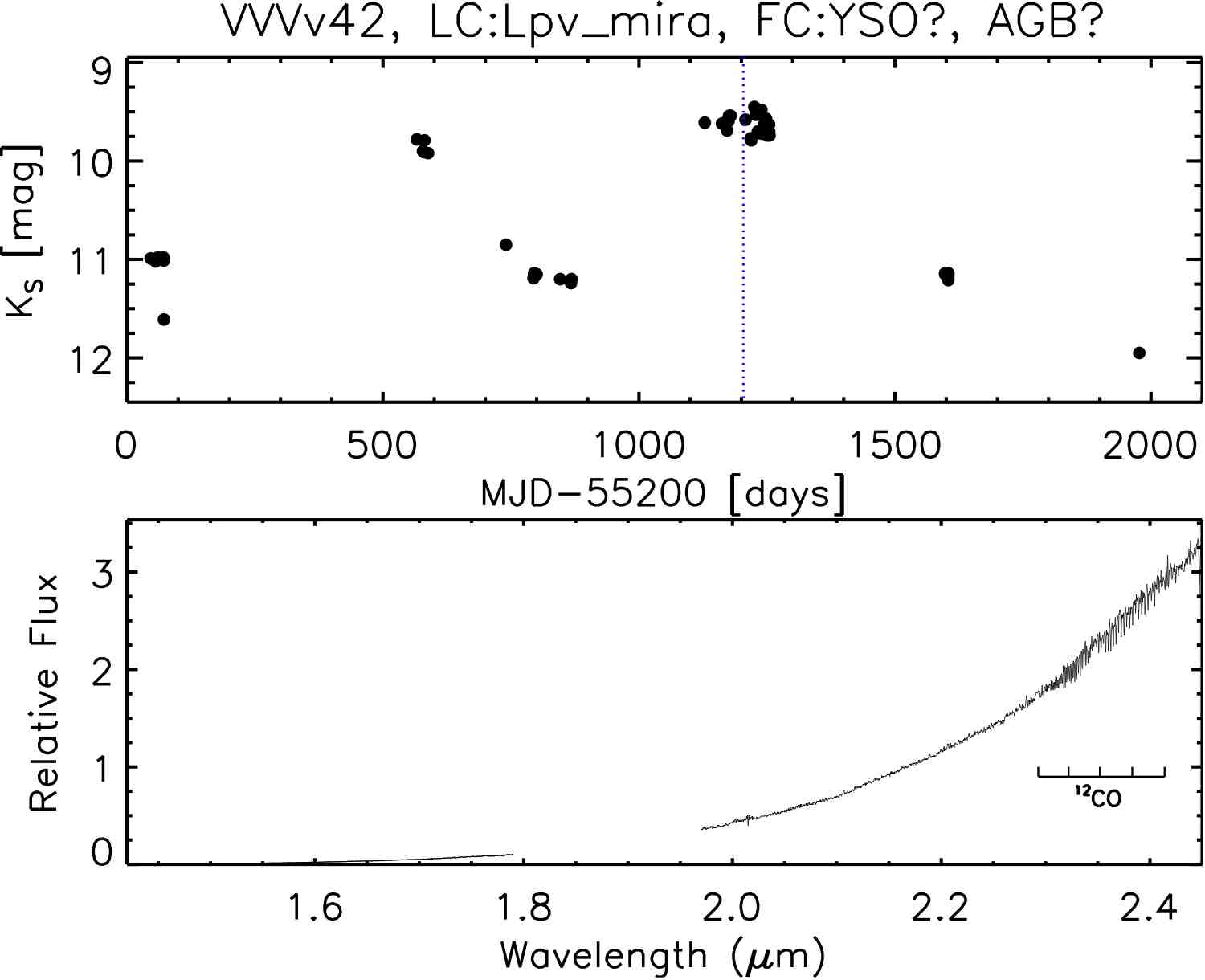}}
\caption{FIRE spectra and $K_{\rm s}$ light curves.}
\label{apen:fig2}
\end{figure*}

\begin{figure*}
\centering
\resizebox{0.75\textwidth}{!}{\includegraphics{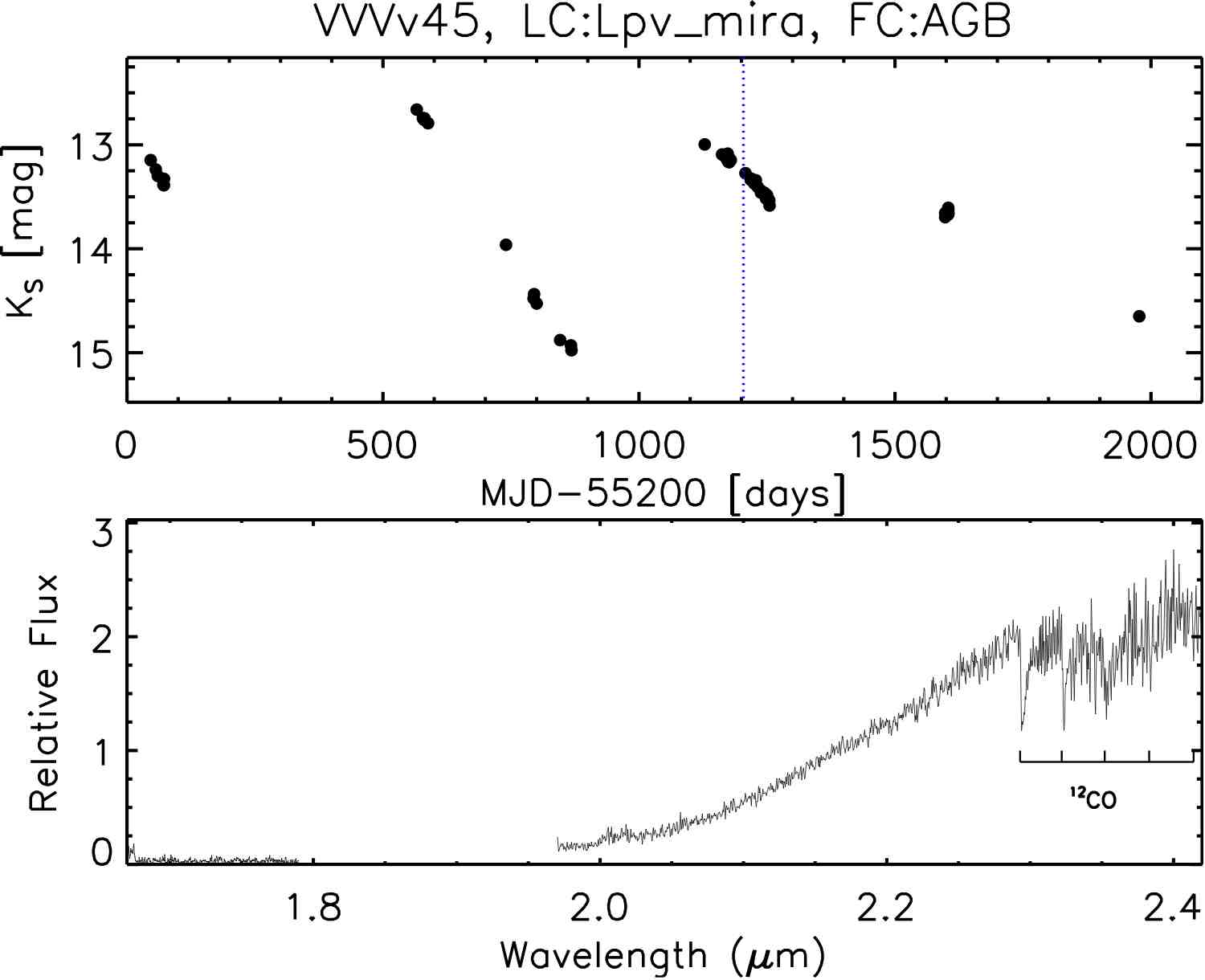}}\\
\resizebox{0.75\textwidth}{!}{\includegraphics{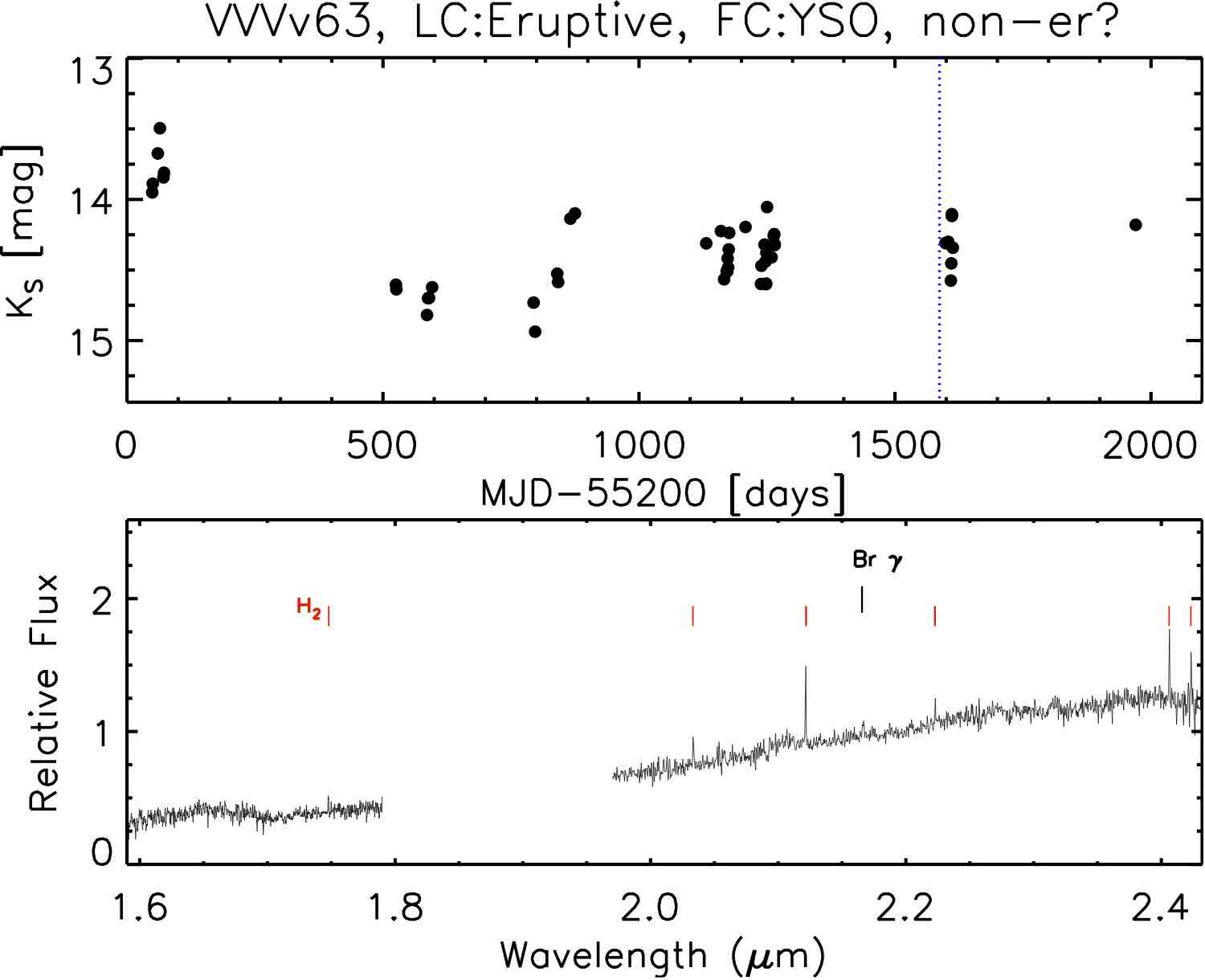}}
\caption{FIRE spectra and $K_{\rm s}$ light curves.}
\label{apen:fig3}
\end{figure*}

\begin{figure*}
\centering
\resizebox{0.75\textwidth}{!}{\includegraphics{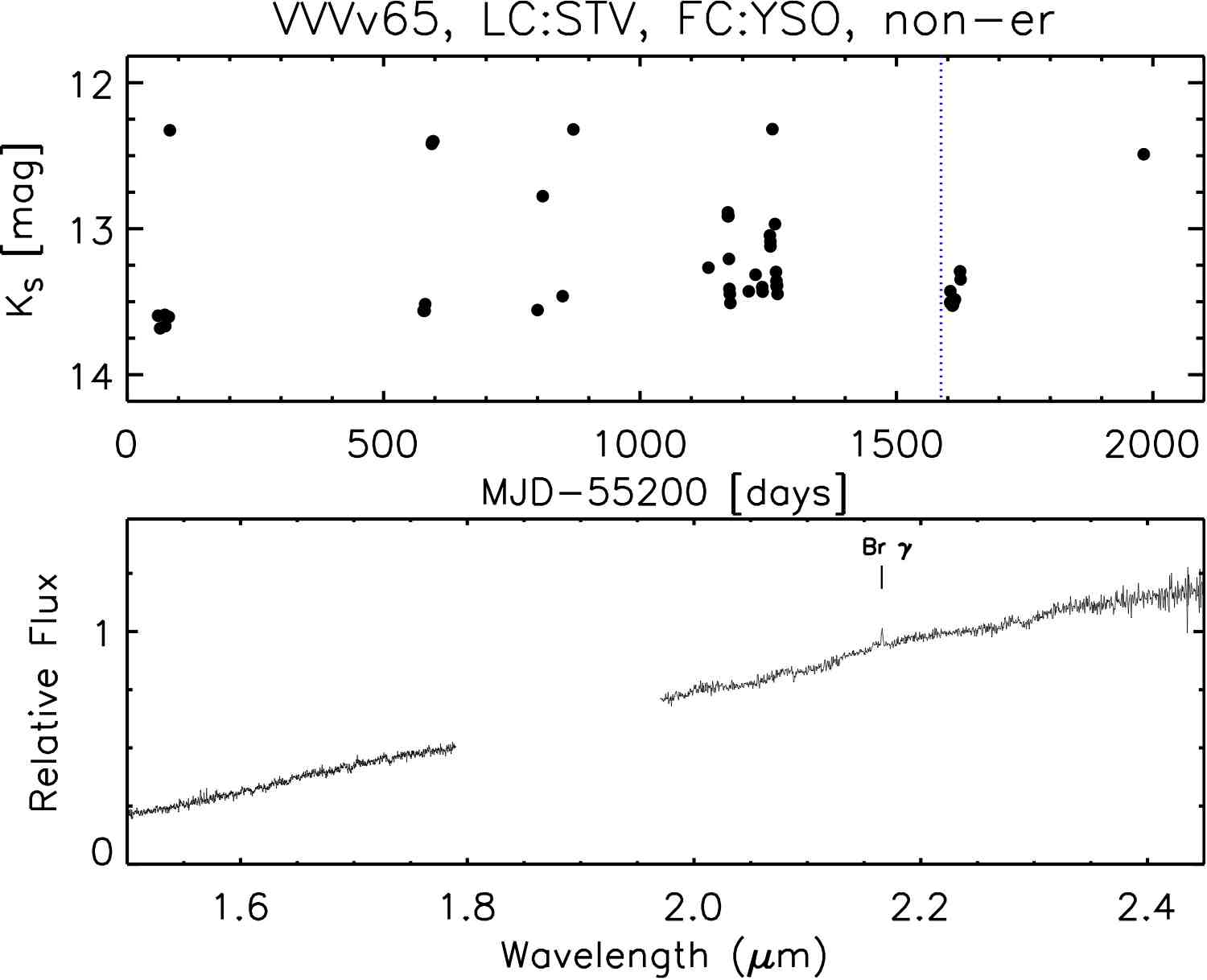}}\\
\resizebox{0.75\textwidth}{!}{\includegraphics{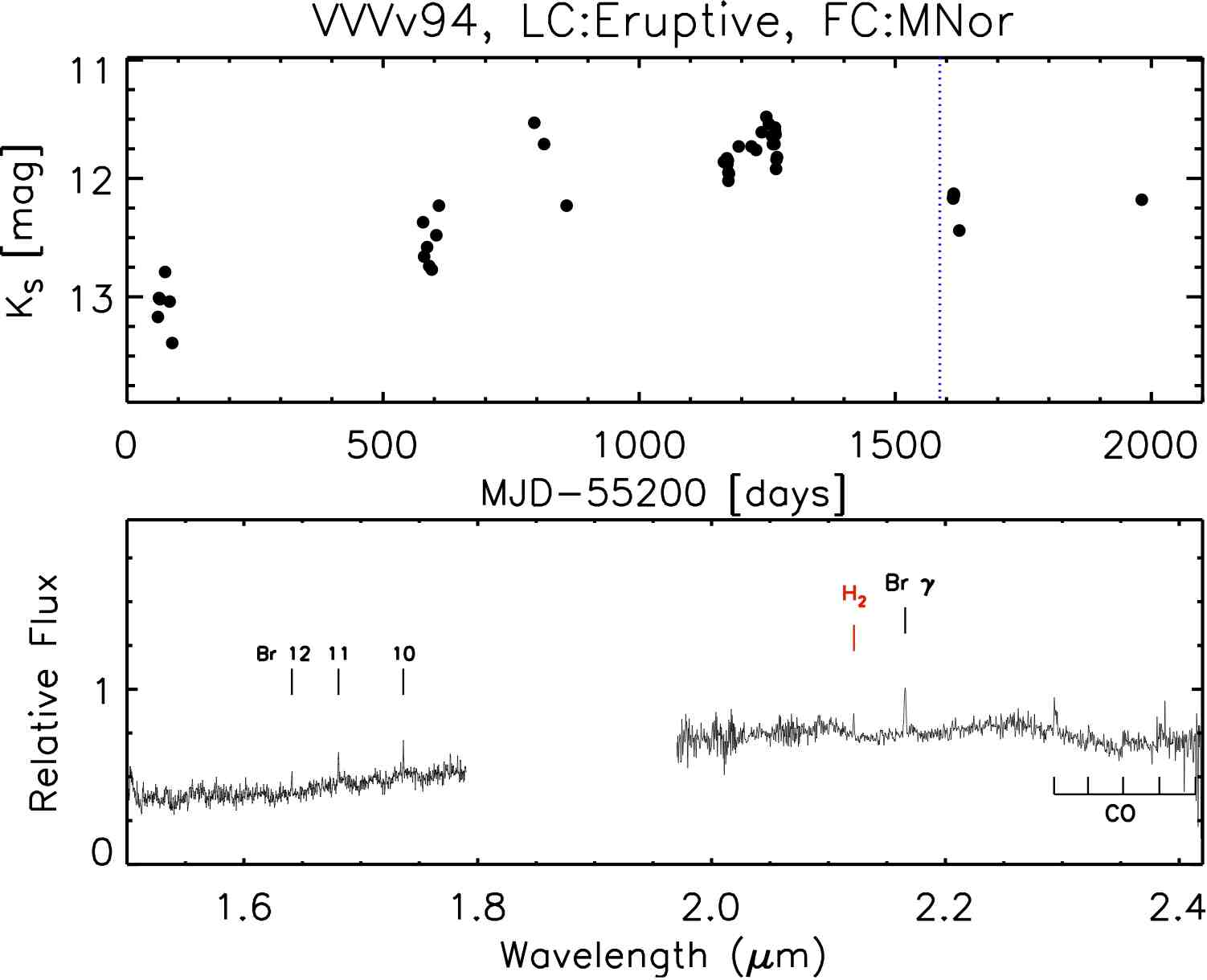}}
\caption{FIRE spectra and $K_{\rm s}$ light curves.}
\label{apen:fig4}
\end{figure*}

\begin{figure*}
\centering
\resizebox{0.75\textwidth}{!}{\includegraphics{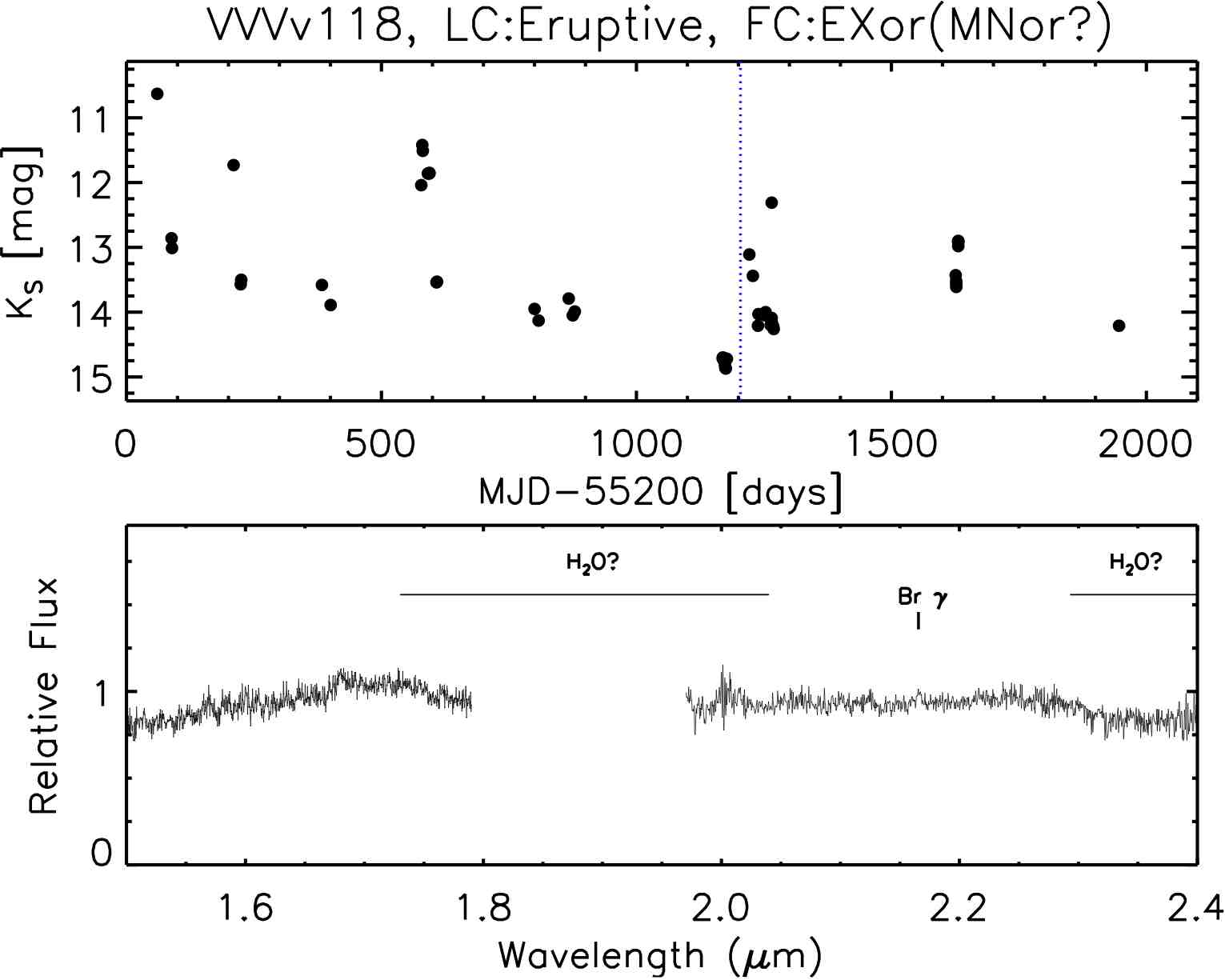}}\\
\resizebox{0.75\textwidth}{!}{\includegraphics{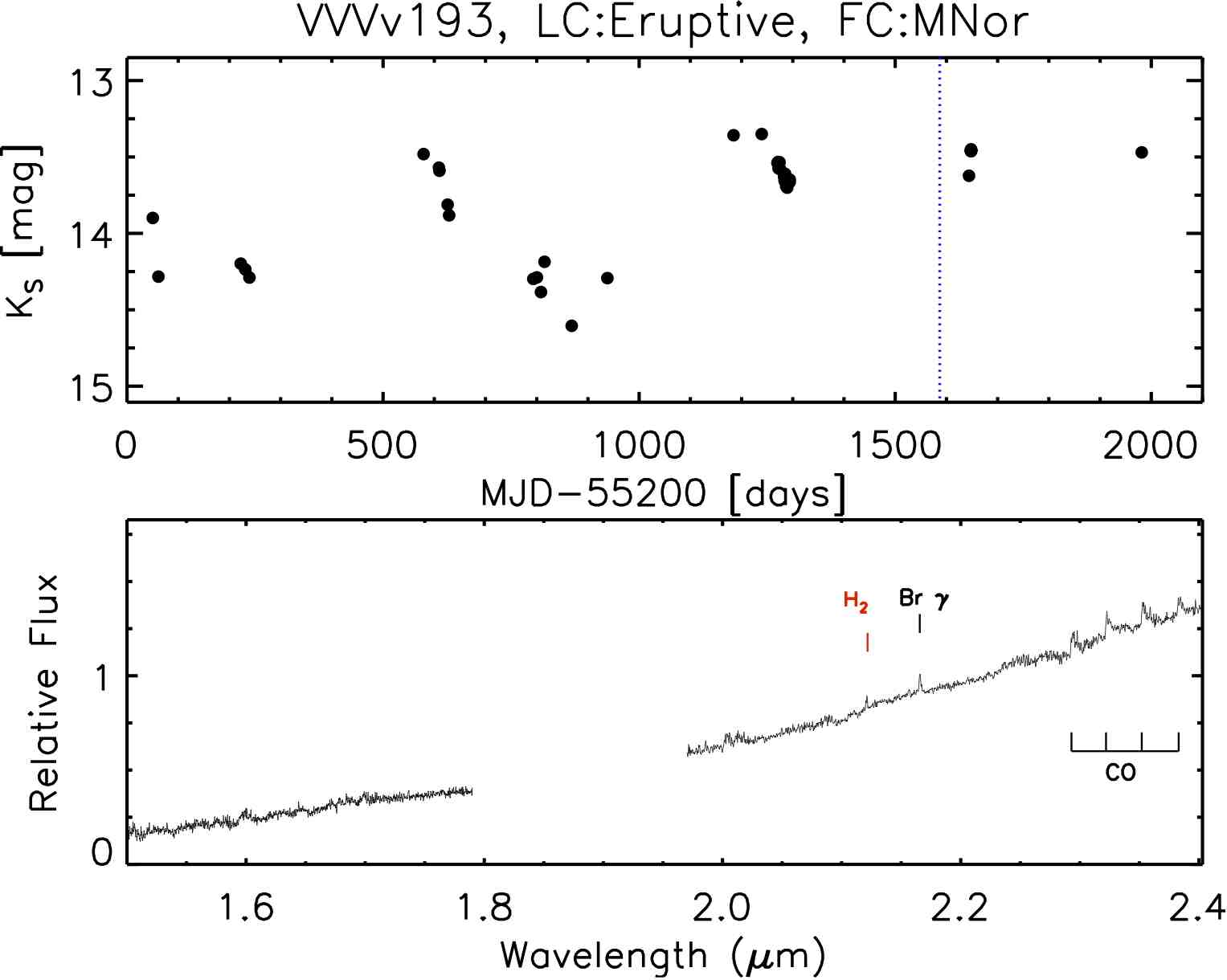}}
\caption{FIRE spectra and $K_{\rm s}$ light curves.}
\label{apen:fig5}
\end{figure*}

\begin{figure*}
\centering
\resizebox{0.75\textwidth}{!}{\includegraphics{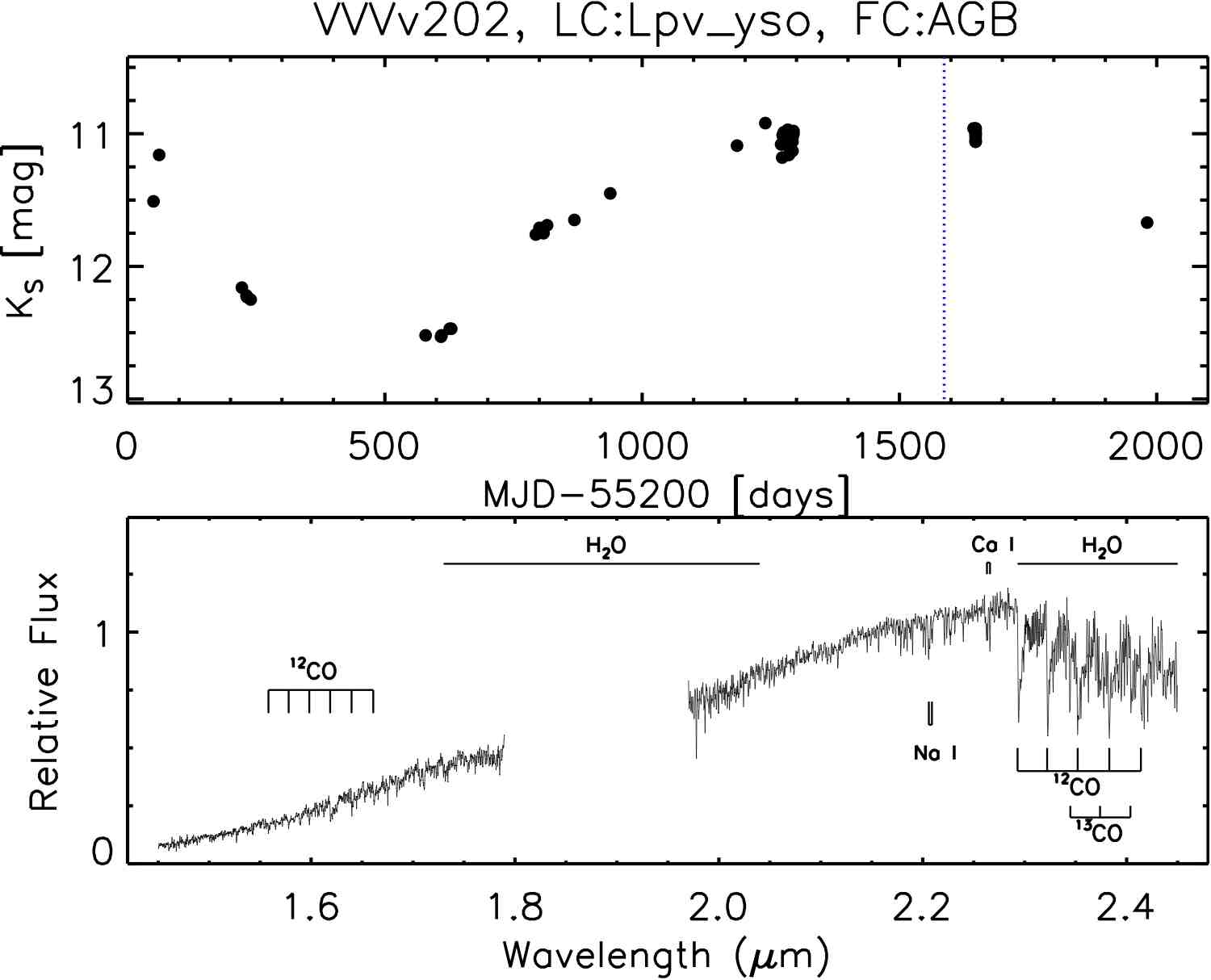}}\\
\resizebox{0.75\textwidth}{!}{\includegraphics{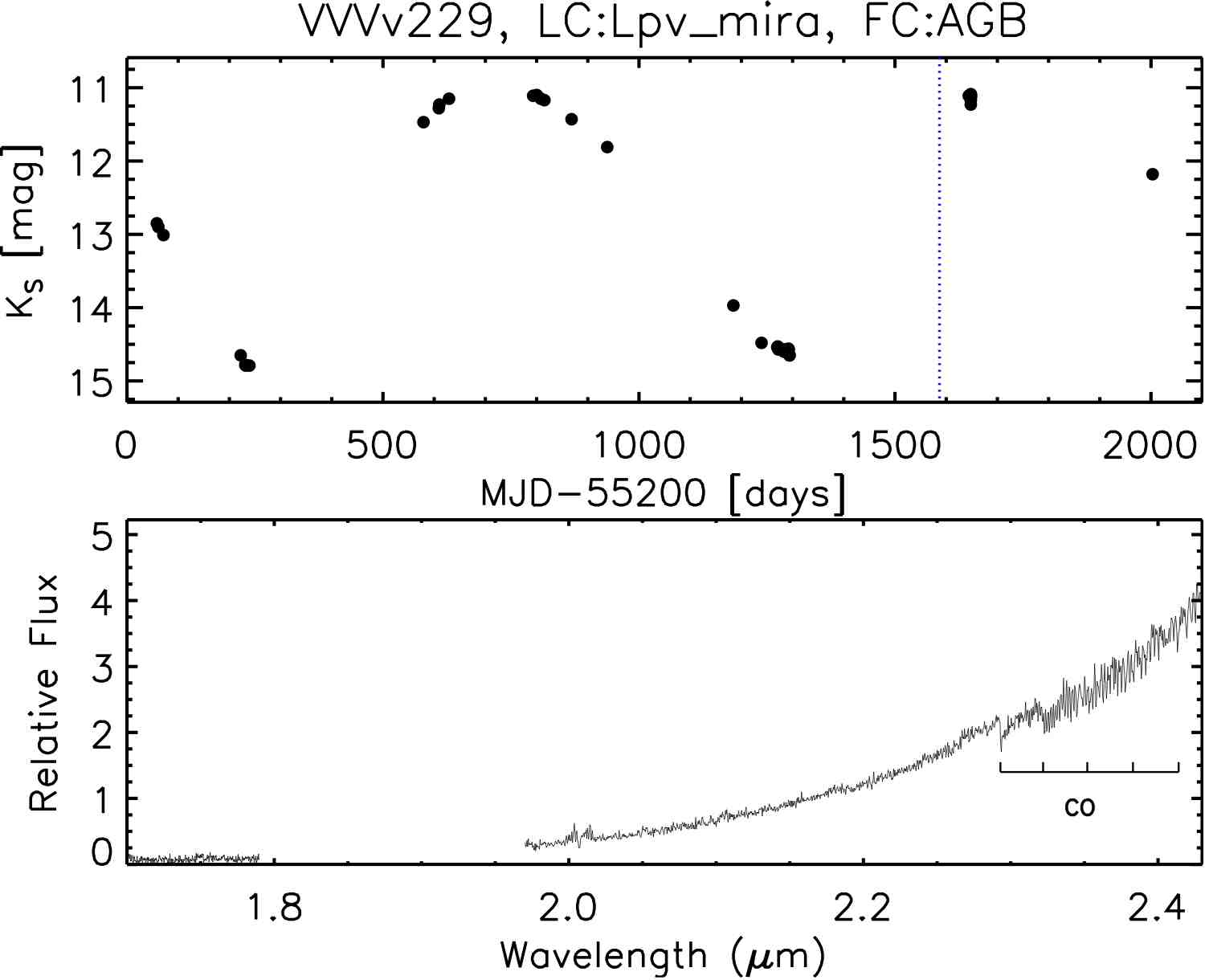}}
\caption{FIRE spectra and $K_{\rm s}$ light curves.}
\label{apen:fig6}
\end{figure*}

\begin{figure*}
\centering
\resizebox{0.75\textwidth}{!}{\includegraphics{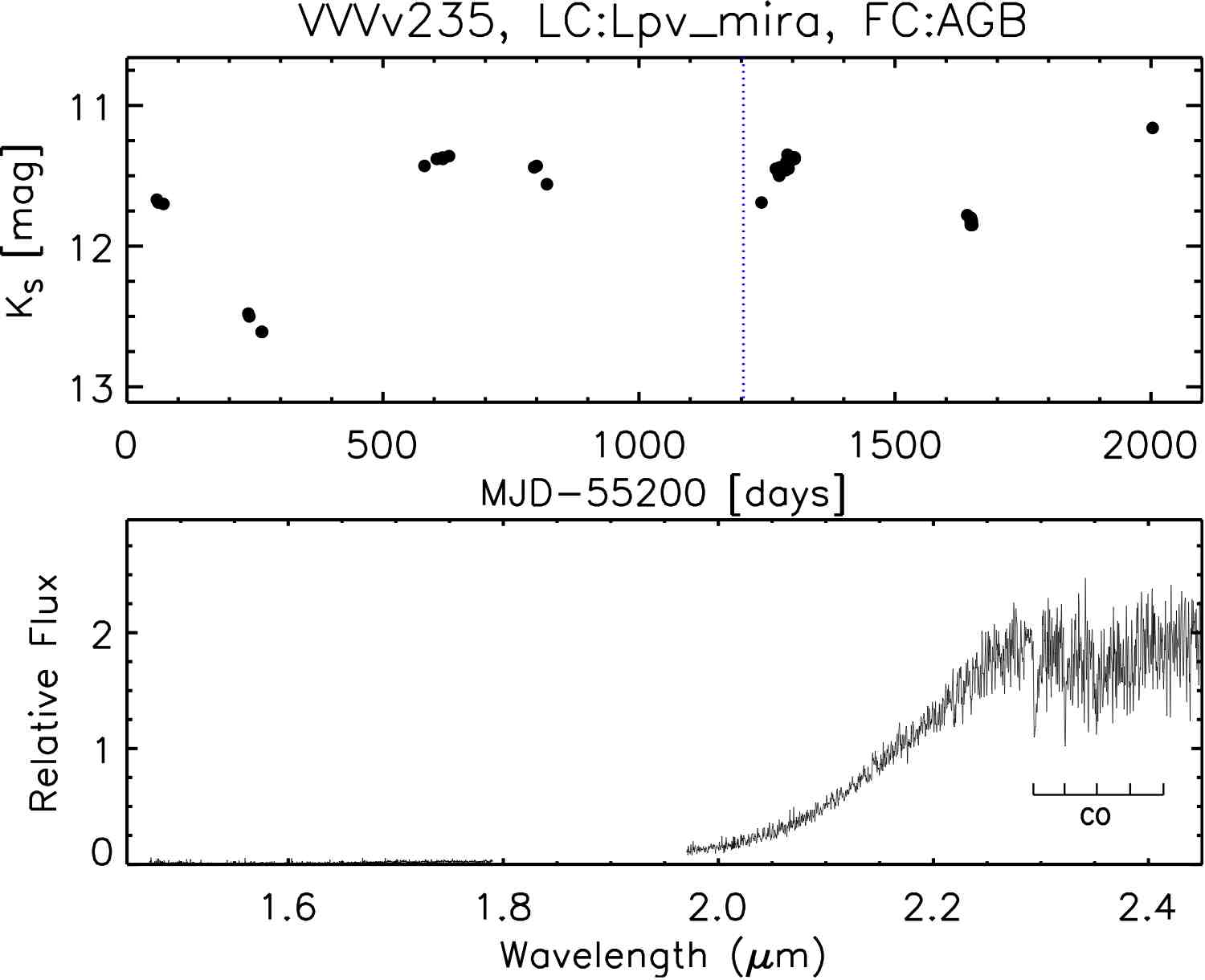}}\\
\resizebox{0.75\textwidth}{!}{\includegraphics{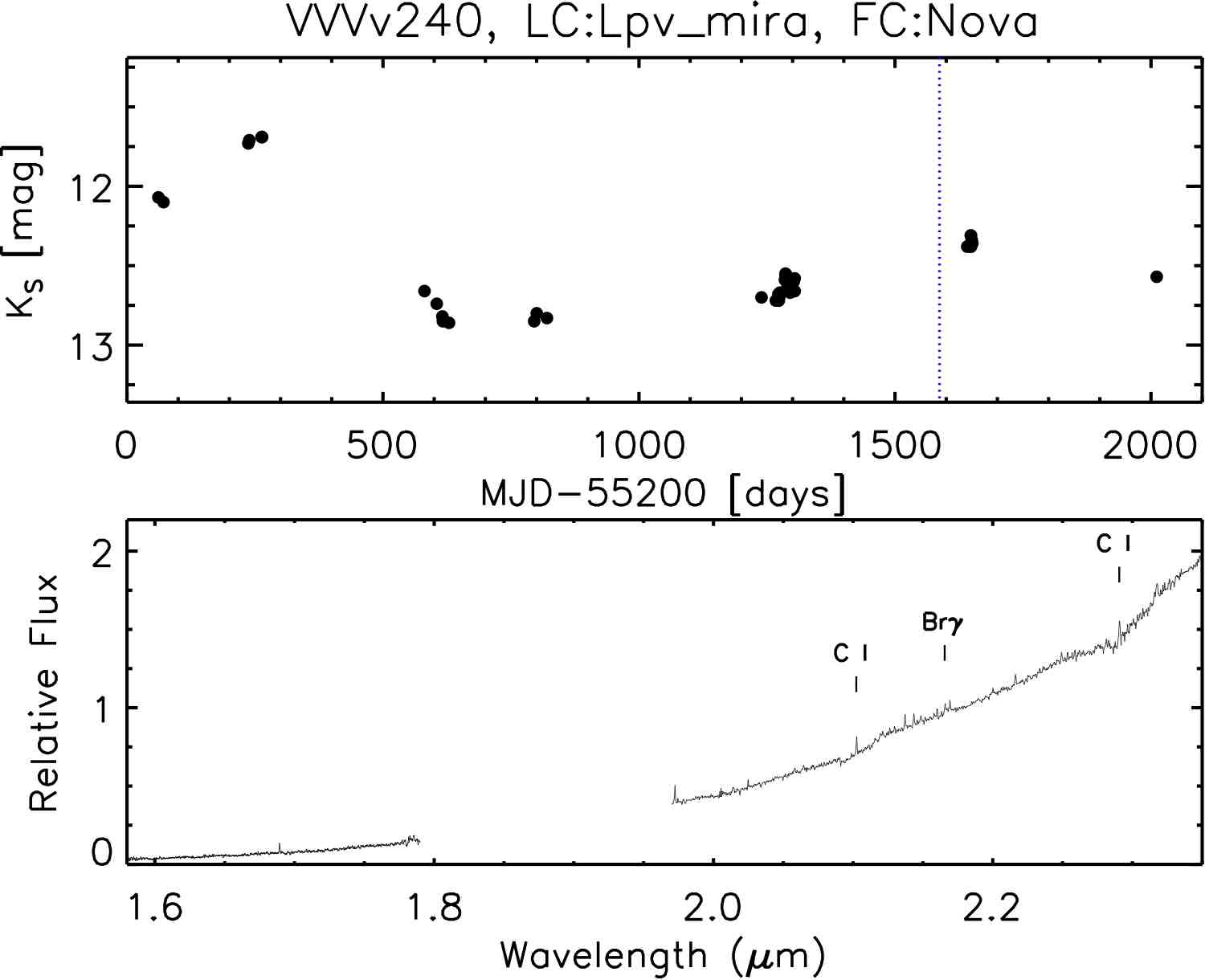}}
\caption{FIRE spectra and $K_{\rm s}$ light curves.}
\label{apen:fig7}
\end{figure*}

\begin{figure*}
\centering
\resizebox{0.75\textwidth}{!}{\includegraphics{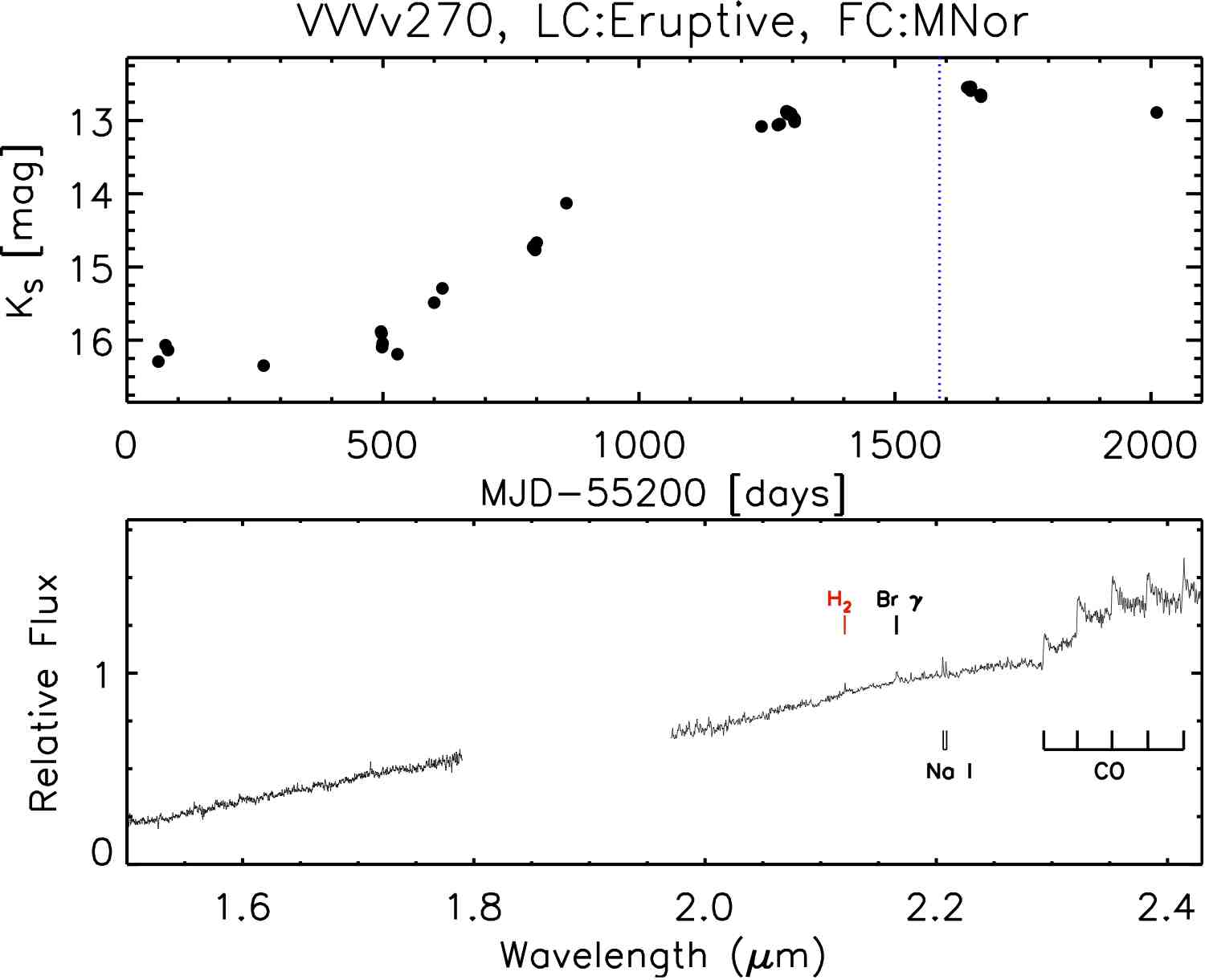}}\\
\resizebox{0.75\textwidth}{!}{\includegraphics{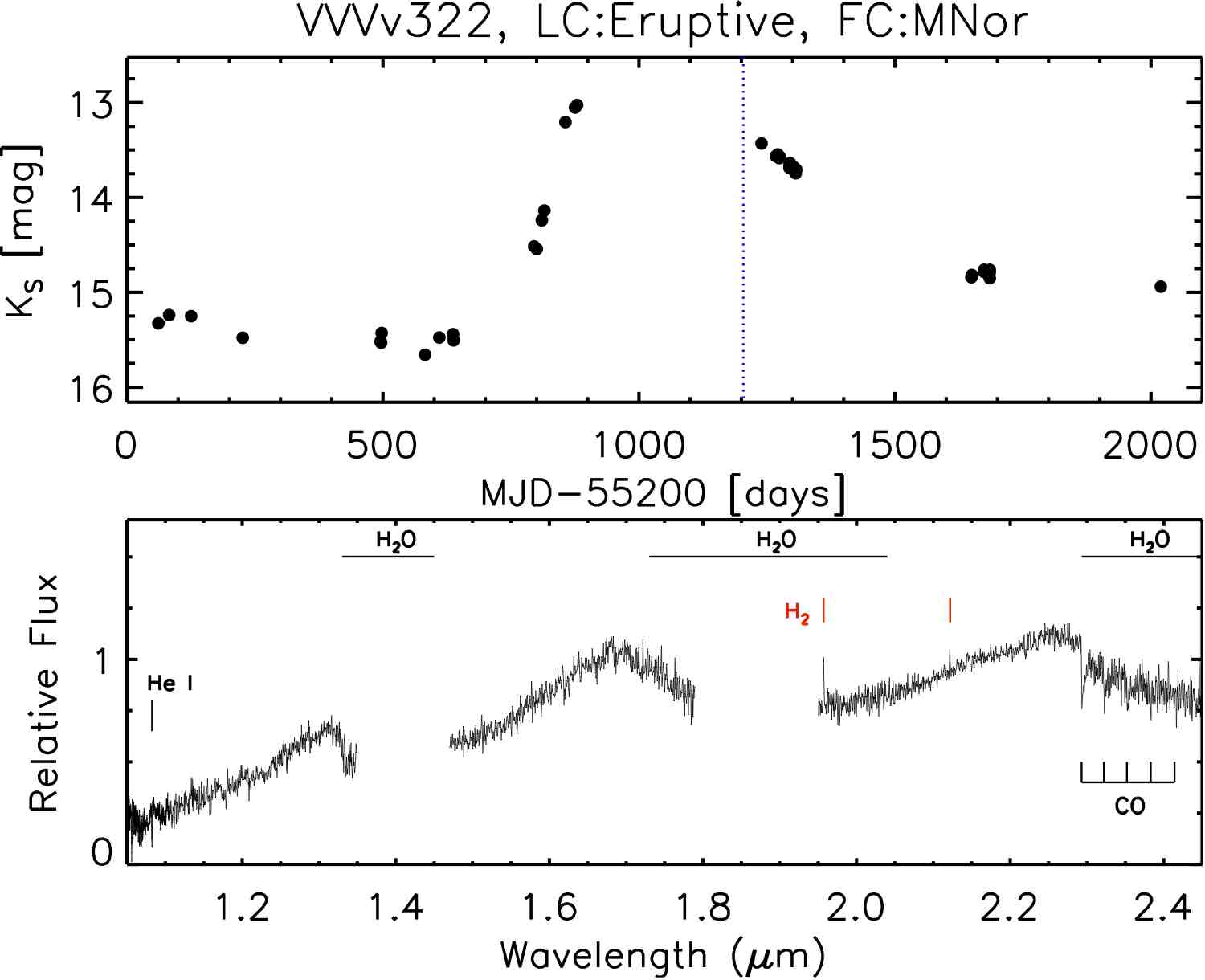}}
\caption{FIRE spectra and $K_{\rm s}$ light curves.}
\label{apen:fig8}
\end{figure*}

\begin{figure*}
\centering
\resizebox{0.75\textwidth}{!}{\includegraphics{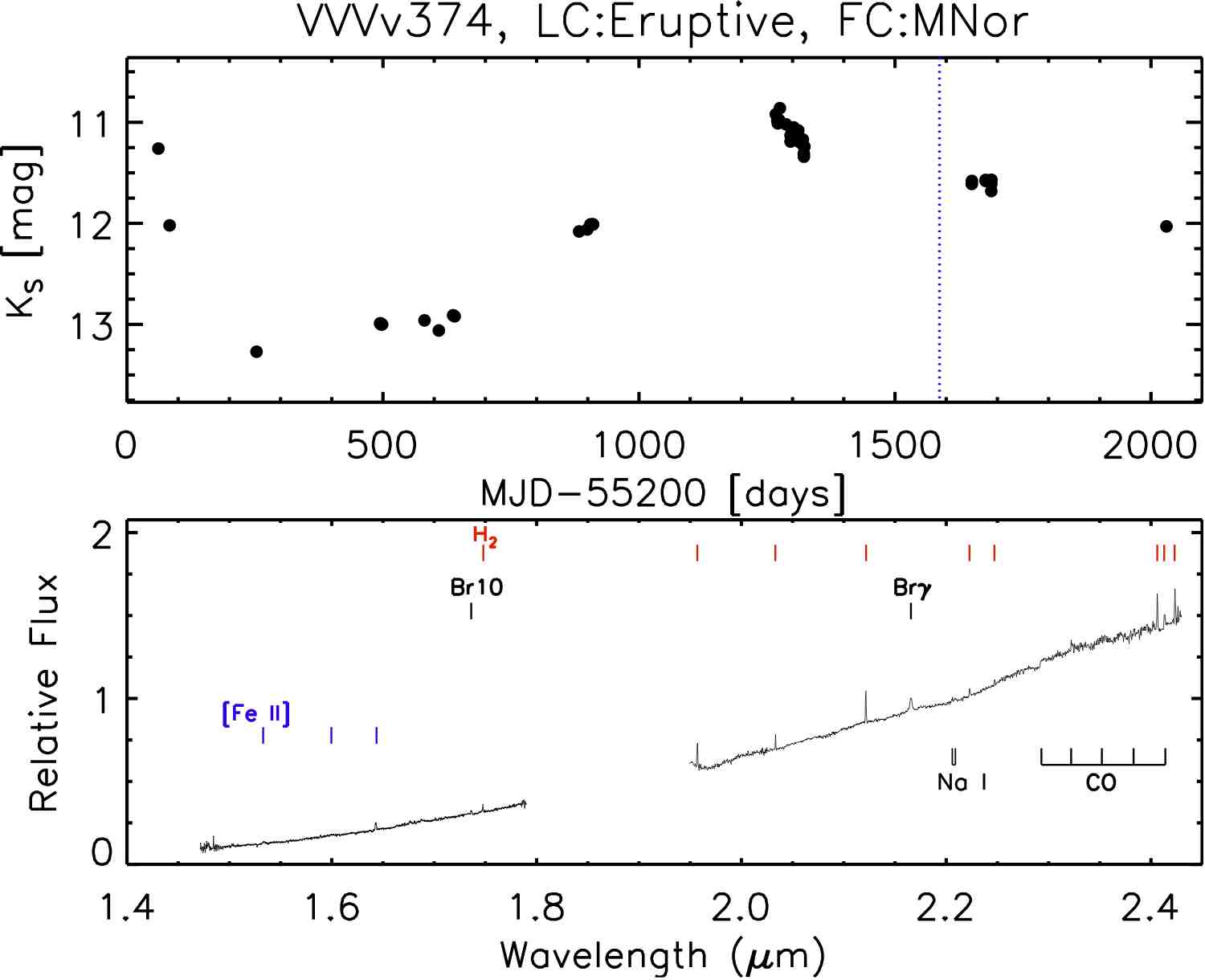}}\\
\resizebox{0.75\textwidth}{!}{\includegraphics{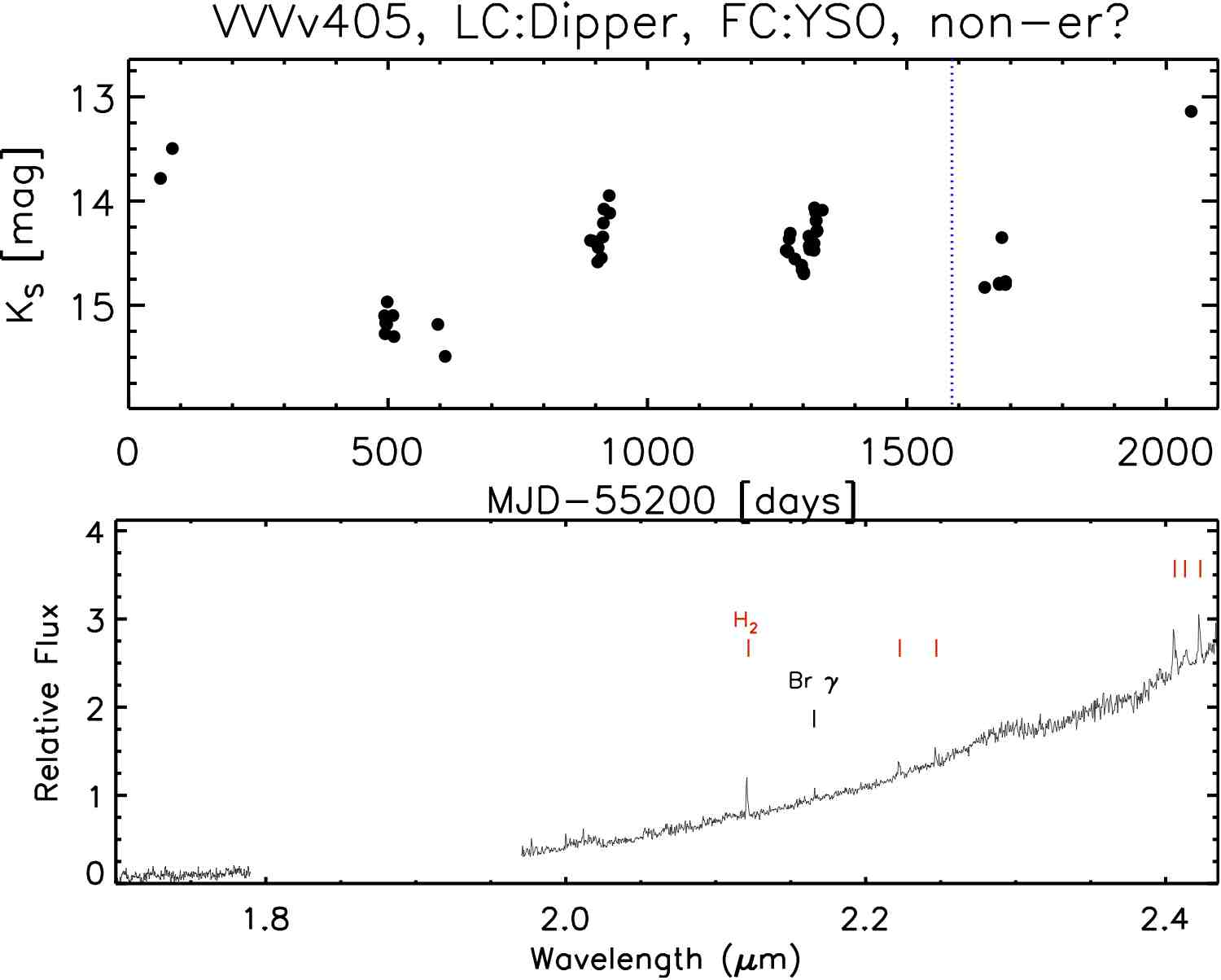}}
\caption{FIRE spectra and $K_{\rm s}$ light curves.}
\label{apen:fig9}
\end{figure*}

\begin{figure*}
\centering
\resizebox{0.75\textwidth}{!}{\includegraphics{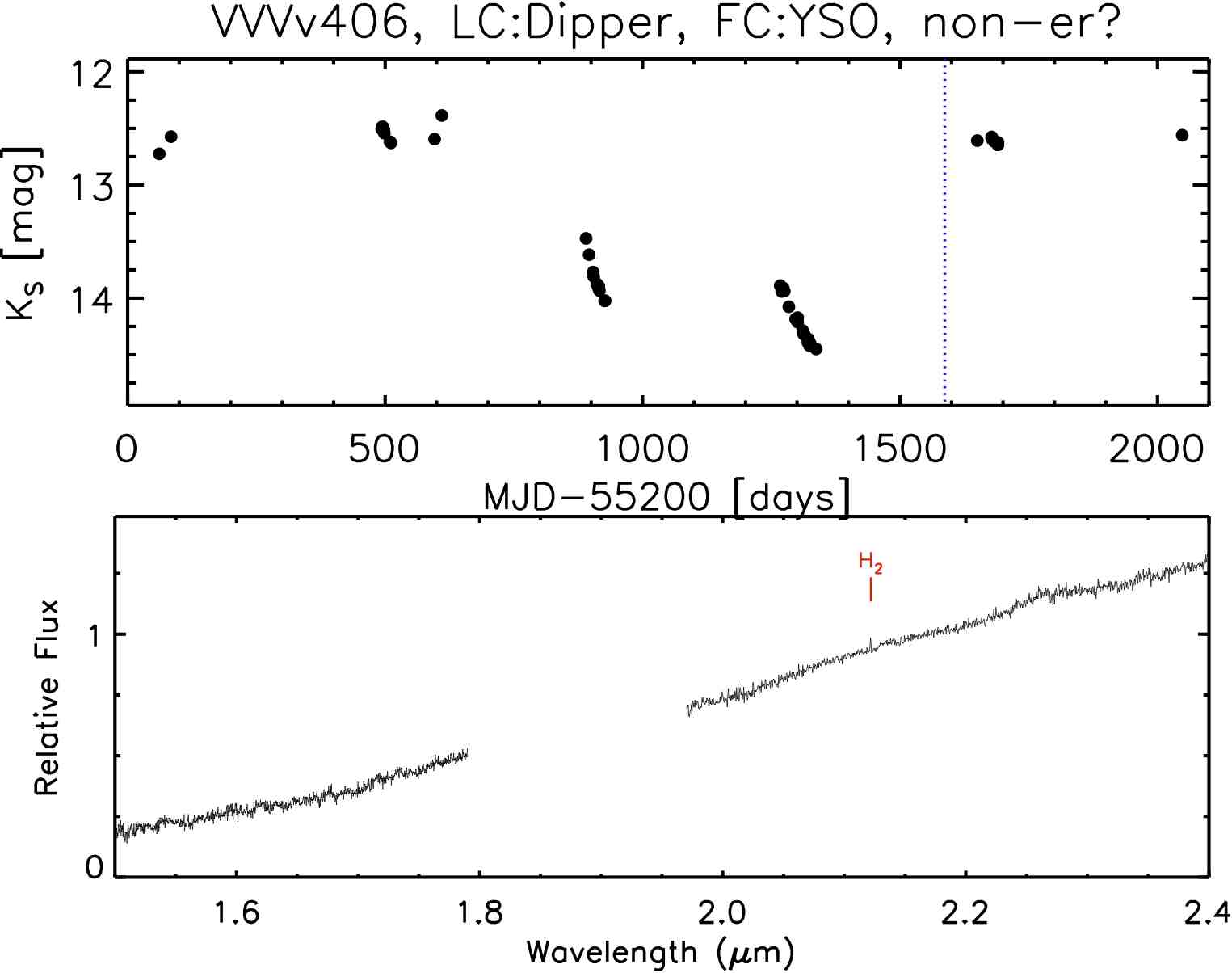}}\\
\resizebox{0.75\textwidth}{!}{\includegraphics{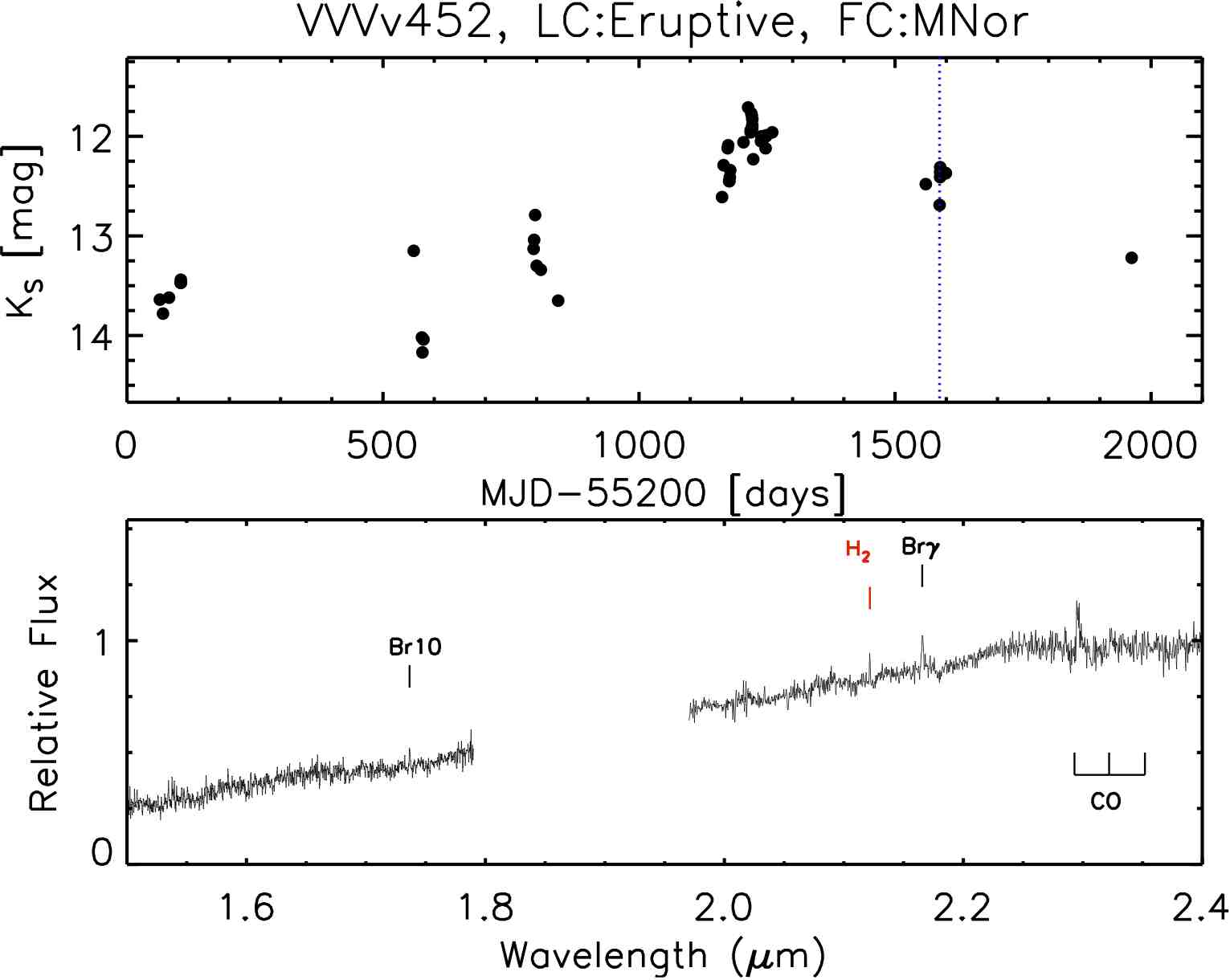}}
\caption{FIRE spectra and $K_{\rm s}$ light curves.}
\label{apen:fig10}
\end{figure*}

\begin{figure*}
\centering
\resizebox{0.75\textwidth}{!}{\includegraphics{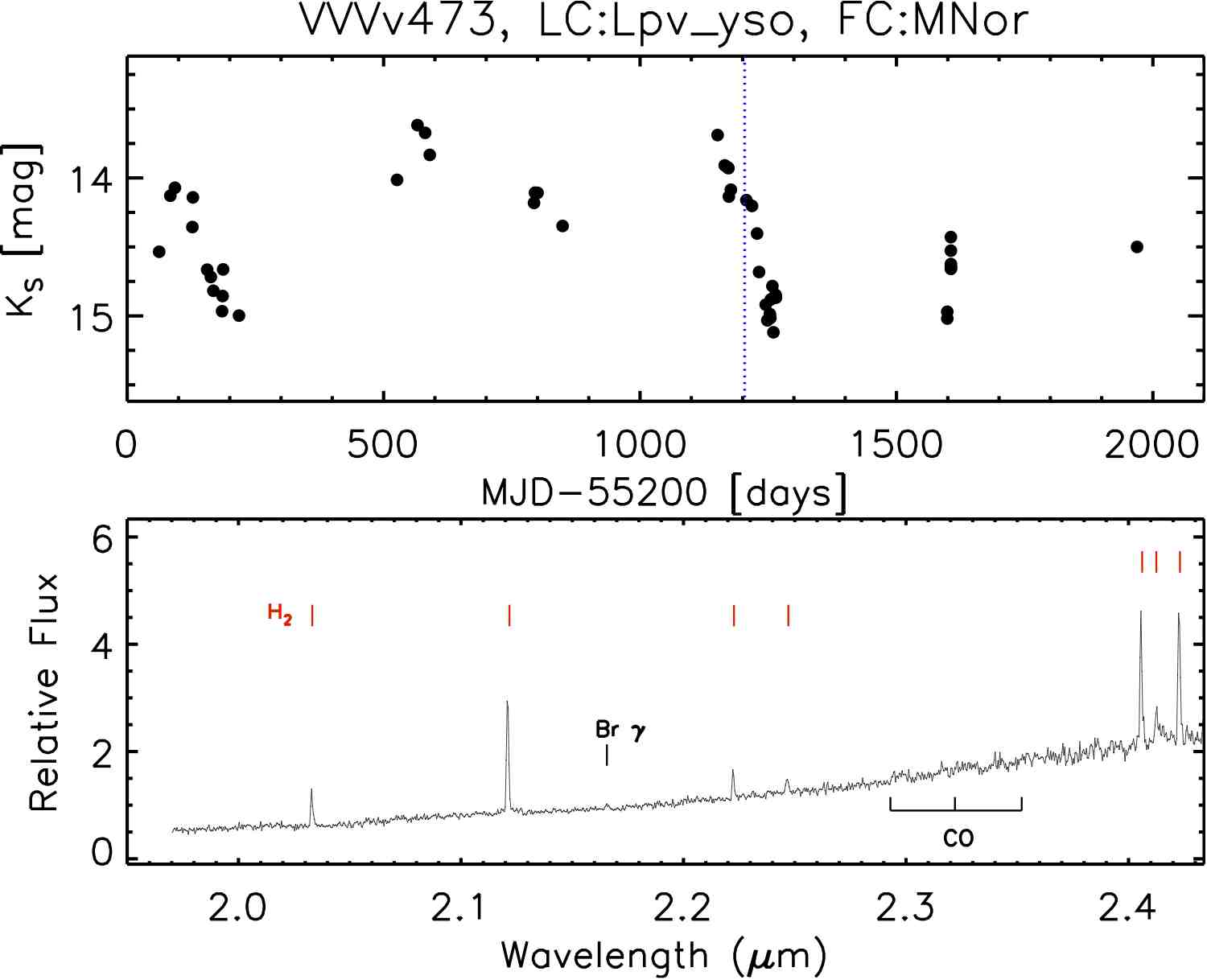}}\\
\resizebox{0.75\textwidth}{!}{\includegraphics{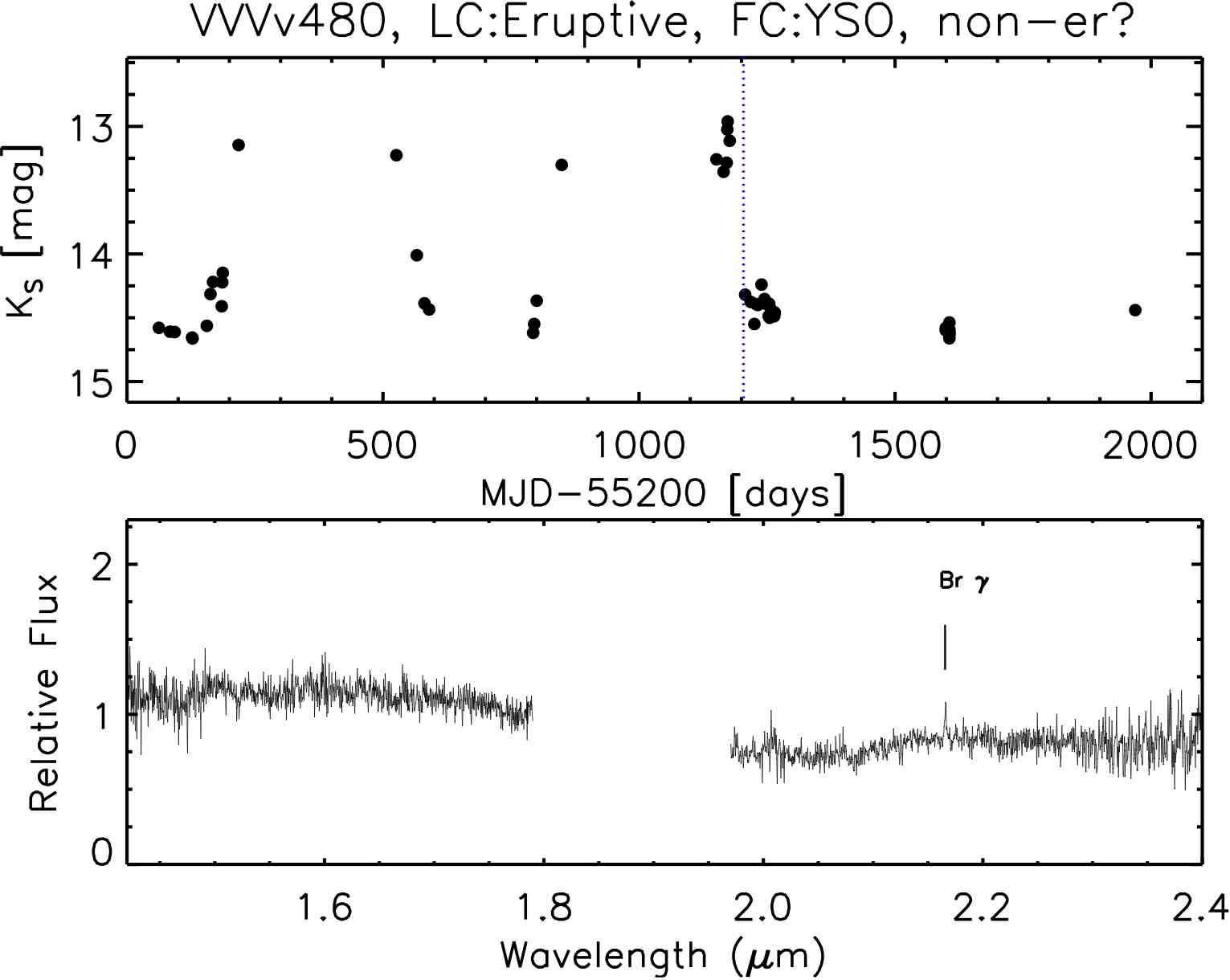}}
\caption{FIRE spectra and $K_{\rm s}$ light curves.}
\label{apen:fig11}
\end{figure*}

\begin{figure*}
\centering
\resizebox{0.75\textwidth}{!}{\includegraphics{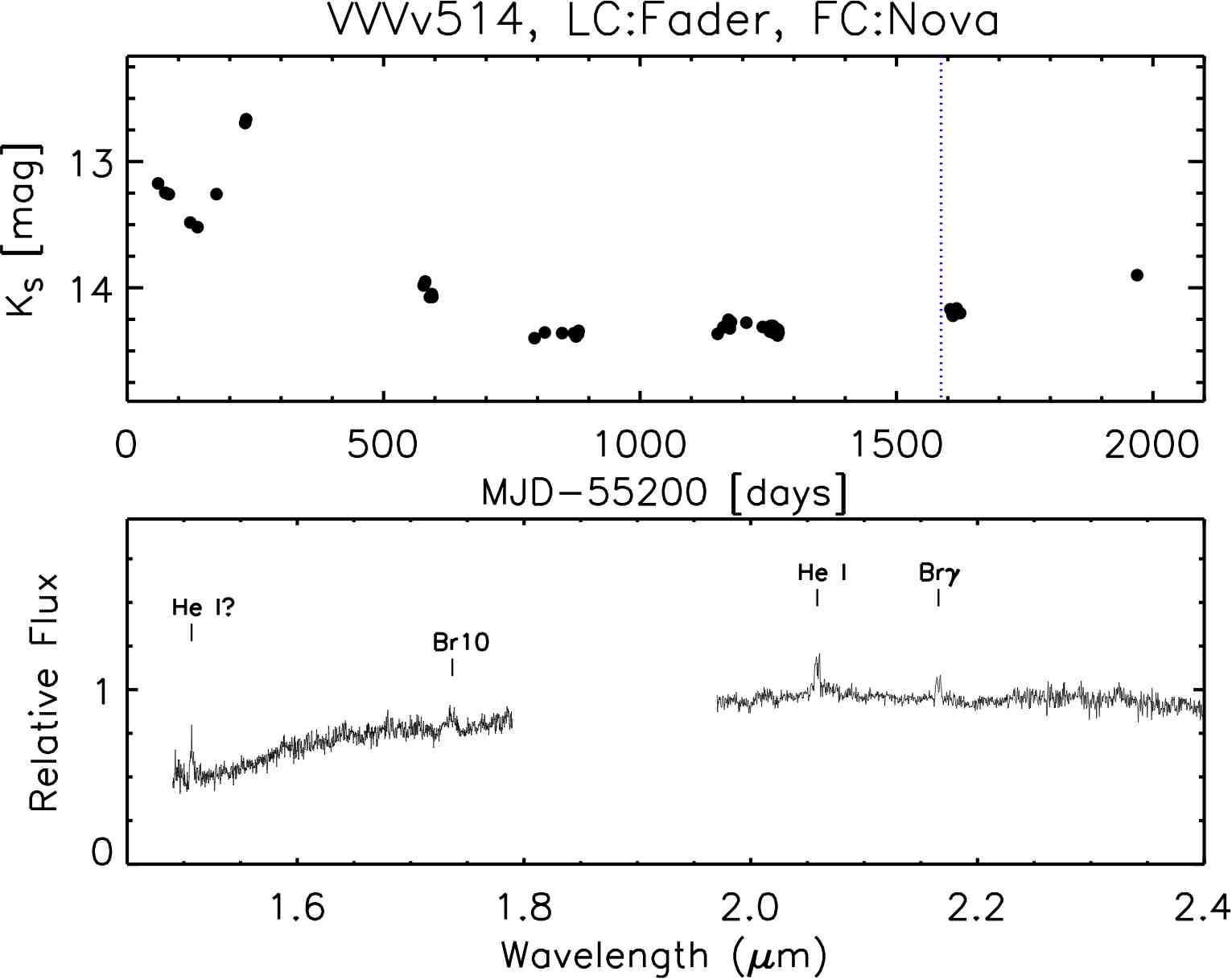}}\\
\resizebox{0.75\textwidth}{!}{\includegraphics{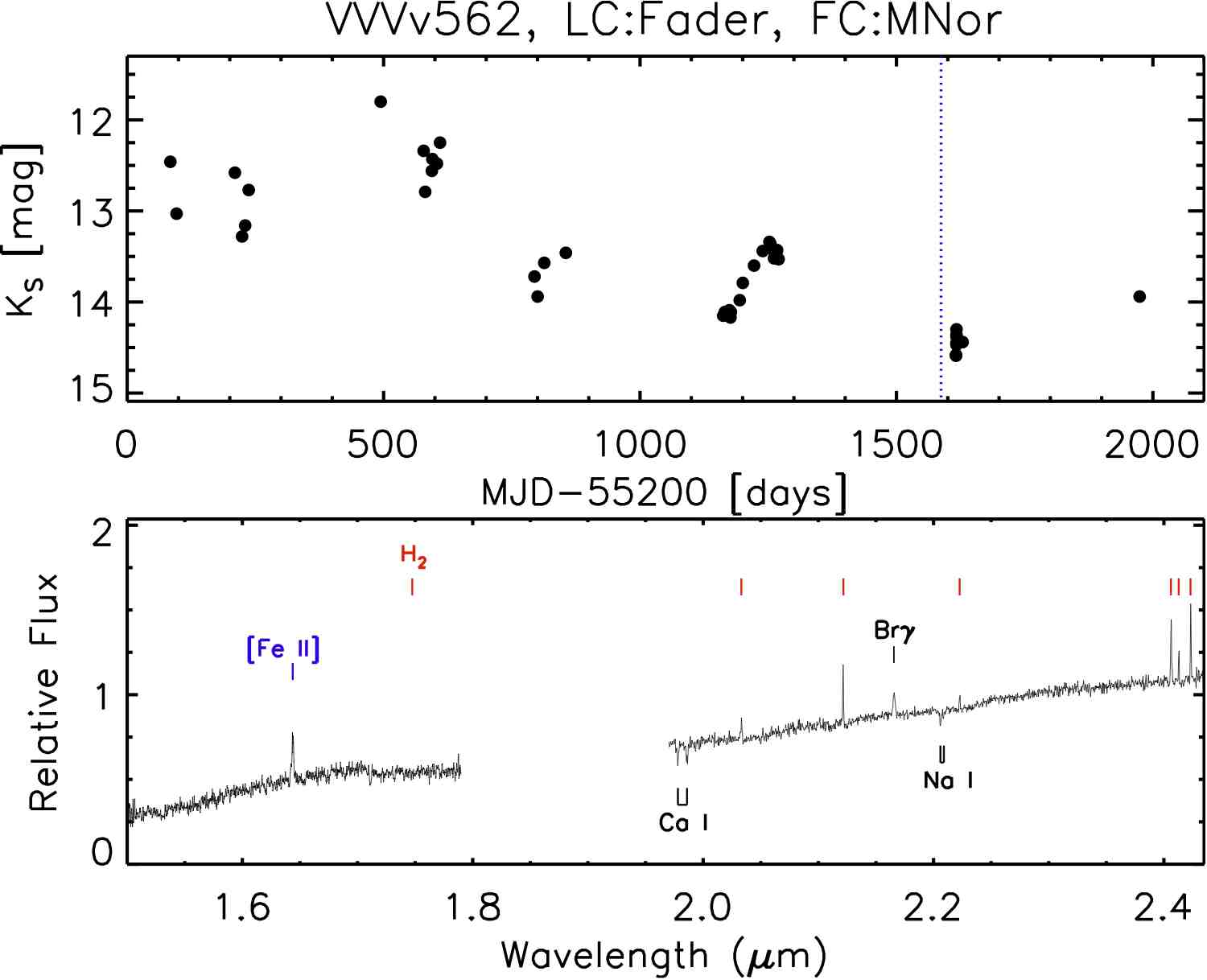}}
\caption{FIRE spectra and $K_{\rm s}$ light curves.}
\label{apen:fig12}
\end{figure*}

\begin{figure*}
\centering
\resizebox{0.75\textwidth}{!}{\includegraphics{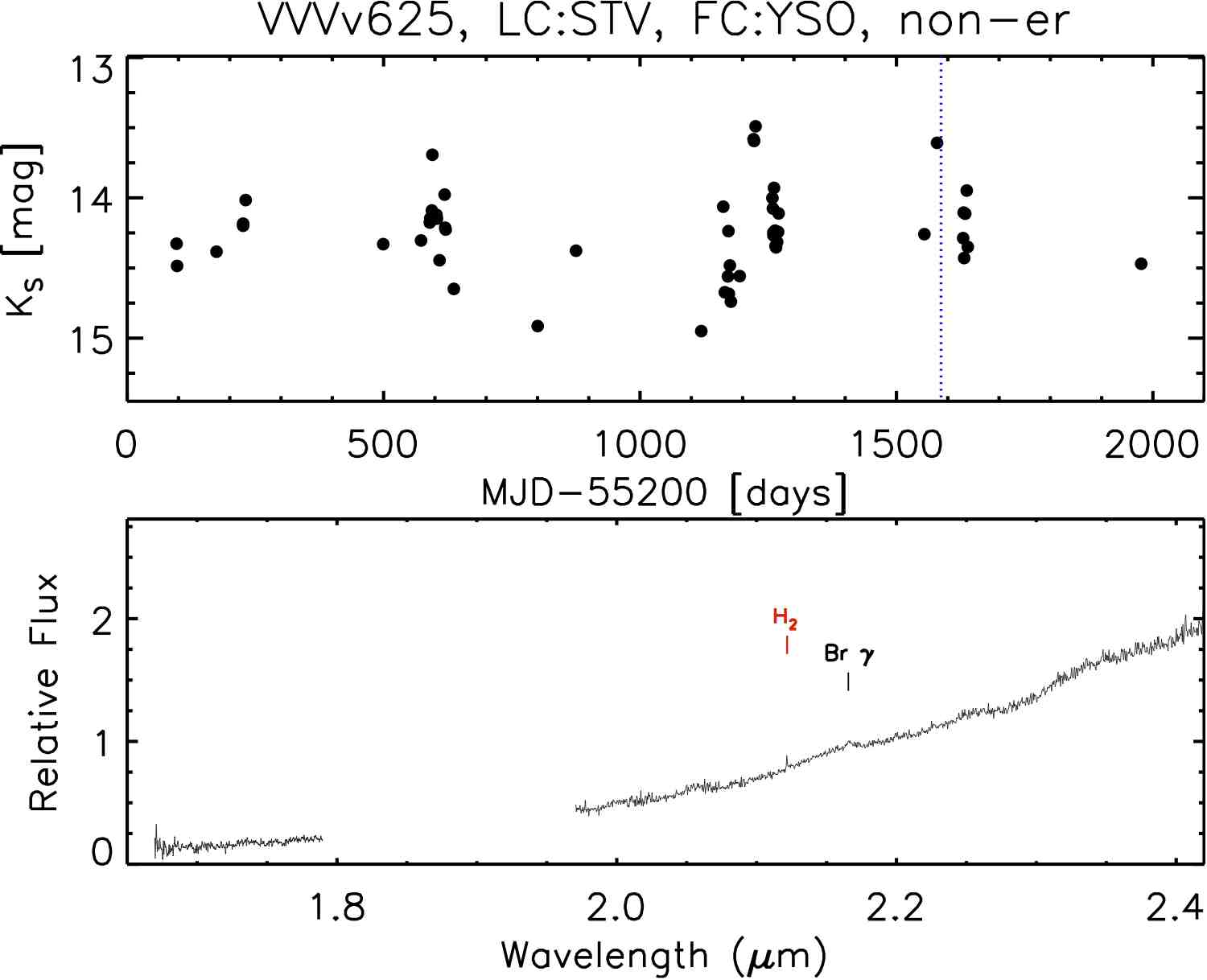}}\\
\resizebox{0.75\textwidth}{!}{\includegraphics{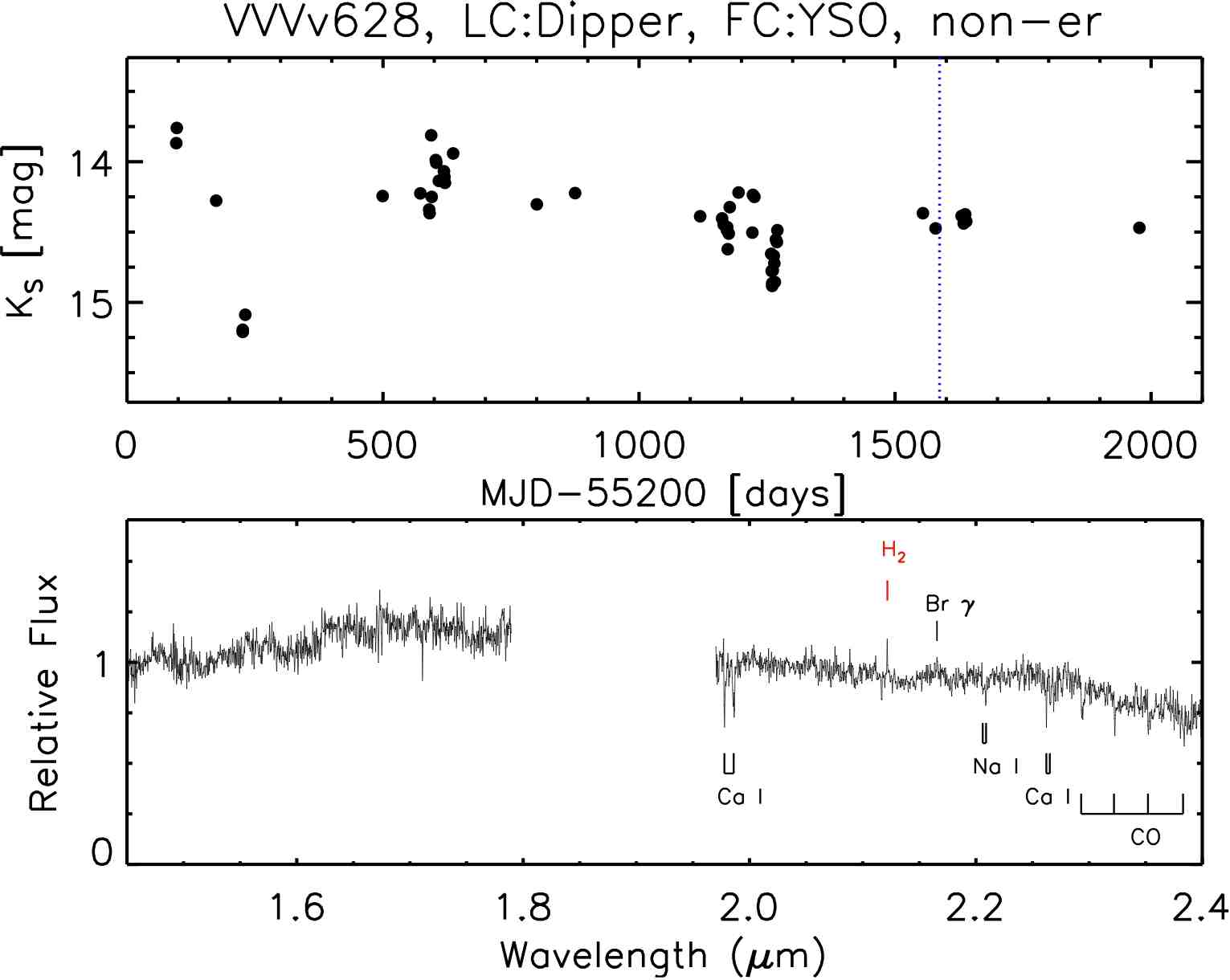}}
\caption{FIRE spectra and $K_{\rm s}$ light curves.}
\label{apen:fig13}
\end{figure*}

\begin{figure*}
\centering
\resizebox{0.75\textwidth}{!}{\includegraphics{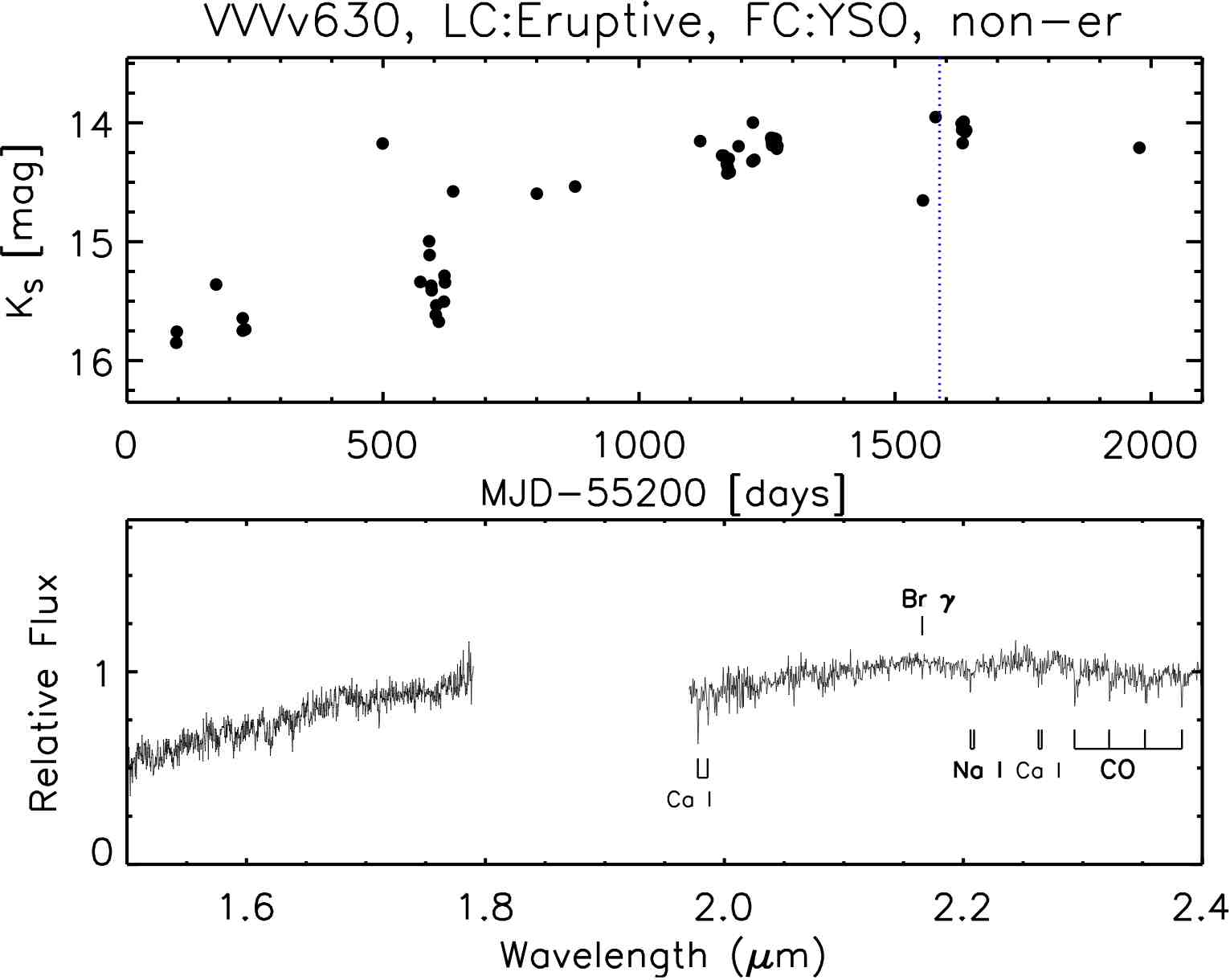}}\\
\resizebox{0.75\textwidth}{!}{\includegraphics{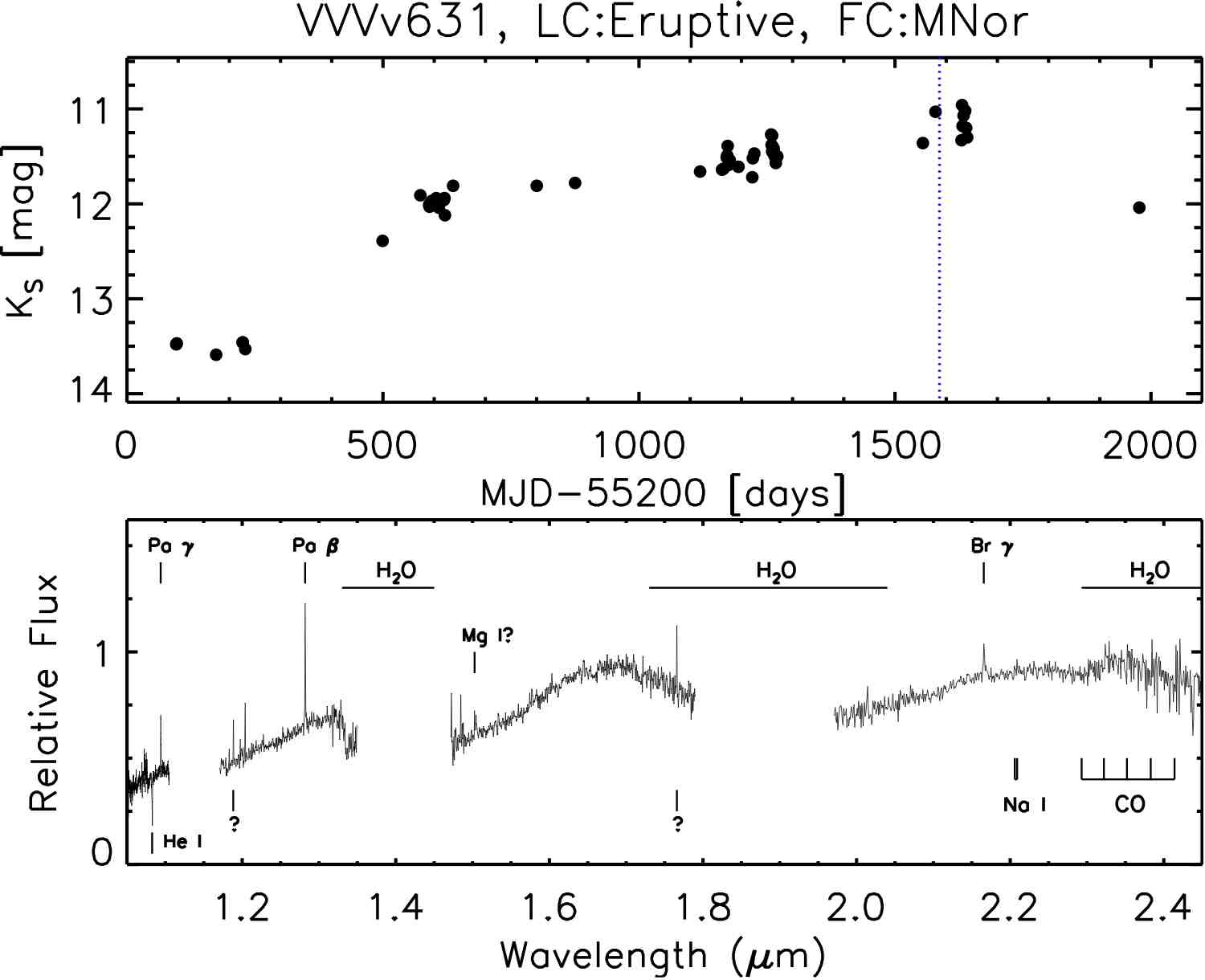}}
\caption{FIRE spectra and $K_{\rm s}$ light curves.}
\label{apen:fig14}
\end{figure*}

\begin{figure*}
\centering
\resizebox{0.75\textwidth}{!}{\includegraphics{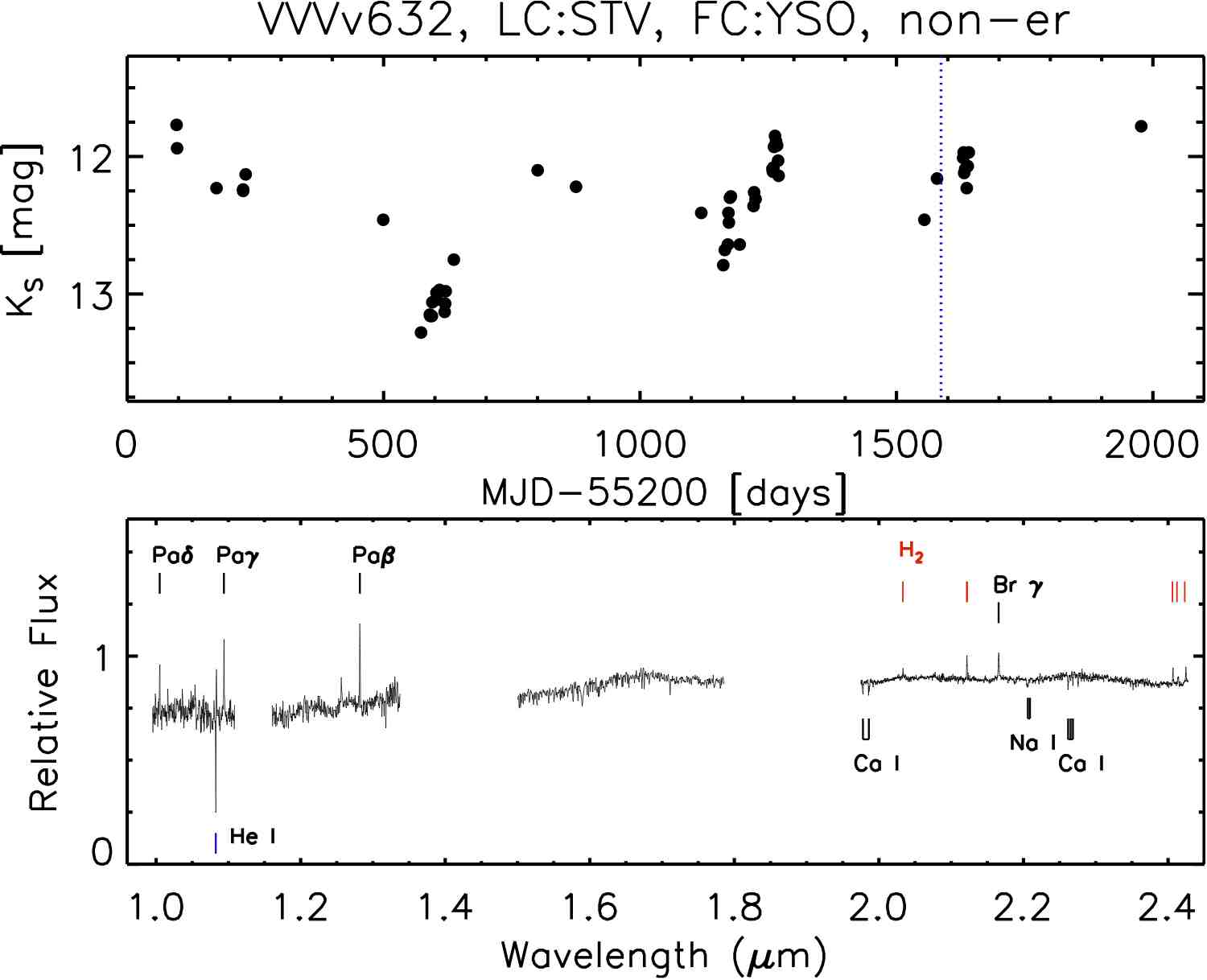}}\\
\resizebox{0.75\textwidth}{!}{\includegraphics{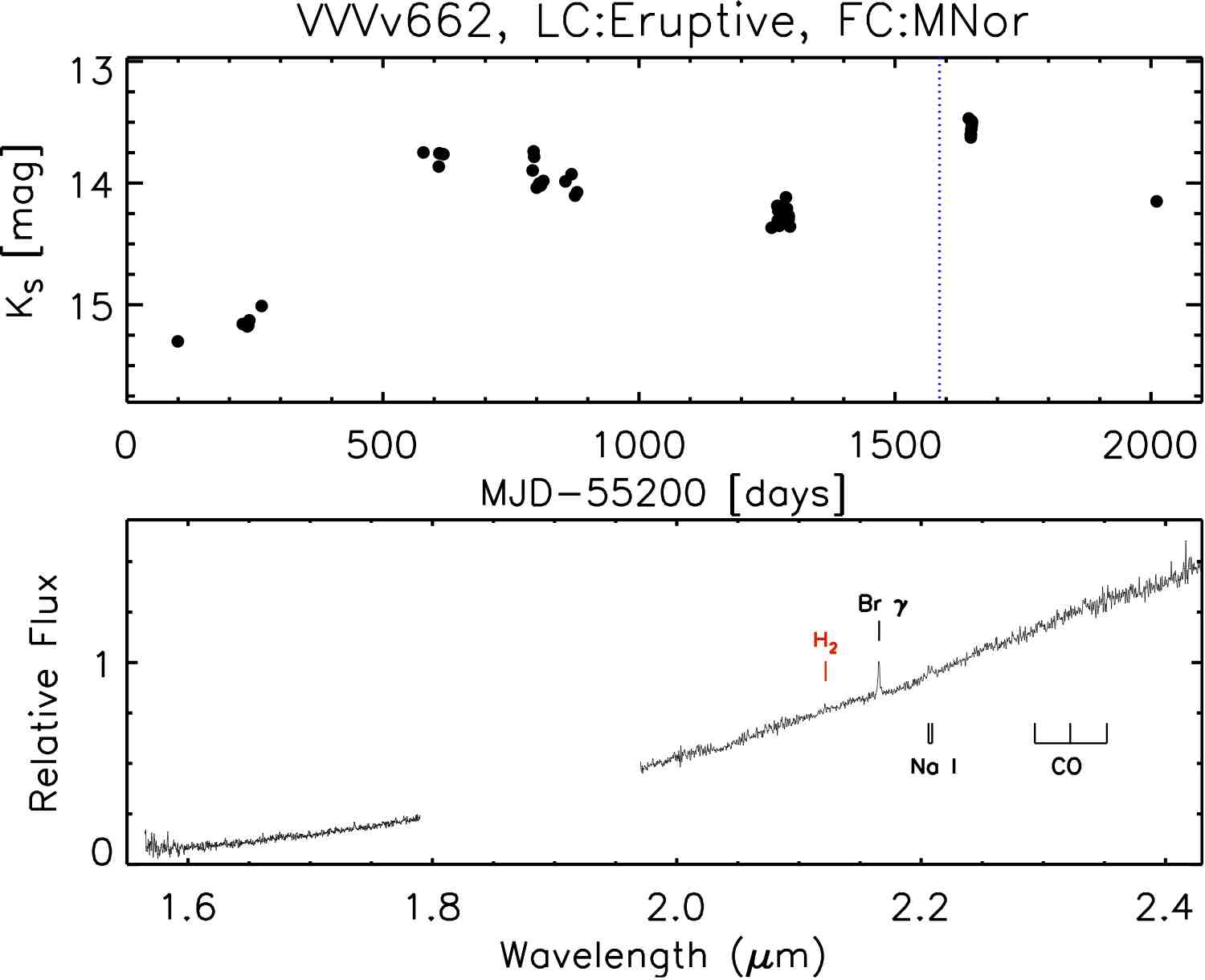}}
\caption{FIRE spectra and $K_{\rm s}$ light curves.}
\label{apen:fig15}
\end{figure*}

\begin{figure*}
\centering
\resizebox{0.75\textwidth}{!}{\includegraphics{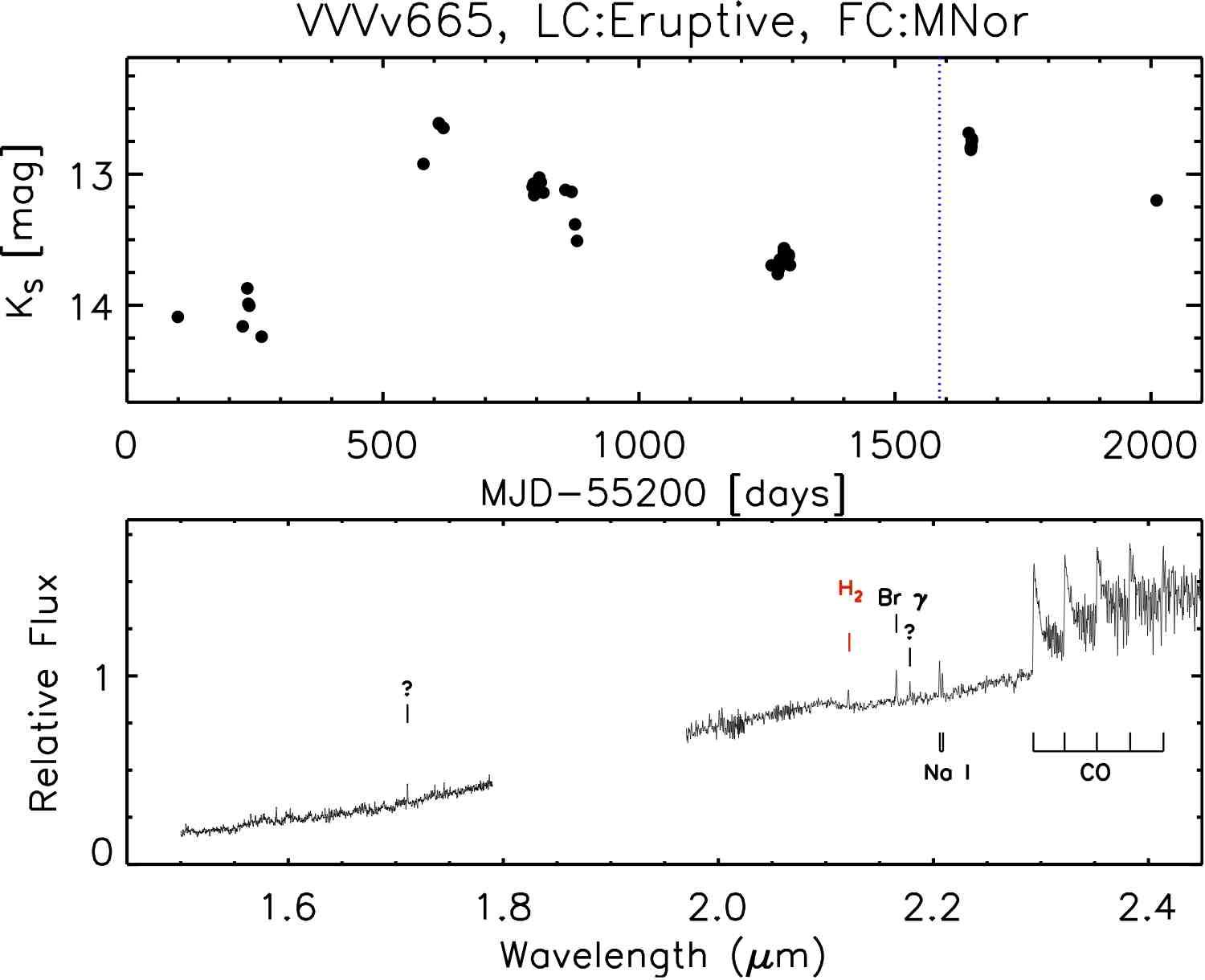}}\\
\resizebox{0.75\textwidth}{!}{\includegraphics{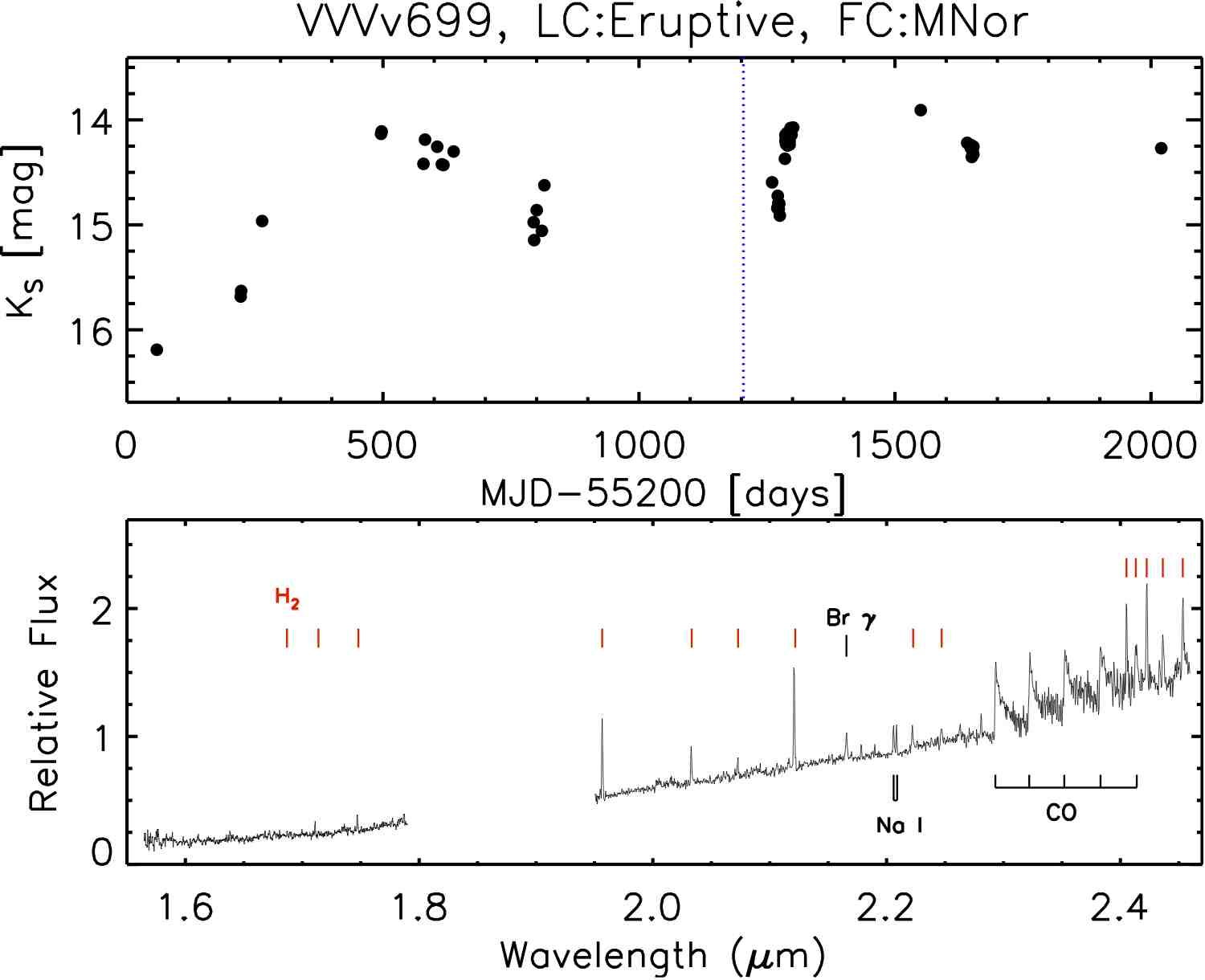}}
\caption{FIRE spectra and $K_{\rm s}$ light curves.}
\label{apen:fig16}
\end{figure*}

\begin{figure*}
\resizebox{0.75\textwidth}{!}{\includegraphics{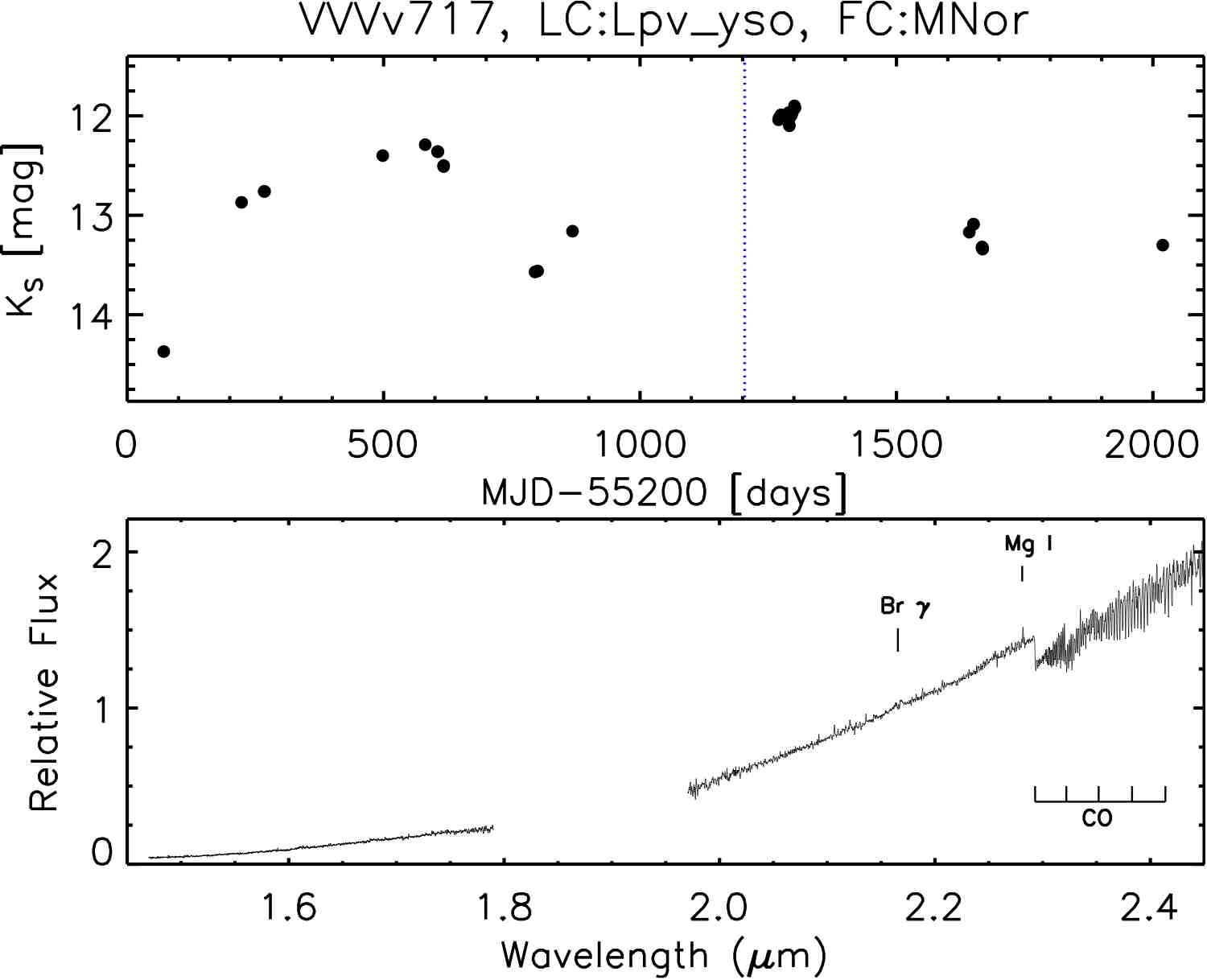}}\\
\resizebox{0.75\textwidth}{!}{\includegraphics{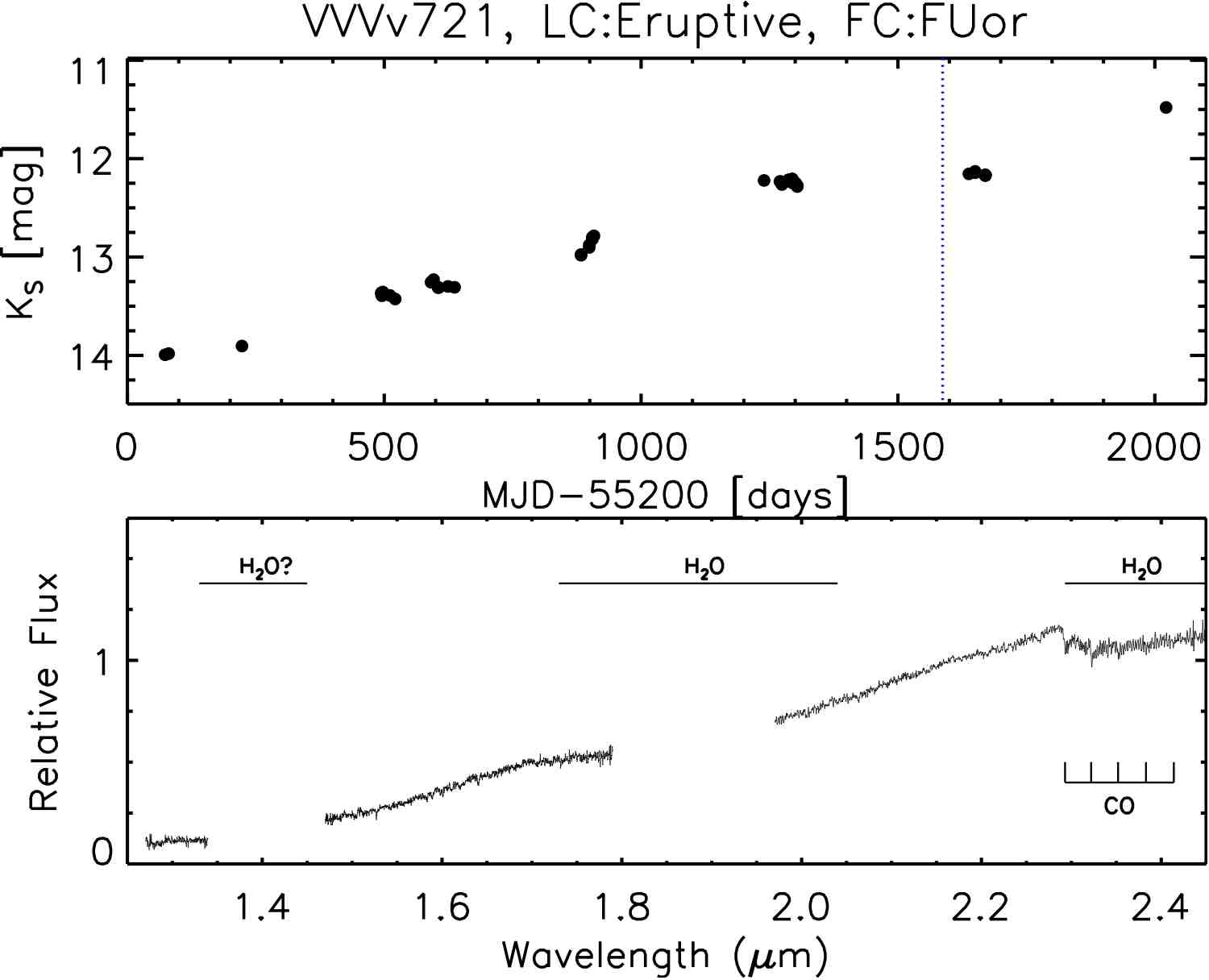}}
\caption{FIRE spectra and $K_{\rm s}$ light curves.}
\label{apen:fig17}
\end{figure*}

\begin{figure*}
\resizebox{0.75\textwidth}{!}{\includegraphics{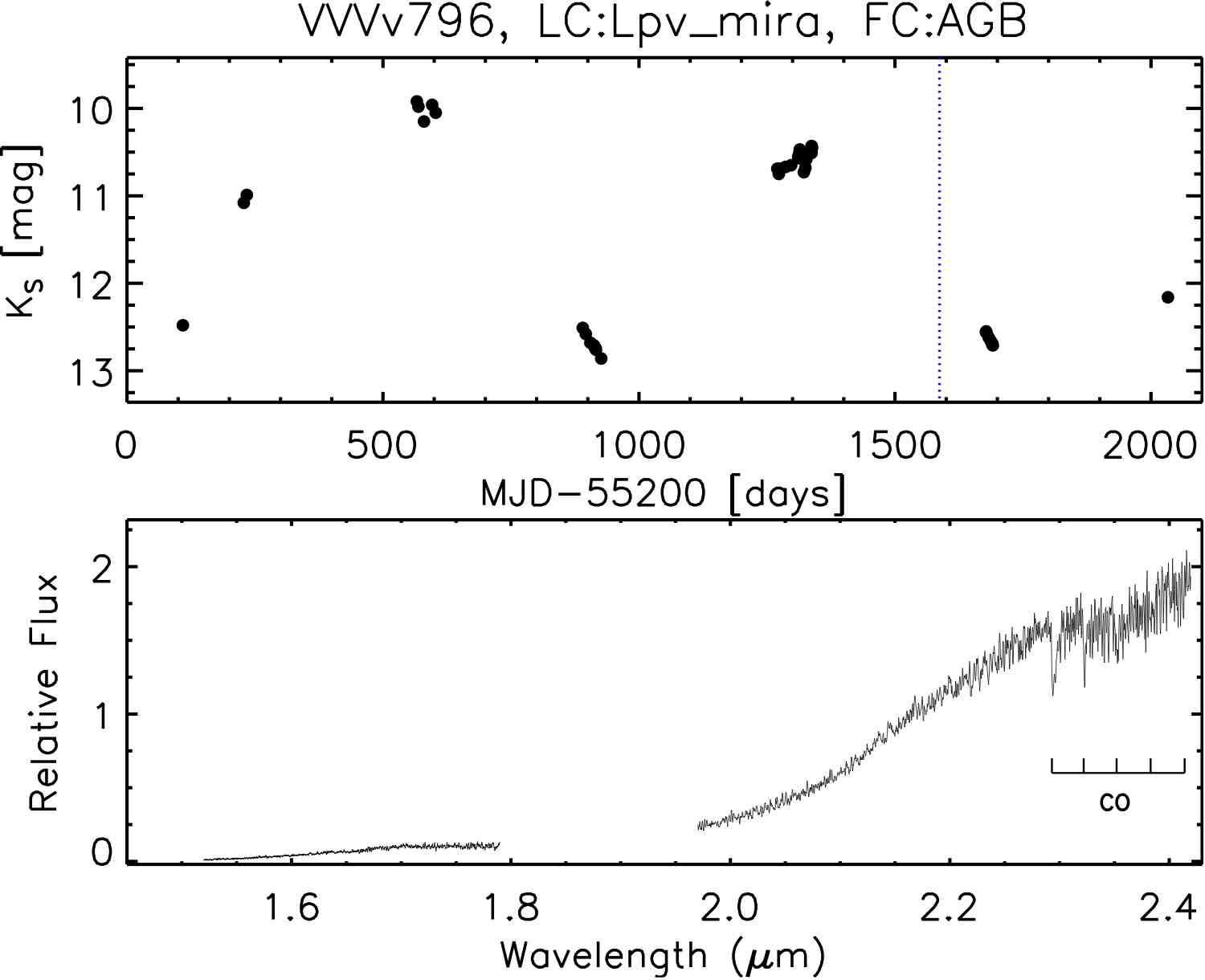}}\\
\resizebox{0.75\textwidth}{!}{\includegraphics{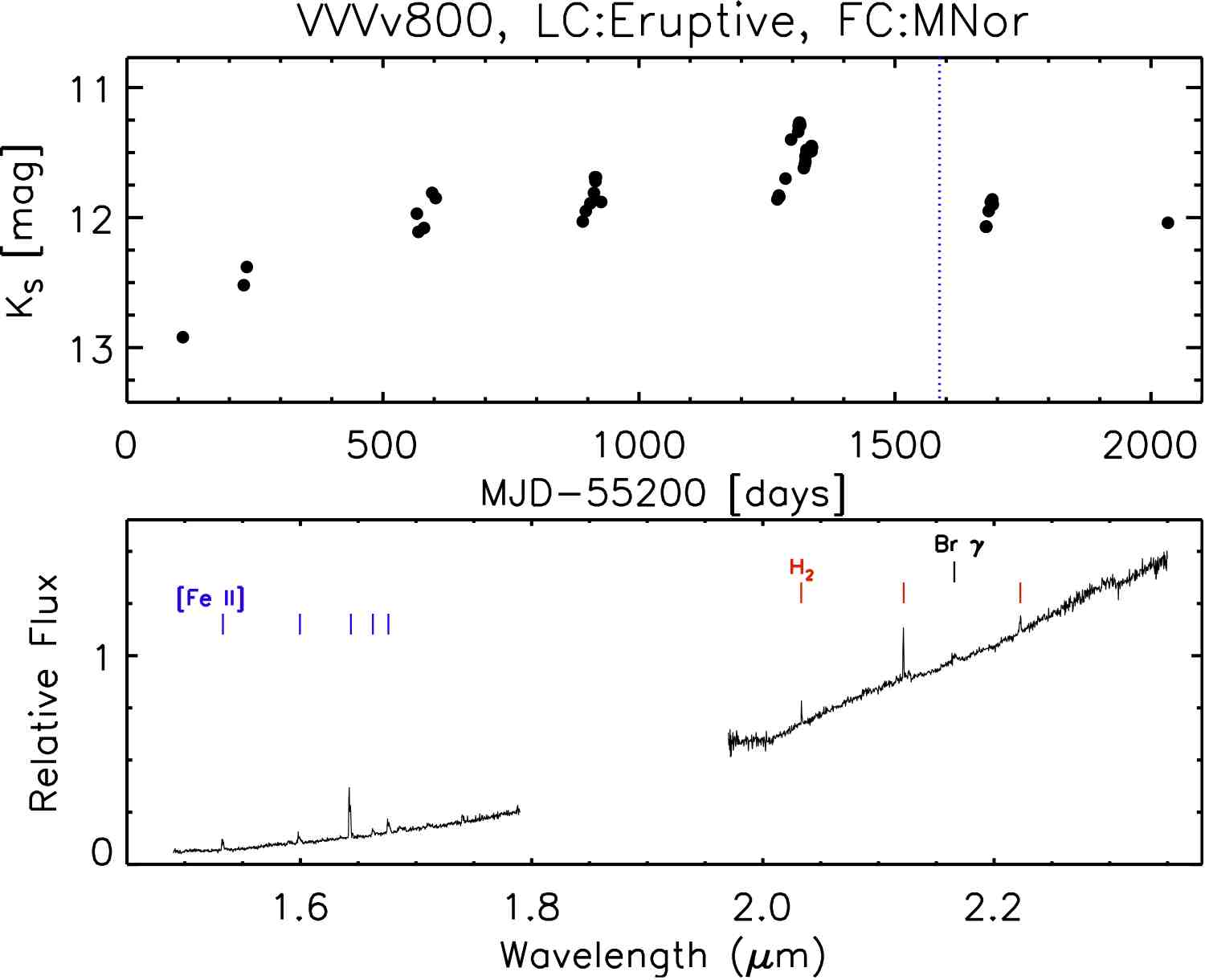}}
\caption{FIRE spectra and $K_{\rm s}$ light curves.}
\label{apen:fig18}
\end{figure*}

\clearpage

\begin{figure*}
\centering
\resizebox{0.75\textwidth}{!}{\includegraphics{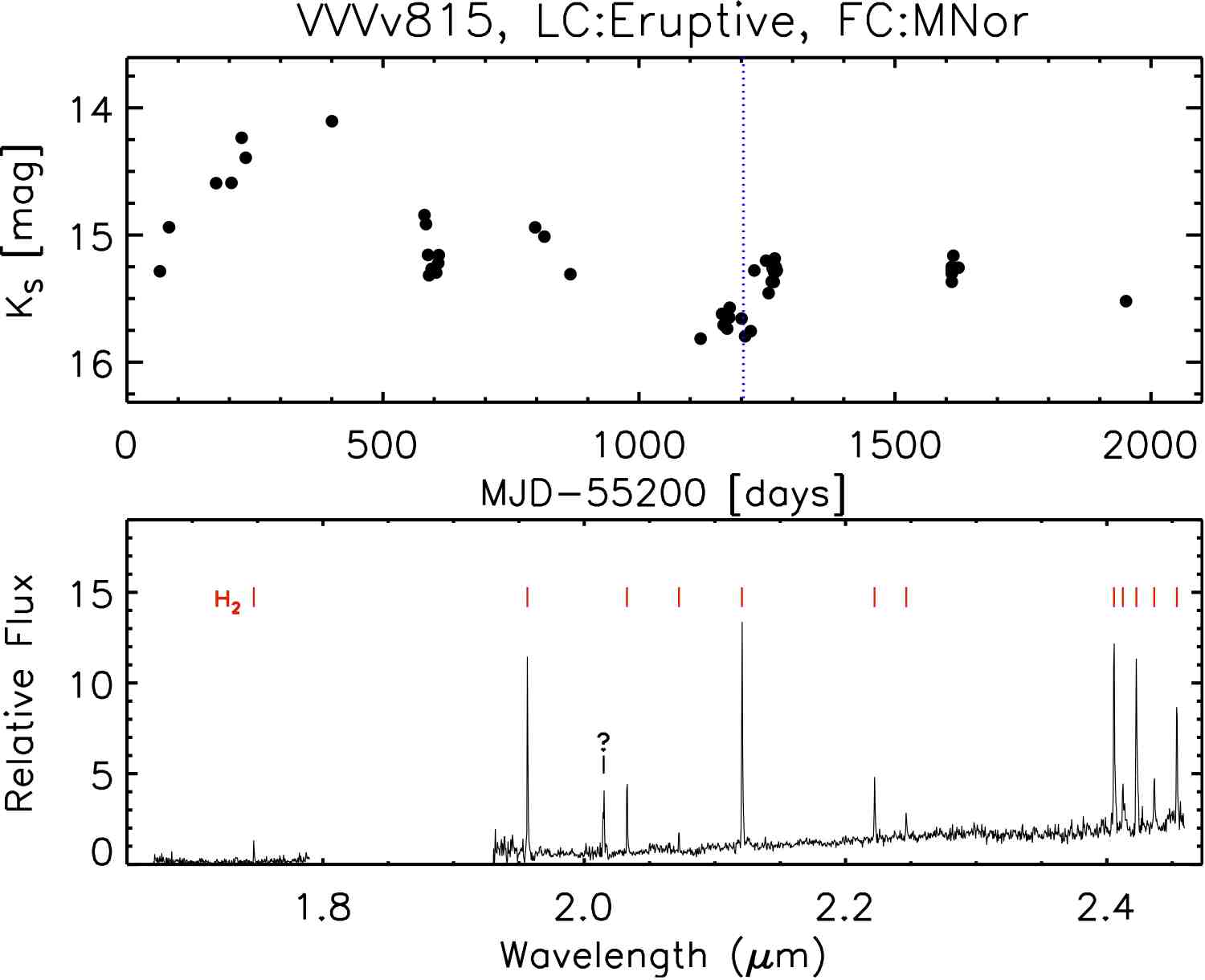}}
\caption{FIRE spectra and $K_{\rm s}$ light curves.}
\label{apen:fig19}
\end{figure*}

\clearpage

\section{Finding Charts}\label{vvv:fcharts}

In this section we present finding charts for the 37 objects analysed in this work. The images are 1\arcmin$\times$1\arcmin~ $K_{s}$ cutouts that were downloaded from the Vista Science Archive and correspond to images from the first epoch of contemporaneous colour photometry in 2010. The designation of each object is shown at the top-left corner of the images.

\begin{figure*}
\centering
\subfloat[12:28:27.97,$-$62:57:13.97]{\resizebox{0.24\textwidth}{!}{\includegraphics{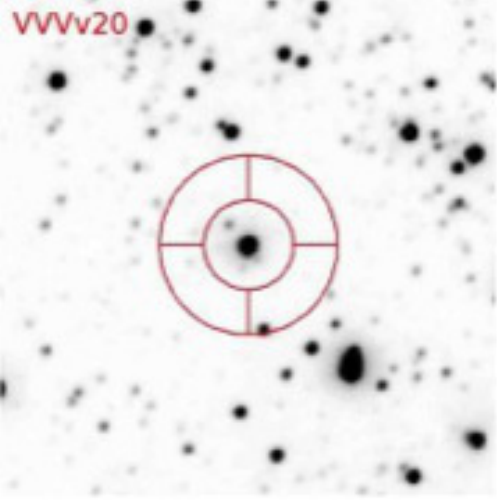}}}
\subfloat[12:35:14.37,$-$62:47:15.63]{\resizebox{0.24\textwidth}{!}{\includegraphics{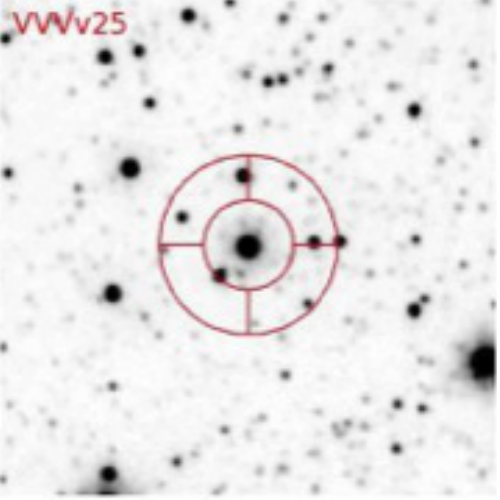}}}
\subfloat[12:43:57.15,$-$62:54:45.09]{\resizebox{0.24\textwidth}{!}{\includegraphics{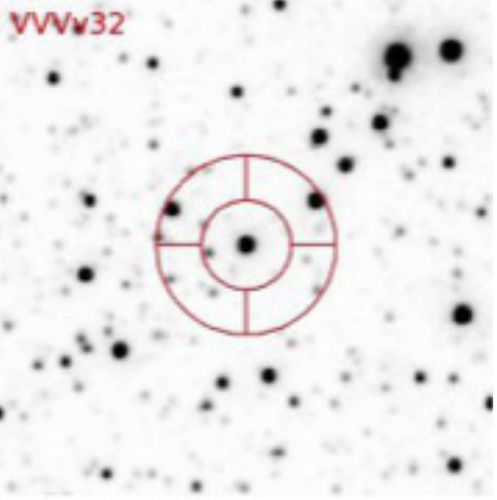}}}
\subfloat[13:09:34.64,$-$62:49:32.52]{\resizebox{0.24\textwidth}{!}{\includegraphics{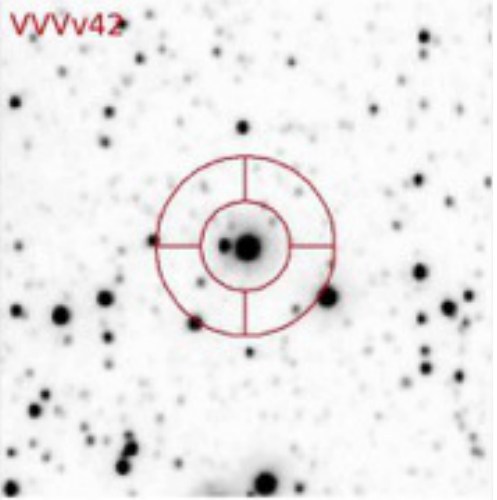}}}\\
\subfloat[13:11:43.07,$-$62:48:54.77]{\resizebox{0.24\textwidth}{!}{\includegraphics{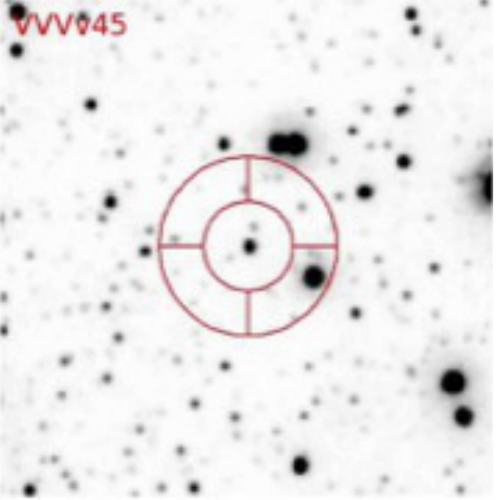}}}
\subfloat[13:46:20.48,$-$62:25:30.81]{\resizebox{0.24\textwidth}{!}{\includegraphics{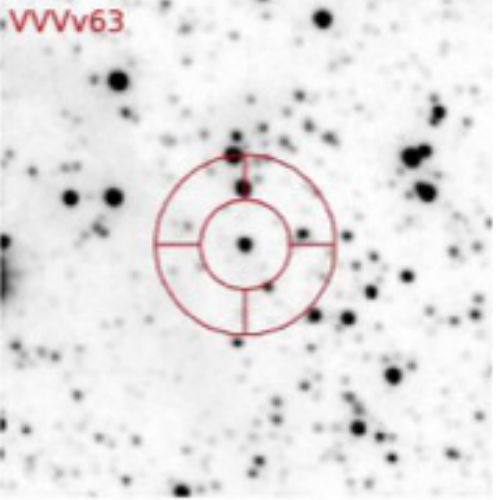}}}
\subfloat[13:47:51.09,$-$62:42:37.46]{\resizebox{0.24\textwidth}{!}{\includegraphics{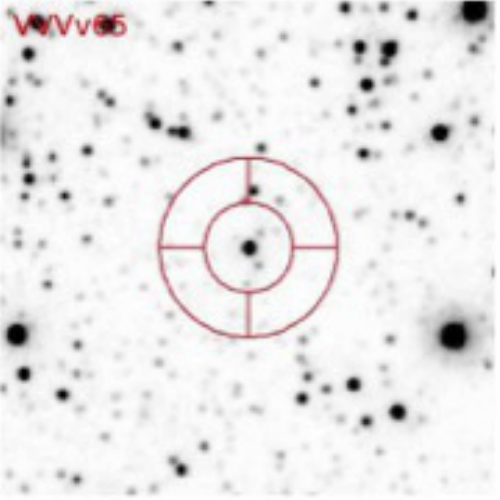}}}
\subfloat[14:22:57.76,$-$61:05:47.03]{\resizebox{0.24\textwidth}{!}{\includegraphics{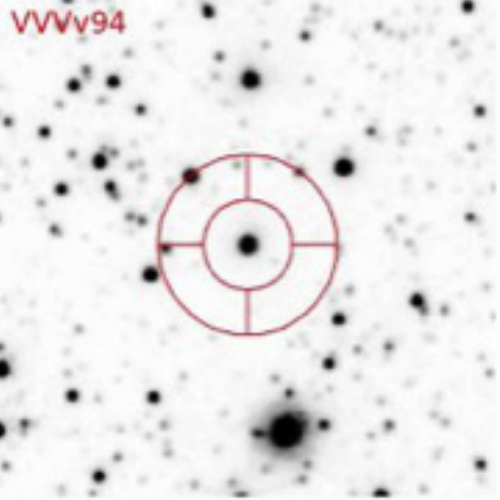}}}\\
\subfloat[14:51:20.97,$-$60:00:27.40]{\resizebox{0.24\textwidth}{!}{\includegraphics{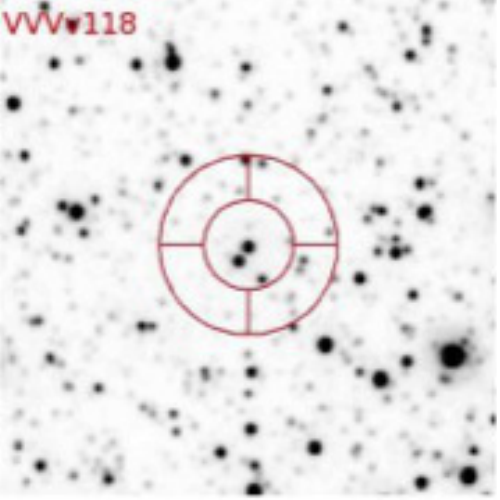}}}
\subfloat[15:49:14.34,$-$54:34:23.66]{\resizebox{0.24\textwidth}{!}{\includegraphics{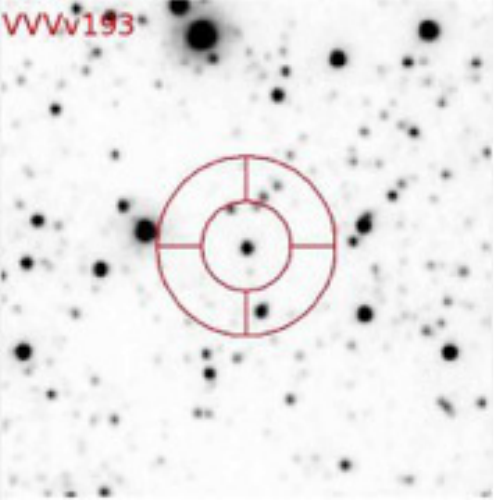}}}
\subfloat[15:54:26.41,$-$54:08:29.40]{\resizebox{0.24\textwidth}{!}{\includegraphics{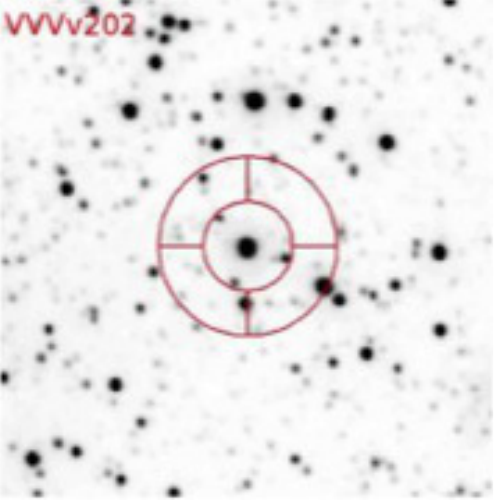}}}
\subfloat[16:04:24.48,$-$53:01:14.01]{\resizebox{0.24\textwidth}{!}{\includegraphics{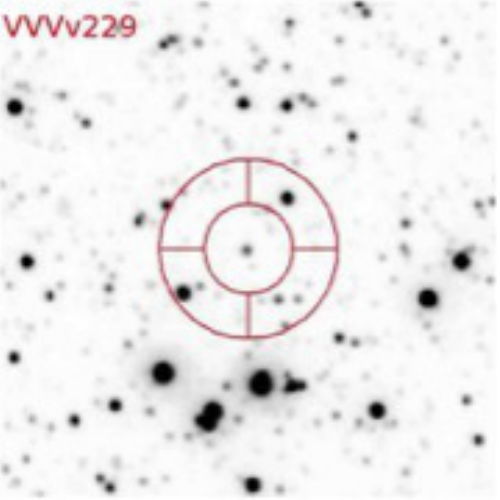}}}\\
\subfloat[16:09:35.53,$-$51:54:14.08]{\resizebox{0.24\textwidth}{!}{\includegraphics{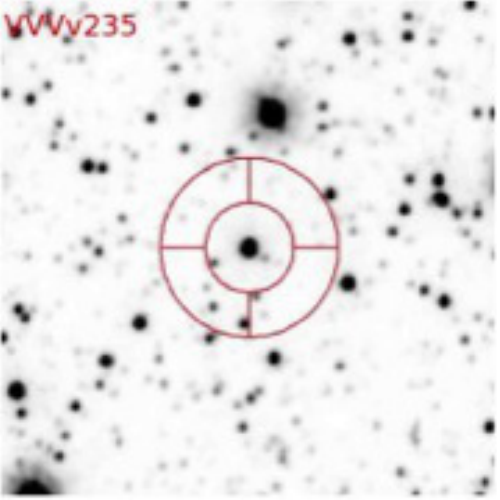}}}
\subfloat[16:13:03.44,$-$51:41:51.91]{\resizebox{0.24\textwidth}{!}{\includegraphics{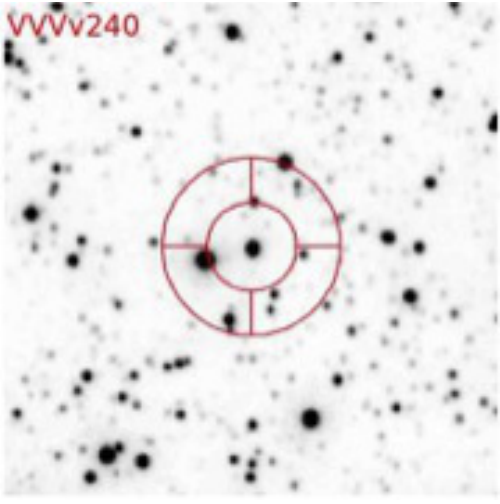}}}
\subfloat[16:23:27.14,$-$49:44:43.96]{\resizebox{0.24\textwidth}{!}{\includegraphics{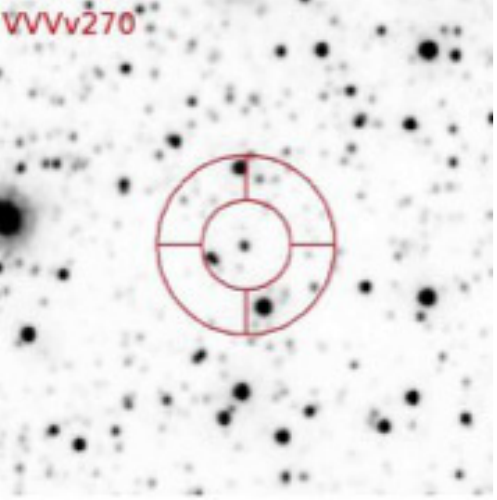}}}
\subfloat[16:46:24.57,$-$45:59:21.04]{\resizebox{0.24\textwidth}{!}{\includegraphics{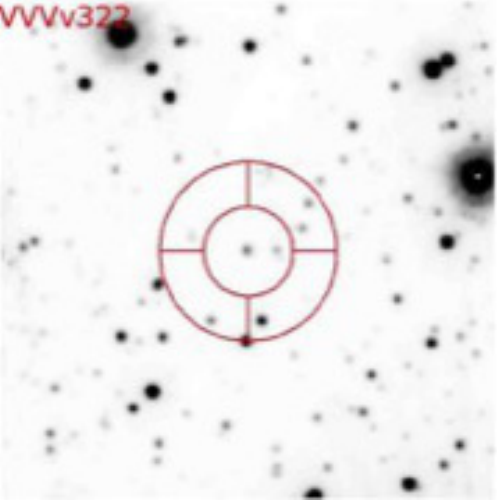}}}\\
\caption{1\arcmin$\times$1\arcmin~ $K_{\rm s}$ images of VVV high amplitude variables analysed in this work. The object designation is shown in the top left corner of the images, whist the right ascension and declination of the object is shown in the individual subcaptions. In the images north is up and east is to the left.}
\label{fctest}
\end{figure*}

\begin{figure*}
\centering
\subfloat[16:58:33.99,$-$42:49:55.25]{\resizebox{0.24\textwidth}{!}{\includegraphics{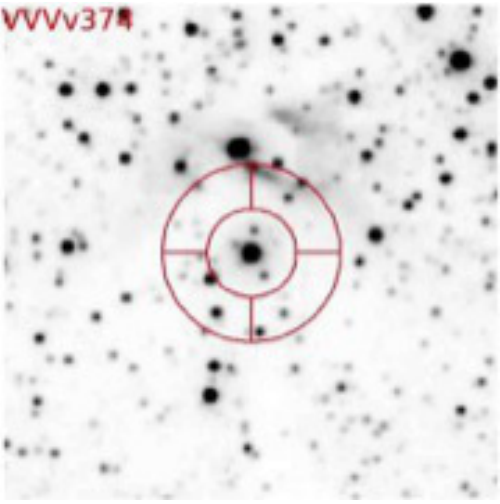}}}
\subfloat[17:09:38.62,$-$41:38:51.81]{\resizebox{0.24\textwidth}{!}{\includegraphics{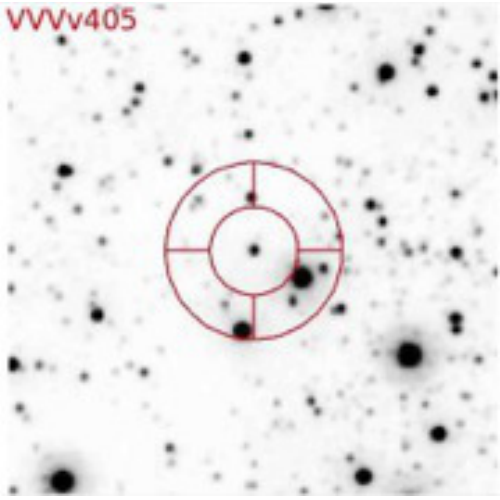}}}
\subfloat[17:09:57.47,$-$41:35:48.87]{\resizebox{0.24\textwidth}{!}{\includegraphics{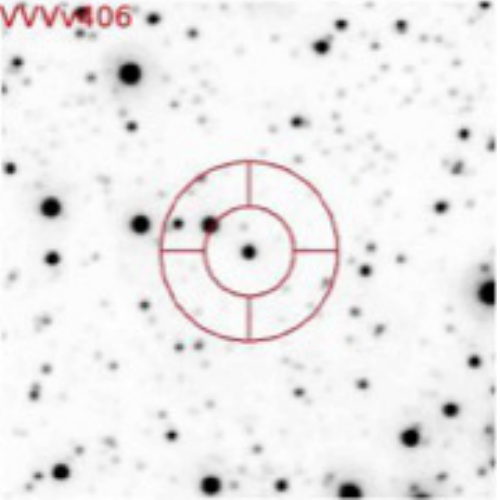}}}
\subfloat[12:41:58.06,$-$62:13:42.90]{\resizebox{0.24\textwidth}{!}{\includegraphics{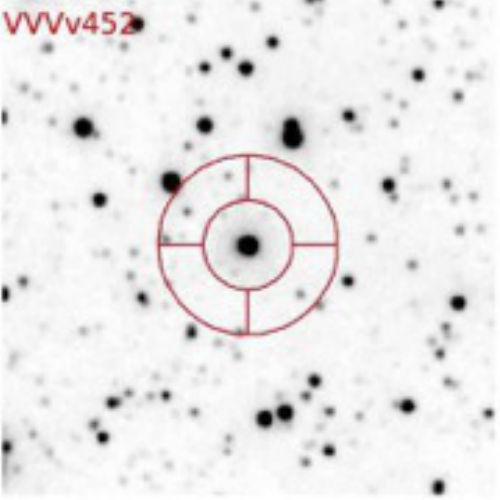}}}\\
\subfloat[13:10:57.49,$-$62:35:22.34]{\resizebox{0.24\textwidth}{!}{\includegraphics{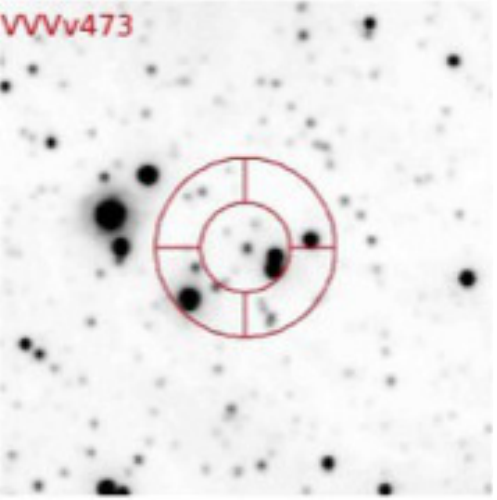}}}
\subfloat[13:16:50.32,$-$62:23:41.61]{\resizebox{0.24\textwidth}{!}{\includegraphics{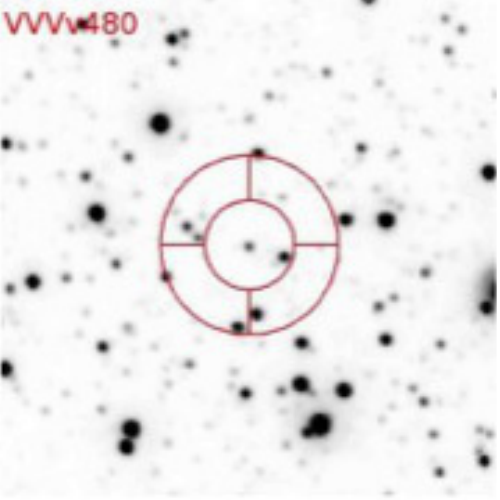}}}
\subfloat[14:00:45.37,$-$61:33:39.95]{\resizebox{0.24\textwidth}{!}{\includegraphics{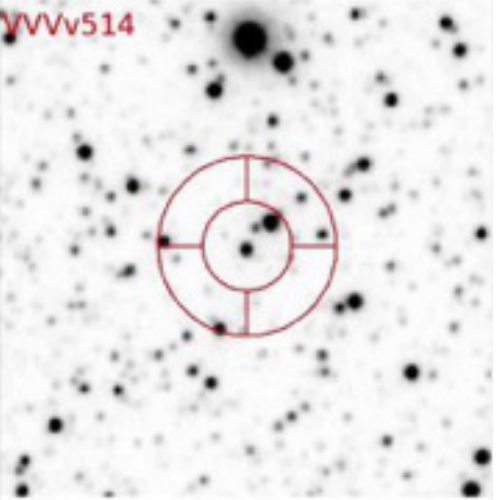}}}
\subfloat[14:53:33.59,$-$59:10:21.73]{\resizebox{0.24\textwidth}{!}{\includegraphics{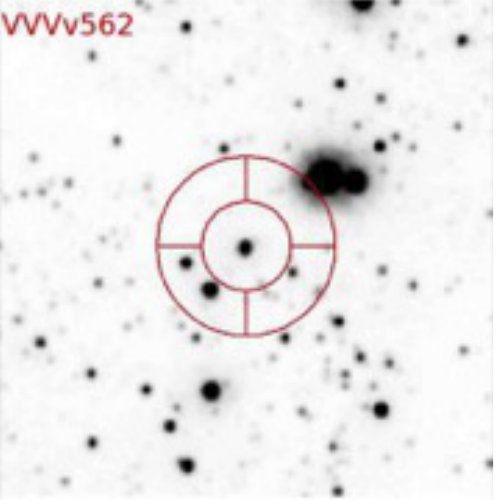}}}\\
\subfloat[15:43:17.95,$-$54:06:47.29]{\resizebox{0.24\textwidth}{!}{\includegraphics{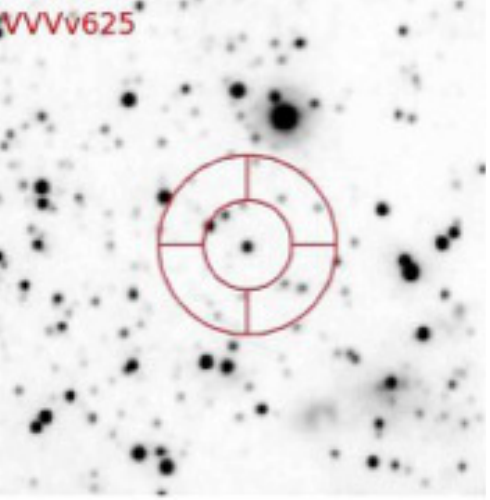}}}
\subfloat[15:44:49.54,$-$54:07:52.08]{\resizebox{0.24\textwidth}{!}{\includegraphics{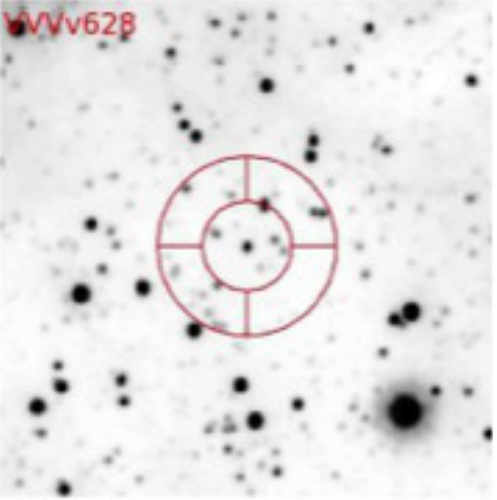}}}
\subfloat[15:44:56.13,$-$54:07:03.18]{\resizebox{0.24\textwidth}{!}{\includegraphics{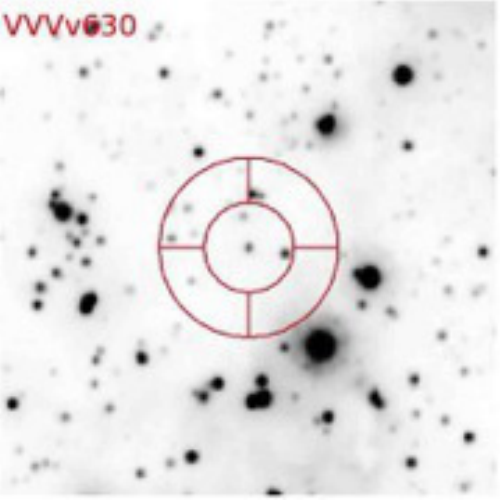}}}
\subfloat[15:45:18.36,$-$54:10:36.87]{\resizebox{0.24\textwidth}{!}{\includegraphics{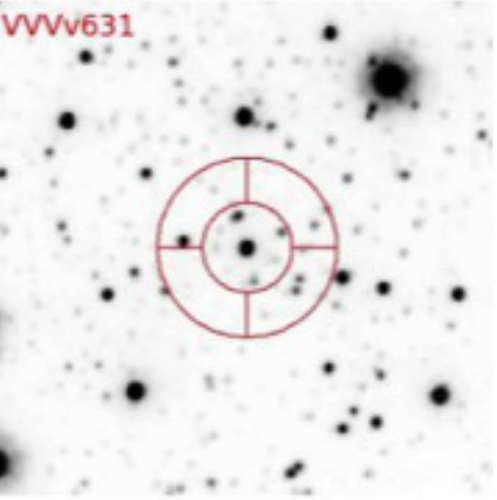}}}\\
\subfloat[15:44:09.80,$-$53:56:27.78]{\resizebox{0.24\textwidth}{!}{\includegraphics{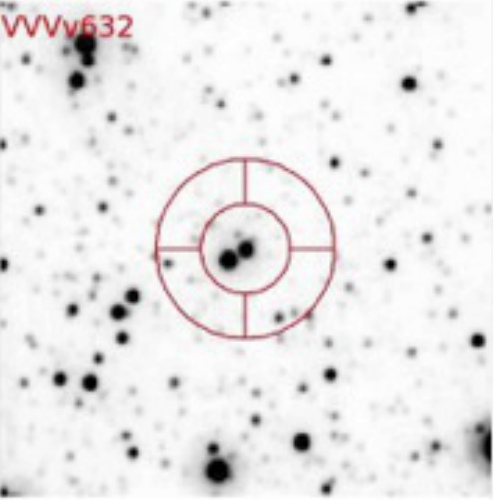}}}
\subfloat[16:10:26.82,$-$51:22:34.13]{\resizebox{0.24\textwidth}{!}{\includegraphics{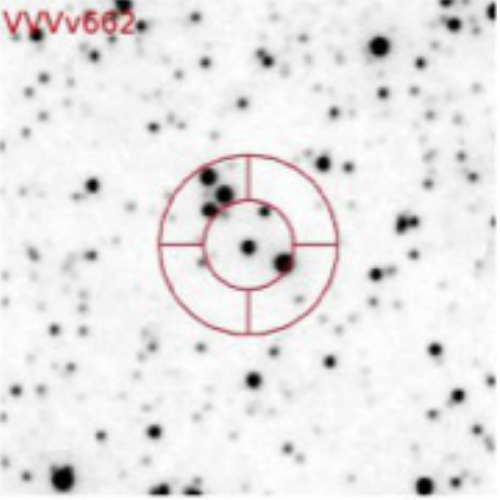}}}
\subfloat[16:09:57.70,$-$50:48:09.42]{\resizebox{0.24\textwidth}{!}{\includegraphics{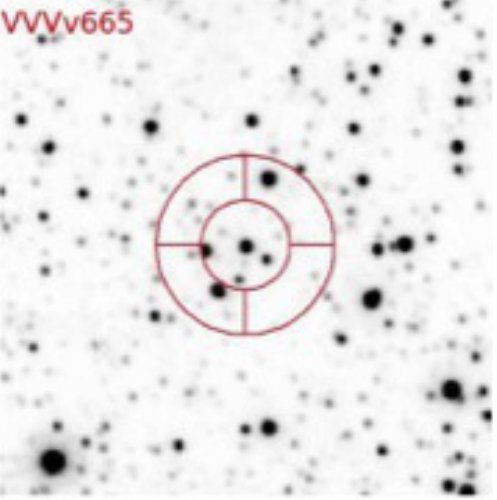}}}
\subfloat[16:23:44.34,$-$48:54:55.29]{\resizebox{0.24\textwidth}{!}{\includegraphics{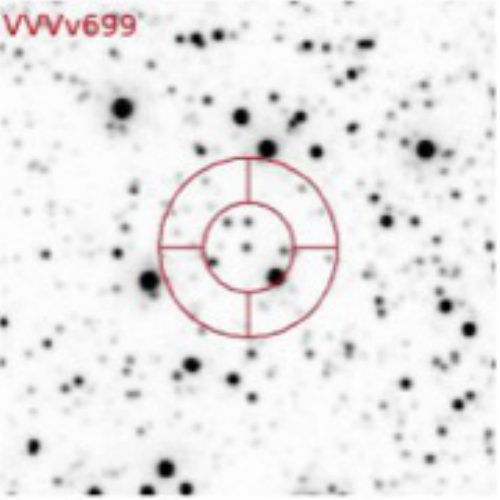}}}\\
\caption{1\arcmin$\times$1\arcmin~ $K_{\rm s}$ images of VVV high amplitude variables analysed in this work. The object designation is shown in the top left corner of the images, whist the right ascension and declination of the object is shown in the individual subcaptions. In the images north is up and east is to the left.}
\label{fctest1}
\end{figure*}

\begin{figure*}
\centering
\subfloat[16:36:05.56,$-$46:40:40.61]{\resizebox{0.24\textwidth}{!}{\includegraphics{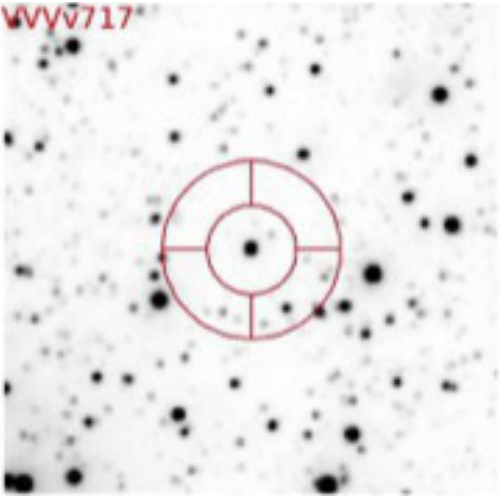}}}
\subfloat[16:39:48.77,$-$45:48:47.96]{\resizebox{0.24\textwidth}{!}{\includegraphics{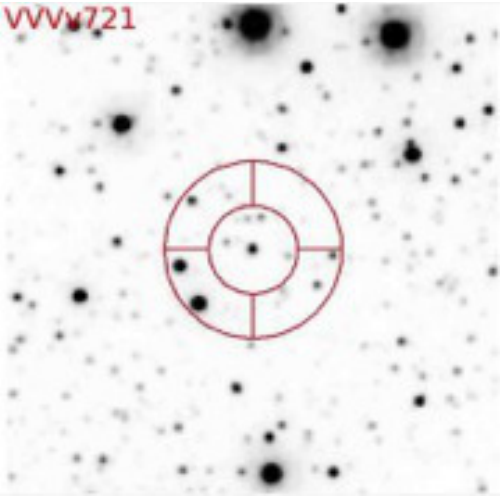}}}
\subfloat[17:12:07.43,$-$38:41:26.86]{\resizebox{0.24\textwidth}{!}{\includegraphics{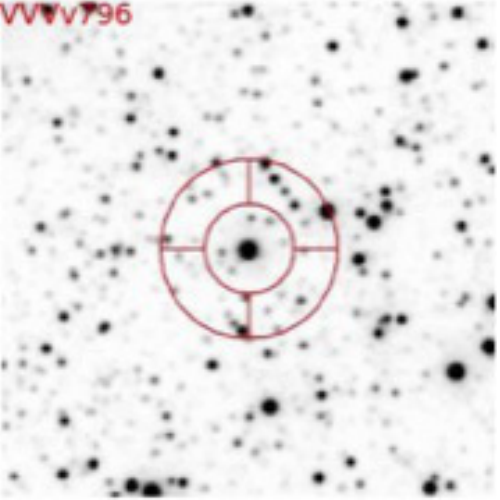}}}
\subfloat[17:12:46.04,$-$38:25:24.63]{\resizebox{0.24\textwidth}{!}{\includegraphics{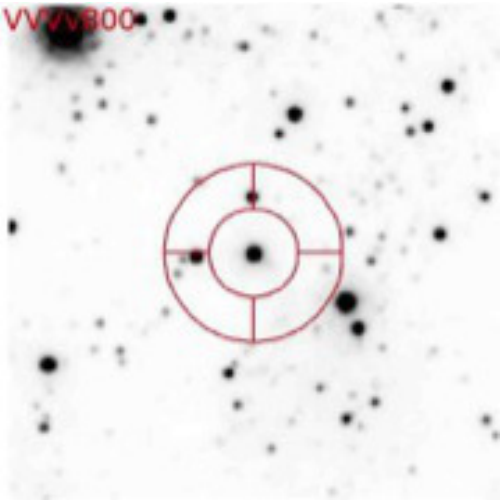}}}\\
\subfloat[14:26:04.95,$-$60:41:16.81]{\resizebox{0.24\textwidth}{!}{\includegraphics{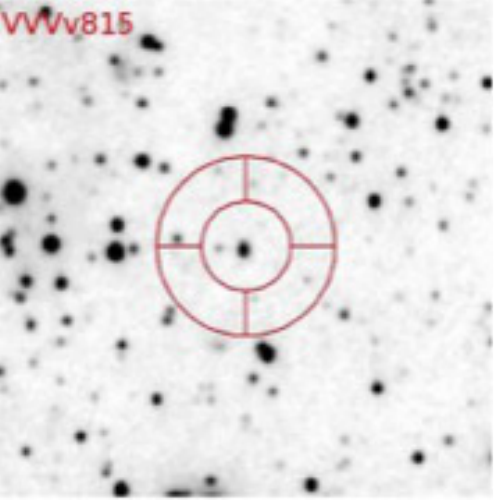}}}
\caption{1\arcmin$\times$1\arcmin~ $K_{\rm s}$ images of VVV high amplitude variables analysed in this work. The object designation is shown in the top left corner of the images, whist the right ascension and declination of the object is shown in the individual subcaptions. In the images north is up and east is to the left.}
\label{fctest2}
\end{figure*}


\clearpage

\label{lastpage}
\end{document}